# Equivalent circuit modeling of electron-hole recombination in semiconductor and mixed ionic-electronic conductor based devices

Davide Moia[1]

Max Planck Institute for Solid State Research, Heisenbergstr. 1, 70569, Stuttgart, Germany

**Abstract**

The understanding and optimization of solar energy conversion and light emitting devices can greatly benefit from equivalent circuit models describing their response. However, a general model of electron-hole recombination in semiconductors is currently missing. This study presents equivalent circuit models of radiative and non-radiative electron-hole recombination based on their linearized analytical treatment. These are integrated in a circuit model of complete devices that is equivalent to the linearized drift-diffusion equations in one dimension. The analysis of the model shows that, for most situations involving semiconductors without mobile ions, approximated models that do not account for local electrostatics are sufficient to describe the contribution of recombination to the device response. The influence of local electrostatics on non-radiative trap-mediated and Auger recombination becomes essential in mixed conducting devices, and it should be included explicitly in equivalent circuit models. Resistors used for traditional semiconductors are indeed a special case of the general model of non-radiative recombination, for which transistors gated by the local change in electrostatic potential implement a more accurate representation. For mixed conducting devices, such as halide perovskite solar cells, appropriate simplifications of the complete model provide analytical solutions describing the bulk and interfacial polarization effects that influence the small signal electrostatics, recombination currents and overall impedance. The resulting analysis is relevant to a wide range of materials and devices used for solar energy conversion as well as other optoelectronic and photo-electrochemical applications.

**Introduction**

This work presents a circuit model that is analytically equivalent to the linearized drift-diffusion equations for semiconductor and mixed ionic-electronic conductor-based devices. The analysis identifies suitable circuit elements describing electron-hole recombination based on the linearized expressions of radiative and non-radiative recombination rates and integrates such elements into a transmission line model. The derivation of practically relevant approximations of such model is presented. These can be helpful in the experimental investigation of devices such as solar cells.

Equivalent circuit models used in conjunction with impedance spectroscopy measurements and other time or frequency domain (opto)electronic techniques represent a powerful approach to the study of semiconductors and mixed ionic-electronic conductors. [1–4] Transmission line equivalent circuit models are a physically meaningful starting point for the description of the small perturbation response due to transport and storage of charges, applicable to one-dimensional devices close to equilibrium. [3,5–8] For devices such as solar cells and light-emitting diodes under operation, the electron and hole populations

[1] moia.davide@gmail.com
Current address: Fluxim AG, Katharina-Sulzer-Platz 2, 8400 Winterthur, Switzerland

are not at equilibrium, requiring appropriate models of the thermal generation and recombination of electronic charge carriers. Resistive elements are commonly used for this purpose. [9] While simple, this approach may not always bear physical correspondence with situations where multiple processes influence the overall electron-hole recombination rate in the semiconductor.

In mixed ionic-electronic conductors, such as hybrid metal-halide perovskites, an additional layer of complexity is introduced by the influence of mobile ionic defects on the recombination dynamics. [10–12] For example, investigation of perovskite solar cells and other mixed conducting devices revealed unphysically large values of capacitance and negative capacitance (inductive behavior) in their impedance at low frequencies, features that have been interpreted based on frequency dependent electronic processes, such as recombination. [13–19]

Ionic-to-electronic current amplification effects can be responsible for such behavior, whereby energy barriers associated with electron transfer reactions change due to ion redistribution, producing changes in electronic current occurring over "ionic time scales". [13,20–22] Such an effect concerns recombination processes as well as charge injection at interfaces, giving rise to additional low-frequency impedance features and influencing the apparent capacitive and inductive behavior of the device. Attempts to translate such phenomena in equivalent circuit model terms have been made. [13,17,20,23–25] The use of bipolar transistors was introduced to implement the "gating" of the electronic current by the interfacial potential changes associated with the response of an ionic circuit branch. The resulting model could describe much of the optoelectronic behavior of hybrid perovskite solar cells. [13] Another approach consists in fitting empirical R-C elements as well as inductive elements to the low frequency response. While these have no physical meaning, they can be related to the ionic influence on the electronic current. [13,17,23,26] A surface polarization model and a modified polarization model have been suggested as analytical approximations of the drift-diffusion model describing perovskite solar cells. [27–29] The approach succeeded at reproducing impedance spectra with one or two low-frequency features that have been observed experimentally, by considering the role of ionic as well as of electronic charge carriers in the frequency dependent electrostatic and recombination behavior.

Despite this progress, the connection between currently proposed equivalent circuit models and the analytical description of electron-hole recombination is still missing. For complete devices including mixed ionic-electronic conductors, the search for appropriate circuit models addressing multiple low frequency impedance features observed in experiments is still an open question. On a general level, progress in the development of accessible models that can facilitate experimental data analysis is needed.

This work formalizes the use of resistors and bipolar transistors to describe electron-hole recombination in semiconductors (section A in the Results and Discussion). Such description, integrated within a transmission line, results in an analytically accurate model that allows the analysis of bulk and interfacial recombination in one-dimensional semiconducting devices with or without mobile ions (section B). The model can be simplified into more accessible versions to match the properties of different device structures and bias conditions (section C). The results exemplify the connection between the complete linearized drift-diffusion model of the problem and simplified circuit models, such as the one proposed in Ref. [13]. Calculated impedance spectra using the proposed models (section D) point to the relevance of these findings for the description of devices based on hybrid perovskites, as well as other mixed conductors, but also for traditional semiconductors with negligible ionic conductivity.

## I. Background

This section reviews some basic concepts related to electronic charge carrier equilibrium and non-equilibrium in semiconductors that will be useful in the remainder of this study. Applying the equilibrium condition to electron-hole thermal generation and recombination in semiconductors leads to the mass-action law

$n_{\mathrm{eq}}p_{\mathrm{eq}} = K_B$ .   Eq. 1

Here, $n_{\mathrm{eq}}$ and $p_{\mathrm{eq}}$ correspond to the equilibrium electron and hole concentrations and $K_B$ indicates the mass-action constant for the electronic charge carriers (often expressed as the square of the intrinsic electronic charge concentration, $n_i^2$). Under such conditions, the recombination rate $R$ counterbalances the "thermal" generation rate $G_{th}$. That is, while $R \neq 0$ and $G_{th} \neq 0$, the net rate of recombination $U = R - G_{th}$ is zero. The equilibrium between the electrons and the holes populations obeys Fermi-Dirac statistics involving a single Fermi level $E_F$ (or electrochemical potential $\tilde{\mu}_{\mathrm{e}^-}$).

The application of light or voltage bias takes the semiconductor out-of-equilibrium, where the electron and hole populations ($n$ and $p$) obey two separate quasi-equilibria. The steady-state situation can be described by the modified mass-action law,

$np = n_i^2 \exp\left[\left(V_p - V_n\right)/V_{th}\right]$ .   Eq. 2

Here, $V_{th}$ is the thermal voltage, defined based on Boltzmann's constant, temperature and the elementary charge as $k_B T/q$. The two quasi-electrochemical potentials for electrons and holes correspond to $qV_n$ and $qV_p$. Here, the use of the symbol $V$ emphasizes their relationship with the "voltages" at the nodes within the equivalent circuit model, as adopted in References [13,30]. These potentials are also indicated with $\tilde{\mu}_p$and $\tilde{\mu}_n$ in the literature, or expressed in terms of quasi-Fermi energies ($E_{Fp}$ and $E_{Fn}$). In Equation 2, the values of $n$ and $p$ differ from the equilibrium case depending on $q(V_p - V_n)$, which is commonly referred to as quasi-Fermi levels splitting ($QFLS$).

When different routes for recombination are at play, the total net recombination rate can be expressed as

$U = \sum_k U_k = \sum_k (R_k - G_{th,k})$ ,  Eq. 3

where $R_k$ and $G_{th,k}$ are the recombination and the thermal generation contributions of the *k*-th process. Analytical expressions of the net recombination rate associated with radiative, Shockley-Read-Hall (SRH, trap-mediated) and Auger recombination processes (Figure 1a) are available [31,32]:

$U_{rad} = k_{rad}(np - n_i^2)$   Eq. 4

$U_{SRH} = \frac{np - n_i^2}{\tau_n(p+p_1) + \tau_p(n+n_1)}$   Eq. 5

$U_{Aug} = \gamma_n\left(n^2 p - n_{\mathrm{eq}}^2 p_{\mathrm{eq}}\right) + \gamma_p\left(np^2 - n_{\mathrm{eq}} p_{\mathrm{eq}}^2\right)$   Eq. 6

The parameters $k_{rad}$, $\gamma_n$ and $\gamma_p$ refer to the radiative rate constant and Auger coefficients of the material, respectively. The terms $\tau_n$ and $\tau_p$ are carrier capture lifetimes that depend on trap concentration, as well as on the capture cross-section and the carrier thermal velocity, while the parameters $n_1$ and $p_1$ are defined based on the trap energy position within the bandgap $E_T$. [33,34]

In a solar cell, the functional form of Equations 4–6 dictates the dependence of the net recombination current in the active material when a potential $V_{app}$ applied to the device leads to a position dependent non-equilibrium condition ($QFLS \neq 0$). This is reflected in the dependence of the total current density flowing in the device, $J_{tot}$, on the applied potential, often discussed in terms of the diode equation:

$$J_{tot} = J_0(e^{\frac{V_{app}}{mV_{th}}} - 1)\ . \quad \text{Eq. 7}$$

In Equation 7, $J_0$ is a saturation current, while $m$ is the ideality factor, which encapsulates information on the position and the type of recombination process that dominates $J_{tot}$ (Figure 1c). [32] In devices including mixed conductors, such information is further convoluted with the influence of mobile ionic defects on recombination. [23,35,36]

**II. Results and discussion**

## A. Equivalent circuit model of recombination processes

The analytical description of recombination presented in the previous section allows the derivation of corresponding equivalent circuit models, through a small signal analysis of the relevant functions (Figure 1a and b). All quantities are presented using a notation ($V = \bar{V} + v$, $J = \bar{J} + j$ or $n = \bar{n} + \tilde{n}$) that identifies the steady-state ($\bar{V}$, $\bar{J}$, $\bar{n}$) and the (small) deviation from steady-state ($v, j, \tilde{n}$). The latter can be caused for example by the ac voltage applied to the device when recording impedance spectra. The lower case is used also to identify differential circuit elements used in the model (*e.g.* small perturbation resistors and capacitors as $r$ and $c$, as well as conductance and transconductance terms as $g$).

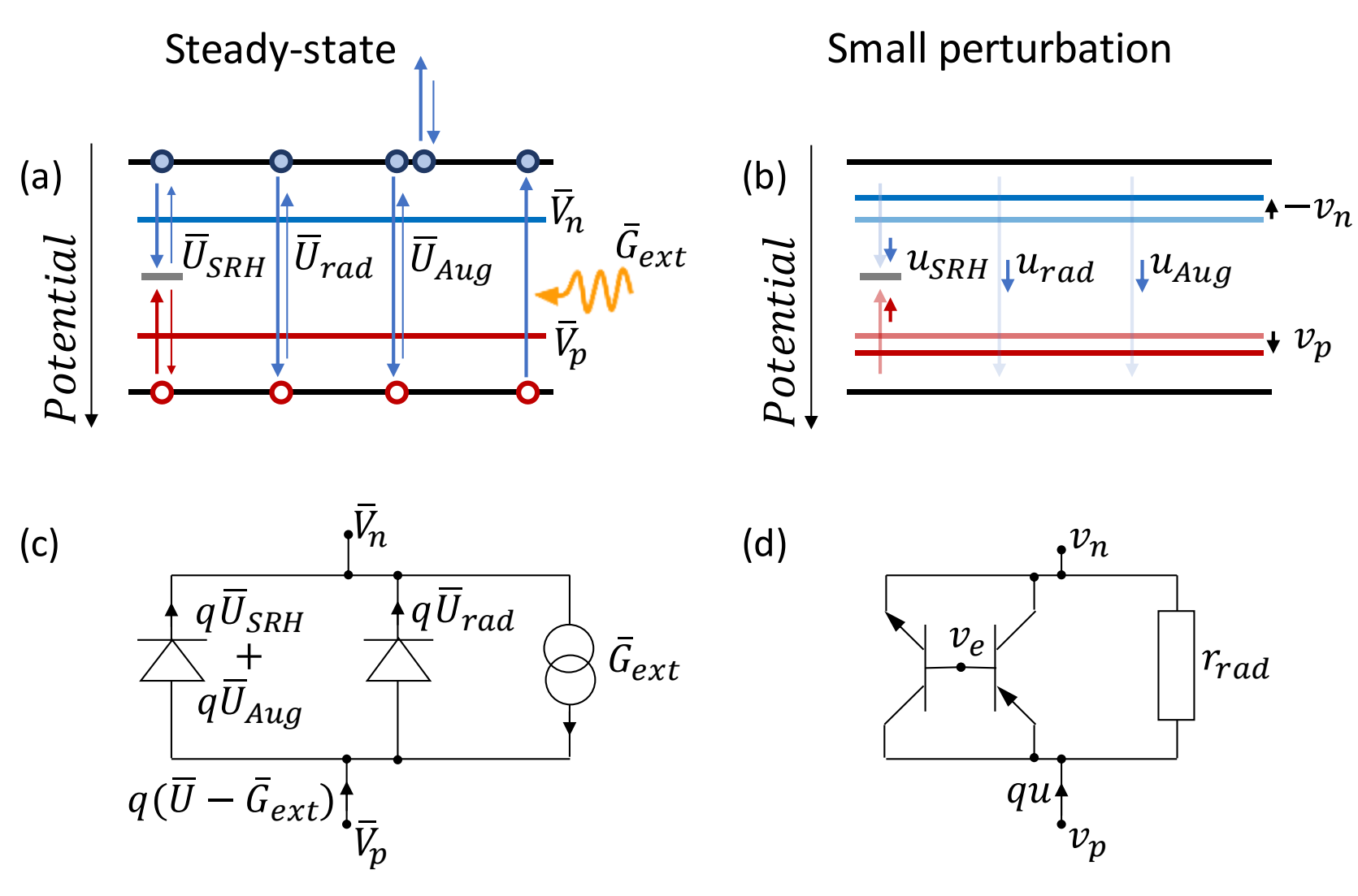

Figure 1. (a, b) Energy diagrams representing (a) the steady state and (b) the small perturbation local recombination and thermal generation processes occurring in a semiconductor under bias (see text). (c, d) Equivalent circuit models describing such processes for (c) the steady-state and (d) the small perturbation regimes. The latter considers only the small electrical perturbation (as it is the case in an impedance experiment) and it involves a resistor $r_{rad}$ in parallel to an *npn* and a *pnp* bipolar transistor pair. The $q\bar{U}$ and $qu$ terms refer to local net recombination currents per unit volume and the $G_{ext}$ term to the, in this case constant, generation rate per unit volume due to illumination. The ideality factor of the radiative recombination diode is $m_{rad} = 1$. For the Auger and SRH diode, $m$ depends on the operation conditions. A "$Potential$" y-axis is used in (a, b) to indicate the connection between the potentials in the diagram and in the circuits below (such $Potential$ is connected to the partial free enthalpy of electronic charges [37], often referred to as energy, $E$).

The total small signal net recombination current per unit volume can be expressed, based on Equation 3, as $qu = \sum_k qu_k$. In view of deriving an appropriate equivalent circuit description of the recombination terms, the analysis focuses on the relation between each contribution, $qu_k$, and the relevant potentials. The latter are the change in electrochemical potentials of electrons and holes, $v_n$ and $v_p$, and the change in electrostatic potentials, $v_e$ (Figure 1b), represented as potentials of a small perturbation circuit network (Figure 1d). By linearizing Equation 4 and focusing on the small perturbation regime, one obtains:

$$u_{rad} = k_{rad}\bar{n}\bar{p}\left(\frac{\tilde{n}}{\bar{n}} + \frac{\tilde{p}}{\bar{p}}\right). \qquad \text{Eq. 8}$$

The change in carrier concentration can be expressed based on the small signal chemical potential of electrons, $v_{\mu,n} = v_e - v_n$, and holes, $v_{\mu,p} = v_p - v_e$. It follows that $\tilde{n} = \bar{n}(v_e - v_n)/V_{th}$ and $\tilde{p} = \bar{p}(v_p - v_e)/V_{th}$, yielding:

$$qu_{rad} = \frac{v_p - v_n}{r_{rad}}. \qquad \text{Eq. 9}$$

Here, $r_{rad} = \frac{V_{TH}}{qk_{rad}\bar{n}\bar{p}}$ is interpreted as a resistor (units of $\Omega\ cm^3$) connected between the nodes associated with $v_p$ and $v_n$ (Figure 1d). Importantly, the change in electrostatic potential $v_e$ does not appear in Equation 9, as radiative recombination scales linearly with the product of the electron and hole concentration.

For the Shockley-Read-Hall recombination rate, linearization of Equation 5 yields:

$$u_{SRH} = \frac{\tau_n\bar{p}(\bar{p}+p_1)+\tau_p\bar{p}n_1+\tau_p n_i^2}{\left[\tau_n(\bar{p}+p_1)+\tau_p(\bar{n}+n_1)\right]^2}\tilde{n} + \frac{\tau_p\bar{n}(\bar{n}+n_1)+\tau_n\bar{n}p_1+\tau_n n_i^2}{\left[\tau_n(\bar{p}+p_1)+\tau_p(\bar{n}+n_1)\right]^2}\tilde{p}. \qquad \text{Eq. 10}$$

This expression accurately describes $u_{SRH}$, if the trapping and de-trapping processes involved in the SRH model are at quasi-equilibrium (valid for time scales that are slower than the dominant time constant controlling such processes). Again, to describe Equation 10 in terms of equivalent circuit model, the small signal changes in electronic charge carrier concentrations ($\tilde{n}$ and $\tilde{p}$) are expressed as a function of the relevant small signal electron and hole chemical potentials. In this case, the change in electrostatic potential $v_e$ is involved in the resulting expression of the small perturbation net recombination current per unit volume.

$$qu_{SRH} = \frac{q}{V_{th}}\left\{\frac{\bar{n}\bar{p}[\tau_n(\bar{p}+p_1)+\tau_p n_1]}{\left[\tau_n(\bar{p}+p_1)+\tau_p(\bar{n}+n_1)\right]^2}(v_e - v_n) + \frac{\tau_n\bar{p}n_i^2}{\left[\tau_n(\bar{p}+p_1)+\tau_p(\bar{n}+n_1)\right]^2}(v_p - v_e) + \right.$$
$$\left. + \frac{\bar{n}\bar{p}[\tau_p(\bar{n}+n_1)+\tau_n p_1]}{\left[\tau_n(\bar{p}+p_1)+\tau_p(\bar{n}+n_1)\right]^2}(v_p - v_e) + \frac{\tau_p\bar{n}n_i^2}{\left[\tau_n(\bar{p}+p_1)+\tau_p(\bar{n}+n_1)\right]^2}(v_e - v_n)\right\} \qquad \text{Eq. 11}$$

Equation 11 is arranged to emphasize the contribution of four terms. These, consistent with the description in Ref. [13], can be interpreted as the current contributions due to transconductance terms of *npn* and *pnp* bipolar transistors (with current gain $\beta \to \infty$). The transistors are appropriately connected to the nodes associated with the small signal electrostatic and quasi-electrochemical potentials as shown in Figure 1d (see section 1 of the Supplemental Material [38] for more details on this representation). On this basis, Equation 11 can be expressed as follows:

$$qu_{SRH} = g_{rec,n}(v_e - v_n) - g_{gen,n}\,(v_e - v_p) + g_{rec,p}(v_p - v_e) - g_{gen,p}(v_n - v_e)\,. \qquad \text{Eq. 12}$$

Here, $g_{rec,n}$ and $g_{rec,p}$ are electron-hole recombination transconductance terms, while $g_{gen,n}$ and $g_{gen,p}$ are (thermal) generation transconductance terms. Each of these parameters is dependent on the specific steady-state condition (the values of $\bar{n}$ and $\bar{p}$), as shown in Equation 11. In the analytical treatment the symbol $g$ represents a transconductance per unit volume (units of A $V^{-1}$ $cm^{-3}$), while in the circuit models discussed below it refers to a discretized transconductance with A $V^{-1}$ $cm^{-2}$ units, similarly to Ref. [13]. The treatment above provides an analytical justification to the use of the bipolar transistor in the modeling of small perturbation local net recombination described with the Shockley-Read-Hall rate.

The expressions describing transconductance terms can be simplified, depending on the relative magnitudes of $\bar{n}$ and $\bar{p}$, and on the trap's properties. If the capture lifetimes for electrons and holes are comparable, the analysis in Appendix A shows that for the case of shallow traps, and for the case of deep traps with $\tau_n\bar{p} = \tau_p\bar{n}$, one obtains $g_{rec,n} \approx g_{rec,p}$. If such values are referred to as $g_{rec}$, and thermal generation can be neglected, it is possible to simplify the expression of $qu_{SRH}$ in Equation 12 to obtain $qu_{SRH} = g_{rec}(v_p - v_n)$, where $v_e$ is no longer involved. This encourages the definition of a recombination resistance:

$$r_{SRH} = 1/g_{rec}\ . \qquad \text{Eq. 13}$$

In all other cases, the transistor representation may be simplified by considering only the current contribution of the transistor type (*npn* or *pnp*) associated with the recombination limiting carrier type ($n$ or $p$). This corresponds to the minority carrier, if $\tau_n$ and $\tau_p$ are of similar magnitude, while different trends are obtained based on Equation 11 if these parameters are very different from each other. These aspects are discussed in section 2 of the Supplemental Material, [38] where the expressions for the generation transconductance terms are also presented (these have negligible contribution under light and/or forward voltage bias).

Finally, linearization of the Auger net recombination expression leads to:

$$qu_{Aug} = q(2\gamma_n\bar{n}\bar{p} + \gamma_p\bar{p}^2)\tilde{n} + q(2\gamma_p\bar{n}\bar{p} + \gamma_n\bar{n}^2)\tilde{p} \qquad \text{Eq. 14}$$

The small signal net recombination once again depends on each of the carrier concentrations. Similarly to the case of the trap-mediated SRH recombination rate, the change in electrostatic potential appears explicitly in the functional form and an analogous approach to the one used for $u_{SRH}$ can be applied. However, in this case, the generation of charge carriers is a constant term and does not appear in the small perturbation treatment (see section 3 of the Supplemental Material [38]):

$$qu_{Aug} = \frac{q}{V_{TH}}\left[2\gamma_n\bar{n}\bar{p}(v_e - v_n) + \gamma_n\bar{n}^2(v_p - v_e) + 2\gamma_p\bar{n}\bar{p}(v_p - v_e) + \gamma_p\bar{p}^2(v_e - v_n)\right] =$$

$$= g_{rec,n}(v_e - v_n) + g_{rec,p}(v_p - v_e)\,. \qquad \text{Eq. 15}$$

In summary, a pair of *npn* and *pnp* transistors with appropriate thermal generation (SRH) and recombination (SRH + Auger) transconductance values combined with a radiative resistor (Figure 1d) is an analytically accurate description of the local small signal net recombination. Importantly, such conclusion is general for any semiconductor (with or without mobile ions), given the fact that the proposed analogy between the transistor equations and the non-radiative SRH and Auger recombination rates is valid at a fundamental level.

The total small signal net recombination per unit volume can be expressed as

$$qu = \sum_k g_{rec,k} v_{rec,k} \ , \qquad \text{Eq. 16}$$

or more explicitly as

$$qu = g_{rec,rad}(v_p - v_n) + g_{rec,n}(v_e - v_n) - g_{gen,n}\,(v_e - v_p) + g_{rec,p}(v_p - v_e) - g_{gen,p}(v_n - v_e)\ , \qquad \text{Eq. 17}$$

where each contribution involves the product of a recombination (trans)conductance, $g_{rec,k}$, and a recombination voltage (driving force), $v_{rec,k}$. Here, the radiative conductance per unit volume is defined as $g_{rec,rad} = r_{rad}^{-1}$, while the recombination transconductance for SRH and Auger processes are combined in $g_{rec,n}$ and $g_{rec,p}$.

### B. Generalized transmission line equivalent circuit model

The recombination elements described above can be included within a small perturbation model of a one-dimensional device, where the local charge carrier dynamics is coupled to long range effects. [2,39] Figure 2a shows the equivalent circuit model for a device where a mixed ionic-electronic conducting active layer is sandwiched between ion-blocking contacts. The circuit consists of electronic rails for electrons and holes, one ionic rail for the one mobile ionic species considered in this case, and the electrostatic rail. Each rail is drawn along a position axis, $x$. This circuit allows the evaluation of the changes in (quasi-)electrochemical potentials of electrons, holes and ions ($v_n$, $v_p$ and $v_{ion}$) due to the applied small perturbation ($v_{app}$) as function of position, encapsulating properties related to the transport, reaction and storage of charges in the material. As for the change in the electrostatic potential ($v_e$), this is determined at each position by an interplay between dielectric contributions and the local change in net charge.

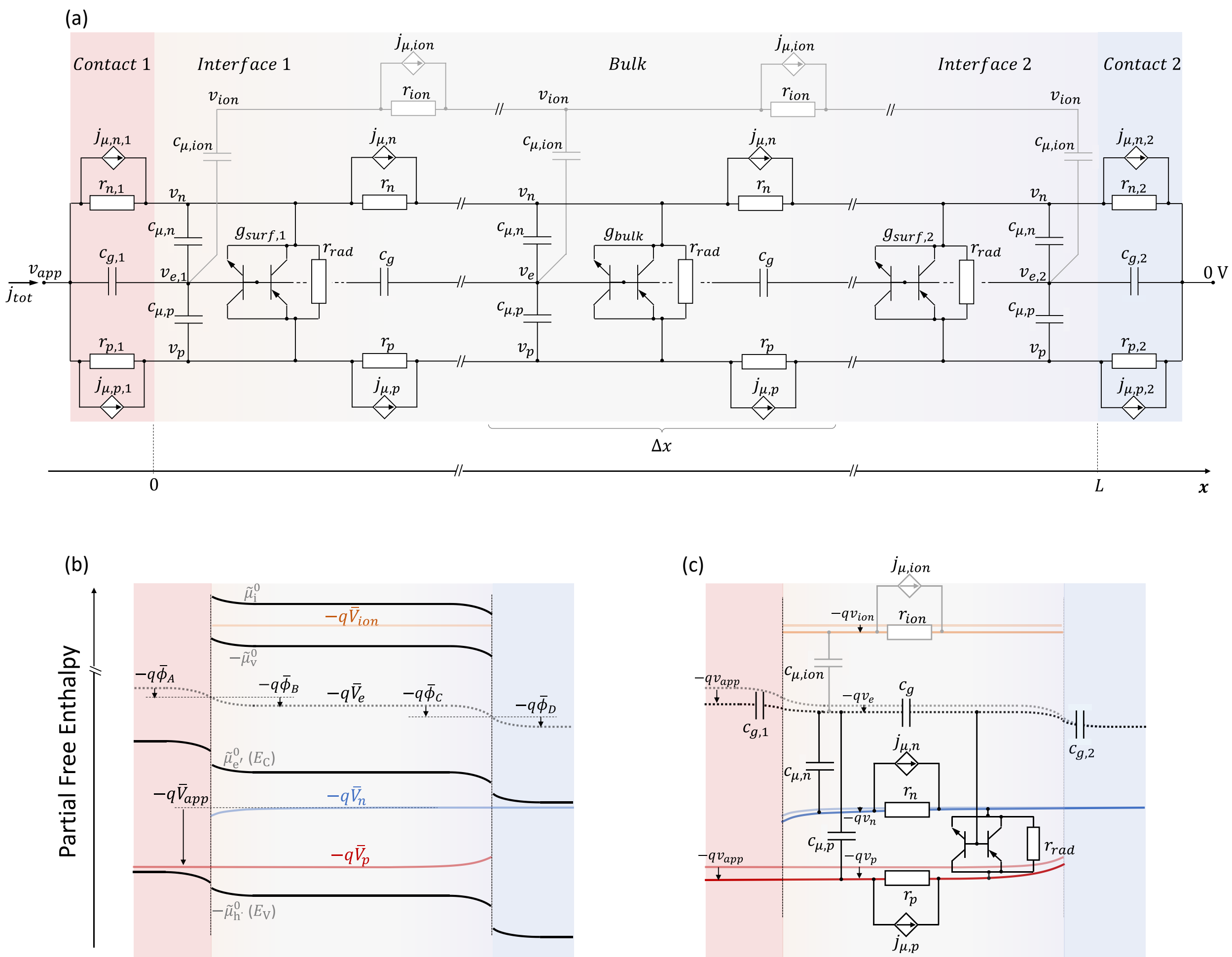

Figure 2. (a) Equivalent circuit model for a mixed ionic-electronic conductor (electrons, holes and one mobile ionic species are considered) between ion-blocking contacts (contact 1 and 2). The model is valid for devices with selective contacts (for each contact either of the electron or hole resistance is very large). The $x$ axis represents the qualitative position in the device, with $L$ being the active layer thickness. Within the "slice" $\Delta x$ of the mixed conductor, the model used for the bulk properties is shown (see text). (b) Generalized energy diagram [37] for a mixed conductor based solar cell between selective contacts ($r_{n,1} \to \infty$ and $r_{p,2} \to \infty$) under bias (*e.g.* under light and/or forward voltage $\bar{V}_{app}$) showing the steady-state potentials described in the text. (c) Schematics illustrating the construction of the equivalent circuit in (a) based on the relation between the small signal potentials in the energy diagram and their representation in the circuit. The relevant voltages at the circuit's nodes are not the absolute potential values, but their changes with respect to the steady-state.

Before describing the circuit in more detail, it is helpful to discuss the steady-state properties of such model device, which are used to derive the value of the relevant circuit elements. Given an applied voltage bias ($\bar{V}_{app}$) and electron-hole generation profile due to light absorption ($G_{ext}(x)$), the steady-state solution is obtained by solving the Nernst-Planck-Poisson problem (NPP or drift-diffusion model). This yields the values of the electrochemical potential for the electronic ($\bar{V}_n, \bar{V}_p$) and for the mobile ionic charges ($\bar{V}_{ion}$), as well as the value of the electrostatic potential ($\bar{V}_e$, here associated with the electronic vacuum level profile), as a function of position. These are schematically represented in an ionic and electronic energy diagram in Figure 2b, [37] for the case of a solar cell with a mixed conducting active layer and under forward bias. In the diagram, the difference in the value of $\bar{V}_e$ at the left and the right boundary of the full device corresponds to $\phi_{bi} - \bar{V}_{app}$, where $\phi_{bi}$ is the built-in potential of the complete stack (difference in work function between the two contact layers). The profile of $\bar{V}_e$ depends on the charge distribution in the device, which is dictated by the concentration of fixed charges and of mobile charge carriers, and by the ability of the latter to screen electric fields. Such screening occurs over a distance on the order of the Debye length, $L_D$ (for small perturbations of the local potential). In the case of a mixed conductor with one monovalent mobile ionic species, $L_D$ is defined as

$$L_D = \sqrt{\frac{\epsilon V_{th}}{q \sum_{j=n,p,ion} n_{j,bulk}}}\ , \qquad \text{Eq. 18}$$

where $\epsilon$ is the dielectric constant of the material and $n_{j,bulk}$ is the bulk charge concentration of carrier $j$ ($j = n, p, ion$). The value of $L_D$ is dominated by the mobile majority charge carrier(s). Variations in the steady-state profile of $\bar{V}_e$ are largely confined to space charge regions of width on the order of the material's $L_D$ or wider, depending on whether the majority carrier is accumulated or depleted at such interface. [40] The space charge potentials developing in the contact and active layer interfacial regions are indicated in Figure 2b as $\bar{\phi}_A, \bar{\phi}_B, \bar{\phi}_C$ and $\bar{\phi}_D$. The electrochemical potentials $\bar{V}_n, \bar{V}_p$ and $\bar{V}_{ion}$ complete the steady-state picture. Their position with respect to their corresponding standard potentials $\tilde{\mu}^0$ and their slope provide information on the local steady-state charge concentrations and current components in the device.

Based on the steady-state solution, the circuit elements relevant to the small perturbation problem can be derived (see Methods for details). The electrochemical resistors [7] and the voltage controlled current sources [39] ($r_j$, and $j_{\mu,j}$, respectively, with subscript $j = n, p, ion$) describe the linearized charge transport equation. The small signal current flowing in the resistors is driven by changes in the relevant quasi-electrochemical potential gradient (gradient in $v_j$ along the charge carrier rail). The current flowing in the controlled sources is sensitive to changes in the charge carrier chemical potential, according to $j_{\mu,j} = g_{\mu,j} v_{\mu,j}$, where $g_{\mu,j} = \frac{\bar{J}_j}{V_{th}}$ represents a transport transconductance ($\bar{J}_j$ is the steady-state current associated with species $j$). The dielectric capacitors ($c_g$) account for the short range dielectric properties, while the chemical capacitors [7] ($c_{\mu,j}, j = n, p, ion$) describe storage of each of the mobile charged species in the material. The generation-recombination elements connected between the rails associated with electrons and with holes complete the circuit describing the active layer (transistors with transconductance $g_{rec}$ and $g_{gen}$, and radiative recombination resistors $r_{rad}$). The current flowing through each of these differential circuit elements is driven by the changes in the electrochemical and electrostatic potentials (see schematics in Figure 2c).

The elements in the circuit implement a discretized version of the continuum differential problem. Their value is position dependent (*e.g.* $r_n = r_n(x)$), based on the local steady-state solution. In Figure 2a, such position dependence is not shown for the small signal circuit elements and potentials, for simplicity. Only for the SRH recombination parameters, the circuit emphasizes that these are, in general, different at the interface with the contacts ($g_{surf}$) compared with the bulk ($g_{bulk}$) of the active layer. This is due to differences in the values of $\tau_n, \tau_p, n_1$ and $p_1$, besides $n$ and $p$, as well as to possible recombination contributions involving electronic charges in the contacts.

This study focuses on devices with ion-blocking contacts. These are described in Figure 2a with resistors ($r_{p,1}, r_{n,1}, r_{p,2}$ and $r_{n,2}$), voltage controlled current sources ($j_{\mu,p,1}, j_{\mu,n,1}, j_{\mu,p,2}$ and $j_{\mu,n,2}$), and with the geometric capacitors, $c_{g,1}$ and $c_{g,2}$. These parameters refer to the transport in and capacitance of the space charge regions at the interfaces on the contact layers' side. This is an approximation which is valid only for selective contacts (*i.e.* for each contact layer either $r_n \to \infty$ and $g_{\mu,n} \to 0$ or $r_p \to \infty$ and $g_{\mu,p} \to 0$). Accurate treatment of both electrons and holes in the contacts requires recombination elements also in these regions, including the explicit description of the electrostatic potential in transmission line terms, as shown in the circuit in Figure S3. Such circuit, is referred to as 'complete model' below. An additional series resistance is expected in practical cases, and it is omitted in Figure 2, for simplicity. If the steady-state ionic current is zero ($\bar{J}_{ion} = 0$), all $g_{\mu,ion} = 0$ and $j_{\mu,ion}$ current source elements can be omitted from the model. Ion penetration in the contacts could also be described by extending the ionic rail throughout the device stack. Finally, the same circuit model without the ionic rail describes devices based on semiconductors without mobile ions.

Note that a complete treatment of the electronic and ionic charge carrier chemistry may include any possible redox reaction through which not only immobile (as in the traditional SRH treatment) but also mobile ionic defects interact with electrons and holes. This influences the electronic recombination dynamics, but also establishes more complex (quasi-)equilibria involving the electronic and ionic defects. In the small signal picture, these reactions can be represented via circuit elements connecting the electronic and ionic rails. [2,30] This work neglects their contributions, while a more detailed discussion of the non-equilibrium defect chemistry of mixed conductors will be subject of a separate study.

The complex impedance of the circuit in Figure 2a is calculated according to:

$$Z(\omega) = \frac{v_{app}}{j_{tot}(\omega)}. \qquad \text{Eq. 19}$$

Here, $v_{app}$ is the amplitude of an applied sinusoidal (small) voltage perturbation at angular frequency $\omega$, and $j_{tot}(\omega)$ is the (complex) current density flowing in the circuit at the node of the applied potential (Figure 2a). The value of $j_{tot}(\omega)$ can be evaluated, for example, by solving the system of linear equations obtained from Kirchhoff's current law applied to all nodes of the circuit minus one. The value of $Z(\omega)$ obtained using the circuit in Figure 2a corresponds to the impedance of the modelled system at the steady-state condition used to derive the values of the circuit elements. Due to the complexity of the analytical solution to this problem, approximated versions of the model are desirable. Possible approaches to this question are presented in the next section.

### C. Role of electrostatic potential and approximated models for bulk and interfaces

The contribution of electrostatic effects to the device's electrical response is often described through the overall geometric capacitor of the device in parallel to the remaining components of the equivalent circuit model. However, as evident from the model in Figure 2a, this approximation is not applicable in the general case. In particular, the model highlights the following important points: local changes in the electrostatic potential, $v_e$, can have a first order influence (1) on the non-radiative recombination rate (transistor elements) and (2) on the electronic charge transport currents ($j_{\mu,j}$ current sources), requiring an explicit account of their gating effect. (3) Due to the connection of the electrostatic rail to the chemical capacitors in the circuit, charge carriers that are high in concentration (high chemical capacitance $c_{\mu,j}, j = n, p, ion$) are expected to influence $v_e$ the most. It is useful to discuss the relevance of (1), (2) and (3) in practical situations. The case of devices with efficient electronic charge transport and ion blocking contacts with opposite selectivity is considered in detail in the following section (relevant to high performance solar cells). In this case, changes in applied potential $v_{app}$ are reflected in changes in $(v_p - v_n)$ throughout the device. Such condition might not apply in general to measurements performed on solar cells, especially when far from open circuit conditions. [41,42]

### *C.1 Semiconductors without mobile ions*

First, let us consider the importance of treating explicitly the electrostatic potential for the description of recombination in semiconducting devices with no mobile ions. For this, the transmission line model in Figure 2a, but without the ionic rail, is the relevant and analytically accurate equivalent circuit model describing the small signal behavior.

As discussed above, the transistor-based circuit couples the electron-hole recombination current to the electrostatic potential. Such coupling becomes negligible, and the transistor can be well approximated with a suitable $r_{SRH}$ resistor, only in the case of shallow traps, and in the case of deep traps for $\tau_n \bar{p} \approx \tau_p \bar{n}$. Interestingly, a similar approximation can be made in situations involving deep traps and $\tau_n \bar{p} \neq \tau_p \bar{n}$, based on the following argument. If the electron and hole concentrations are very different from each other, *e.g.* $\bar{p} \ll \bar{n}$, it is reasonable to expect also $c_{\mu,p} \ll c_{\mu,n}$. Given the low impedance associated with the large $c_{\mu,n}$, under the small perturbation regime and for $\omega > 0$, the approximation $v_e \approx v_n$ can be used. It follows that the single transistor indeed behaves once again like a resistor (in this example, $r_{SRH} = 1/g_{rec,p}$, see Figure S4).

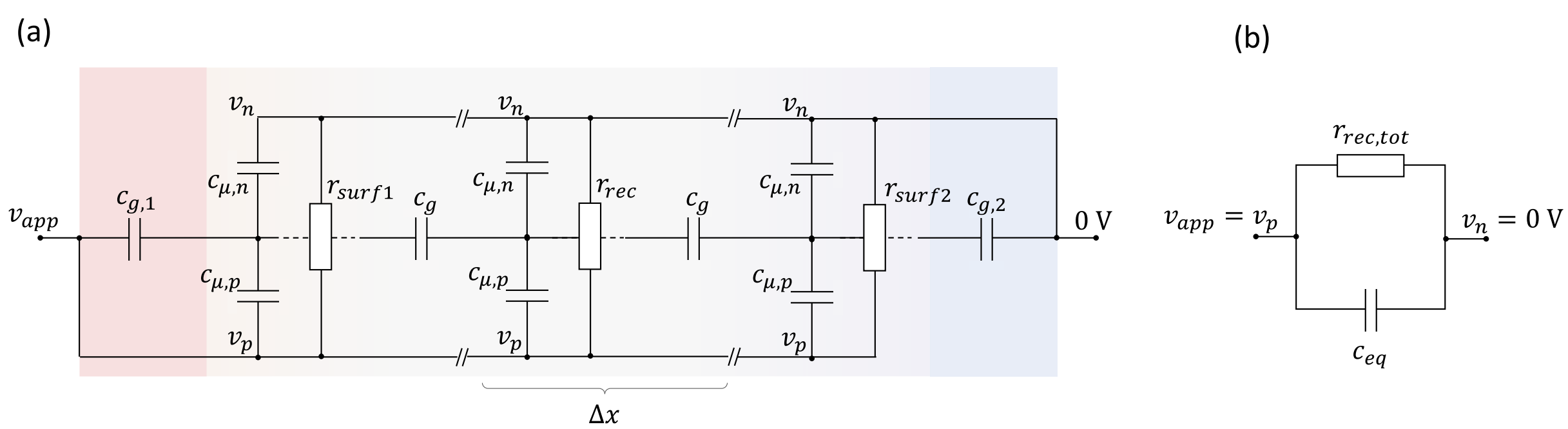

Figure 3. (a) Simplified version of the transmission line model for a device based on a semiconductor without mobile ions, assuming that all recombination processes can be modeled as resistors, the rate of electronic transport is significantly faster than the rate of recombination, and ideal contact selectivity (holes for contact 1 and electrons for contact 2). (b) Simplified version of (a), where $r_{rec,tot} = [(\sum r_{rec}^{-1}) + r_{surf,1}^{-1} + r_{surf,2}^{-1}]^{-1}$, while $c_{eq}$ is the equivalent capacitance of the circuit in (a) which includes the electrostatic and the chemical contributions (see Appendix B).

The use of bipolar transistors to describe the SRH recombination rate in semiconducting optoelectronic devices, while formally accurate, can be reasonably approximated with recombination resistors when looking at many practical situations (see model in Figure 3a, where efficient electronic transport and selective contacts are also considered). Only in the case of deep traps, $\bar{p} \approx \bar{n}$ and $\tau_n \bar{p} \neq \tau_p \bar{n}$, is the transistor description needed. Influence of the change in local electrostatic potential on the recombination current is expected in such cases despite negligible ion transport in the device, an aspect that deserves future investigation.

An analytical solution to the impedance of the circuit in Figure 3a is available (see Appendix B). This can be interpreted as a simplified zero-dimensional circuit model consisting of an equivalent capacitance ($c_{eq}$) and a total recombination resistance ($r_{rec,tot}$) connected in parallel (Figure 3b). A similar approach has been discussed in previous studies on solar cells, including the treatment of electronic transport limitations and of contributions from the contacts. [43–45]

The term $c_{eq}$ is often interpreted as the parallel of a total electrostatic capacitance $c_{g,tot} = \left(c_{g,1}^{-1} + c_{g,2}^{-1} + \sum c_g^{-1}\right)^{-1}$ and a total electronic chemical capacitance $c_{\mu,eon,tot} = \sum \left(c_{\mu,n}{}^{-1} + c_{\mu,p}{}^{-1}\right)^{-1}$. The discussion in Appendix B shows that this is a good approximation in many cases and that the definition of a capacitive term $\hat{c} = c_{eq} - c_{g,tot}$ can be used to extract information on the effective electronic contribution to the capacitance ($\hat{c}$, like $c_{g,tot}$ and $c_{\mu,eon,tot}$, describes the whole device). Such contribution can be chemical in nature ($\hat{c} \approx c_{\mu,eon,tot}$), while interfacial space charge contributions can become significant if $c_{\mu,n}$ and $c_{\mu,p}$ are very different (*e.g.* low injection condition in a doped semiconductor-based device).

### *C.2 Semiconductors with mobile ions*

The following discussion focuses on devices where a mixed ionic-electronic conductor is used as active layer of a solar cell, such as the ones based on halide perovskites. The analytical treatment of the full model is complex, even when considering ideal selective contacts and fast electronic transport (see section 6.2 in the Supplemental Material [38]).

Different approximations of the model can be derived depending on which charge carrier(s) play a significant role in the determination of the electrostatics. To address this question quantitatively, it is helpful to refer to the Debye length defined for each carrier type $j$ as $L_{D,j} = \sqrt{\frac{\epsilon V_{th}}{q n_{j,bulk}}}$. Because $L_{D,j}$ refers to the individual carrier screening length, $L_{D,j} < L$ or $L_{D,j} > L$ indicates whether carrier $j$ contributes significantly to the determination of the electrostatic landscape in the active layer or not.

In many cases of interest, the ionic charge carrier concentration is large enough, such that $L_{D,ion} \ll L$. As a result, whenever the recombination transistor elements cannot be reduced to resistors, the ionic situation is key to the electronic response, and the electrostatic rail needs to be considered explicitly (see point (1) above). Assuming $L_{D,ion} \ll L$, two scenarios are discussed depending on whether also for at least one of the electronic charge carriers $L_{D,j} < L$, $j = n, p$ (see also discussion in Ref.  [29]). This is equivalent to stating that not only the ionic chemical capacitance but also at least one of the chemical capacitances associated with electronic charges can be large enough to influence the local changes in electrostatic potential (point (3) above).

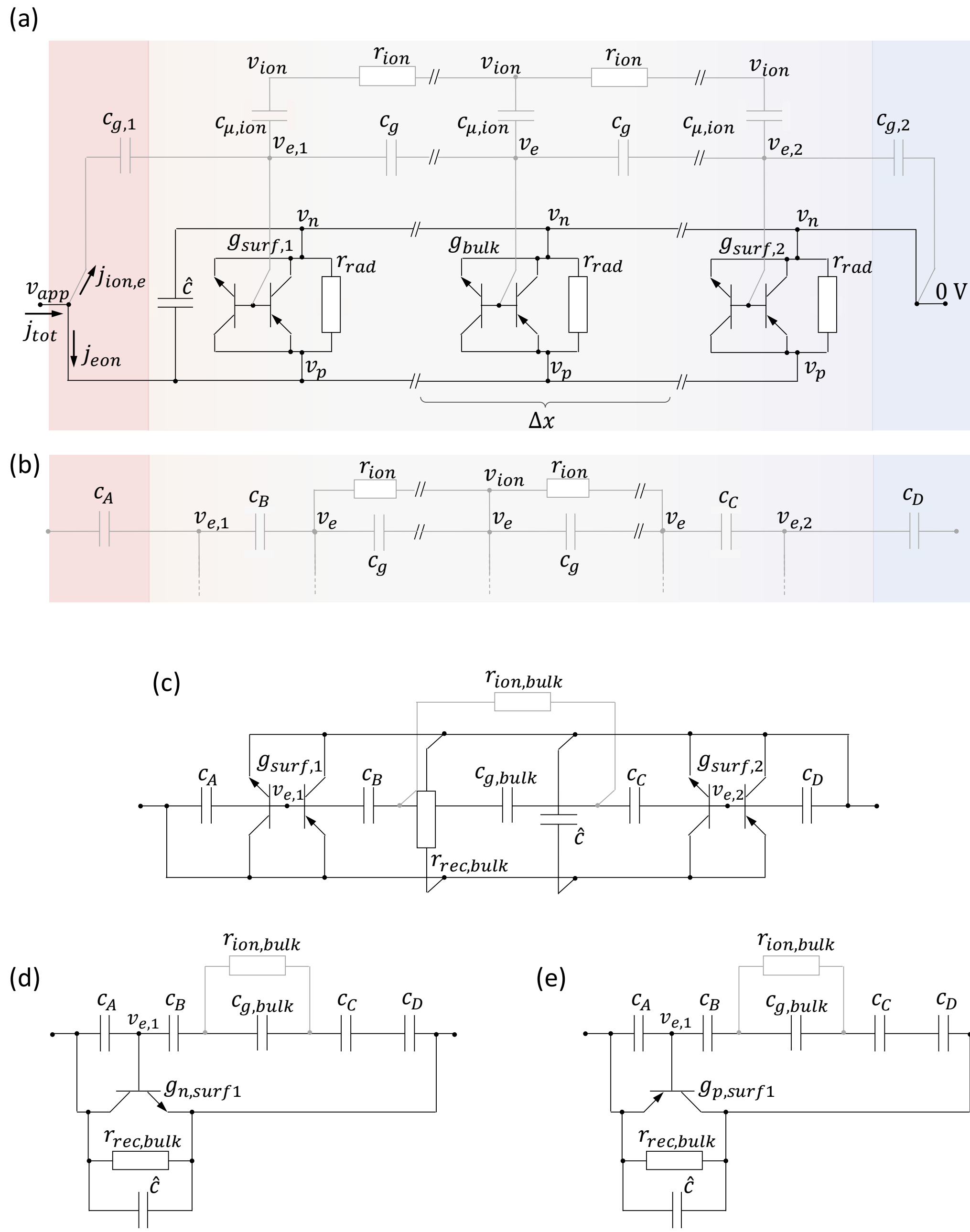

Figure 4. Ionic conductor approximation (IC) for the treatment of the electrostatic potential in a mixed conducting solar cell. (a) Simplified model obtained from Figure 2a, by neglecting the electronic chemical capacitors ($L_{D,n} \gg L$ and $L_{D,p} \gg L$). Infinite electronic charge carrier mobility limit and ideal contact selectivity (holes for contact 1 and electrons for contact 2) are assumed. (b) Further approximation of the electrostatic and ionic rails (see text and Appendix C). In (c–e), the network of $r_{ion}$ and $c_g$ elements in (b) is summarized in $r_{ion,bulk}$ (bulk ionic resistance) and $c_{g,bulk}$ (bulk geometric capacitance). The bulk non-radiative and radiative recombination elements are combined within a resistor $r_{rec,bulk}$ (see text). The models include: (c) all four transistors describing the recombination at the two interfaces; (d, e) only one (dominant) transistor for the cases where recombination due to (d) electrons or (e) holes minority carriers at the interface with the hole injecting contact (contact 1) dominates the total recombination current.

***IC (Ion Conductor) approximation*** *(valid for $L_{D,ion} \ll L$, $L_{D,n} \gg L$ and $L_{D,p} \gg L$).* In a first scenario, the steady-state electronic charge concentrations are small enough so that their corresponding Debye lengths are larger than the active layer thickness (*i.e.* their ability to screen electric fields is negligible over length scales in the order of $L$). Equivalently, this means that the total chemical capacitance of the active layer associated with either electrons or holes is small compared with its geometric capacitance ($c_{\mu,n}$ and $c_{\mu,p}$ present a large impedance and can be neglected). This allows one to evaluate the electrostatic-ionic behavior using an "ion conductor" approximation (IC), independently from the electronic properties of the device.

The total current density can be written as the sum of an ionic-electrostatic component and of an electronic component ($j_{ion,e}$ and $j_{eon}$, respectively, see Figure 4a), $j_{tot}(\omega) = j_{ion,e}(\omega) + j_{eon}(\omega)$. Based on Equation 19, this leads to

$$Z(\omega) = \frac{v_{app}}{j_{eon}(\omega)+j_{ion,e}(\omega)} = \left[Z_{eon}(\omega)^{-1} + Z_{ion,e}(\omega)^{-1}\right]^{-1}, \qquad \text{Eq. 20}$$

where the electronic and the ionic-electrostatic impedance are defined as $Z_{eon}(\omega) = \frac{v_{app}}{j_{eon}(\omega)}$ and $Z_{ion,e}(\omega) = \frac{v_{app}}{j_{ion,e}(\omega)}$, respectively.

Figure 4a shows the resulting equivalent circuit model, for a situation where the condition $v_p(x,\omega) - v_n(x,\omega) = v_{app}$ (fast electronic transport) is also valid. By neglecting the $g_{gen}$ terms (device under forward and/or light bias), and by integrating both SRH and Auger terms in one set of recombination transconductance parameters, one obtains:

$$Z_{eon}(\omega) = \frac{L}{r_{rad}} + i\omega\hat{c} + \int_0^L \left[\frac{v_e(x,\omega)}{v_{app}} g_{rec,n}(x) + \left(1 - \frac{v_e(x,\omega)}{v_{app}}\right) g_{rec,p}(x)\right] dx\,. \qquad \text{Eq. 21}$$

Such expression includes a frequency independent radiative recombination impedance, and a frequency dependent impedance associated with the SRH and Auger recombination current. The $\hat{c}$ element, connected between the nodes corresponding to $v_p$ and $v_n$, accounts for the electronic capacitive contribution, similarly to the discussion in the previous section. Indeed, while $c_{\mu,n} \ll c_{\mu,ion}$ and $c_{\mu,p} \ll c_{\mu,ion}$ based on the assumption in this first scenario, $\hat{c}$ can become relevant at high frequencies, for situations where its value approaches the capacitance due to the electrostatic term (Appendix B).

Replacing the ionic-electrostatic circuit branch in Figure 4a with the one displayed in Figure 4b is a further approximation leading to simple expressions for $Z_{ion,e}(\omega)$ and $Z_{eon}(\omega)$, and therefore $Z(\omega)$ (Equation C1–6). In Figure 4b, the small perturbation potential dropping across the space charge capacitors $c_B$ and $c_C$ refer to the changes in the values of $\phi_B$ and $\phi_C$, respectively. Also, $c_{g,1}$ and $c_{g,2}$ are renamed as $c_A$ and $c_D$ to reflect the nomenclature of potentials in Figure 2b. The replacement of $c_{\mu,ion}$ with short circuits ($v_e \approx v_{ion}$) is valid for large enough values of such capacitance and for non-zero angular frequencies. Note that a solution is available also for the circuit in Figure 4a, but only when considering a flat band situation (section 6.3 of the Supplemental Material [38]).

This approach allows one to explicitly account for the frequency dependence of recombination at all positions in the bulk of the active layer and at the surfaces, extending the concept of ionic-to-electronic current amplification introduced in Ref. [13] to processes occurring across the whole device. Note that in its simplest form ($Z_{ion,e}$ evaluated with the circuit in Figure 4b), the model does not account for recombination across the space charges explicitly (more on this in the Methods).

Finally, further simplifications to the model can be obtained in special cases. Figure 4c shows a circuit with focus on the interfacial behavior, and where the spatial dependence of bulk properties is no longer explicitly described. Here, the $r_{ion}$ and $c_g$ bulk elements in Figure 4c have been replaced with a single parallel circuit involving the total bulk ionic resistance and geometric capacitance ($r_{ion,bulk}$ and $c_{g,bulk}$). In addition, the bulk non-radiative recombination and radiative recombination elements are combined within a resistor $r_{rec,bulk}$. This approximation can be relevant to real devices, either because of the conditions discussed in the previous section (shallow traps, or $\tau_n\bar{p} \approx \tau_p\bar{n}$ throughout the bulk, see Appendix A), or because of negligible dependence of the total bulk recombination current on the small signal electrostatic potential.

Regarding the interfacial transistor elements, the circuit in Figure 4c shows some differences with the 4-transistor circuit proposed in Ref. [13]. These are related to the description of local surface recombination/generation, included in this work, and of electronic charge transfer reactions across interfaces, neglected here while considered in Ref. [13]. The interested reader is referred to section 7 of the Supplemental Material for a detailed analysis. [38] One observation from such discussion is that the model for an injection limited situation is in fact analogous to the model of a device where the dominant recombination at an interface involves the charge carrier type injected from the contact at such interface as minority carrier. These conditions lead to low frequency inductive behavior, as it will be shown in the next section.

When evaluating the circuit in Figure 4c for most practical situations, one transistor (the one with the largest value of $g_{rec}$) is sufficient to describe the small signal electronic response of the device under bias. In Figure 4d and 4e such simplification is displayed, for the case where the recombination of the minority carriers electrons or holes, respectively, at the hole injecting interface dominates the electrical response (the models described in Refs [13,46] are recovered, see Appendix C for impedance evaluation). The elements $r_{rec,bulk}$ and $\hat{c}$ can still be included to complete the model, similarly to Figure 4a.

In all models shown in Figure 4, the low frequency time constant of the circuit describes the space charge polarization process, whereby redistribution of mobile ions in the bulk of the active layer and electronic charges in the contacts charge/discharge the interfacial capacitors. This time constant can be expressed as $\tau^{\perp} = r_{ion,bulk}c^{\perp}$, where $c^{\perp}$ is the series of $c_A, c_B, c_C$ and $c_D$, and it represents the characteristic time of the slow ionic and electronic currents in the device operating in the IC approximation regime. [13,27,30,46–48] Importantly $\tau^{\perp}$ depends on the properties of the active layer ($r_{ion,bulk}$, $c_B$ and $c_C$) and also of the contacts ($c_A$ and $c_D$).

***MC-i (Mixed Conductor with mobile ions as majority carriers) approximation*** *(valid for $L_{D,ion} \ll L$, $L_{D,ion} \ll L_{D,p}$ and $L_{D,ion} \ll L_{D,n}$).* The second scenario relaxes the assumption of low electronic charge concentration. It can describe also situations where one or both electronic carrier types contribute significantly to the determination of the electrostatics in the active layer, while mobile ions remain the majority carriers. This condition is relevant to the modified surface polarization model presented by Clarke *et al.* [28,29] Such large steady-state concentrations for at least one of the electronic charge carriers are obtained if large enough (voltage and/or light) bias is applied to the mixed conducting device. Note that, unlike semiconductors with no mobile ions, the condition $\bar{p} \neq \bar{n}$ in the bulk of mixed conductors is the norm rather than the exception, even at large biases (see section 8 in the Supplemental Material [38]). The condition $L_{D,p} > L$ and/or $L_{D,n} > L$ may be met also for small bias, if very thick samples are considered (large $L$), or if the charge carrier equilibrium due to material and device preparation leads to significant values of $\bar{n}$ or $\bar{p}$ in the bulk (*e.g.* intrinsic or extrinsic doping). In equivalent circuit modeling terms, this means that the electron and/or the hole chemical capacitance is large, so that the ionic-electrostatic circuit impedance and the function $v_e(x,\omega)/v_{app}$ cannot be calculated separately from the electronic circuit, as it was the case in the IC approximation.

Aiming to obtain a model with an analytical solution also in this case, the transmission line shown in Figure 2a is simplified as follows. Chemical capacitors associated with the neutral component (see Ref. [2] and Figure S15) that couple either electrons or holes with the ionic rail are defined:

$$c_n^\delta = \frac{c_{\mu,ion}c_{\mu,n}}{c_{\mu,ion}+c_{\mu,n}} \qquad \text{Eq. 22}$$

$$c_p^\delta = \frac{c_{\mu,ion}c_{\mu,p}}{c_{\mu,ion}+c_{\mu,p}} \qquad \text{Eq. 23}$$

Note that the voltage dropping across $c_{\mu,j}$ capacitors ($j = n, p, ion$) is the small signal chemical potential of the individual *j*-th charge. On the other hand, the voltage dropping across $c_n^\delta$ or $c_p^\delta$ refers to the small signal quasi-chemical potential of the neutral species (defined by the ionic and *j*-th electronic charge carrier), which is related to changes in local stoichiometry ($\delta$).

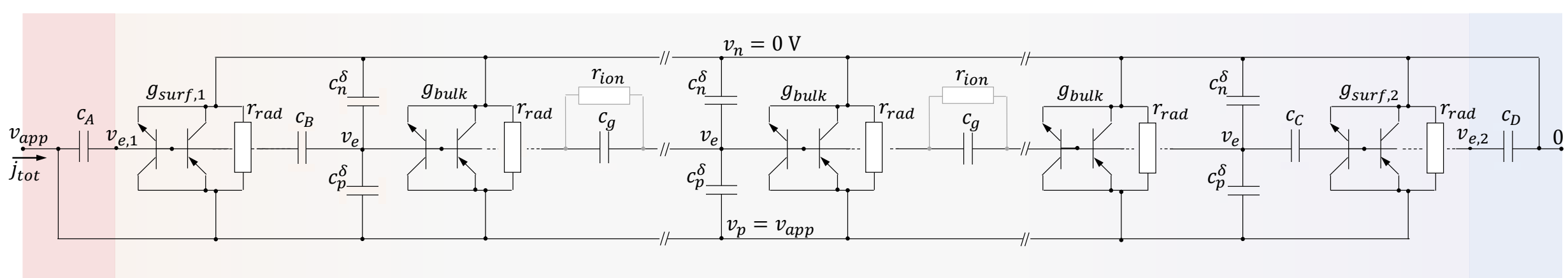

Figure 5. Mixed conductor with mobile ionic majority carriers approximation (MC-i) for a solar cell with ideal selective and ion-blocking contacts. The model can account for the effect of large electronic charge concentrations via the chemical capacitors $c_n^\delta$ and $c_p^\delta$. The circuit is obtained from Figure 2a, with the approximation $v_e \approx v_{ion}$ in the bulk (valid for angular frequencies $\omega > 0$), fast electronic transport, and by introducing the space charge capacitors $c_B$ and $c_C$ in the active layer, as discussed for Figure 4b.

Assuming that the conditions $c_{\mu,ion} \gg c_{\mu,n}$ and $c_{\mu,ion} \gg c_{\mu,p}$ are still valid, the approximation $c_n^\delta \approx c_{\mu,n}$ and $c_p^\delta \approx c_{\mu,p}$, as well as $v_e \approx v_{ion}$ can be used. This treatment accounts for the mixed conducting properties of the active layer and for the fact that mobile ions are assumed to be majority carriers (MC-i approximated model). Such simplification, for the case of a solar cell with selective contacts and fast electronic transport, is displayed in Figure 5.

While separating the total current into an ionic-electrostatic component and an electronic component is no longer possible, a reasonably simple expression of $v_e(x,\omega)/v_{app}$ that also accounts for electronic contribution to the electrostatic potential through the $c_n^\delta$ and $c_p^\delta$ terms is obtained (Equation C8). Equation C13 yields the overall impedance $Z(\omega)$ of the circuit (MC-i approximation). Its expression reduces to the results obtained for Figure 4 for the low electronic charge concentration case (IC solution). Note that the $\hat{c}$ element is no longer needed in the circuit, as approximations for the electronic capacitive contributions are explicitly considered in Figure 5 through $c_n^\delta$ and $c_p^\delta$. The approach used to obtain the MC-i circuit can be generalized to treat situations where the chemical capacitance associated with any one charge carrier (ionic or electronic) is dominant (*e.g.* MC-n: $L_{D,n} \ll L$, $L_{D,n} \ll L_{D,ion}$ *and* $L_{D,n} \ll L_{D,p}$).

In Figure 5, the changes in electrostatic potential in the active layer are dictated not only by the ionic charging of interfacial capacitors, as seen in the IC approximation scenario, but also by the ionic and electronic charging of bulk chemical capacitors. The latter process is associated with the ambipolar diffusion of electronic and ionic charges (chemical diffusion of the neutral component), which leads to stoichiometric polarization effects in the bulk. While mostly established for mixed conducting devices in the dark under local equilibrium, [40,49–51] stoichiometric polarization effects have been discussed also in the context of oxide-based high-temperature photoelectrochemical devices under local non-equilibrium (*e.g.* light bias). [52,53]

The characteristic time constant of this bulk diffusion process, $\tau^{\delta}$, is associated with the product of a chemical resistance $r^{\delta}$ and a chemical capacitance $c^{\delta}$. In Figure 5, these correspond to $r_{ion,bulk}$ (assuming negligible electronic transport resistance) and to $c_n^{\delta} + c_p^{\delta}$ integrated over the bulk of the active layer, respectively. [54] For the case of stoichiometric polarization close to equilibrium, a proportionality prefactor of 1/12 is needed in the calculation of $\tau^{\delta}$, due to the evolution of the potential profile during the polarization process. [2] Here, such prefactor is expected to depend on the specific device and steady-state situation considered. Note that, if the impedance associated with electronic transport cannot be neglected, the $r_j$ and $g_{\mu,j}$ terms ($j = n, p$) as well as the recombination elements, are also involved in the determination of $\tau^{\delta}$.

Importantly, the time constant $\tau^{\delta}$ scales with the concentration of electronic charges (included in the chemical capacitance terms), and with the active layer thickness squared. [2,30] It follows that $\tau^{\delta}$ is very fast in thin film mixed conducting solar cells with low electronic charge concentrations. Because of that, and since the electronic contribution to the electrostatics is negligible, stoichiometric polarization effects are not visible in the electrical response for low bias and intrinsic active layers. Under large bias, the values of $c_n^{\delta}$ and/or $c_p^{\delta}$ (and therefore $\tau^{\delta}$) increase. If $L_{D,p} < L$ and/or $L_{D,n} < L$, the ambipolar diffusion dynamics (charging of the chemical capacitors) gives rise to a significant change in $v_e$. This modulates recombination voltage driving forces and the corresponding currents, introducing an additional low-frequency feature in the impedance spectrum (see next section).

The time constant $\tau^{\perp}$ is expected to also change with the bias condition. Especially in situations where the value of $c^{\perp}$ is limited by the presence of a depletion interfacial capacitor in the device, its dependence on bias is expected to be milder than for $c^{\delta}$. The relation $\tau^{\perp} > \tau^{\delta}$ is then valid only for low enough $c^{\delta}$ (intrinsic active layer under low bias and thin films). In addition, because charging of $c_n^{\delta}$ and $c_p^{\delta}$ occurs also during the space charge polarization process, these elements contribute to the determination of $\tau^{\perp}$ in the MC-i approximation (Appendix C). Such contribution is negligible only if the value of $c^{\delta}$ integrated along the device thickness is significantly lower than the interfacial capacitance terms. It follows that the parameters determining the characteristic times $\tau^{\delta}$and $\tau^{\perp}$ depends on the properties of the active layer and of the contacts (including their selectivity), as well as on the applied bias.

### D. Interpretation and discussion of calculated impedance spectra

The impedance of devices at equilibrium or under bias can be calculated using the equivalent circuit models presented in the previous sections. Here, the proposed models are validated based on the comparison with drift-diffusion simulations. In all cases, solar cells with a mixed conducting active layer and ion-blocking contacts are considered, with input parameters that are relevant to halide perovskite devices (see Methods, and sections 10 and 11 of the Supplemental Material [38]).

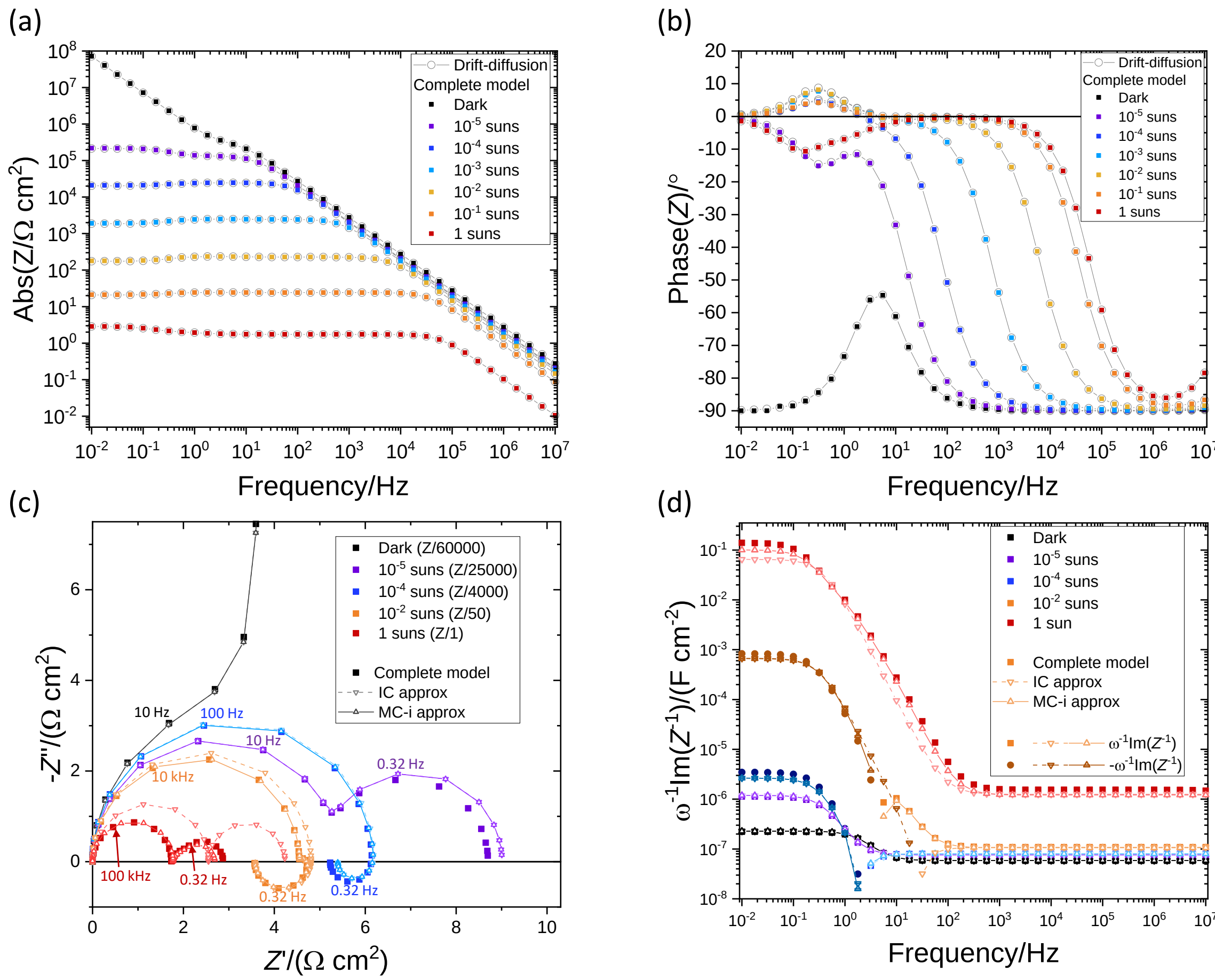

Figure 6. Impedance of a solar cell based on a semiconducting active layer with mobile ions calculated using either drift-diffusion or the transmission line transistor based equivalent circuit models presented in this study. Bode plots of the (a) magnitude and (b) phase of the impedance evaluated at open circuit in the dark and for different bias light intensities. The data obtained using the equivalent circuit model in Figure S3 (complete model) are compared with drift-diffusion simulation results. (c) and (d) show the Nyquist and apparent capacitance-frequency spectra of selected data from (a, b), where the results obtained with the complete transmission line and the approximated IC and MC-i models are compared.

In Figure 6, Bode plots show the calculated impedance of a solar cell with a mixed conducting active layer under open circuit conditions in the dark and for different steady-state bias light intensities. The impedance obtained from drift-diffusion simulations using the method described in Refs [13,46] is compared with the impedance calculated using the complete model (Figure S3). Both calculated results refer to the same steady-state drift-diffusion solutions. The datasets obtained with the two different methods match with good accuracy, confirming the analytical correspondence of the proposed equivalent circuit model with the linearized drift-diffusion equations.

The input parameters selected for this example (see Table S3) are such that recombination integrated across the active layer bulk is slightly larger than the recombination occurring at the interfaces, validating the model in its one-dimensional form. Moreover, the selected example highlights various transitions in the low frequency impedance as a function of bias, as shown in the Nyquist plot and apparent capacitance-frequency spectra in Figure 6c and d. These data emphasize several of the key factors determining the small perturbation electrical response of the device, including:

- Ionic-to-electronic current amplification effects at the interfaces and in the bulk, giving rise to one or two low frequency impedance features associated with either capacitive or inductive behavior.
- The contribution of ionic conduction to the impedance at low electronic charge concentrations.
- Transition from impedance spectra that can be explained with the IC approximation (see data for dark, $10^{-5}$ and $10^{-4}$ suns bias) to situations where the MC-i approximation is more appropriate ($10^{-2}$ suns and 1 sun data). This transition is evident from the emergence of a second low frequency feature, and it occurs at bias conditions leading to enhanced high frequency capacitance due to electronic contribution in the active layer.
- Changes in shape of the low frequency features at different bias due to changes in the relative magnitude of the steady-state concentration of electrons and holes in the bulk (see $10^{-2}$ suns vs 1 sun data).
- Effect of the magnitude of $\bar{V}_{app}$ relative to $\phi_{bi}$ and inversion in the majority carrier type at interfaces (relevant to situations where surface recombination dominates, this is not the case in Figure 6).

Further discussion of these factors with reference to the data in Figure 6, as well as of the influence of electronic transport limitations, is reported in section 12 of the Supplemental Material. [38]

In Figure 7, the analysis of individual representative cases is discussed, with focus on the significance and validity of the IC and MC-i models. The discussion of this figure exemplifies a key message deriving from Equation 16, and applied to the case of mixed-conducting solar cells with efficient electronic transport: impedance spectra can be interpreted based on the evolution of $v_e$ with frequency in the region of the device hosting the dominant recombination current, and on which recombination (trans)conductance $g_{rec,k}$ is dominant. Such region can be localized at surface 1, at surface 2 or can be extended throughout the bulk of the active layer. The same input device parameters are used to calculate all impedance spectra in the figure, and only the capture lifetime of electrons and holes at the interfaces and in the bulk are varied to access different responses. Here, $\tau_n = \tau_p$ is used, for simplicity. Section 2 of the Supplemental Material [38] discusses the influence of the relative magnitude of $\tau_n$ and $\tau_p$ on the type of the recombination limiting carrier, and therefore on the dominant transconductance and recombination current.

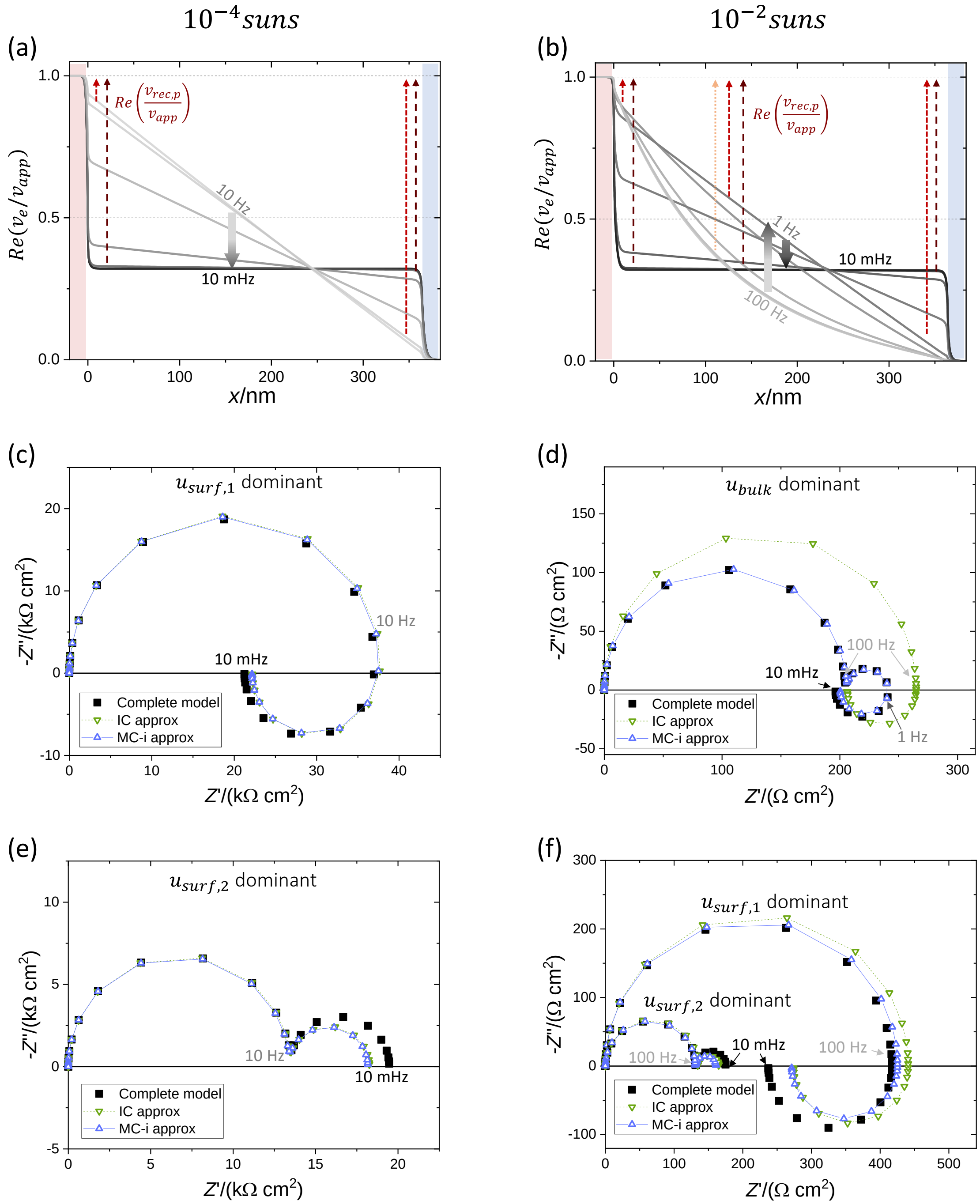

Figure 7. Impedance of mixed conducting solar cells under bias at open circuit explained based on the frequency dependent profile of the small signal electrostatic potential $v_e$. The left column and the right column refer to light intensities of $10^{-4}$ and $10^{-2}$ suns, respectively. (a) and (b) show the real part of the $v_e/v_{app}$ function extracted from the solution of the complete model as a function of position in the device and for different frequencies. (c–f) Nyquist plot showing the impedance calculated with either the complete model based on the drift-diffusion steady-state solution, or with the IC or MC-i approximated circuit models (see Methods for details). The same input parameters are used to obtain the steady-state condition in all cases (see section 11 of the Supplemental Material [38]), except the carrier capture lifetimes in the active layer

and at interfaces 1 and 2. Specifically, $\tau_n = \tau_p$ in each region, and their value is varied to select situations where the recombination current at either interface 1 or 2, or in the bulk is dominant. Data in (a) and in (b) refer to the input parameters used for calculations shown in (c) and (d), respectively.

The $v_e$ profiles obtained from the solution to the complete model allow the identification of the changes in recombination voltages. These are $v_{rec,n} = v_e - v_n \approx v_e$ and $v_{rec,p} = v_p - v_e \approx v_{app} - v_e$ for the *npn* and *pnp* transistors, respectively (assuming fast electronic transport). Because in all cases considered here $g_{rec,p} \gg g_{rec,n}$ (holes are minority carriers across the active layer), only $v_{rec,p}$ is highlighted in Figure 7a and b (see arrows indicating $Re(v_{rec,p}/v_{app})$ at specific positions and for different frequencies). Once again, the driving force for radiative recombination is independent of position and of the frequency of the applied perturbation for cases involving efficient electronic transport and ideal contacts with opposite selectivity ($v_{rec,rad} = v_p - v_n \approx v_{app}$).

The frequency dependent $Re(v_e/v_{app})$ data for the $10^{-4}$ suns case highlight an essentially linear $v_e$ profile within the bulk of the active layer, with its slope (electric field) varying from a finite value to approximately zero for decreasing frequency. This corresponds to the transition from the short-range dielectric response of the device stack to the response due to long range ionic redistribution (space charge polarization with time constant $\tau^{\perp}$) at high and low frequencies, respectively. Note that the value of $v_e$ in the bulk at low frequency is lower than $v_{app}/2$ because of the asymmetry in the interfacial capacitors (in this case $c_B \approx c_C$ and $c_A < c_D$).

Figure 7c and e show the resulting impedance depending on whether the electronic charge carriers' lifetime is shortest at interface 1 or 2, respectively. In the former case, a negative capacitance (or inductive behavior) is observed, due to the increase in recombination voltage $v_{rec,p}$ (decrease in impedance) with decreasing frequencies at interface 1 (see Figure 7a). The opposite dependence of $v_{rec,p}$ is observed for interface 2, leading to a low frequency feature with capacitive behavior when recombination at such interface dominates. The data highlight that recombination of the rate limiting electronic carrier in proximity to the interface where either such carrier type or the opposite carrier type is injected leads to negative or positive capacitance behavior, respectively.

When performing the same analysis for the data associated with high light intensities ($10^{-2}$ suns), an additional regime is found. A non-linear $Re(v_e/v_{app})$ profile in the active layer (Figure 7b) dictates the recombination driving forces and currents in the bulk and at interfaces at high frequencies (*e.g.* 100 Hz). This result can be assigned to the electronic charging of the chemical capacitors $c_n^{\delta}$ and $c_p^{\delta}$, which occurs already at fast time scales. As discussed in the previous section, such an effect happens at any bias conditions, but only for situations involving large enough steady-state electronic charge concentrations it influences $v_e$ significantly. In other words, when $c_n^{\delta}$ and/or $c_p^{\delta}$ cannot be neglected, the bulk of the device cannot be approximated as electroneutral at high frequencies. As a result, the capacitance observed at high frequencies is expected to deviate from the geometric contribution, as also discussed in the previous section and in Appendix B. If transport of electrons and holes is sluggish, the above trend occurs only at frequencies that are lower than the rate of electronic transport (see discussion of Figures S24–S28).

Electroneutrality and a linear profile of $Re(v_e/v_{app})$ in the bulk are established only at frequencies that are low enough so that ionic charges can redistribute within the bulk. Once again, this polarization describes the ambipolar diffusion of ionic and electronic charges in the bulk of the active layer and occurs over time scales in the order of $\tau^{\delta}$. At even lower frequencies, the space charge polarization described above occurs also in this case.

Figure 7d displays the implications of such evolution in $v_e$ in terms of the impedance spectral shape, for the case where bulk recombination is dominant. The resulting impedance highlights two clear features at low frequencies, one related with the bulk diffusion process and one related with the space charge polarization with characteristic frequencies of $\sim\left(2\pi\tau^{\delta}\right)^{-1}$ and $\sim(2\pi\tau^{\perp})^{-1}$, respectively. In this specific example, the ambipolar diffusion process leads to a decrease in recombination driving force for decreasing frequencies and to an increase in impedance highlighted by a feature with positive capacitance occurring in the 100 – 1 Hz frequency range. The increase in $v_{rec,p}$ caused by space charge polarization at even lower frequencies (1 – 0.01 Hz) leads to a negative capacitive feature.

Interestingly, when the dominant recombination contribution is at either interface of the device, only one low frequency feature (associated with space charge polarization) is evident (Figure 7f), even at large bias conditions. This is consistent with the minimal variation in electrostatic potential (and in $v_{rec,p}$) at the interfaces during the bulk polarization process associated with ambipolar diffusion.

For cases where both polarization processes are visible in the impedance spectrum, two features with two positive, two negative, or one positive and one negative capacitive behavior are possible. Figures S16–S20 show that, whether the minority or the majority electronic charge carrier is limiting recombination in the bulk determines the nature of the feature associated with stoichiometric polarization (positive or negative capacitive behavior, respectively). [29] The discussion in section 12.2 of the Supplemental Material [38] emphasizes that a more complex picture describes the situation where electronic transport is limiting the device current and the overall impedance of the device. Gating of charge transport current (due to the $j_{\mu}$ elements) by the changes in the charge carrier chemical potential can lead to "transport limited" low-frequency impedance features. Importantly, because such gating effect is sensitive to the local $v_e$ in the region that is rate limiting to charge transport, it is not necessarily confined to the position hosting the dominant recombination process in the device (*e.g.* the presence of two low-frequency feature does not imply dominant bulk recombination).

Figures 7c–f compare the impedance data obtained with the complete model with the IC and MC-i approximated models described above. While for the low electronic charge concentration case the IC and MC-i models essentially coincide and well describe the device response, at large bias only the MC-i model returns a satisfactory spectral shape. Interestingly, if $c_n^{\delta}$ and $c_p^{\delta}$ are small compared with the interfacial capacitors, then the IC model can also reproduce the low frequency feature associated with space charge polarization, despite the presence of the ambipolar diffusion feature (*e.g.* $10^{-2}$ suns data in Figure 6c, and Figure S16). If, however, $c_n^{\delta}$ and $c_p^{\delta}$ contribute significantly also to the space charge polarization process, such correspondence is no longer there (*e.g.* 1 sun data in Figure 6c, $10^{-2}$ suns data in Figure 7d, Figure S17–20). Therefore, even though including the $\hat{c}$ element in the IC approximation allows for an acceptable description of the capacitive behavior at high frequencies (see Figure 6d), the influence of $c_n^{\delta}$ and $c_p^{\delta}$ on the magnitude and frequency dependence of the $v_e$ profile at low frequencies can only be captured with the MC-i approximation (see Appendix C). Some discrepancies in the dataset obtained with the complete and the approximated models are present (*e.g.* low frequency impedance in Figure 7e and f) and are expected to be due to the method used to estimate the recombination voltage at the interfaces (see Methods).

The equivalent circuit modeling approach presented in this work facilitates the study of devices for which electron-hole recombination processes play an important role in the electrical response. The analysis emphasizes that the frequency dependence of the driving force associated with the dominant recombination process largely determines the impedance of the device at low frequencies. In the context of halide perovskite solar cells, this also means that mobile ions can be an unlikely researcher's "ally" when investigating the nature of such dominant recombination process.

The transistor-based treatment of recombination integrated within a transmission line model may provide a more general physical interpretation of solar cell behavior, including the discussion of parameters such as the device's ideality factor, and the analysis of other time- and frequency-domain measurements. Relating such information to physical properties (*e.g.* $\phi_{bi}$, $L_D$, charge carrier concentrations, recombination lifetimes) should be subject of future studies. Finally, the model can be extended to account for other aspects (*e.g.* multiple mobile ionic species, non-ideal contacts, redox-reactions, explicit treatment of traps and of (de-)trapping processes), extending its relevance to a broad range of photo-electrochemical devices.

## Conclusions

Linearization of the radiative and non-radiative recombination rates in semiconductors reveals that resistors and transistors, respectively, are the accurate equivalent circuit elements representing these processes in the small perturbation regime. Integrating such elements in a transmission line results in a circuit that is analytically equivalent to the linearized drift-diffusion model. The equivalent circuit model emphasizes the dependence of recombination on the local changes in electrostatic potential, a general effect relevant to semiconducting materials with or without mobile ions. Such influence can be ignored in most cases involving semiconductors with negligible mobile ionic concentrations, leading to an approximated description of recombination through simple resistor elements. On the other hand, in presence of large mobile ion concentrations, as it is the case in halide perovskite solar cells, the influence of ion redistribution on the electrostatic potential results in frequency dependent recombination currents that largely dictates the electronic response of the device at low frequencies. The analysis of ionic-to-electronic current amplification effects occurring at interfaces [13] is extended to the bulk recombination and for situations where the electronic charge also influences the electrostatic potential. The discussion of stoichiometric polarization effects, in addition to the well-established space charge polarization, occurring in mixed conducting devices is addressed for the non-equilibrium case, with simplified analytical solutions that can aid experimental data analysis and fitting. The method and models proposed in this work provide a general platform for the study of the electrical response of semiconducting and mixed ionic-electronic conducting devices out-of-equilibrium.

## Methods

**Drift-diffusion and complete equivalent circuit model calculations**. Impedance calculations performed using drift-diffusion simulations are based on the method described in Refs [13,46], using the Driftfusion software. [55] The impedance calculations of all equivalent circuit models described in this work are performed using MATLAB codes that include the system of equations associated with Kirchhoff's current law at all nodes minus one (see section 10 and 15 of the Supplemental Material [38]). Data presented in the main text involve calculations in the frequency range 10 MHz – 10 mHz, with four data points per decade, performed assuming devices at room temperature. For the determination of the values of the equivalent circuit elements in the discretized model, information on the steady-state situation is needed.

For the data obtained using the complete model (Figure 6 and 7), the steady-state solution is obtained from the Driftfusion software. The same approach is expected to apply to other simulation tools that can provide the position dependent charge concentrations and electrostatic potential. The circuit model shown in Figure S3 is used, where the transmission line is extended to the contact layers. This accounts for generation and recombination processes occurring in contacts 1 and 2 (for calculations relevant to solar cells, these are referred to as the hole transport material, HTM, and the electron transport material, ETM, respectively), as well as for non-ideal contact selectivity. The elements are defined on a mesh that includes the position of the voltage nodes $x_{i+1/2}$ ($i = 1, 2, \dots N$) and the position associated with the elements connecting neighboring voltage nodes $x_i$ ($i = 1, 2, \dots N+1$) (see section 13 of the Supplemental Material for details on the discretized model [38]). Below, the method to evaluate the equivalent circuit elements is described, following previously reported analysis [7,39].

To model *charge transport* of charge carrier $j$, the relation describing the total (particle) current density as a function of the carrier concentration and mobility ($n_j$ and $u_j$, respectively) and the gradient of the relevant electrochemical potential is considered:

$$J_j = -qu_j n_j \frac{dV_j}{dx}. \qquad \text{Eq. 24}$$

The relation is valid for electronic charges and monovalent ions. Differentiation of Equation 24 yields

$$j_j = -qu_j \left[\tilde{n}_j \frac{d\bar{V}_j}{dx} + \bar{n}_j \frac{d\widetilde{V}_j}{dx}\right]. \qquad \text{Eq. 25}$$

Since $\frac{d\widetilde{V}_j}{dx} = \frac{dv_j}{dx}$ and, in the small signal regime, $\tilde{n}_j \approx \bar{n}_j \frac{v_{\mu,j}}{V_{th}}$, Equation 25 can be expressed as

$$j_j = \frac{\bar{J}_j}{V_{th}} v_{\mu,j} - \sigma_j \frac{dv_j}{dx}. \qquad \text{Eq. 26}$$

The second term in Equation 26 describes the small signal particle current due to the steady-state local partial conductivity $\sigma_j = qu_j\bar{n}_j$, and the change in electrochemical potential gradient, $\frac{dv_j}{dx}$. This current component corresponds, in discretized form, to the current flowing through an electrochemical resistor defined at position $x_i$, $r_{i,j} = \Delta x_{i+1/2}/\sigma_{i,j}$, connected between the voltage nodes $x_{i-1/2}$ and $x_{i+1/2}$ ($\Delta x_{i+1/2} = x_{i+1/2} - x_{i-1/2}$). The first term in Equation 26 describes the small signal particle current density due to changes in chemical potential of carrier $j$. By defining the transport transconductance $g_{\mu,j} = \frac{\bar{J}_j}{V_{th}}$, such current component can be expressed as $j_{\mu,j} = g_{\mu,j} v_{\mu,j}$. In the discretized model of Figure 2a, $j_{\mu,j}$ is implemented as a voltage controlled current source. Note that a controlled source is required here, because $j_{\mu,j}$ flows in the charge carrier rail but it is proportional to $v_{\mu,j}$, which is the potential difference between nodes of such rail and of the electrostatic rail. Evaluation of $j_{\mu,j}$ flowing from node $x_{i-1/2}$ to $x_{i+1/2}$ requires calculating the average of $v_{\mu,j}$ at these two positions. For this reason, two current sources are used to represent this current component at each interval in Ref. [39]. In this work, a single current source controlled by the average of $v_{\mu,j}$ at position $x_{i-1/2}$ and $x_{i+1/2}$ is used at each $x_i$ position for simplicity.

To model *charge storage*, the local chemical capacitance per unit volume associated with charge carrier $j$ is defined as $c_{\mu,n_j} = \frac{q\tilde{n}_j}{v_{\mu,j}}$. In absence of significant trapping and assuming a dilute defect situation, the discretized chemical capacitance associated with node $x_{i+1/2}$ can be expressed as

$$c_{i+1/2,\mu,n_j} = \Delta x_i \frac{q\bar{n}_j}{V_{th}}, \qquad \text{Eq. 27}$$

where $\Delta x_i = x_{i+1} - x_i$.

Finally, the *electrostatic capacitors* are defined based on the local geometric capacitance as

$$c_{i,g} = \frac{\epsilon_i}{\Delta x_{i+1/2}}, \quad \text{Eq. 28}$$

where $\epsilon_i$ is the local dielectric constant.

**Approximated equivalent circuit models.** The equivalent circuit elements used in calculations involving the IC and MC-i approximated models in Figure 4 and 5 are evaluated based on the steady-state solution described above. For the bulk parameters (*e.g.* $r_{ion,bulk}$, $g_{rec,bulk}$, $c_n^\delta$ and $c_p^\delta$), a single value is used and estimated by considering the charge concentrations in the middle of the active layer. This approximation represents a source of error in the impedance calculation, as the transconductance profiles vary especially within the space charge regions in the active layer with possible local maxima (*e.g.* in general $\bar{p} \neq \bar{n}$ in the bulk, while the condition $\bar{p} = \bar{n}$, and therefore large values of $g_{rec}$, can occur in a localized region of the space charges). For the interfacial parameters, values of recombination transconductance are obtained by integrating their value over the junction layers (in this case 2 nm in thickness) used in the Driftfusion software to describe the interfaces. [55] Interfacial capacitors $c_A$, $c_B$, $c_C$ and $c_D$ are defined based on the steady-state concentration of the majority carrier in the relevant layer's bulk and on the space charge potential. [27,56] For example, the space charge capacitance associated with contact 1 (HTM for the case of a solar cell) is evaluated based on:

$$c_A = sign(\bar{\phi}_A)\sqrt{\frac{q\epsilon_{HTM}N_A}{2V_{th}}}\frac{1-e^{-\frac{\bar{\phi}_A}{V_{th}}}}{\sqrt{e^{-\frac{\bar{\phi}_A}{V_{th}}}+\frac{\bar{\phi}_A}{V_{th}}-1}} \quad \text{Eq. 29}$$

where $\epsilon_{HTM}$ and $N_A$ are the dielectric constant and the acceptor doping density in the HTM (the latter is assumed to correspond to the hole concentration in the contact's bulk). Equation 29 can be used to evaluate the other interfacial capacitors too, by replacing the relevant space charge potential, dielectric constant and bulk majority carrier concentration (*e.g.* $p$ in the HTM, $n_{ion}$ in the active layer and $n$ in the ETM, for the mixed conducting solar cells in this work). Note that $\bar{\phi}_A$ is defined to be positive for space charge situations leading to depletion of majority carriers in the HTM. For the calculation of $c_B$, the relevant space charge potential should be included in Equation 29 with a negative sign, as depletion of the positive ionic species occurs for negative values of $\bar{\phi}_B$, as defined in Figure 2b. Because significant changes in the steady-state electrostatic potential occur within the junction layers used in the drift-diffusion simulations, the use of a single estimate for each space charge potential in the calculation of the interfacial capacitors can be a source of discrepancy between the approximated and the complete models. In this study, the space charge potentials are calculated considering the potential in each of the layers' bulk and the potential at the outer boundary of the junction layer (interface between each junction and the respective contact layer). This approach is valid for negligible steady-state electric field values in the bulk of the active layer, as it is expected for the IC and MC-i approximations.

## Appendix A. Steady-state electronic charge concentration and recombination transconductance terms

Table 1. Expressions for the recombination transconductance terms describing SRH recombination in a semiconductor under light and/or electrical bias. Situations involving similar ($\bar{n} \approx \bar{p}$) or very different ($\bar{n} \ll \bar{p}$ or $\bar{n} \gg \bar{p}$) electronic charge concentrations are discussed (referred to as high- or low-injection respectively, see text) for the deep or the shallow trap cases. All $g_{rec}$ terms have the units of A V$^{-1}$ cm$^{-3}$. These simplified expressions are valid only if $\tau_n$ and $\tau_p$ have comparable values. Under such conditions, and assuming forward bias (or light bias) leading to $QFLS > 0$, dominant term(s) for each condition are highlighted in yellow. The energy level diagram shows the position of the quasi-Fermi levels ($E_{Fn}$ and $E_{Fp}$) and of the trap energy ($E_T$) with respect to the conduction and valence band edges ($E_C$ and $E_V$).

| $E$: $E_C$, $E_{Fn}$, $E_T$, $E_{Fp}$, $E_V$ | $\bar{n} \approx \bar{p}$ High injection | $\bar{n} \ll \bar{p}$ Low injection (p-type) | $\bar{n} \gg \bar{p}$ Low injection (n-type) |
|---|---|---|---|
| **Shallow trap ($E_T$ close to $E_V$)** | $p_1 \gg \bar{p}$; $g_{rec,n} \approx \frac{q\bar{n}\bar{p}}{V_{th}\tau_n p_1}$; $g_{rec,p} \approx \frac{q\bar{n}\bar{p}}{V_{th}\tau_n p_1}$ | $p_1 \gg \bar{p}$; $g_{rec,n} \approx \frac{q\bar{n}\bar{p}}{V_{th}\tau_n p_1}$; $g_{rec,p} \approx \frac{q\bar{n}\bar{p}}{V_{th}\tau_n p_1}$ | $p_1 \gg \bar{n}$; $g_{rec,n} \approx \frac{q\bar{n}\bar{p}}{V_{th}\tau_n p_1}$; $g_{rec,p} \approx \frac{q\bar{n}\bar{p}}{V_{th}\tau_n p_1}$ |
| **Deep trap** | $p_1 \ll \bar{p}$, $n_1 \ll \bar{n}$; $g_{rec,n} \approx \frac{q\tau_n\bar{n}}{V_{th}(\tau_n+\tau_p)^2}$; $g_{rec,p} \approx \frac{q\tau_p\bar{n}}{V_{th}(\tau_n+\tau_p)^2}$ | $p_1 \ll \bar{p}$, $p_1 \ll \bar{n}$, $n_1 \ll \bar{n}$; $g_{rec,n} \approx \frac{q\bar{n}}{V_{th}\tau_n}$; $g_{rec,p} \approx \frac{q\tau_p\bar{n}^2}{V_{th}\tau_n^2\bar{p}}$ | $p_1 \ll \bar{p}$, $n_1 \ll \bar{n}$, $n_1 \ll \bar{n}$; $g_{rec,n} \approx \frac{q\tau_n\bar{p}^2}{V_{th}\tau_p^2\bar{n}}$; $g_{rec,p} \approx \frac{q\bar{p}}{V_{th}\tau_p}$ |
| **Shallow trap ($E_T$ close to $E_C$)** | $n_1 \gg \bar{p}$; $g_{rec,n} \approx \frac{q\bar{n}\bar{p}}{V_{th}\tau_p n_1}$; $g_{rec,p} \approx \frac{q\bar{n}\bar{p}}{V_{th}\tau_p n_1}$ | $n_1 \gg \bar{p}$; $g_{rec,n} \approx \frac{q\bar{n}\bar{p}}{V_{th}\tau_p n_1}$; $g_{rec,p} \approx \frac{q\bar{n}\bar{p}}{V_{th}\tau_p n_1}$ | $n_1 \gg \bar{n}$; $g_{rec,n} \approx \frac{q\bar{n}\bar{p}}{V_{th}\tau_p n_1}$; $g_{rec,p} \approx \frac{q\bar{n}\bar{p}}{V_{th}\tau_p n_1}$ |

Table 1 provides the approximated expression of the recombination transconductance parameters for the different possible situations relevant to the trap-mediated SRH recombination in a semiconductor, where light and/or electrical bias leads to $QFLS > 0$. These include situations where the steady-state concentrations of electrons and holes are comparable ($\bar{n} \approx \bar{p}$), commonly referred to as "high-injection" in semiconductor physics, or where the two carriers differ significantly in concentration (where $\bar{n} \gg \bar{p}$ or $\bar{n} \ll \bar{p}$), referred to as "low-injection". Note that the high- and low-injection terminology may be misleading in the case of mixed conductors, where the condition $\bar{p} \neq \bar{n}$ holds for most situations, even for devices under large bias (section 8 of the Supplemental Material [38]). The analysis in Table 1 is carried out for the cases where either a shallow or a deep trap is at play. Here, a trap is defined as shallow or deep depending on the relative position of the trap energy and of the electronic majority carrier(s) quasi-Fermi level. That is, if either $n_1$ or $p_1$ is greater than both $n$ and $p$, the trap is shallow and it is deep otherwise (the trap energy can be deep and yet not be between the quasi-Fermi levels). In the table, the recombination transconductance terms that are expected to dominate the expression of $u_{SRH}$ for each situation are highlighted. Importantly, similar values of $\tau_n$ and $\tau_p$ are assumed in the approximations shown in Table 1, while different expressions are obtained based on Equation 11 if these parameters are very different from each other, also affecting the dominant transconductance term. This point is discussed in section 2 of the Supplemental Material, [38] where the expressions for the generation transconductance terms are also shown (these have negligible contribution under light and/or forward voltage bias).

As evident from Table 1, for all cases involving a shallow trap, $g_{rec,n} = g_{rec,p}$ justifying the definition of a recombination resistance $r_{SRH}$ according to Equation 13 in the main text. For the case of deep traps, the terms $g_{rec,n}$ and $g_{rec,p}$ are in general different, implying that the complete transistor description is needed. When considering $\bar{n} \approx \bar{p}$ with a deep trap, $g_{rec,n}$ and $g_{rec,p}$ can be of similar magnitude. For the special case of $\tau_n \bar{p} = \tau_p \bar{n}$, one obtains $g_{rec,n} = g_{rec,p}$, and a recombination resistance $r_{SRH} = 1/g_{rec}$ can be adopted instead of the transistor model in this case, too. Table 2 summarizes these considerations.

Table 2. Expressions for the radiative and trap-mediated recombination terms for a semiconductor with $QFLS > 0$ under small perturbation conditions, and their corresponding equivalent circuits. For the SRH transconductance, $\tau_n$ and $\tau_p$ are assumed to be of similar magnitude. If this is not the case, the full expression in Equation 11 needs to be used.

| **Recombination term** | **Equivalent circuit** |
|---|---|
| Radiative recombination<br>$qu_{rad} = \frac{v_p - v_n}{r_{rad}}$ $\quad$ $r_{rad} = \frac{V_{th}}{qk_{rad}\bar{n}\bar{p}}$ | 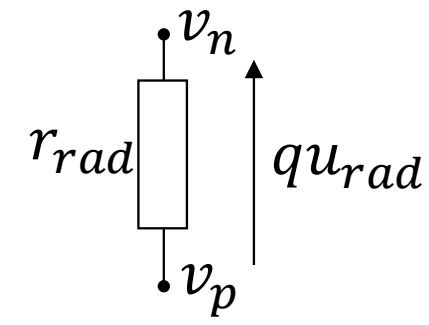 |
| SRH (deep trap, low injection, n-type)<br>$qu_{SRH}(\bar{n} \gg \bar{p}) \approx g_{rec,p}(v_p - v_e)$ $\quad$ $g_{rec,p} = \frac{q\bar{p}}{V_{th}\tau_p}$ | 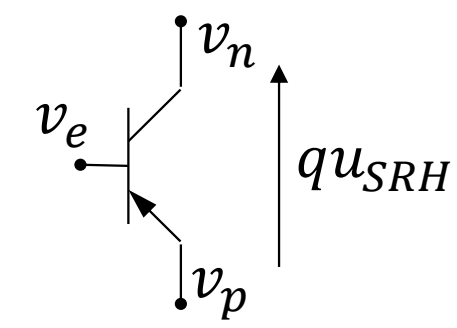 |
| SRH (deep trap, low injection, p-type)<br>$qu_{SRH}(\bar{p} \gg \bar{n}) \approx g_{rec,n}(v_e - v_n)$ $\quad$ $g_{rec,n} = \frac{q\bar{n}}{V_{th}\tau_p}$ | 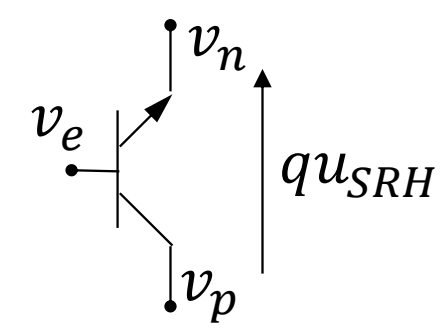 |
| SRH (deep trap, high injection)<br>$qu_{SRH}(\bar{n} \approx \bar{p}) \approx g_{rec,p}(v_p - v_e) + g_{rec,n}(v_e - v_n)$<br>$g_{rec,p} = \frac{q\bar{p}\tau_p}{V_{th}(\tau_n+\tau_p)^2}$ $\quad$ $g_{rec,n} = \frac{q\bar{n}\tau_n}{V_{th}(\tau_n+\tau_p)^2}$ | 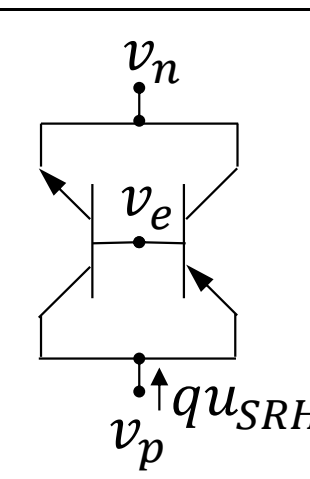 |
| SRH (deep trap, $\tau_n\bar{p} \approx \tau_p\bar{n}$)<br>$qu_{SRH}(\tau_n\bar{p} \approx \tau_p\bar{n}) \approx \frac{v_p - v_n}{r_{SRH}}$ $\quad$ $r_{SRH} = \frac{4V_{th}\tau_n}{q\bar{n}}$ | 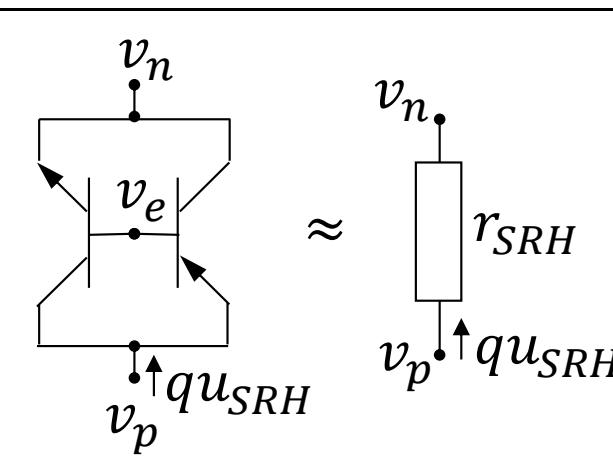 |
| SRH (shallow trap, *e.g.* $E_T$ close to $E_C$)<br>$qu_{SRH}(n_1 \gg \bar{p}, \bar{n}) \approx \frac{v_p - v_n}{r_{SRH}}$ $\quad$ $r_{SRH} = \frac{V_{th}\tau_p n_1}{q\bar{n}\bar{p}}$ | 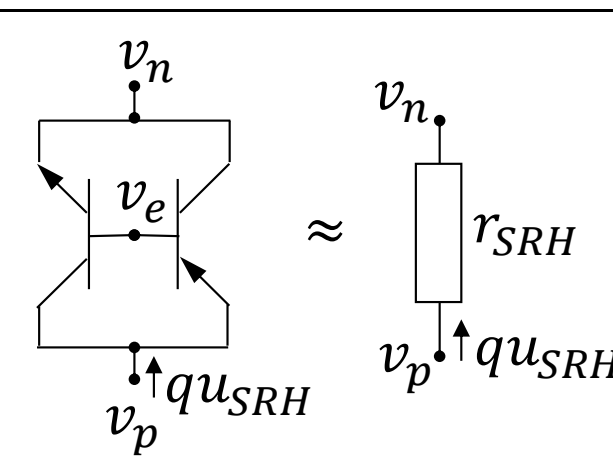 |

**Appendix B. Equivalent capacitance and impedance approximations for solar cells without mobile ions**

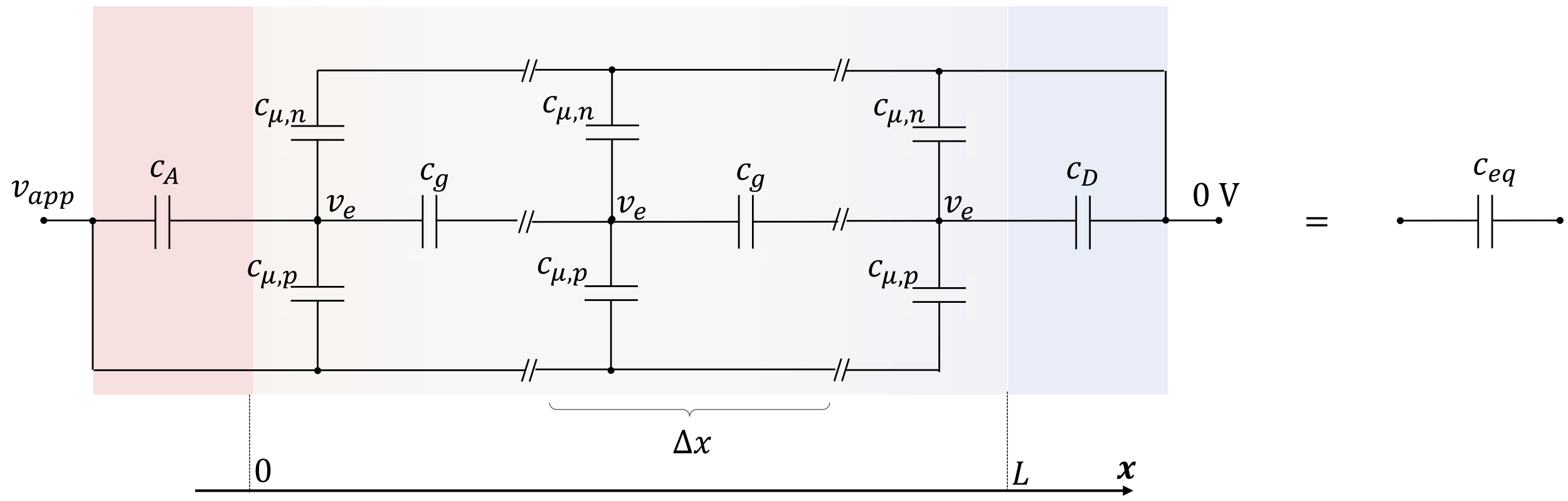

Figure 8. Capacitive network associated with a semiconducting solar cell (no mobile ions) assuming selective contacts (with opposite selectivity) and fast transport ($r_{p,1}, r_{n,2}, r_p,\ r_n \approx 0$, and $r_{p,1}, r_{n,2} \to \infty$,). The analytical expression of the equivalent capacitance $c_{eq}$ depends on both the electrostatic and on chemical contributions (see text).

Figure 8 shows the capacitive transmission line associated with a semiconducting device with no mobile ions and ideal contacts with opposite electronic selectivity. By considering the differential value of the electrical elements $c'_\mu = \frac{dc_\mu}{dx}$, $c'_g = \left(\frac{d(c_g)^{-1}}{dx}\right)^{-1}$, the equivalent capacitance can be expressed as (see derivation in section 6.1 of the Supplemental Material [38]):

$$c_{eq} = \{2c_A c_D c'_{\mu,n} c'_{\mu,p} + \left[c_A c_D \left({c'_{\mu,n}}^2 + {c'_{\mu,p}}^2\right) + c'_{\mu,n} c'_{\mu,p} (c_A + c_D)\left(c'_{\mu,n} + c'_{\mu,p}\right) L\right] \cosh[\kappa L] + \kappa\left[c_A c'_{\mu,n}\left(c'_g c'_{\mu,n} + c_D c'_{\mu,p} L\right) + c'_g c'_{\mu,p}\left(c_D c'_{\mu,p} + c'_{\mu,n}\left(c'_{\mu,n} + c'_{\mu,p}\right) L\right)\right] \sinh[\kappa L]\} \left\{(c_A + c_D)\left(c'_{\mu,n} + c'_{\mu,p}\right)^2 \cosh[\kappa L] + c'_g \kappa^3 \left(c_A c_D + c'_g\left(c'_{\mu,n} + c'_{\mu,p}\right)\right) \sinh[\kappa L]\right\}^{-1}$$

(B1)

The total impedance of the circuit in Figure 3a can be obtained based on this result as:

$$Z = \left(r_{rec,tot}^{-1} + i\omega c_{eq}\right)^{-1} \qquad \text{(B2)}$$

The modification to this problem to include situations that differ from the flat-band case and the discussion of the evaluation of $c_{eq}$ in mixed conducting devices at high-frequencies are presented in section 6.1 of the Supplemental Material [38]. As discussed in the main text, an effective electronic capacitance can be expressed as $\hat{c} = c_{eq} - c_{g,tot}$ (parallel of $c_{g,tot}$ and $\hat{c}$ corresponds to $c_{eq}$).

## Appendix C. IC and MC-i approximations to the transmission line problem

The complete differential problem associated with Figure 2a has a complex solution, even when considering selective contacts and fast electronic transport (see section 6.2 of the Supplemental Material [38]). Simplified versions of the circuit model are considered here, for which more accessible analytical solutions are available.

### *Ionic conductor approximation (IC model)*

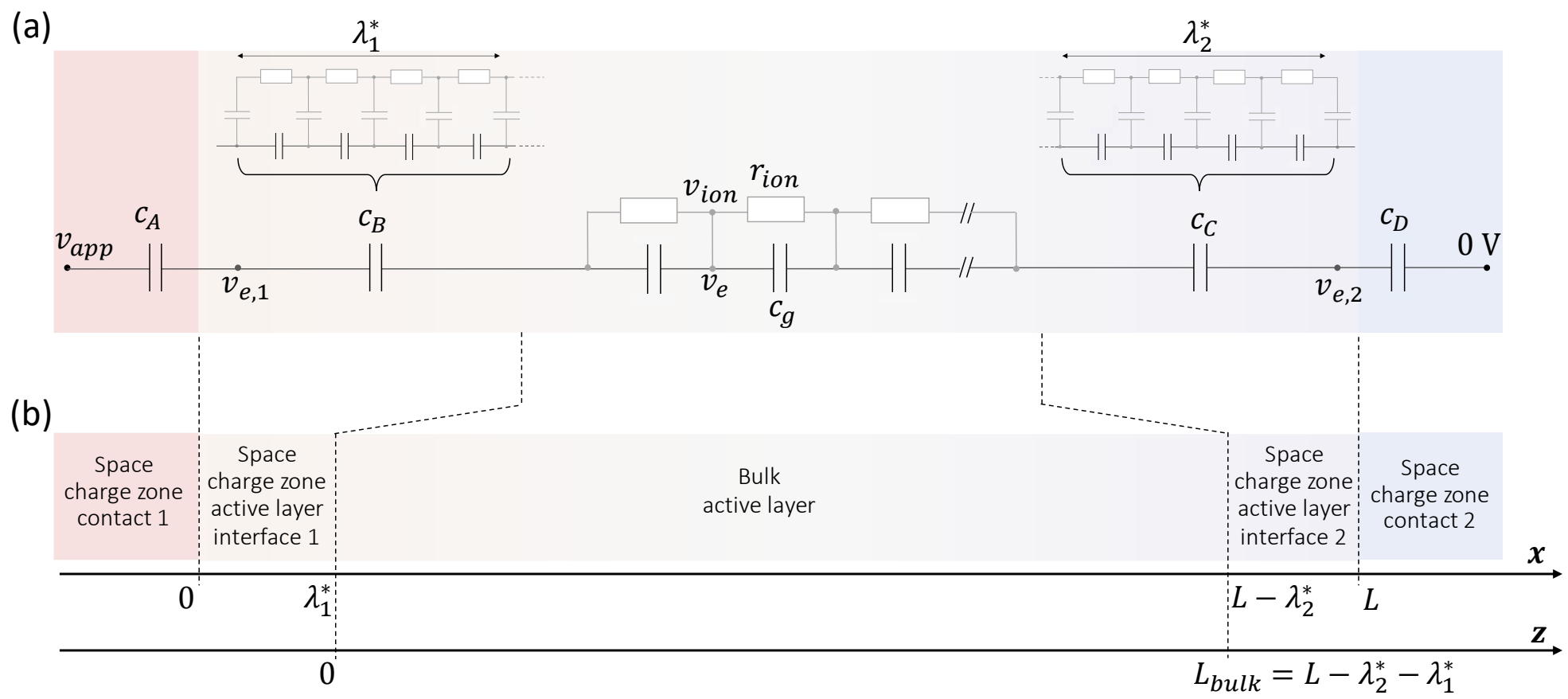

Figure 9. (a) Approximated circuit determining the changes in electrostatic potential $v_e$ in a mixed conducting device with ion-blocking contacts (IC approximation, see text). (b) Schematics of the full device highlighting the relation between the circuit elements in (a) and the different regions in the contacts and in the active layer. The space charge widths in the active layer for interface 1 and interface 2 are indicated as $\lambda_1^*$ and $\lambda_2^*$, respectively. The position variable $z$ is introduced to describe the bulk of the device.

Figure 9 shows the transmission line circuit used for the determination of the small signal electrostatic potential $v_e$ in a mixed conducting device under the ion conductor approximation (IC model, Figure 4). The circuit essentially represents an ionic conductor between ion-blocking contacts (for a detailed discussion, see section 6.3 of the Supplemental Material [38]). The widths of the space charge widths at interfaces 1 and 2 on the active layer side are indicated as $\lambda_1^*$ and $\lambda_2^*$, respectively (Figure 9b). These correspond to the Debye length $L_D$ in case of Gouy-Chapman situations (accumulation or small depletion of the majority carrier), while wider values may be expected for significant ionic depletion (Mott-Schottky situations). The model focuses on the changes in the values of $\phi_B$ and $\phi_C$ (see Figure 2c), while it no longer explicitly describes the position dependence of $v_e$ within the space charge zones. Furthermore, because in the bulk $v_e \approx v_{ion}$ based on the assumption of large ionic chemical capacitance compared with the electronic counterparts, $c_{\mu,ion}$ elements can be replaced with short circuits, as shown in Figure 4b and Figure 9a (valid for non-zero angular frequencies, and only for the small perturbation analysis).

By defining a position variable describing the bulk as $z = x - \lambda_1^*$ and the bulk thickness as $L_{bulk} = L - \lambda_1^* - \lambda_2^*$, the solution $\frac{v_e(z)}{v_{app}}$ for $0 < z < L_{bulk}$ is obtained.

$$\frac{v_e(z)}{v_{app}} = \frac{c_1}{c_1+c_2} \frac{1+i\omega r_{ion,bulk}\left(c_{g,bulk}+\frac{c_2 z}{L_{bulk}}\right)}{1+i\omega r_{ion,bulk}\frac{c_1 c_2}{c_1+c_2}} \qquad \text{(C1)}$$

Here $c_1 = \frac{c_A c_B}{c_A + c_B}$, $c_2 = \frac{c_C c_D}{c_C + c_D}$, $c_{g,bulk} = \frac{c'_g}{L_{bulk}}$ and $r_{ion} = L_{bulk} r'_{ion}$ (all differential parameters are assumed constant in the bulk).

The transfer function in Equation C1 provides the value of $v_e(x) = v_e(z + \lambda_1^*)$ for $0 < z < L_{bulk}$ given a (small) $v_{app}$. The value of $v_e(x)$ is also defined at the interfaces of the active layer with the contacts as:

$$v_{e,1} = v_e(x = 0) = v_e(z = 0)\frac{c_B}{c_A + c_B} + v_{app}\frac{c_A}{c_A + c_B} \quad \text{(C2)}$$

$$v_{e,2} = v_e(x = L) = v_e(z = L_{bulk})\frac{c_C}{c_D + c_C} \quad \text{(C3)}$$

The impedance of the circuit in Figure 9a can be expressed as:

$$Z_{ion,e} = \frac{1}{i\omega\frac{c_1 c_2}{c_1 + c_2}} + \frac{r_{ion,bulk}}{1 + i\omega r_{ion,bulk} c_{g,bulk}} = \frac{1 + i\omega r_{ion,bulk}\left(\frac{c_1 c_2}{c_1 + c_2} + c_{g,bulk}\right)}{i\omega\frac{c_1 c_2}{c_1 + c_2}(1 + i\omega r_{ion,bulk} c_{g,bulk})} \quad \text{(C4)}$$

By including the expressions for $\frac{v_e(z)}{v_{app}}$ in Equation 21 in the main text and the recombination contributions from the interfaces, one obtains the solution for the electronic impedance $Z_{eon}(\omega) = \frac{v_{app}}{j_{eon}(\omega)}$:

$$Z_{eon}(\omega) = \left\{\frac{L}{r_{rad}} + i\omega\hat{c} + \frac{v_{e,1}}{v_{app}} g_{rec,n,surf1} + \left(1 - \frac{v_{e,1}}{v_{app}}\right) g_{rec,p,surf1} + \frac{v_{e,2}}{v_{app}} g_{rec,n,surf2} + \left(1 - \frac{v_{e,2}}{v_{app}}\right) g_{rec,p,surf2} + \int_0^{L_{bulk}} \left[\frac{v_e(z,\omega)}{v_{app}} g_{rec,n} + \left(1 - \frac{v_e(z,\omega)}{v_{app}}\right) g_{rec,p}\right] dz\right\}^{-1}, \quad \text{(C5)}$$

where the value of $\hat{c}$ can be evaluated using Equation B1. The overall impedance of the device is calculated using:

$$Z(\omega) = \left[Z_{ion,e}^{-1} + Z_{eon}^{-1}(\omega)\right]^{-1} \quad \text{(C6)}$$

In Figure 4c, d and e in the main text, further simplification to the model are shown, where the influence of the electrostatic potential on the bulk recombination is not explicitly implemented. For these circuits, the total impedance can still be evaluated based on Equation C6, where:

- Equation C4 still allows the determination of $Z_{ion,e}$
- Equation C5 is simplified to the following expression

$$Z_{eon}(\omega) = \left\{r_{rec,bulk}^{-1} + i\omega\hat{c} + \frac{v_{e,1}}{v_{app}} g_{rec,n,surf1} + \left(1 - \frac{v_{e,1}}{v_{app}}\right) g_{rec,p,surf1} + \frac{v_{e,2}}{v_{app}} g_{rec,n,surf2} + \left(1 - \frac{v_{e,2}}{v_{app}}\right) g_{rec,p,surf2}\right\}^{-1} \quad \text{(C7)}$$

### *Mixed conductor with ionic majority carriers approximation (MC-i)*

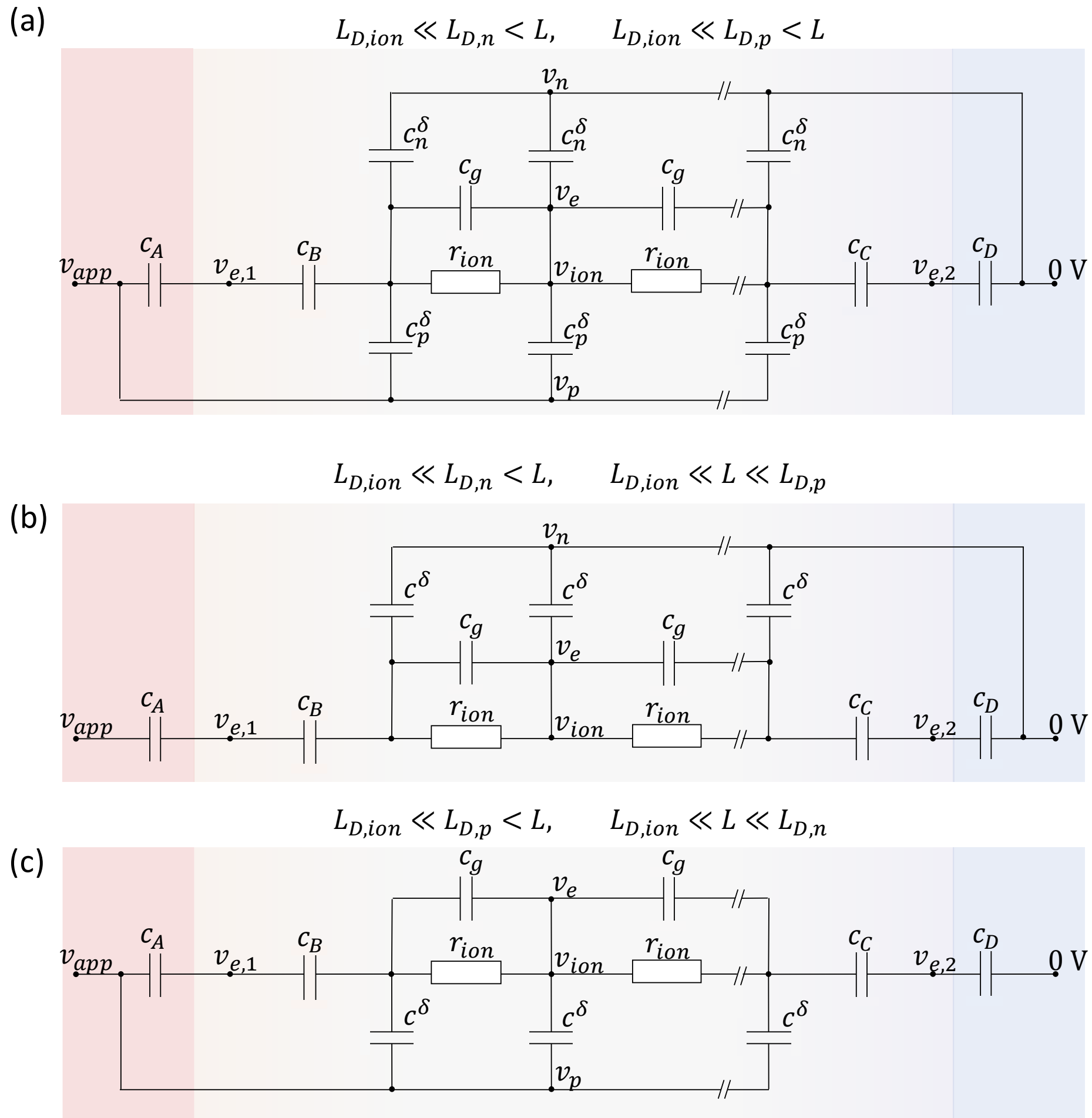

Figure 10. (a) MC-i equivalent circuit model in Figure 5, without the recombination circuit elements. (b) and (c) show the simplified versions of (a) for situations where only (b) electrons or (c) holes influence significantly the electrostatic potential.

The MC-i approximation accounts for the fact that the electronic charge concentrations are large enough to influence the electrostatic landscape. Figure 10a shows the circuit network in Figure 5, without the recombination elements. By solving the relevant differential problem coupled with boundary conditions (see section 6.4 of the Supplemental Material [38]) one obtains:

$$\frac{v_e(z)}{v_{app}} = \frac{c_p^{\delta\prime}}{c_p^{\delta\prime}+c_n^{\delta\prime}} +$$

$$+\frac{\left(1-\frac{c_2}{c_1}\frac{c_p^{\delta\prime}}{c_n^{\delta\prime}}\right)\cosh(\kappa z)\left(1+i\omega r'_{ion}c'_g\right)\kappa^2+i\omega r_{ion}\left[\left(c_n^{\delta\prime}+c_p^{\delta\prime}\right)\left(\cosh[\kappa(L_{bulk}-z)]-\cosh[\kappa z]\right)+c_2\kappa\left(\sinh[\kappa(L_{bulk}-z)]-\frac{c_p^{\delta\prime}}{c_n^{\delta\prime}}\sinh[\kappa z]\right)\right]}{\left(1+\frac{c_p^{\delta\prime}}{c_n^{\delta\prime}}\right)\left\{i\omega r'_{ion}\left[\left(c_n^{\delta\prime}+c_p^{\delta\prime}\right)\left(\cosh[\kappa L_{bulk}]-1\right)+c_2\kappa\sinh[\kappa L_{bulk}]\right]+\left(1+i\omega r'_{ion}c'_g\right)\kappa^2\left(1+\frac{c_2}{c_1}\cosh[\kappa L_{bulk}]+\frac{c_n^{\delta\prime}+c_p^{\delta\prime}}{\kappa c_1}\sinh[\kappa L_{bulk}]\right)\right\}}$$

(C8)

where the differential component chemical capacitors (*e.g.* $c_p^{\delta'} = \frac{dc_p^\delta}{dx}$) as well as the parameter $\kappa = \sqrt{\frac{i\omega(c_p^{\delta'}+c_n^{\delta'})r'_{ion}}{1+i\omega r'_{ion}c'_g}}$ are introduced.

If only one of the two electronic charge carriers is present with large concentration, further simplification to the transmission lines can be applied. The resulting circuit considers only one type of chemical capacitors that involve mobile ionic defects and either electrons (Figure 10b) or holes (Figure 10c). By defining $\kappa = \sqrt{\frac{i\omega c^{\delta'} r'_{ion}}{1+i\omega r'_{ion}c'_g}}$, where $c^{\delta'} = c_n^{\delta'}$ or $c^{\delta'} = c_p^{\delta'}$, depending on the electronic majority carrier, slightly simpler solutions are obtained in these cases. For $\bar{p} \ll \bar{n}$ (Figure 10b), one finds

$$\frac{v_e(z)}{v_{app}} = \frac{\cosh[\kappa(L_{bulk}-z)]}{\cosh(\kappa L_{bulk})} \frac{1+i\omega r'_{ion}\left\{c'_g+\frac{c_2}{\kappa}\tanh[\kappa(L_{bulk}-z)]\right\}}{\left(1+\frac{c_2}{c_1}\right)+\frac{c^{\delta'}}{\kappa c_1}\tanh(\kappa L_{bulk})+i\omega r'_{ion}\left[c'_g\left(1+\frac{c_2}{c_1}\right)+\left(\frac{c_2}{\kappa}+\frac{c^{\delta'}c'_g}{\kappa c_1}\right)\tanh(\kappa L_{bulk})\right]} \quad \text{(C9)}$$

$$\frac{v_e(z)}{v_{app}}(\omega \to 0) \approx \frac{1}{1+\frac{c_2}{c_1}+\frac{c^{\delta'}}{c_1}L_{bulk}} = \frac{c_1}{c_1+c_2+c^{\delta'}L_{bulk}} \quad \text{(C10)}$$

For the $\bar{p} \gg \bar{n}$ situation (Figure 10c):

$$\frac{v_e(z)}{v_{app}} = \frac{\cosh(\kappa z)+i\omega r'_{ion}\left[c'_g\cosh(\kappa z)+\frac{c_1}{\kappa}\sinh(\kappa z)\right]}{-\left[1+\frac{c_1}{c_2}\right]\cosh(\kappa L_{bulk})-\frac{c^{\delta'}}{\kappa c_2}\sinh(\kappa L_{bulk})+i\omega r'_{ion}\left\{-c'_g\left[1+\frac{c_1}{c_2}\right]\cosh(\kappa L_{bulk})+\left(-\frac{c_1}{\kappa}-\frac{c^{\delta'}c'_g}{\kappa c_2}\right)\sinh(\kappa L_{bulk})\right\}} + 1 \quad \text{(C11)}$$

$$\frac{v_e(z)}{v_{app}}(\omega \to 0) \approx 1 - \frac{1}{1+\frac{c_1}{c_2}+\frac{c^{\delta'}}{c_2}L_{bulk}} = \frac{c_1+c^{\delta'}L_{bulk}}{c_1+c_2+c^{\delta'}L_{bulk}} \quad \text{(C12)}$$

Importantly, the changes in electrostatic potential in the active layer on slow enough perturbation is not only determined by the value of $c_1$ and $c_2$, but also by the chemical capacitance between the electronic and the ionic rails $c^{\delta'}L_{bulk}$.

The impedance of the whole device can then be calculated as follows:

$$Z(\omega) = \left\{\left(1-\frac{v_{e,1}}{v_{app}}\right)i\omega c_A + \frac{L}{r_{rad}} + \frac{v_{e,1}}{v_{app}}g_{rec,n,surf1} + \left(1-\frac{v_{e,1}}{v_{app}}\right)g_{rec,p,surf1} + \frac{v_{e,2}}{v_{app}}g_{rec,n,surf2} + \left(1-\frac{v_{e,2}}{v_{app}}\right)g_{rec,p,surf2} + \int_0^{L_{bulk}}\left[\frac{v_e(z,\omega)}{v_{app}}g_{rec,n} + \left(1-\frac{v_e(z,\omega)}{v_{app}}\right)\left(g_{rec,p}+i\omega c_p^\delta\right)\right]dz\right\}^{-1} \quad \text{(C13)}$$

Here, $v_{e,1}$ and $v_{e,2}$ are calculated by substituting $\frac{v_e(z)}{v_{app}}$ from Equation C8 (or its simplified forms when relevant, Equations C9 and C11) in Equations C2 and C3.

**Acknowledgements**

Prof. Joachim Maier, Dr. Rotraut Merkle and Dr. Piers Barnes are gratefully acknowledged for reading and providing useful feedback on this manuscript. D.M. is grateful to the Alexander von Humboldt foundation for financial support.

Supplemental Material

# Equivalent circuit modeling of electron-hole recombination in semiconductor and mixed ionic-electronic conductor based devices

Davide Moia[1]

Max Planck Institute for Solid State Research, Heisenbergstr. 1, 70569, Stuttgart, Germany

## Contents

[1] moia.davide@gmail.com
Current address: Fluxim AG, Katharina-Sulzer-Platz 2, 8400 Winterthur, Switzerland

## 1. On the use of bipolar transistors as equivalent circuit models for recombination

In this work, similarly to Ref. [1], bipolar transistors are used as equivalent circuit elements to describe recombination in semiconductors. As shown in the main text, the expression of the linearized Shockley-Read-Hall (SRH) net recombination is analogous, in terms of its functional form, to the relation describing the small perturbation current in *npn* and *pnp* transistors. The expression involves transconductance terms (here indicated with $g_{rec}$ and $g_{gen}$) and the potential difference associated with the two junctions in the component:

$$j_{npn} = g_{rec,n}(v_B - v_E) - g_{gen,n}(v_B - v_C) \qquad \text{Eq. S1}$$

$$j_{pnp} = g_{rec,p}(v_E - v_B) - g_{gen,p}(v_C - v_B) \qquad \text{Eq. S2}$$

Here $v_B, v_C$ and $v_E$ correspond to the base, collector and emitter small perturbation potentials of *npn* or *pnp* transistors. The $j_{npn}$ and $j_{pnp}$ currents are defined positive when they flow from the collector to the emitter and from the emitter to the collector, respectively.

The bipolar transistor is a non-linear circuit element. Formally, when discussing equivalent circuit models associated with a linearized description of a system, only linear elements should be included. In the case of the transistor, this would involve the use of voltage controlled current sources (Figure S1a). Such circuit emphasizes a key property of the transistor that distinguishes it from other bipoles (*e.g.* resistors and capacitors): the current flowing from the collector to the emitter (*e.g.* in an *npn* transistor) is not dependent on the voltage difference between these two terminals only. Equations S1 and S2 show that such current depends on the voltage difference $(v_B - v_E)$ and $(v_B - v_C)$, where the voltage of a third terminal (the base) is also involved.

In this work, for simplicity in representation, the symbol of the non-linear component is used in the equivalent circuit models to describe also the small signal behavior.

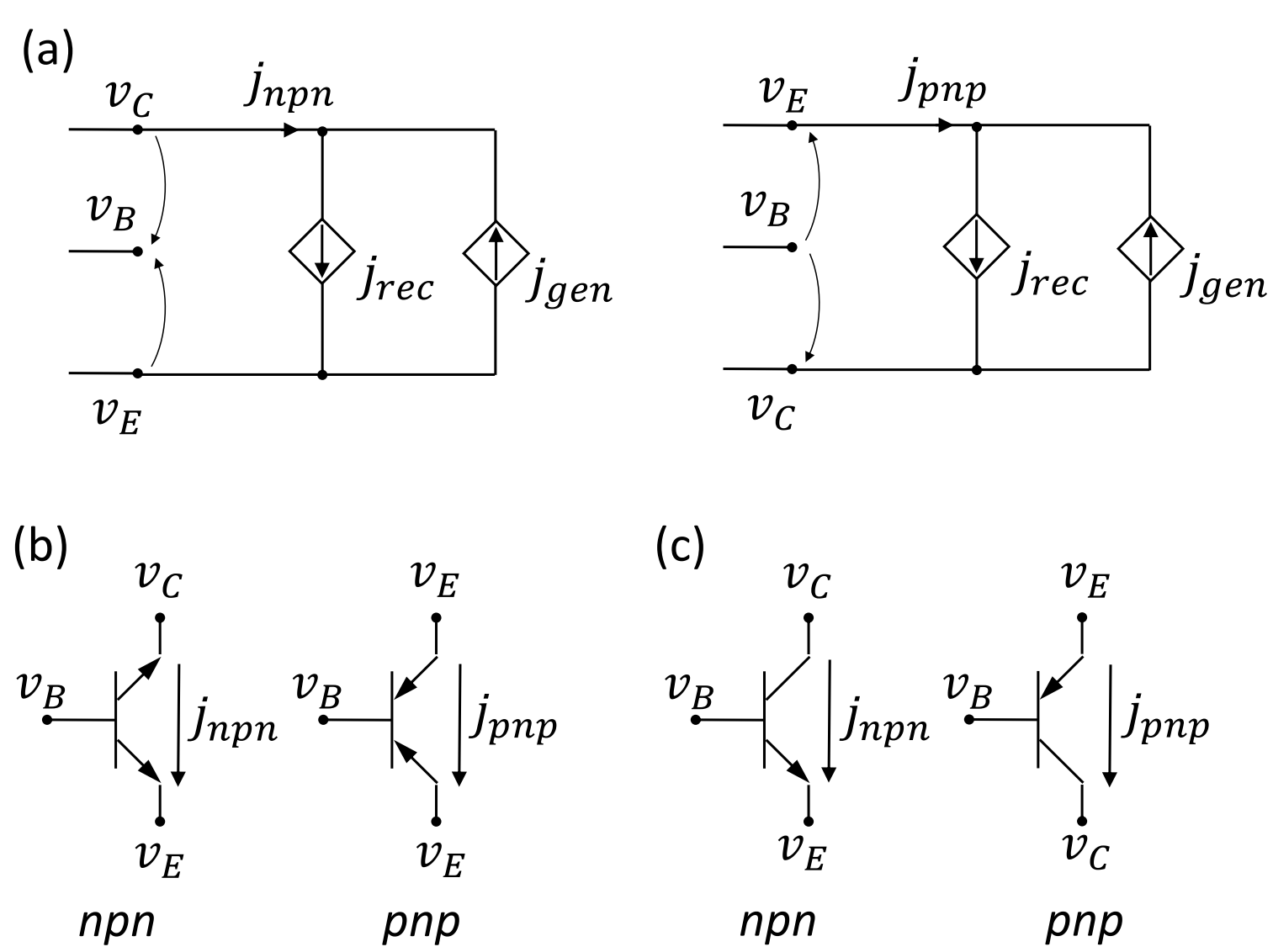

Figure S1. (a) Small signal circuit model of a bipolar transistor, involving a voltage controlled current source. Symbols for the bipolar transistors (b) in Ref. [1] and (c) in this work, used also for the small signal analysis.

In Ref. [1], a modified symbol of the bipolar transistor is used (Figure S1b), to emphasize that the current flowing in the equivalent transistor model can be positive or negative, depending on whether recombination or thermal generation of electronic charges is dominant. In this work, the more conventional symbol is used (Figure S1c). This allows the identification of the recombination current direction. Also, such asymmetric symbol is consistent with the asymmetry in the dependence of the transconductance terms on the applied steady-state electrochemical potentials (see Equation 11 in the main text), a feature that is shared with most physical transistor components. Finally, the analytical treatment in the main text considers the transconductance terms to be defined per unit volume, so that a net recombination current per unit volume term $qu$ (instead of a current density per unit area, $j$) flows through these elements (see main text). In the discretized model, transconductance terms for each transistor are multiplied by the width $\Delta x$ of the device slice that they model, yielding units of A $V^{-1}$ $cm^{-2}$.

Another important difference between the models described in the main text and the ones in Ref. [1] is that in the circuits shown in Figure 2, 4a and 5 the base of the transistors is connected to the electrostatic potential rail. This aspect deserves some clarification. When operating a physical bipolar transistor, the potential of the base can be controlled with an external bias. Such bias effectively controls the electrochemical potential of the majority carriers in the base. These are holes in the *p* region of a *npn* device or electrons in the *n* region of a *pnp* device. The applied changes in potential of the base correspond to changes in the electrostatic potential in this region, only if the *p*-doped base operates in low-injection regime. Physical bipolar transistor components are commonly used in such regime, for example in analog amplification circuits. It is worth noting that, if the device is operated under high-injection regime, the changes in the electrostatic potential in the base no longer follow the applied potential (and the electrochemical potential of the holes, *i.e.* the hole quasi-Fermi level $E_{Fp}$). Indeed, to respond to the (large) change in concentration of the electrons that are injected from the emitter, the hole concentration changes too. To describe such situations in a physical transistor, models other than Equation S1 and S2 need to be considered.

In this work, the transistor symbol is used simply based on the correspondence of its basic relation (Equation S1 and S2) with the relation obtained from the linearized recombination current equations in the main text. Therefore, even though connecting the base terminal to nodes of the electrostatic potential rail in a circuit model may appear unphysical, this is in practice convenient as it reflects the analytical treatment. As already specified in the text, the current amplification factor of the transistors used in this study is infinity, implying that no current flows into the base contact. In this way, the electronic current that flows through the transistor is gated by the potential differences (recombination and thermal generation potentials) that also involve the base voltage ($v_e$), while making sure that no current flows between the electrostatic and the electronic rails.

## 2. Approximations for the recombination and generation transconductance

The approximated values of thermal generation and recombination transconductance based on Equation 11 in the main text, and for situations involving different electronic charge concentrations and different trap energies, are shown in Table S1.

Table S1. Expressions for the transconductance terms describing trap-mediated recombination and thermal generation for situations involving different electron and hole charge concentrations and for deep or shallow trap levels. All $g$ terms have the units of A V$^{-1}$ cm$^{-3}$. The dominant term(s) assuming a forward bias (or light bias) situation for each condition are highlighted in yellow, as also shown in the main text (only valid if $\tau_n$ and $\tau_p$ have comparable values. If this is not the case the complete expression in Equation 11 needs to be used). The energy level diagram show the position of the quasi-Fermi and trap energies on a partial free enthalpy axis ($E$).

| 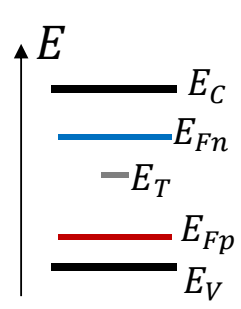 | $\bar{n} \approx \bar{p}$<br>High injection<br>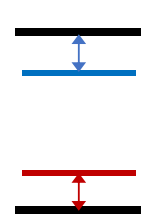 | $\bar{n} \ll \bar{p}$<br>Low injection (p-type)<br>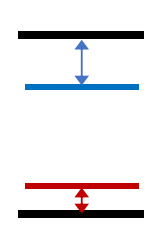 | $\bar{n} \gg \bar{p}$<br>Low injection (n-type)<br>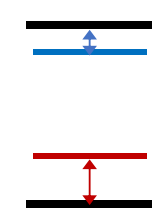 |
|---|---|---|---|
| **Shallow trap ($E_T$ close to $E_V$)** | 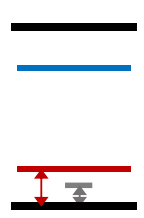<br>$p_1 \gg \bar{p}$<br>$g_{rec,n} \approx \frac{q\bar{n}\bar{p}}{V_{th}\tau_n p_1}$<br>$g_{gen,n} \approx \frac{q\bar{p}n_i^2}{V_{th}\tau_n p_1^2}$<br>$g_{rec,p} \approx \frac{q\bar{n}\bar{p}}{V_{th}\tau_n p_1}$<br>$g_{gen,p} \approx \frac{q\tau_p\bar{n}n_i^2}{V_{th}\tau_n^2 p_1^2}$ | 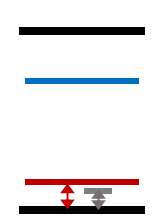<br>$p_1 \gg \bar{p}$<br>$g_{rec,n} \approx \frac{q\bar{n}\bar{p}}{V_{th}\tau_n p_1}$<br>$g_{gen,n} \approx \frac{q\bar{p}n_i^2}{V_{th}\tau_n p_1^2}$<br>$g_{rec,p} \approx \frac{q\bar{n}\bar{p}}{V_{th}\tau_n p_1}$<br>$g_{gen,p} \approx \frac{q\tau_p\bar{n}n_i^2}{V_{th}\tau_n^2 p_1^2}$ | 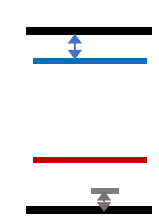<br>$p_1 \gg \bar{n}$<br>$g_{rec,n} \approx \frac{q\bar{n}\bar{p}}{V_{th}\tau_n p_1}$<br>$g_{gen,n} \approx \frac{q\bar{p}n_i^2}{V_{th}\tau_n p_1^2}$<br>$g_{rec,p} \approx \frac{q\bar{n}\bar{p}}{V_{th}\tau_n p_1}$<br>$g_{gen,p} \approx \frac{q\tau_p\bar{n}n_i^2}{V_{th}\tau_n^2 p_1^2}$ |
| **Deep trap** | 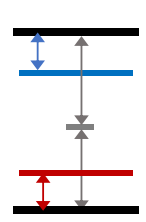<br>$p_1 \ll \bar{p}$<br>$n_1 \ll \bar{n}$<br>$g_{rec,n} \approx \frac{q\tau_n\bar{n}}{V_{th}(\tau_n+\tau_p)^2}$<br>$g_{gen,n} \approx \frac{q\tau_n n_i^2}{V_{th}(\tau_n+\tau_p)^2\bar{n}}$<br>$g_{rec,p} \approx \frac{q\tau_p\bar{n}}{V_{th}(\tau_n+\tau_p)^2}$<br>$g_{gen,p} \approx \frac{q\tau_p n_i^2}{V_{th}(\tau_n+\tau_p)^2\bar{n}}$ | 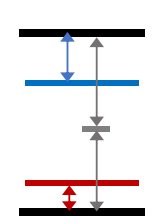<br>$p_1 \ll \bar{p}$<br>$p_1 \ll \bar{n}$<br>$n_1 \ll \bar{n}$<br>$g_{rec,n} \approx \frac{q\bar{n}}{V_{th}\tau_n}$<br>$g_{gen,n} \approx \frac{qn_i^2}{V_{th}\tau_n\bar{p}}$<br>$g_{rec,p} \approx \frac{q\tau_p\bar{n}^2}{V_{th}\tau_n^2\bar{p}}$<br>$g_{gen,p} \approx \frac{q\tau_p\bar{n}n_i^2}{V_{th}\tau_n^2\bar{p}^2}$ | 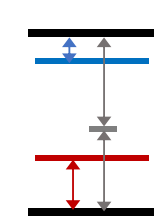<br>$p_1 \ll \bar{p}$<br>$n_1 \ll \bar{n}$<br>$n_1 \ll \bar{n}$<br>$g_{rec,n} \approx \frac{q\tau_n\bar{p}^2}{V_{th}\tau_p^2\bar{n}}$<br>$g_{gen,n} \approx \frac{q\tau_n\bar{p}n_i^2}{V_{th}\tau_p^2\bar{n}^2}$<br>$g_{rec,p} \approx \frac{q\bar{p}}{V_{th}\tau_p}$<br>$g_{gen,p} \approx \frac{qn_i^2}{V_{th}\tau_p\bar{n}}$ |
| **Shallow trap ($E_T$ close to $E_C$)** | 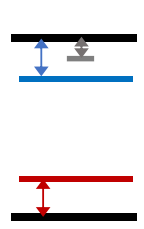<br>$n_1 \gg \bar{p}$<br>$g_{rec,n} \approx \frac{q\bar{n}\bar{p}}{V_{th}\tau_p n_1}$<br>$g_{gen,n} \approx \frac{q\tau_n\bar{p}n_i^2}{V_{th}\tau_p^2 n_1^2}$<br>$g_{rec,p} \approx \frac{q\bar{n}\bar{p}}{V_{th}\tau_p n_1}$<br>$g_{gen,p} \approx \frac{q\bar{n}n_i^2}{V_{th}\tau_p n_1^2}$ | 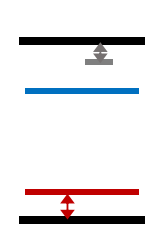<br>$n_1 \gg \bar{p}$<br>$g_{rec,n} \approx \frac{q\bar{n}\bar{p}}{V_{th}\tau_p n_1}$<br>$g_{gen,n} \approx \frac{q\tau_n\bar{p}n_i^2}{V_{th}\tau_p^2 n_1^2}$<br>$g_{rec,p} \approx \frac{q\bar{n}\bar{p}}{V_{th}\tau_p n_1}$<br>$g_{gen,p} \approx \frac{q\bar{n}n_i^2}{V_{th}\tau_p p_1^2}$ | 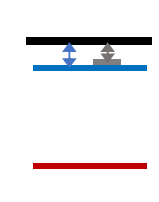<br>$n_1 \gg \bar{n}$<br>$g_{rec,n} \approx \frac{q\bar{n}\bar{p}}{V_{th}\tau_p n_1}$<br>$g_{gen,n} \approx \frac{q\tau_n\bar{p}n_i^2}{V_{th}\tau_p^2 n_1^2}$<br>$g_{rec,p} \approx \frac{q\bar{n}\bar{p}}{V_{th}\tau_p n_1}$<br>$g_{gen,p} \approx \frac{q\bar{n}n_i^2}{V_{th}\tau_p n_1^2}$ |

If the condition $\tau_n \approx \tau_p$ does not apply, the approximations in Table S1 are not valid and the complete expression in Equation 11 should be used. Figure S2 shows the profiles of $g_{rec,n}$ and $g_{rec,p}$ as a function of $\tau_n$ assuming $\tau_p = 1\ \mu s$, for the case of deep, mid-gap trap level (Figure S2a, b, c) and for a shallow trap close to the conduction band minimum (Figure S2d, e and f). The left, middle and right panels refer to different relative magnitudes of $\bar{n}$ and $\bar{p}$.

It is clear from these trends that:

- For the case $\bar{n} \approx \bar{p}$ and if $\tau_n \neq \tau_p$, $g_{rec,n} \neq g_{rec,p}$ is obtained and the two-transistor model is required
- If $\tau_n$ and $\tau_p$ are of similar magnitude, the recombination transconductance of the minority carrier dominates
- The recombination transconductance of the majority carrier becomes dominant if its lifetime is much longer (by a factor corresponding to the concentration ratio) compared to the minority carrier's
- $g_{rec,n} = g_{rec,p}$ applies for the shallow trap case and the resistor approximation can be used also if $\tau_n \neq \tau_p$. However, if the lifetime of the carrier populating the band closer to the trap level is much longer than the lifetime of the carrier populating the band far from the trap energy, the recombination becomes limited by the former carrier and a transistor description is required for this case too.

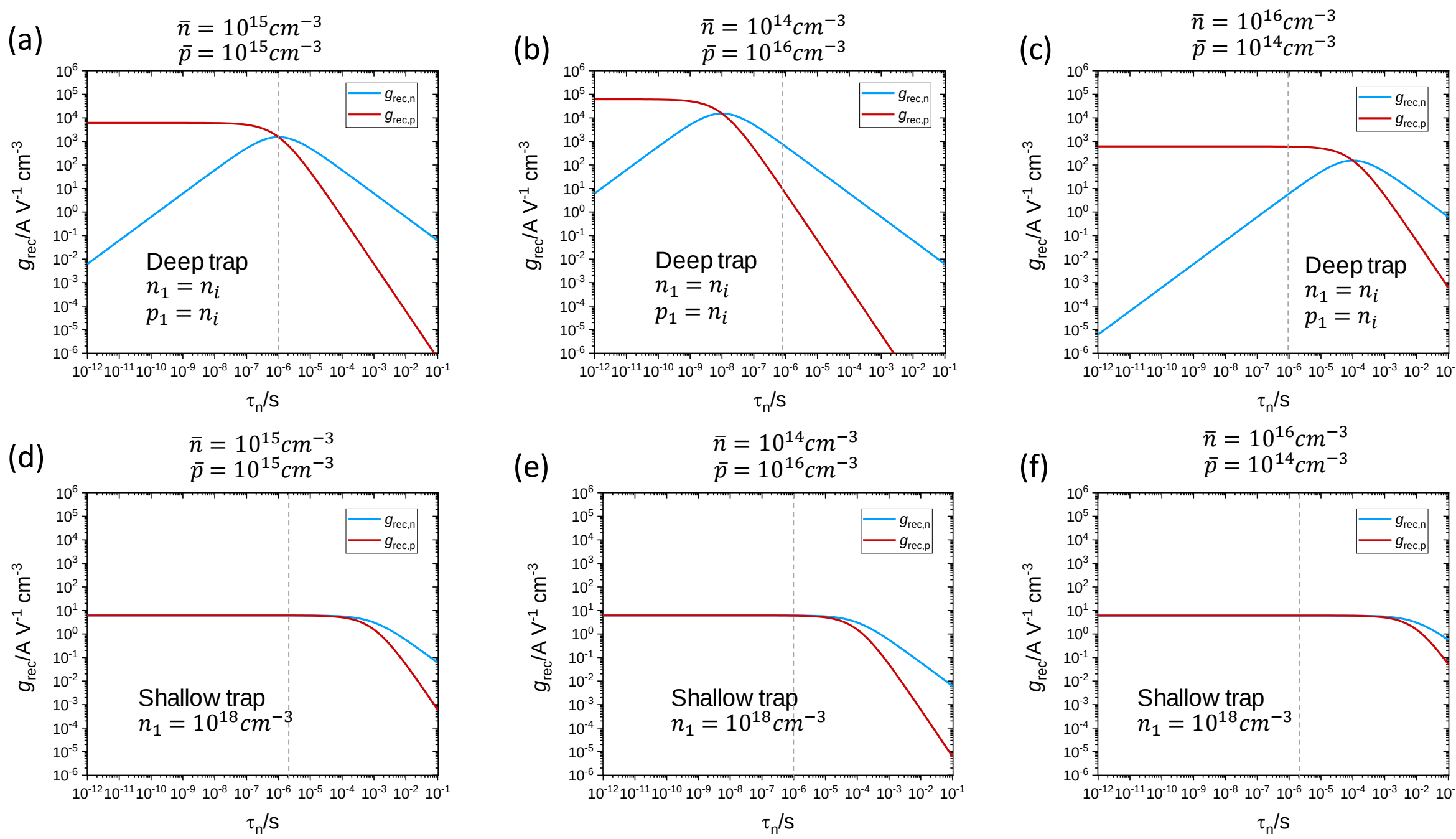

Figure S2. Profiles of the recombination transconductance for electrons and holes (*npn* and *pnp* transistors) according to Equation 11 and 12 in the main text. The data are plotted vs the electron lifetime. The cases of a mid-gap trap (a, b, c) and of a shallow trap close to the conduction band edge (d, e, f) are considered, assuming a hole lifetime of 1 $\mu$s. For the shallow trap, a value of $n_1 = 10^{18}\ cm^{-3}$ is used, which corresponds to a trap energy level that lies approximately 60 meV below the conduction band of the semiconductor (effective density of states $N_C = N_V = 10^{19}\ cm^{-3}$ is assumed). Situations where $\bar{n} = \bar{p}$ (a, d), $\bar{n} \ll \bar{p}$ (b, e) and $\bar{n} \gg \bar{p}$ (c, f) are considered.

As described in the main text, the transconductance parameters are derived from linearization of the expression for the SRH rate at quasi-equilibrium. The transistor representation is therefore appropriate to describe the small signal SRH recombination only at time scales that are slower than the trapping and de-trapping dynamics, described by the parameters $\tau_n$ and $\tau_p$. In situations where such time constants are slower than the time scale of the measurement, a model that describes trapping and de-trapping events explicitly may be required. This is relevant to semiconductors without mobile ions, where interaction of free carriers with traps represents a slow contribution to their response relative to other electronic processes. Note that also for semiconductors with mobile ions, such as hybrid perovskites, slow variations in the spectroscopic and optoelectronic response, often associated with ion transport effects, have been discussed in terms of trapping processes. [2]

### 3. Small perturbation model of Auger recombination

Equation 15 in the main text illustrates the small perturbation functional form of the Auger recombination as follows:

$$u_{Auger} = \frac{q}{V_{TH}}\left[2\gamma_n\bar{n}\bar{p}(v_e - v_n) + \gamma_n\bar{n}^2\left(v_p - v_e\right) + 2\gamma_p\bar{n}\bar{p}\left(v_p - v_e\right) + \gamma_p\bar{p}^2(v_e - v_n)\right] = g_{rec,n}(v_e - v_n) + g_{rec,p}(v_p - v_e) \qquad \text{Eq. S3}$$

The approximated expression for the transconductance under different injection regimes is shown in Table S2. Also in this case, the recombination transconductance of the minority carriers dominates (assuming $\gamma_n$ and $\gamma_p$ to be of similar order).

As explained in the main text, the recombination transconductance of SRH and Auger terms can be combined in a single pair of $g_{rec,n}$ and $g_{rec,p}$ terms, while only SRH processes contribute to the thermal generation transconductance terms.

Table S2. Expressions for the $g_{rec,n}$ and $g_{rec,p}$ terms associated with Auger recombination for situations involving different electronic charge concentrations.

| $\bar{n} \approx \bar{p}$ | $\bar{n} \ll \bar{p}$ | $\bar{n} \gg \bar{p}$ |
|---|---|---|
| 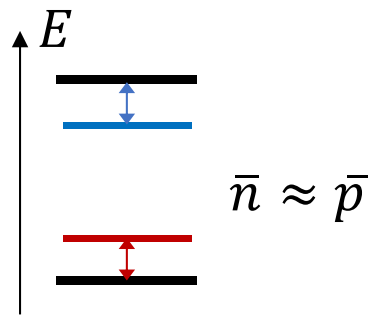 | 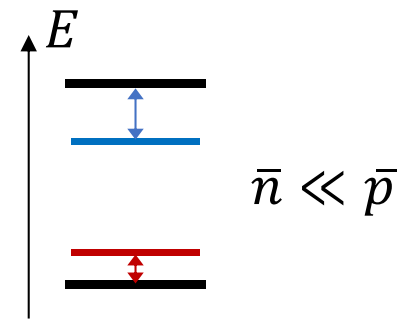 | 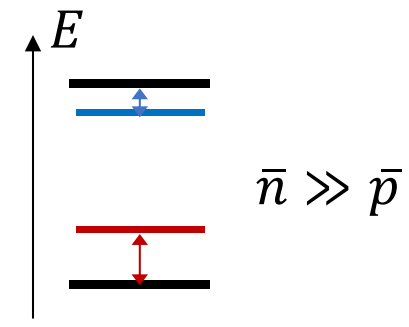 |
| $g_{rec,n} = \frac{q\bar{n}^2}{V_{TH}}(2\gamma_n + \gamma_p)$ | $g_{rec,n} = \frac{q\bar{p}^2}{V_{TH}}\gamma_p$ | $g_{rec,n} = \frac{q\bar{n}\bar{p}}{V_{TH}}2\gamma_n$ |
| $g_{rec,p} = \frac{q\bar{n}^2}{V_{TH}}(2\gamma_p + \gamma_n)$ | $g_{rec,p} = \frac{q\bar{n}\bar{p}}{V_{TH}}(2\gamma_p + \gamma_n)$ | $g_{rec,p} = \frac{q\bar{n}^2}{V_{TH}}\gamma_n$ |

## 4. Complete model for devices with ion blocking contacts

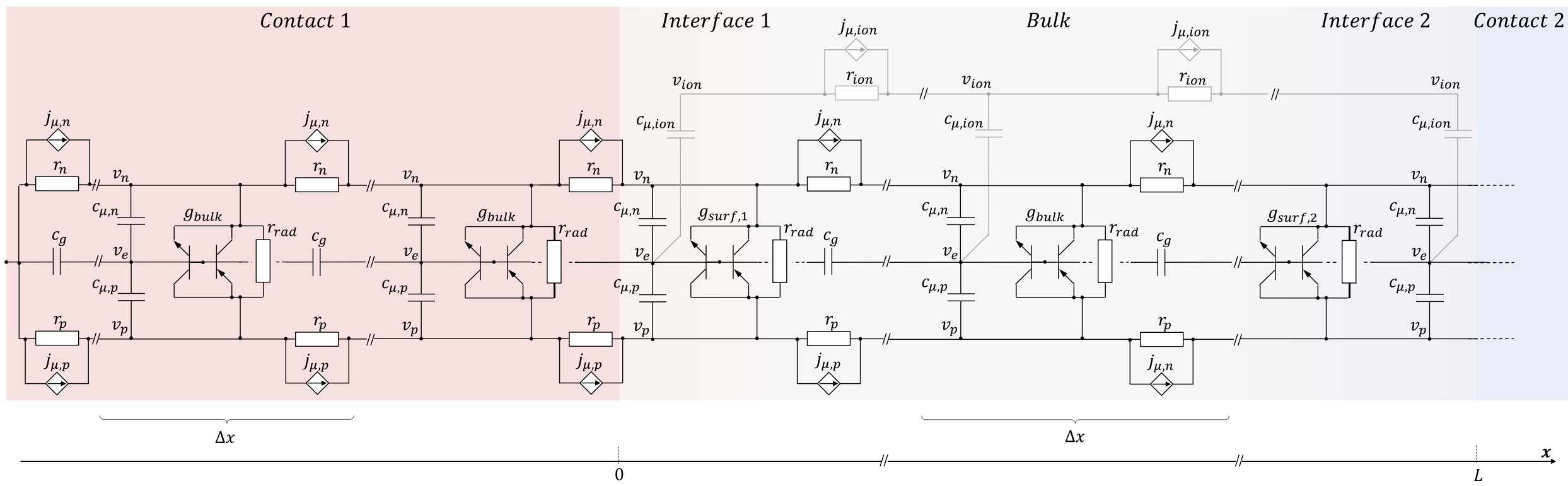

Figure S3. Transmission line model where the electron and hole rails are treated explicitly also in the ion-blocking contacts (here only contact 1 is shown, the model for contact 2 is constructed analogously).

Figure S3 illustrates the full transmission line model for situations where the ion-blocking contacts are not selective, and where the transport, storage and recombination of electrons and holes, as well as the electrostatic potential, are included within the discretized model and in the calculation of the impedance. This model is used for the calculations shown in the main text referred to as "complete model". The value of each circuit element is evaluated based on the steady-state solution obtained from drift-diffusion simulations (see Methods section).

## 5. Approximations to the transistor recombination circuit model in a semiconductor

In a semiconductor under the low-injection regime, it is possible to neglect one of the two transistors in the description of $u_{SRH}$ (Figure S4a and b). For example, if $\bar{n} \gg \bar{p}$, it is expected that $c_{\mu,n} \gg c_{\mu,p}$ and therefore that the local changes in electrostatic potential are largely determined by the changes in the majority carrier electrochemical potential. The approximation illustrated in Figure S4c including a short circuit instead of the capacitor $c_{\mu,n}$ is valid for non-zero frequencies, as here the transistors are assumed to have infinite impedance at the base contact. The response of the obtained configuration (collector connected with the base) corresponds to the one of a diode. Under the small perturbation regime, this can be replaced by a resistor (Figure S4d)

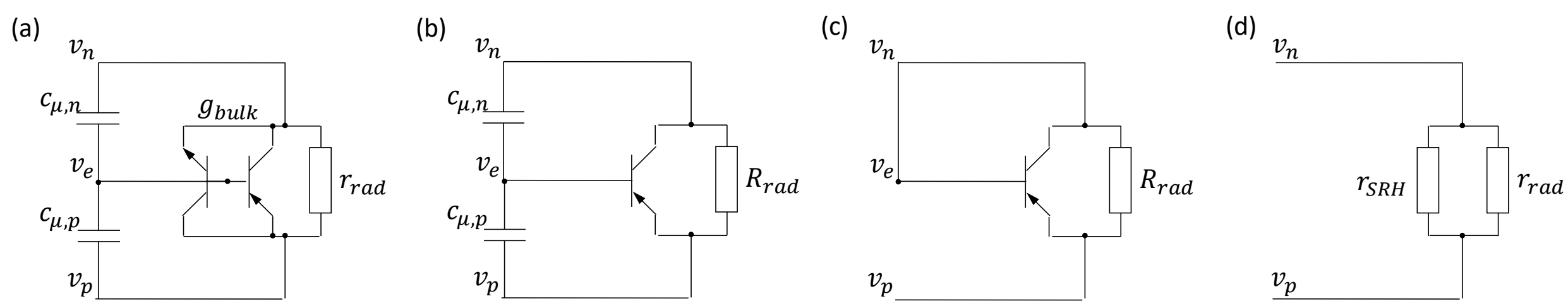

Figure S4. Sequence of approximations applicable to the (a) recombination equivalent circuit model in an *n*-type semiconductor under low-injection ($\bar{n} \gg \bar{p}$). (b) The recombination of the holes minority carriers is well described by the *pnp* transistor (see Table S1). (c) $c_{\mu,n} \gg c_{\mu,p}$ implies that $v_e \approx v_n$. (d) The transistor in diode configuration reduces to a resistor in the small perturbation regime.

## 6. Differential problems to the approximated circuit models

The analytical solutions to the differential impedance of selected equivalent circuit models discussed in the main text are presented below. These are obtained by, first, solving the relevant differential equations and, secondly, solving the system of boundary conditions. The latter step is performed using the Mathematica software.

### *6.1 Equivalent capacitance in transmission line circuits*

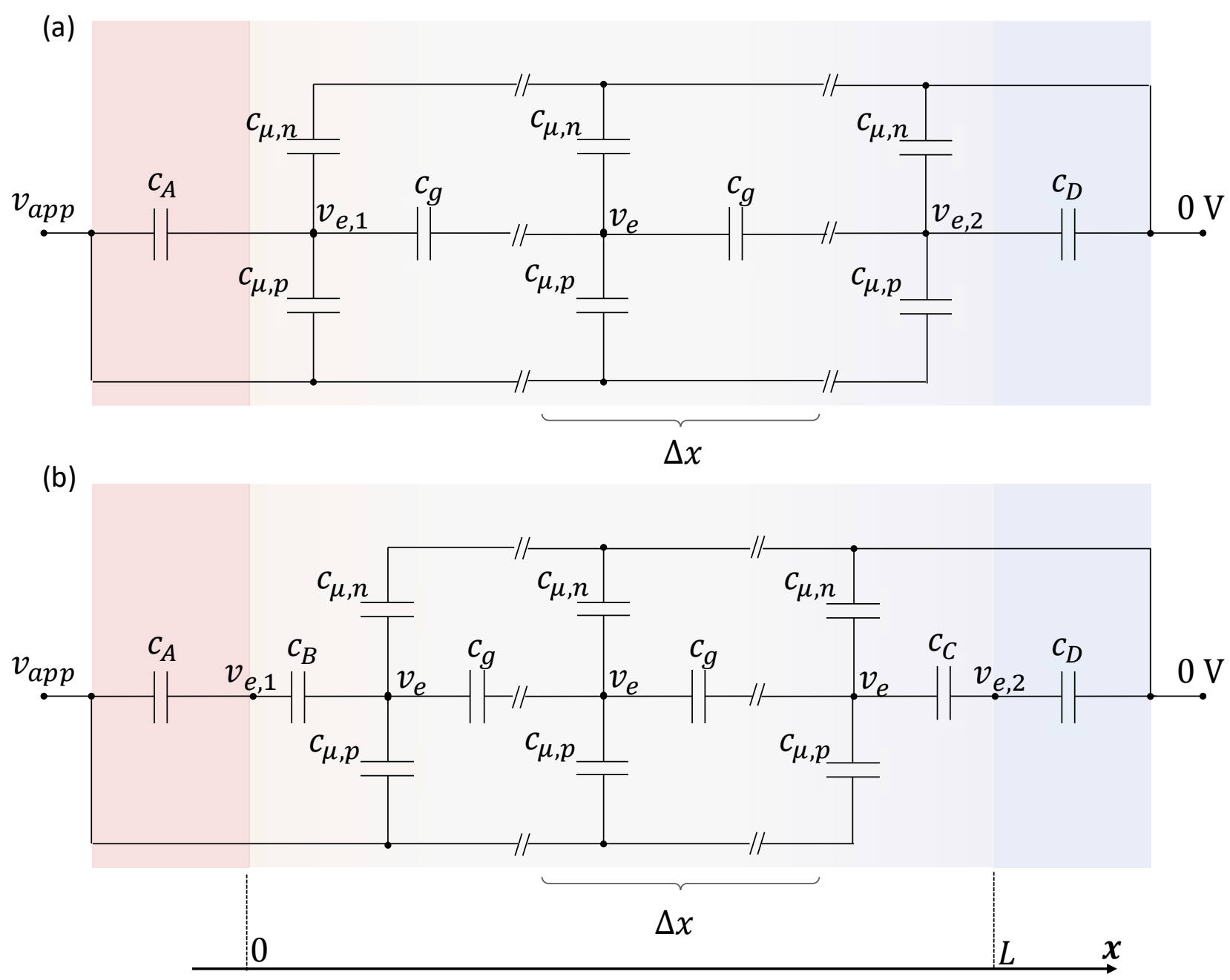

Figure S5. Capacitive network associated with a semiconducting solar cell (no mobile ions) assuming selective contacts (with opposite selectivity) and fast transport ($r_{p,1}, r_{n,2}, r_p,\ r_n \approx 0$, and $r_{p,1}, r_{n,2} \to \infty$). The analytical expression of the equivalent capacitance $c_{eq}$ depends on both the electrostatic and the chemical contributions (see text). The circuit in (a) is an accurate representation of the physical system. However, a simple analytical solution to (a) is possible only for a homogeneous active layer, where the specific differential chemical and electrostatic capacitors are constant throughout the layer thickness. This implies a flat-band situation. The circuit in (b) is a possible approximation whereby the capacitors $c_B$ and $c_C$ account for space charge capacitors at the interfaces with the contacts in the active layer.

The dependence of the equivalent capacitance of a semiconducting device on the electrostatic, space charge and chemical capacitance contributions is presented in this section. The focus is the case of a semiconducting device with no mobile ions and ideal contacts with opposite electronic charge carrier selectivity (see network in Figure S5a). To find the value of the equivalent capacitance $c_{eq}$ associated with such circuit, the relevant differential problem for the small signal electrostatic potential, $v_e$, is:

$$\frac{d^2 v_e}{dx^2} = -\frac{c'_{\mu,p}}{c'_g} v_{app} + \frac{c'_{\mu,p}+c'_{\mu,n}}{c'_g} v_e \ . \qquad \text{(S4)}$$

Here, the differential value of the electrical elements is considered i.e. $c'_\mu = \frac{dc_\mu}{dx}$, $c'_g = \left(\frac{d(c_g)^{-1}}{dx}\right)^{-1}$.
Under dilute conditions and for negligible trapping, $c'_\mu$ is an indicator of the charge carrier concentration. $c'_g$ corresponds to the dielectric constant as function of position.

For the simple case of position independent $c'_{\mu,p}$, $c'_{\mu,n}$ and $c'_g$, a solution in the form $v_e(x) = Ae^{\kappa x} + Be^{-\kappa x} + C$ is expected, where $\kappa = \sqrt{\frac{c'_{\mu,p}+c'_{\mu,n}}{c'_g}}$.

By applying the boundary conditions for the conservation of the electric displacement at the interfaces

$$-c'_g \frac{dv_e}{dx}(x=0) = c_A[v_{app} - v_e(x=0)] \qquad \text{(S5)}$$

$$-c'_g \frac{dv_e}{dx}(x=L) = c_D v_e(x=L) \qquad \text{(S6)}$$

the following solution is obtained:

$$\frac{v_e(x)}{v_{app}} = \frac{-c_D c'_{\mu,p}[c_A \sinh[\kappa x] + c'_g \kappa \cosh[\kappa x]] + c_A c'_{\mu,n}[c_D \sinh[\kappa(L-x)] + c'_g \kappa \cosh[\kappa(L-x)]]}{(c'_{\mu,p}+c'_{\mu,n})\left[\left(c_A c_D + (c'_g \kappa)^2\right)\sinh[\kappa L] + (c_D + c_A)c'_g \kappa \cosh[\kappa L]\right]} + \frac{c'_{\mu,p}}{c'_{\mu,p}+c'_{\mu,n}}. \qquad \text{(S7)}$$

The equivalent capacitance of the circuit can be calculated for example as $c_{eq} = \left(1 - \frac{v_e(x=0)}{v_{app}}\right)c_A$, which yields (see Appendix B):

$$c_{eq} = \{2c_A c_D c'_{\mu,n} c'_{\mu,p} + [c_A c_D (c'^{\,2}_{\mu,n} + c'^{\,2}_{\mu,p}) + c'_{\mu,n} c'_{\mu,p}(c_A + c_D)(c'_{\mu,n} + c'_{\mu,p})L]\cosh[\kappa L] + \kappa[c_A c'_{\mu,n}(c'_g c'_{\mu,n} + c_D c'_{\mu,p} L) + c'_g c'_{\mu,p}(c_D c'_{\mu,p} + c'_{\mu,n}(c'_{\mu,n} + c'_{\mu,p})L)]\sinh[\kappa L]\}\left\{(c_A + c_D)(c'_{\mu,n} + c'_{\mu,p})^2 \cosh[\kappa L] + c'_g \kappa^3 \left(c_A c_D + c'_g(c'_{\mu,n} + c'_{\mu,p})\right)\sinh[\kappa L]\right\}^{-1}. \qquad \text{(S8)}$$

The total impedance of the circuit in Figure 3 can be obtained based on this result as:

$$Z = \left(r^{-1}_{rec,tot} + i\omega c_{eq}\right)^{-1}. \qquad \text{(S9)}$$

The functional form of Equation S8 can be used to treat the network in Figure S5b, where space charge capacitors $c_B$ and $c_C$ are introduced to describe situations that differ from the flat-band case discussed above. In this case, the electronic majority carrier accumulation or depletion at each interface determines the value of these capacitors. By defining the total interfacial capacitors $c_1 = \frac{c_A c_B}{c_A + c_B}$ and $c_2 = \frac{c_C c_D}{c_C + c_D}$, and the thickness of the bulk $L_{bulk} = L - \lambda_1^* - \lambda_2^*$ ($\lambda_1^*$ and $\lambda_2^*$ are the space charge widths of the two interfaces) one finds:

$$c_{eq} = \{2c_1 c_2 c'_{\mu,n} c'_{\mu,p} + [c_1 c_2 (c'^{\,2}_{\mu,n} + c'^{\,2}_{\mu,p}) + c'_{\mu,n} c'_{\mu,p}(c_1 + c_2)(c'_{\mu,n} + c'_{\mu,p})L_{bulk}]\cosh[\kappa L_{bulk}] + \kappa[c_1 c'_{\mu,n}(c'_g c'_{\mu,n} + c_2 c'_{\mu,p} L_{bulk}) + c'_g c'_{\mu,p}(c_2 c'_{\mu,p} + c'_{\mu,n}(c'_{\mu,n} + c'_{\mu,p})L_{bulk})]\sinh[\kappa L_{bulk}]\}\left\{(c_1 + c_2)(c'_{\mu,n} + c'_{\mu,p})^2 \cosh[\kappa L_{bulk}] + c'_g \kappa^3 \left(c_1 c_2 + c'_g(c'_{\mu,n} + c'_{\mu,p})\right)\sinh[\kappa L_{bulk}]\right\}^{-1}. \qquad \text{(S10)}$$

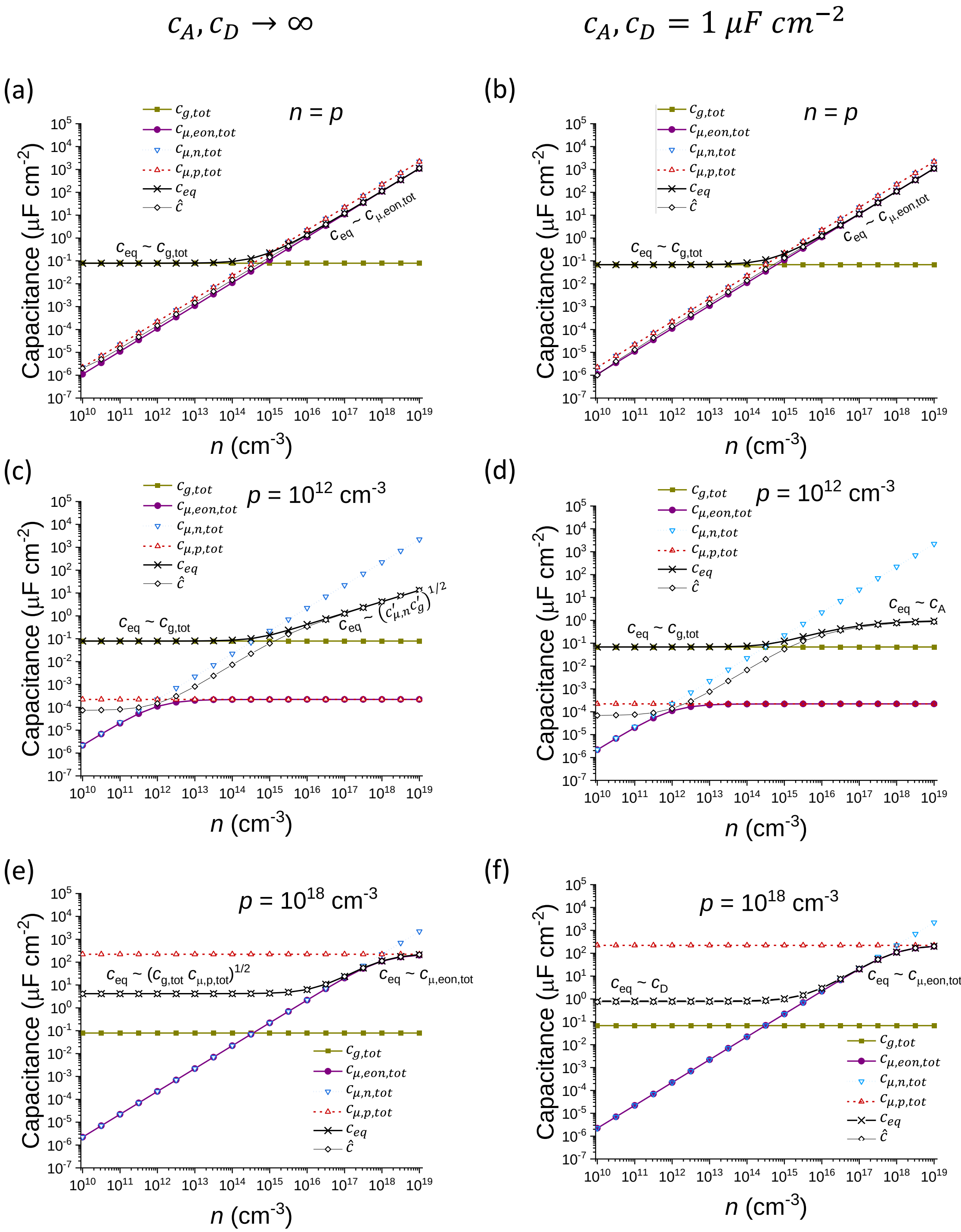

Figure S6. Values of the total chemical, geometric and space charge capacitors relevant to the circuit in Figure S5a, plotted as a function of the electron concentration $n$. The equivalent capacitance is shown, based on Equation S8, and the effective electronic capacitance is defined as $\hat{c} = c_{eq} - c_{g,tot}$. Each panel refers to a specific condition for the hole concentration $p$: (a) $n = p$, (b) $p = 10^{12}\ cm^{-3}$, (c) $p = 10^{18}\ cm^{-3}$.

Figure S6 shows the value of $c_{eq}$ for different situations, where the electronic charge carrier concentrations are such that the electronic chemical capacitors are either larger or smaller than the geometric capacitance (flat-band case is considered for simplicity). These trends highlight that:

a) $c_{g,tot} \gg c_{\mu,n,tot}, c_{g,tot} \gg c_{\mu,p,tot}$ leads to $c_{eq} \approx c_{g,tot}$
b) $c_{g,tot} \ll c_{\mu,n,tot}, c_{g,tot} \ll c_{\mu,p,tot}$ leads to $c_{eq} \approx \max\left[c_{\mu,eon,tot}, c_1, c_2\right]$
c) $c_{g,tot} \ll c_{\mu,n,tot}, c_{g,tot} \gg c_{\mu,p,tot}$ leads to $c_{eq} \approx \max\left[c_{\mu,eon,tot}, c_1\right]$

Where $c_{g,tot} = \frac{c_g'}{L}$, $c_{\mu,n,tot} = c_{\mu,n}' L$, $c_{\mu,p,tot} = c_{\mu,p}' L$ and $c_{\mu,eon,tot} = \frac{c_{\mu,n}' c_{\mu,p}'}{c_{\mu,n}' + c_{\mu,p}'} L$.

The data in Figure S6 show that when the interfacial capacitance dominates (case (b)), this can be limited by either the contact capacitance or the space charge capacitance in the active layer. For $c_A \to \infty$ and $c_D \to \infty$, $c_{eq} \sim \sqrt{c_{\mu,n}' c_g'} = \frac{\epsilon}{L_{D,n}} = c_{Debye}$, where $c_{Debye}$ is the Debye capacitance associated with the majority carriers (electrons in this example). For finite values of $c_A$ and $c_D$, their values represent the upper limit to $c_{eq}$. Specifically:

$c_{eq} \le c_A$ for $c_{g,tot} \ll c_{\mu,n,tot}, c_{g,tot} \gg c_{\mu,p,tot}$

$c_{eq} \le c_D$ for $c_{g,tot} \gg c_{\mu,n,tot}, c_{g,tot} \ll c_{\mu,p,tot}$.

To generalize the treatment of the capacitive response in semiconductors (but also of mixed conductors for the high frequency case, as discussed in the main text for the IC approximation) it is convenient to define an effective electronic capacitance. This can be expressed as,

$$\hat{c} = c_{eq} - c_{g,tot} , \qquad \text{(S11)}$$

so that the parallel of $c_{g,tot}$ and $\hat{c}$ corresponds to the value of $c_{eq}$. As shown in Figure S6, its value can be approximated with the electronic chemical capacitance ($c_{\mu,eon,tot}$), with the space charge capacitance of one of the contacts ($c_A$ or $c_D$), or with the space charge capacitance in the active layer due to the electronic majority carrier ($c_{Debye}$) depending on the specific situation. Using the full expression in Equations S8 and S10 returns a better approximation for the general case.

In real devices, $c_{\mu,n}'$ and $c_{\mu,p}'$ are not constant throughout the active layer. While Equation S10 cannot be directly applied in such cases, one of the approximations $c_{eq} \approx c_{g,tot}$, $c_{eq} \approx c_{\mu,eon,tot}$, $c_{eq} \approx c_1$ or $c_{eq} \approx c_2$ may be still be acceptable.

The use of Equations S8 and S10 to evaluate $c_{eq}$ and $\hat{c}$ applies also to the analysis of the high frequency capacitance of mixed conducting devices according to the IC approximation, as illustrated in the main text (see Figure 4). It is important to note that the values of $c_B$ and $c_C$ in the models displayed in Figure 4 refer to the interfacial capacitance due to accumulation or depletion of the mobile ionic majority carriers, and not to the space charge capacitors due to electronic charges as discussed in this section. The large concentration of mobile ions in mixed conducting devices (assuming IC or MC-i approximation to be valid) implies that the space charge regions on the active layer sides are much thinner than they would be if electronic carriers only were responsible for the electric field screening. It follows that the value of the interfacial capacitance due to electronic charges ($c_B$ and $c_C$ in Figure S5b and in Equation S10) needed in the calculation of $c_{eq}$ of a mixed conductor at high frequencies is in the order of the electronic majority carrier's Debye capacitance. This is due to the steady-state concentration of such carriers deviating from the bulk value only within a region that is thinner than the carrier's Debye length. For this reason, Equation S8 is used in this study to evaluate the high frequency $c_{eq}$ (and therefore $\hat{c}$) for the calculations involving mixed conducting devices under the IC approximation, even when situations far from flat-band are considered.

### *6.2 Differential problem for the complete transmission line model*

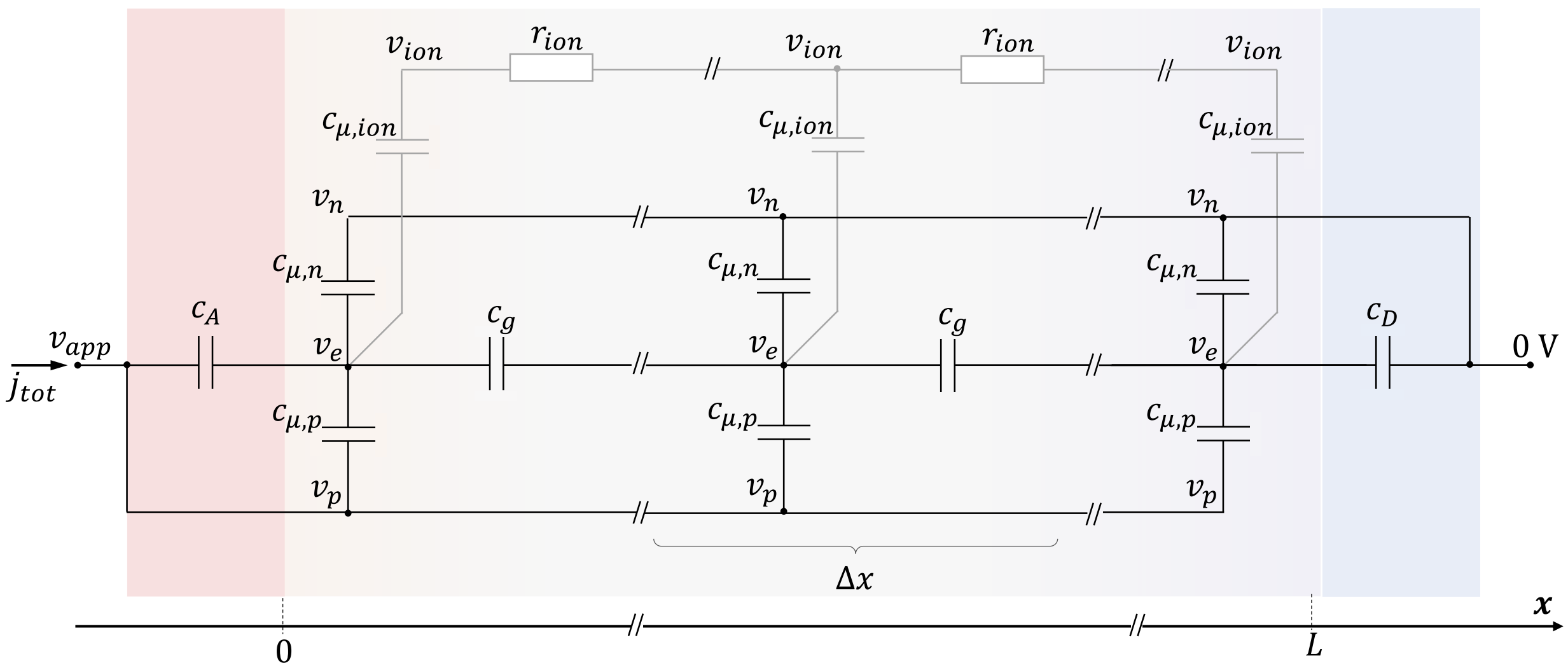

Figure S7. Circuit network involved in the determination of the small signal electrostatic potential in a mixed ionic-electronic conducting solar cell, based on Figure 2, assuming ideal selective contacts (with opposite selectivity) and fast electronic transport ($r_{p,1}, r_{n,2}, r_p,\ r_n \approx 0$, and $r_{p,1}, r_{n,2} \to \infty$,).

Figure S7 shows the circuit model of a mixed conducting solar cell with efficient electronic transport and ion-blocking contacts with opposite selectivity, without the generation-recombination elements. The differential problem describing the functions $v_e(x)$ and $v_{ion}(x)$ referring to this model can be expressed as follows:

$$\begin{cases} \frac{d^2 v_e}{dx^2} = \frac{c'_{\mu,n}+c'_{\mu,p}+c'_{\mu,ion}}{c'_g} v_e - \frac{c'_{\mu,p}}{c'_g} v_{app} - \frac{c'_{\mu,ion}}{c'_g} v_{ion} \\ \qquad \frac{d^2 v_{ion}}{dx^2} = i\omega c'_{\mu,ion}\, r'_{ion}\, (v_{ion} - v_e) \end{cases} \qquad \text{(S12)}$$

Similarly to above, the differential value of the electrical elements is considered i.e. $c'_\mu = \frac{dc_\mu}{dx}$, $r'_{ion} = \frac{dr_{ion}}{dx}$, $c'_g = \left(\frac{d(c_g)^{-1}}{dx}\right)^{-1}$.

From the first equation, an expression of $v_{ion}$ is obtained

$$v_{ion} = -\frac{c'_g}{c'_{\mu,p}} \frac{d^2 v_e}{dx^2} + \frac{c'_{\mu,n}+c'_{\mu,p}+c'_{\mu,ion}}{c'_{\mu,ion}} v_e - \frac{c'_{\mu,p}}{c'_{\mu,ion}} v_{app}\,, \qquad \text{(S13)}$$

which can be substituted in the second equation to obtain the fourth order differential equation

$$\frac{d^4v_e}{dx^4}+\left(-\frac{c'_{\mu,n}+c'_{\mu,p}+c'_{\mu,ion}}{c'_g}-i\omega c'_{\mu,ion}\,r'_{ion}\right)\frac{d^2v_e}{dx^2}+i\omega c'_{\mu,ion}\,r'_{ion}\,\frac{c'_{\mu,n}+c'_{\mu,p}}{c'_g}v_e-\frac{i\omega c'_{\mu,p}c'_{\mu,ion}}{c'_g}\,r'_{ion}v_{app}=0$$
(S14)

The equation to the eigenvalues for the corresponding homogeneous differential equation follows:

$$\kappa^4+\left(-\frac{c'_{\mu,n}+c'_{\mu,p}+c'_{\mu,ion}}{c'_g}-i\omega c'_{\mu,ion}\,r'_{ion}\right)\kappa^2+i\omega c'_{\mu,ion}\,r'_{ion}\,\frac{c'_{\mu,n}+c'_{\mu,p}}{c'_g}=0 \quad \text{(S15)}$$

which yields:

$$\kappa_1=\sqrt{\frac{\frac{c'_{\mu,n}+c'_{\mu,p}+c'_{\mu,ion}}{c'_g}+i\omega c'_{\mu,ion}\,r'_{ion}+\sqrt{\left(\frac{c'_{\mu,n}+c'_{\mu,p}+c'_{\mu,ion}}{c'_g}+i\omega c'_{\mu,ion}\,r'_{ion}\right)^2-4i\omega c'_{\mu,ion}\,r'_{ion}\frac{c'_{\mu,n}+c'_{\mu,p}}{c'_g}}}{2}} \qquad \text{(S16)}$$

$$\kappa_2=\sqrt{\frac{\frac{c'_{\mu,n}+c'_{\mu,p}+c'_{\mu,ion}}{c'_g}+i\omega c'_{\mu,ion}\,r'_{ion}-\sqrt{\left(\frac{c'_{\mu,n}+c'_{\mu,p}+c'_{\mu,ion}}{c'_g}+i\omega c'_{\mu,ion}\,r'_{ion}\right)^2-4i\omega c'_{\mu,ion}\,r'_{ion}\frac{c'_{\mu,n}+c'_{\mu,p}}{c'_g}}}{2}} \qquad \text{(S17)}$$

The four eigenvalues are therefore $\kappa_1, -\kappa_1, \kappa_2, -\kappa_2$. This leads to the following form for $v_e(x)$ associated with the homogeneous problem.

$$v_e(x)=Ae^{\kappa_1 x}+Be^{-\kappa_1 x}+Ce^{\kappa_2 x}+De^{-\kappa_2 x} \qquad \text{(S18)}$$

Regarding the non-homogeneous problem, this is solved by a solution of the form

$$v_e(x)=Ae^{\kappa_1 x}+Be^{-\kappa_1 x}+Ce^{\kappa_2 x}+De^{-\kappa_2 x}+E \qquad \text{(S19)}$$

The value of $E$ is found by substituting this expression in the original differential problem, leading to:

$$E=\frac{c'_{\mu,p}}{c'_{\mu,n}+c'_{\mu,p}}\,v_{app} \qquad \text{(S20)}$$

The value of the constants $A, B, C, D$ can be found by applying the boundary conditions for the conservation of the electric displacement at the interface between the contact and the active layer and Kirchhoff's current law:

$$\begin{cases} -c'_g\frac{dv_e}{dx}(x=0)=c_A\left[v_{app}-v_e(x=0)\right] \\ -c'_g\frac{dv_e}{dx}(x=L)=c_D v_e(x=L) \\ \left[v_{app}-v_e(x=0)\right]i\omega c_A+\int_0^L\left(v_{app}-v_e(x)\right)i\omega c'_{\mu,p}dx=v_e(x=L)i\omega c_D+\int_0^L v_e(x)i\omega c'_{\mu,n}dx \\ 0=-\frac{1}{r'_{ion}}\frac{dv_{ion}}{dx}(x=0) \end{cases}$$
(S21)

The analytical solution for $v_e(x)$ and $v_{ion}(x)$ is very complex. The IC and MC-i approximations introduced in the main text are described below in detail, with the derivation of simple analytical expressions for $v_e(x)$ and the total impedance of the circuits.

*6.3 Differential problem for the ionic-electrostatic transmission line (IC approximation)*

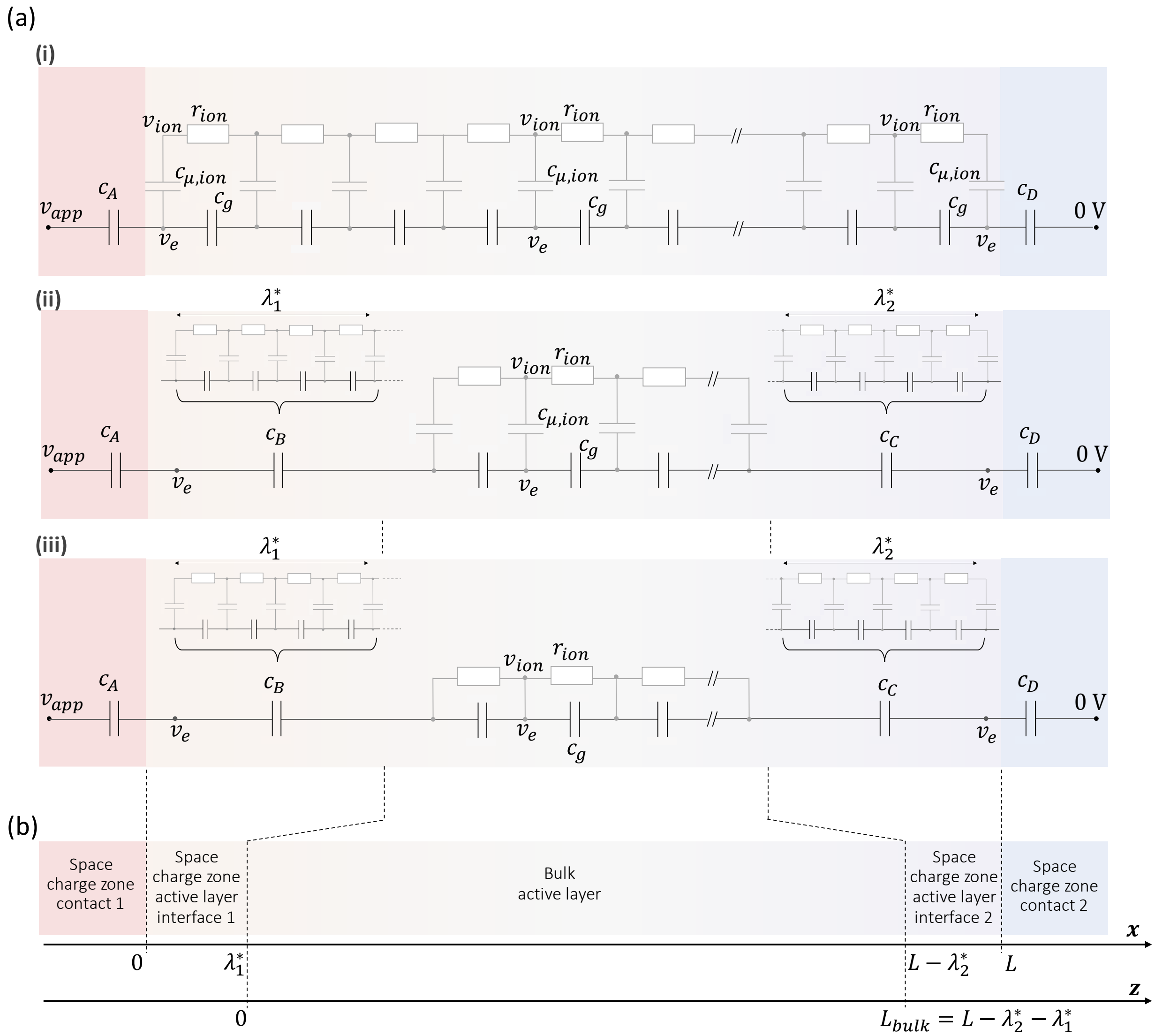

Figure S8. (a) Three levels of approximations for the circuit determining the changes in electrostatic potential $v_e$ in the IC approximation: (i) complete transmission line; (ii) description of the space charges in the mixed conductor through the capacitors $c_B$ and$c_C$; (iii) the ionic chemical capacitors are replaced with short circuits in the bulk, resulting in $v_e = v_{ion}$ (valid only for the small signal and for non-zero applied frequencies). The elements $c_A$ and $c_D$ refer to the contacts' space charge capacitors $c_{g,1}$ and $c_{g,2}$ in Figure 2a. (b) Schematics of the full device highlighting the relation between the circuit elements in (a) and the different regions in the contacts and in the active layer. The space charge widths in the active layer for interface 1 and interface 2 are indicated as $\lambda_1^*$ and $\lambda_2^*$, respectively.

Figure S8a shows transmission line circuits that can be used for the determination of the small signal electrostatic potential $v_e$ in a mixed conducting device (see Figure S8b for details on the structure) under the ionic conductor approximation (IC). The circuits essentially represent an ionic conductor between ion-blocking contacts based on different levels of approximation.

*Approximation (i).* Level (i) corresponds to including the full transmission line circuit within the active layer up to the interfaces, as also shown in Figure 4a. The solution to the bulk transmission line with approximation (i) can be obtained considering a differential version of Kirchhoff's current law, leading to the system of equations:

$$\begin{cases} \frac{d^2 v_e}{dx^2} = -\frac{c'_{\mu,ion}}{c'_g}(v_{ion} - v_e) \\ \frac{d^2 v_{ion}}{dx^2} = i\omega c'_{\mu,ion}\, r'_{ion}\,(v_{ion} - v_e) \end{cases} \tag{S22}$$

Again, the differential value of the electrical elements is considered i.e. $c'_{\mu,ion} = \frac{dc_{\mu,ion}}{dx}$, $r'_{ion} = \frac{dr_{ion}}{dx}$, $c'_g = \left(\frac{d(c_g)^{-1}}{dx}\right)^{-1}$.

This leads to the following form for the functions $v_e(x)$ and $v_{ion}(x)$:

$$v_e(x) = A\kappa^2 e^{\kappa x} + B\kappa^2 e^{-\kappa x} + Cx + D \tag{S23}$$

$$v_{ion}(x) = A\kappa^2\left(1 - \frac{c'_g}{c'_{\mu,ion}}\kappa^2\right)e^{\kappa x} + B\kappa^2\left(1 - \frac{c'_g}{c'_{\mu,ion}}\kappa^2\right)e^{-\kappa x} + Cx + D \tag{S24}$$

Where $\kappa = \sqrt{\frac{c'_{\mu,ion}}{c'_g}(1 + i\omega r'_{ion} c'_g)}$. Considering the presence of $c_A$ and $c_D$, Kirchhoff's current law and the conservation of the electric displacement at the interface between the contact and the active layer leads to the following boundary conditions:

$$\begin{cases} -c'_g \frac{dv_e}{dx}(x=0) = c_A\left[v_{app} - v_e(x=0)\right] \\ -c'_g \frac{dv_e}{dx}(x=L) = c_D v_e(x=L) \\ \left[v_{app} - v_e(x=0)\right] i\omega c_A = v_e(x=L) i\omega c_D \\ 0 = -\frac{1}{r'_{ion}}\frac{dv_{ion}}{dx}(x=0) \end{cases} \tag{S25}$$

From these, the following solution for the electrostatic potential in the active layer is obtained ($0 < x < L$):

$$\frac{v_{e(i)}(x)}{v_{app}} = \frac{\tanh\left(\frac{\kappa L}{2}\right)[1+\cosh(\kappa x)] - \sinh(\kappa x) + \left[\frac{\kappa c'_g}{c_D} + \left(\kappa - \frac{c'_{\mu,ion}}{c'_g \kappa}\right)(L-x)\right](1+i\omega r'_{ion} c'_g)}{2\tanh\left(\frac{\kappa L}{2}\right) + \left(\frac{c'_g \kappa}{c_A} + \frac{c'_g \kappa}{c_D} + \left(\kappa - \frac{c'_{\mu,ion}}{c'_g \kappa}\right)L\right)(1+i\omega r'_{ion} c'_g)} \tag{S26}$$

*Approximation (ii).* By neglecting the faradaic ionic current in the interfacial region, the transmission line circuit associated with the space charge region can be approximated with a single capacitor ($c_B$ and $c_C$ for interface 1 and 2, respectively, see (ii) in Figure S8a). [4] Each space charge has a width equal to $\lambda^*$, which corresponds to the Debye length $L_D$ in case of Gouy-Chapman situations, while wider values may be expected for ionic depletion (Mott-Schottky situations). The resulting model allows the description of the interfacial electrostatics under small signal perturbation, with focus on the changes in the values of $\phi_B$ and $\phi_C$ (see Figure 2b). While the use of circuit (i) allows the extraction of $v_e$ at all positions, the approximation in (ii) no longer explicitly describes the position dependence of $v_e$ within the space charge zones.

When considering approximation (ii), the analytical problem described in (i) can still be used, but with the boundary conditions above no longer applying to the boundaries of the active layer with the contacts but rather to the boundaries of the bulk region with the space charge regions in the active layer.

A similar expression to Eq. S26 is found, valid only for ($\lambda_1^* < x < L - \lambda_2^*$). By defining the bulk position variable $z = x - \lambda_1^*$ and $L_{bulk} = L - \lambda_1^* - \lambda_2^*$, the solution $\frac{v_e(z)}{v_{app}}$ for $0 < z < L_{bulk}$ becomes

$$\frac{v_{e(ii)}(z)}{v_{app}} = \frac{\tanh\left(\frac{\kappa L_{bulk}}{2}\right)[1+\cosh(\kappa z)]-\sinh(\kappa z)+\left[\frac{\kappa c_g'}{c_D}+\left(\kappa-\frac{c_{\mu,ion}'}{c_g'\kappa}\right)(L_{bulk}-z)\right](1+i\omega r_{ion}' c_g')}{2\tanh\left(\frac{\kappa L_{bulk}}{2}\right)+\left(\frac{c_g'\kappa}{c_A}+\frac{c_g'\kappa}{c_D}+\left(\kappa-\frac{c_{\mu,ion}'}{c_g'\kappa}\right)L_{bulk}\right)(1+i\omega r_{ion}' c_g')}, \quad \text{(S27)}$$

where $c_1 = \frac{c_A c_B}{c_A+c_B}$ and $c_2 = \frac{c_C c_D}{c_C+c_D}$.

It follows that $v_e(x) = v_e(z+\lambda_1^*)$ for $0 < z < L_{bulk}$. The value of $v_e(x)$ is also defined at the interfaces of the active layer with the contacts as:

$$v_{e,1} = v_e(x=0) = v_e(z=0)\frac{c_B}{c_A+c_B} + v_{app}\frac{c_A}{c_A+c_B} \quad \text{(S28)}$$

$$v_{e,2} = v_e(x=L) = v_e(z=L_{bulk})\frac{c_C}{c_D+c_C} \quad \text{(S29)}$$

Approximation (ii) assumes that there is no ionic current contribution to the charging of the space charge capacitors. This leads to significant deviation from the solution obtained for the full differential problem in (i).

*Approximation (iii).* Because in the bulk $v_e \approx v_{ion}$ for large values of $c_{\mu,ion}'$ (large mobile ion concentration), $c_{\mu,ion}'$ elements can be replaced with short circuits, as shown in (iii) in Figure S8a (see also Figure 4b in the main text). Note that this approximation is valid for non-zero angular frequencies, and only for the small perturbation analysis. This leads to the solution (see also Appendix C):

$$\frac{v_{e(iii)}(z)}{v_{app}} = \frac{c_1}{c_1+c_2}\frac{1+i\omega r_{ion,bulk}\left(c_{g,bulk}+\frac{c_2 z}{L_{bulk}}\right)}{1+i\omega r_{ion,bulk}\frac{c_1 c_2}{c_1+c_2}} \quad \text{(S30)}$$

The value of $v_e(x)$ at the interfaces of the active layer with the contacts can still be evaluated using Equations S28 and S29.

If the value of the electrostatic potential in the bulk is not required (no bulk recombination transistor in the model), the bulk circuit shown in (iii) is simplified further. The model shown in Figure 4c is obtained, where one ionic resistor $r_{ion,bulk}$ and one geometric capacitor $c_{g,bulk}$ are used to represent the bulk electrostatic and ionic properties.

The resulting impedance of the ionic and electrostatic rails based on the analytical solutions above can be evaluated as follows. For circuit (i):

$$Z_{ion,e(i)} = \frac{1}{i\omega\left(\frac{1}{C_A}+\frac{1}{C_D}+\frac{L}{c_g'}+\frac{c_{\mu,ion}'\left(2\tanh\left[\frac{L\kappa}{2}\right]-\kappa L\right)}{c_g'^{2}\kappa^3}\right)^{-1}}, \qquad \text{(S31)}$$

where $\kappa = \sqrt{\frac{c_{\mu,ion}'}{c_g'}\left(1+i\omega r_{ion}' c_g'\right)}$.

For circuits (ii), similarly:

$$Z_{ion,e(ii)} = \frac{1}{i\omega\left(\frac{1}{C_1}+\frac{1}{C_2}+\frac{L_{bulk}}{c_g'}+\frac{c_{\mu,ion}'\left(2\tanh\left[\frac{L_{bulk}\kappa}{2}\right]-\kappa L_{bulk}\right)}{c_g'^{2}\kappa^3}\right)^{-1}}. \qquad \text{(S32)}$$

Finally, for circuit (iii):

$$Z_{ion,e(iii)} = \frac{1}{i\omega\frac{C_1C_2}{C_1+C_2}} + \frac{L_{bulk}r_{ion}'}{1+i\omega r_{ion}' c_g'} = \frac{1+i\omega L_{bulk}r_{ion}'\left(\frac{C_1C_2}{C_1+C_2}+\frac{c_g'}{L_{bulk}}\right)}{i\omega\frac{C_1C_2}{C_1+C_2}\left(1+i\omega r_{ion}' c_g'\right)} \qquad \text{(S33)}$$

Based on $\frac{v_e}{v_{app}}(x)$, $Z_{eon}(\omega)$ is calculated (Equation 21 in the main text). Note that the following treatment can be applied to either of the (i), (ii) or (iii) level of approximation above, by selecting the appropriate $\frac{v_e}{v_{app}}$ and $Z_{ion,e}$ expressions above. For model (i), one can write

$$j_{eon(i)}(\omega) = v_{app}\left\{\frac{L}{r_{rad}} + i\omega\hat{c} + \int_0^L \left[\frac{v_e(x,\omega)}{v_{app}} g_{rec,n} + \left(1-\frac{v_e(x,\omega)}{v_{app}}\right) g_{rec,p}\right] dx\right\} \qquad \text{(S34)}$$

While for model (iii) (as well as model (ii)), the recombination current in the bulk can be integrated using the $z$ coordinate, while the surface recombination current can be added separately based on the values of the interfacial potentials $v_{e,1} = v_e(x=0)$ and $v_{e,2} = v_e(x=L)$ (see Equations S28 and S29)

$$j_{eon(iii)}(\omega) = v_{app}\left\{\frac{L}{r_{rad}} + i\omega\hat{c} + \frac{v_{e,1}}{v_{app}} g_{rec,n,surf1} + \left(1-\frac{v_{e,1}}{v_{app}}\right) g_{rec,p,surf1} + \frac{v_{e,2}}{v_{app}} g_{rec,n,surf2} + \left(1-\frac{v_{e,2}}{v_{app}}\right) g_{rec,p,surf2} + \int_0^{L_{bulk}} \left[\frac{v_e(z,\omega)}{v_{app}} g_{rec,n} + \left(1-\frac{v_e(z,\omega)}{v_{app}}\right) g_{rec,p}\right] dz\right\} \qquad \text{(S35)}$$

In both cases, the value of $\hat{c}$ can be evaluated using Equation S10.

By defining $Z_{eon}(\omega) = \frac{v_{app}}{j_{eon}(\omega)}$, one obtains for the overall impedance (see also main text):

$$Z(\omega) = \left[Z_{ion,e}^{-1} + Z_{eon}^{-1}(\omega)\right]^{-1} \qquad \text{(S36)}$$

The accuracy of the three levels of approximations is compared, based on a simplified transmission line problem. Figure S9a and S9b illustrate the calculated real part of the normalized small signal electrostatic potential $\frac{v_e}{v_{app}}(x)$ obtained using approximation (i), (ii) and (iii), and for three different frequencies (an input parameter set representative of halide perovskites is used, see figure caption).

Small deviations from the analytically accurate solution (level (i)) are observed for solution (ii) and (iii). Approximation (ii) reproduces less accurately the solution from (i) compared with approximation (iii). This is especially the case at low frequencies, where charging of the space charge capacitors occurs predominantly via the ionic contribution (which is neglected in (ii)). The resulting ionic-electrostatic impedance is shown in Figure S9c.

To evaluate the accuracy of these approximation, impedance calculations using the circuit in Figure 2a are performed. Figure S9d shows the analytical results for the spectra of $Z(\omega)$ evaluated for the cases (i)–(iii) and for a constant recombination transconductance $g_{rec,n}$ and $g_{rec,p}$ across the active layer. Examples with significant surface recombination at either interface 1 or 2 are also shown. In all cases, the results are compared with the numerical solution from the transmission line in Figure 2a using transport parameters such that the rate of electronic transport is significantly faster than the rate of recombination. Once again, the comparison between model (i) and models (ii) and (iii) is relevant only for the simplified flat-band case. The simplified models can account for the distributed nature of the problem described in the main text. In particular, the IC approximation (iii), which is the one discussed in Figure 4b–e, yields good results when compared with the complete model.

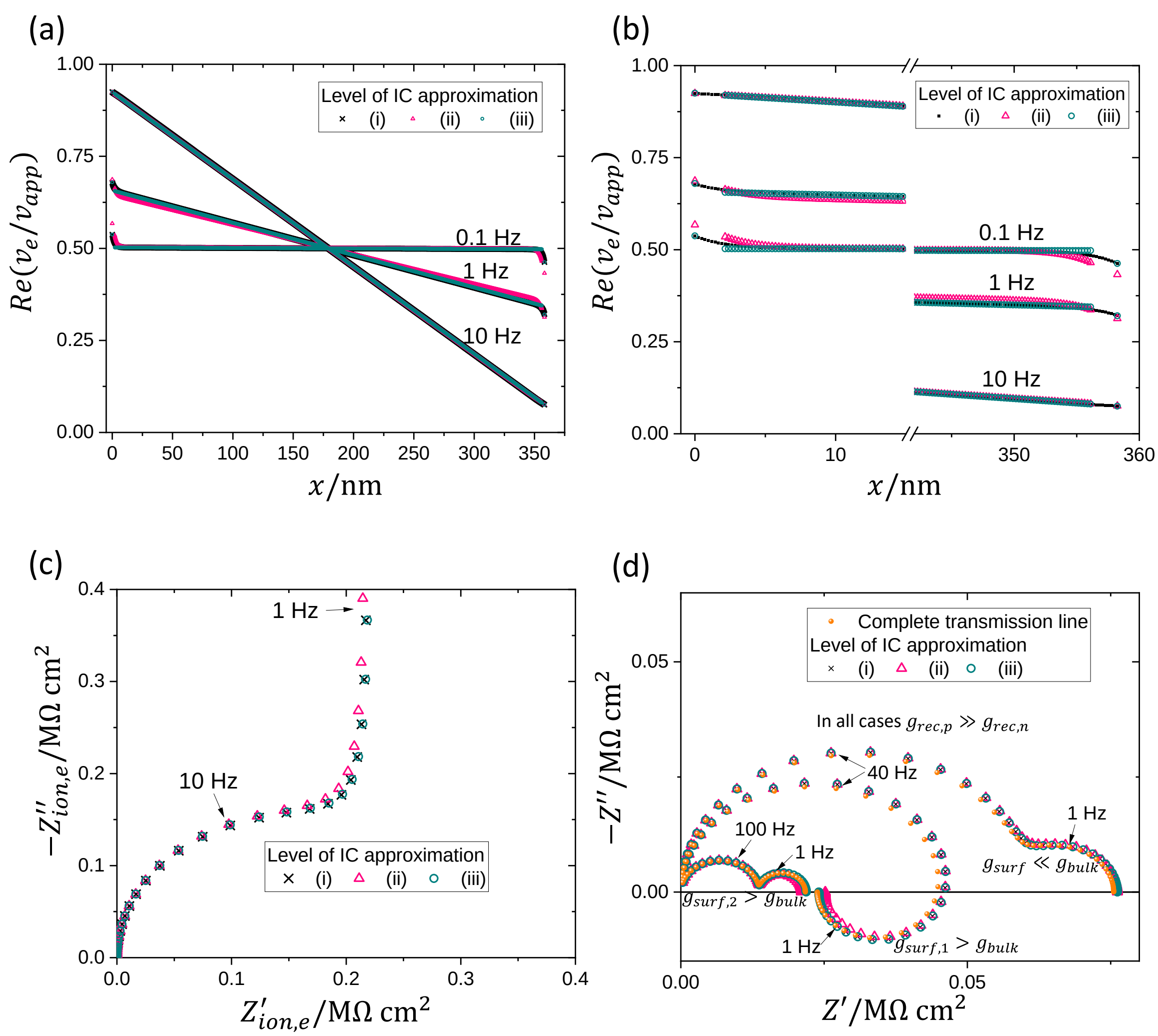

Figure S9. Comparison between different levels of IC approximation (i), (ii) and (iii) discussed in the text for (a) the real part of the small signal electrostatic potential distribution in the active layer of a mixed conducting device with ion-blocking contacts (see models in Figure S8) for the case of $\bar{\phi}_B = \bar{\phi}_C = 0$ V. In (b) the difference between the datasets in proximity of the interfaces is highlighted. Note that for (ii) and (iii) the value of $v_e$ is not defined within the space charge regions. (c) and (d) show Nyquist plot of the impedance spectra $Z_{ion,e}$ and $Z$ obtained with the three approximations. The data in (c) refer to the impedance of the ionic and electrostatic rails in Figure S8. The data in (d) consider the complete impedance where the electronic contribution is included. For this comparison, a constant bulk recombination transconductance (in these cases $g_{rec,p} \gg g_{rec,n}$), negligible transport resistance for the electronic charge carriers and situations where different values for the surface recombination transconductance terms are considered. The data are obtained using $c_A = c_D = 1\ \mu F\ cm^{-2}$, $[\mathrm{V_I^{\cdot}}] = 10^{19}\ cm^{-3}$, $u_{ion} = 10^{-10} cm^2V^{-1}s^{-1}$, $\bar{p} = 10^{12}\ cm^{-3}$, $\bar{n} = 10^{14}\ cm^{-3}$, $\tau_n = \tau_p = 10\ \mu s$. In one case, negligible surface recombination is assumed ($g_{surf} \ll g_{bulk}$), while for the cases of high surface recombination at either interface, $\frac{g_{rec,surf}}{g_{rec,bulk}} \approx 3$. Flat-band condition is assumed for the steady-state at the interfaces. In addition to the analytical results obtained with the three levels of approximation (i)–(iii), the numerical solution from the transmission line in Figure 2a with the same input parameters is also included.

### *6.4 Differential problem for the mixed conductor with ionic majority carrier approximation (MC-i)*

$$L_{D,ion} \ll L_{D,n} < L, \qquad L_{D,ion} \ll L_{D,p} < L$$

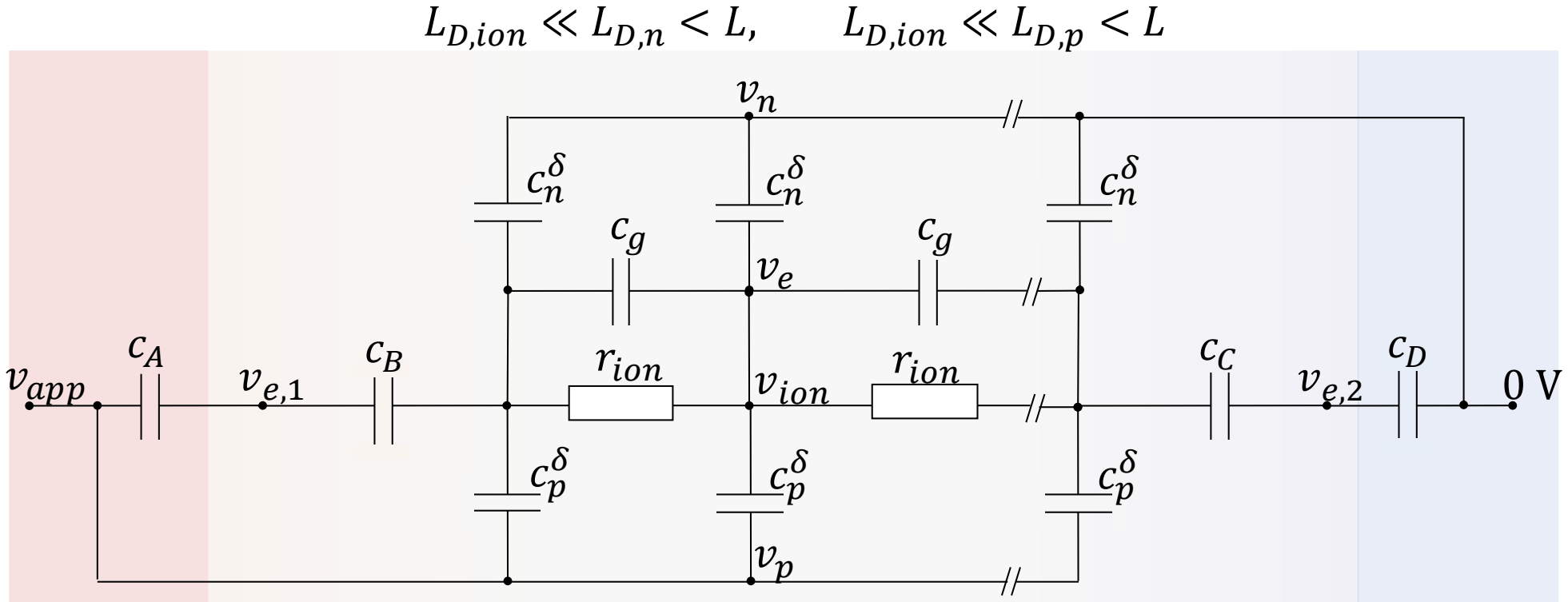

Figure S10. Simplified equivalent circuit model of a solar cell with ideally selective contacts and mobile ions in the active layer. The circuit well approximates the electrostatic, ionic and electronic properties of the device under large bias, where the electronic chemical capacitors cannot be neglected. Only the recombination elements are omitted for simplicity (see main text for their description).

Figure S10 shows the electrostatic and chemical capacitive network combined with the ionic resistors relevant to the mixed conductor with ionic majority carriers (MC-i) approximation (see Figure S15 for the definition of the chemical capacitors $c_n^\delta$ and $c_p^\delta$). To derive the analytical solution for $v_e/v_{app}$ as a function of $z$ ($z = x - \lambda_1^*$, see above) and the impedance of the circuit, the following problem is considered:

$$\frac{d^2 v_e}{dz^2} = -\frac{r'_{ion}}{1+i\omega r'_{ion} c'_g}\left[(v_{app} - v_e) i\omega c_p^{\delta\prime} - v_e i\omega c_n^{\delta\prime}\right] \qquad \text{(S37)}$$

Differential terms for the chemical capacitors are defined as in the previous sections (e.g. $c_p^{\delta\prime} = \frac{dc_p^\delta}{dx}$). The solution to the homogeneous problem is

$$v_e(z) = Ae^{\kappa z} + Be^{-\kappa z}\,, \qquad \text{(S38)}$$

where $\kappa = \sqrt{\frac{i\omega(c_p^{\delta'}+c_n^{\delta'})r'_{ion}}{1+i\omega r'_{ion}c'_g}}$. It follows that the particular solution is

$$v_e(z) = Ae^{\kappa z} + Be^{-\kappa z} + E \qquad \text{(S39)}$$

where $E = \frac{c_p^{\delta'}}{c_p^{\delta'}+c_n^{\delta'}} v_{app}$.

The value of A and B are determined based on the boundary conditions referring to the continuity of the electric displacement at the interface and to the equality of the currents at the two boundaries.

$$\begin{cases} -\left(\frac{r'_{ion}}{1+i\omega c'_g r'_{ion}}\right)^{-1} \frac{dv_e}{dx}(z=0) = i\omega c_1\left[v_{app} - v_e(z=0)\right] \\ i\omega c_1\left[v_{app} - v_e(z=0)\right] + \int_0^{L_{bulk}}\left[v_{app} - v_e(z)\right] i\omega c_p^{\delta'} dz = i\omega c_2 v_e(z = L_{bulk}) + \int_0^{L_{bulk}} v_e(z) i\omega c_n^{\delta'} dz \end{cases}$$

(S40)

This yields the solution:

$$\frac{v_e(z)}{v_{app}} = \frac{c_p^{\delta'}}{c_p^{\delta'}+c_n^{\delta'}} +$$

$$+\frac{\left(1-\frac{c_2}{c_1}\frac{c_p^{\delta'}}{c_n^{\delta'}}\right)\cosh(\kappa z)(1+i\omega r'_{ion}c'_g)\kappa^2 + i\omega r_{ion}\left[\left(c_n^{\delta'}+c_p^{\delta'}\right)(\cosh[\kappa(L_{bulk}-z)]-\cosh[\kappa z]) + c_2\kappa\left(\sinh[\kappa(L_{bulk}-z)] - \frac{c_p^{\delta'}}{c_n^{\delta'}}\sinh[\kappa z]\right)\right]}{\left(1+\frac{c_p^{\delta'}}{c_n^{\delta'}}\right)\left\{i\omega r'_{ion}\left[\left(c_n^{\delta'}+c_p^{\delta'}\right)(\cosh[\kappa L_{bulk}]-1) + c_2\kappa\sinh[\kappa L_{bulk}]\right] + (1+i\omega r'_{ion}c'_g)\kappa\left(\kappa + \frac{c_2}{c_1}\kappa\cosh[\kappa L_{bulk}] + \frac{c_n^{\delta'}+c_p^{\delta'}}{c_1}\sinh[\kappa L_{bulk}]\right)\right\}}.$$

(S41)

## 7. Discussion on transistor based equivalent circuit models for interfacial behavior

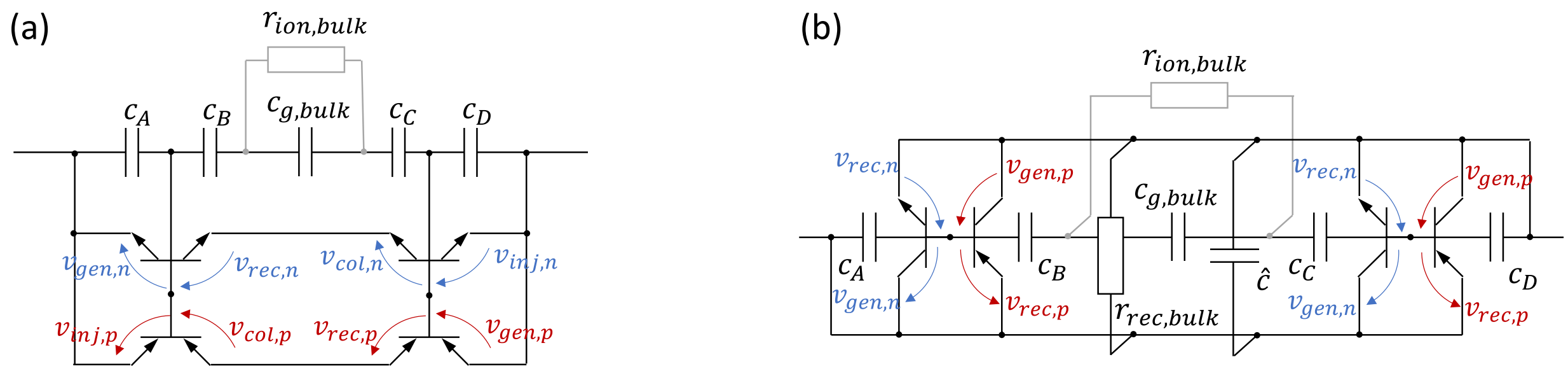

Figure S11. (a) Equivalent circuit model presented in Ref. [1]. (b) Equivalent circuit model described in this work. The orientation of the transistor is representative of the type of electron transfer process: horizontal for transfer across an interface; vertical for a local recombination process.

The equivalent circuit model presented in Ref. [1] is reported in Figure S11a. Here, the interfaces of the mixed conducting solar cell are modelled with a transistor pair (*npn* and *pnp*) controlled by the interfacial change in potential. For either interface, one transistor describes the injection and collection of the carrier that is majority carriers, assumed to be the same as in the respective contact (e.g. holes in proximity of the hole transport material), while the other transistor describes the recombination of minority carriers (e.g. electrons) with the majority carrier in the transport layer. It follows that, in Figure S11a, the transistors represent interfacial processes, specifically electron transfer processes *across* an interface.

In this work, the behavior of *local* electron-hole recombination is investigated. Local recombination in this context refers to processes occurring *without long range transfer*, that is the transistor and resistor elements discussed in the main text refer to recombination occurring locally between carriers. Limiting the description to only recombination (and thermal generation) occurring at the interfaces, the circuit in Figure S11b is obtained (see also Figure 4c in the main text). The apparent difference is that in Figure S11a, the two branches including the series of *npn* transistors and the series of *pnp* transistors are in parallel to each other, while in Figure S11b the pair of *npn* and *pnp* transistors at interface 1 are in parallel to the pair at interface 2. In both equivalent circuit models, it is reasonable to expect that, at any given condition, one of the transistors determines the electronic contribution to the impedance of the model. What becomes evident is that:

- if recombination at either interface is dominant, the two circuits become identical, assuming that at such interface the opposite carrier to the injected species from the contact is the minority carrier (*i.e.* electrons close to the hole transport material, interface 1, or holes close to the electron transport material, interface 2).
- If the *injection* of a carrier from an interface is the dominant contribution to the impedance in Figure S11a, the resulting circuit is identical to the case where *recombination* at the same interface is dominant in Figure S11b, if the same carrier type that is injected is also the minority carrier. Note that this would be the 'wrong' minority carrier at such interface in a common solar cell under operation.

Two further observations become evident from the discussion in this work and the comparison of the models in Figure S11. First, in Figure 2a and Figure S3, modeling of charge transport without considering long range effects is implemented through resistors and voltage controlled current sources. The latter elements include gating effects on the local current by $v_\mu$ (and therefore by $v_e$), which could apply also to the description of injection/collection processes at interfaces. As the relations between small signal voltage and current are different for such elements and for the injection transistors in Figure 11a, differences in the predicted behavior are expected, an aspect that is worth of future investigation.

Secondly, the analysis in the main text indicates that the equivalent circuit model described in Figure 2a and its simplified version in Figure 4c (Figure S11b) have a direct physical connection with the drift-diffusion equations describing the mixed conducting device, assuming only local recombination to occur. The electron transfer across interfaces invoked in Figure S11a and in Ref. [1] may, however, be significant in real devices. A distinction can be made between recombination processes involving transfer across an interface that can still be considered local, and non-local effects. If negligible variation in $v_e$ is present between the positions of the interacting carriers, a recombination transistor gated by the local change in electrostatic potential is an accurate description of the former process. This means that, when describing interfacial processes, the recombination transconductance associated with the relevant transistor could include contributions from recombination between electrons and holes that are in the semiconducting active layer (as discussed in the main text), as well as between electrons and holes that lie at the interface but in different phases (see heterogeneous electron transfer component in Figure S12). In the discretized model, the net recombination per unit volume nomenclature ($u$) is replaced by the net surface recombination current density ($j$).

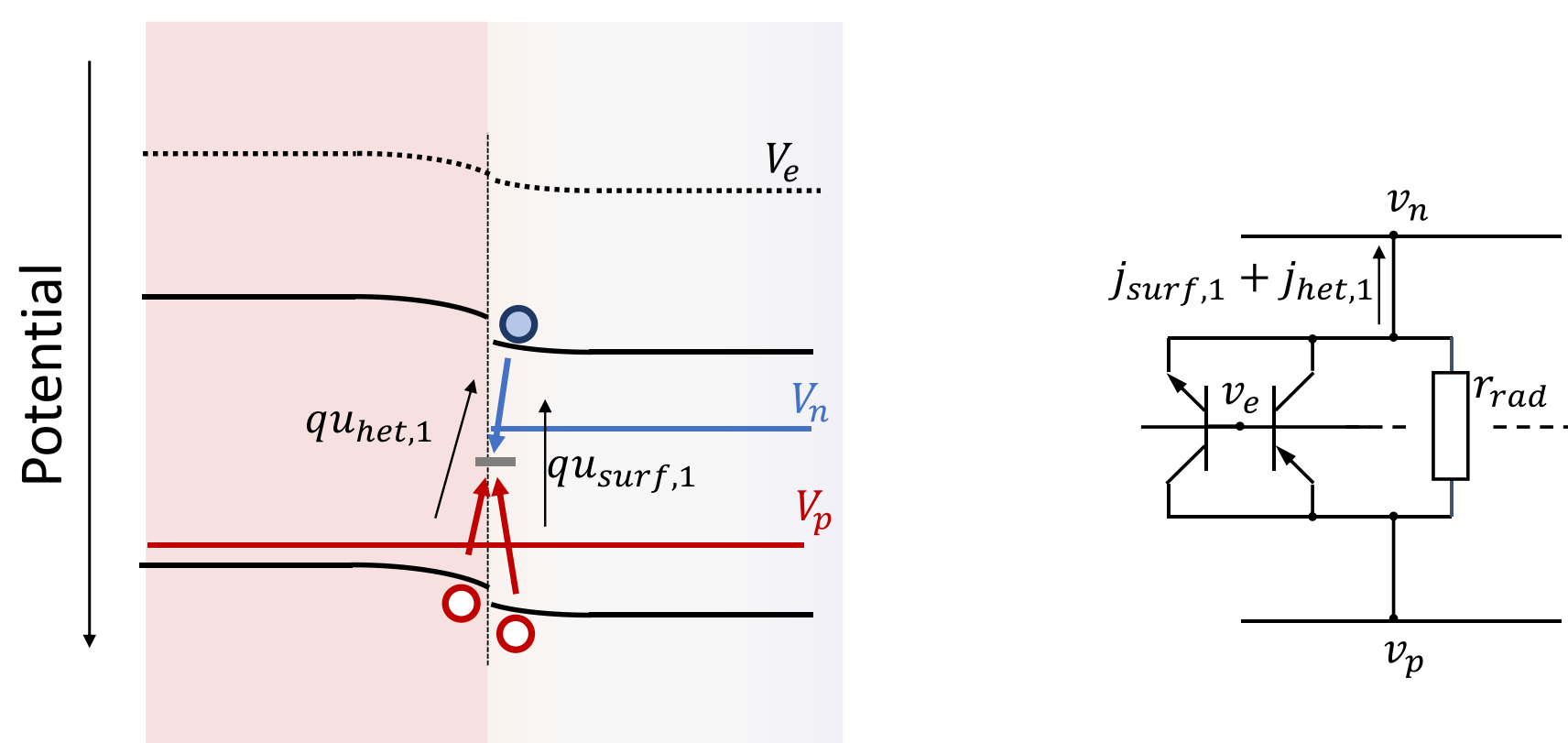

Figure S12. Interfacial recombination can occur via local recombination between electrons and holes in the same phase but also via recombination between electrons and holes in different phases. If for the latter process the same interfacial electrostatic potential affects the concentration of both carriers, its SRH recombination current contribution can be effectively incorporated within the same transistor pair describing local recombination.

## 8. Steady-state electronic charge concentration in mixed conducting solar cells

The electronic steady-state situation in semiconducting devices, such as solar cells, depends on whether ionic defects are mobile or not. In a solar cell device where the active layer is a semiconductor *with no mobile ions*, application of a large enough bias leads to the "high injection regime", whereby $\bar{p} \gg p_{\mathrm{eq}}$ and $\bar{n} \gg n_{\mathrm{eq}}$ ($p_{\mathrm{eq}}$ and $n_{\mathrm{eq}}$ are the hole and electron bulk concentrations at equilibrium). In the bulk, the electroneutrality condition implies $\bar{p} = \bar{n}$.

If the active layer is a semiconductor *with mobile ions* (*e.g.* iodide vacancies $\mathrm{V_I^{\cdot}}$ in iodide perovskites) as majority carriers, these can be compensated by another mobile or immobile defect. The latter case, *e.g.* compensating acceptor dopants $\mathrm{A'}$ or frozen-in intrinsic defects, is considered here. The condition of electroneutrality implies $[\mathrm{V_I^{\cdot}}]_{eq} \approx [\mathrm{A'}]$, if the mixed conductor is in the intrinsic regime. [37,57] In this study, no additional ionic defect disorder equilibrium and/or component exchange reaction with the gas phase are considered. The only boundary conditions determining all charge concentrations are established by the properties of the contacts (see also treatment in Ref. [47]).

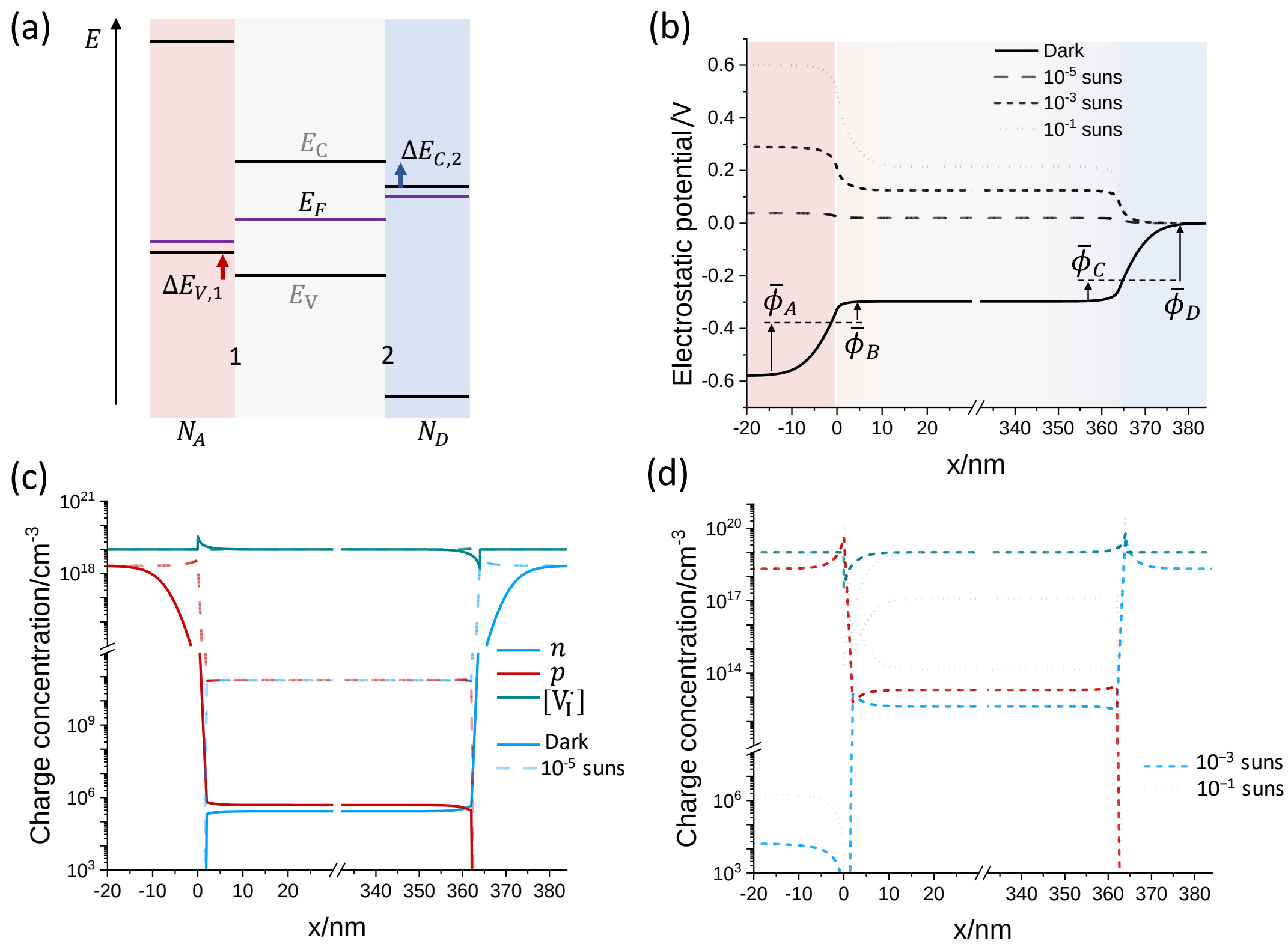

Figure S13. (a) Energy level diagram of a solar cell (before contact between the layers) where the active layer is sandwiched between an acceptor doped HTM and a donor doped ETM ($N_A$ and $N_D$ doping concentrations, respectively). (b) Electrostatic potential and (c, d) charge concentration at open circuit for different light intensities as function of position. For this example of a "symmetrical" solar cell: the active layer is at the intrinsic point, $\bar{p}_{eq} = \bar{n}_{eq}$, $[\mathrm{V_I^{\cdot}}]_{eq} = [\mathrm{A'}] = 10^{19}\ cm^{-3}$ before contact ($\mathrm{A'}$ is a background charge of immobile anions, present across the device, not shown for simplicity); the effective density of states, mobilities and recombination parameters are the same for electrons and holes; the hole transport layer (HTM) and the electron transport layer (ETM) show symmetrical properties in terms of mobilities,

effective density of states, bandgap and dielectric constant (see section 11). In addition, $\Delta E_{V,1} = \Delta E_{C,2} = 0.4\ eV$ and $N_D = N_A = 2.1 \times 10^{18}\ cm^{-3}$. Note that the cations in the contacts are immobile.

By considering the space charge (quasi-)equilibrium with the contacts (see Figure 2b) and assuming fast transport (flat quasi-Fermi levels), the electron and hole concentrations in the bulk obey:

$$\frac{\bar{p}}{\bar{n}} = \frac{N_{C,ETM} N_V N_A}{N_{V,HTM} N_C N_D} e^{\frac{\bar{\phi}_C + \bar{\phi}_D + \frac{\Delta E_{C,2}}{q} - \bar{\phi}_A - \bar{\phi}_B - \frac{\Delta E_{V,1}}{q}}{V_{TH}}} \quad \text{(S42)}$$

where $\Delta E_{V,1}$ and $\Delta E_{C,2}$ are the offsets of the relevant band edges with respect to the active layer (see Figure S13a), $N_C$ and $N_V$ are the effective density of states for the conduction and valence band of the active layer ($N_{C,ETM}$ and $N_{V,HTM}$ have the same meaning but for the ETM and HTM, respectively). $N_A$ and $N_D$ are the acceptor and donor doping of the HTM and ETM, respectively (majority carrier concentration in the contacts is assumed to correspond to the doping density).

Figure S13 considers an example of the steady-state charge carrier distribution in a device at open circuit as function of position for different bias light conditions. The device implements the simplest symmetrical situation for a solar cell with mixed conducting active layer and ion-blocking contacts (see Figure's caption). In the dark and at small bias, the discrepancy from the $\bar{p} = \bar{n}$ situation in the bulk is due to the one source of asymmetry in the system: only one (positively charged) ionic species is mobile and the negatively charged compensating defect is immobile, leading to asymmetry in the space charge situation at the two interfaces, and to differences in the magnitude of $\bar{\phi}_B$ and $\bar{\phi}_C$. If the two space charge regions involve the same amount of charge (ionic defects depleted at one interface accumulate at the other, valid for small electronic charge concentrations), $\bar{\phi}_A$ and $\bar{\phi}_D$ are the same in this example and Equation S42 reduces to $\frac{\bar{p}_{eq}}{\bar{n}_{eq}} = e^{\frac{\bar{\phi}_C - \bar{\phi}_B}{V_{TH}}}$. Because a larger potential drops across a depletion than an accumulation layer on equal stored charge, $\bar{\phi}_B < \bar{\phi}_C$ at equilibrium in the dark, leading to a slightly larger concentration for holes than for electrons. Such mismatch becomes small when the magnitude of $\bar{\phi}_B$ and $\bar{\phi}_C$ is small, in that both an accumulation and a depletion layer with a small space charge potential ($|\bar{\phi}| \ll V_{TH}$) follow the properties of a Debye layer in terms of potential profile and capacitance. In the $10^{-5}$ suns case (Figure S13c), very small values of $\bar{\phi}_B$ and $\bar{\phi}_C$ are obtained ($V_{OC} \approx \phi_{bi}$), leading to $\bar{p} = \bar{n}$ in the bulk of the active layer. For larger light intensities, the sign of $\bar{\phi}_B$ and $\bar{\phi}_C$ becomes negative ($V_{OC} > \phi_{bi}$), implying that ion depletion and ion accumulation occur now at interface 1 and interface 2, respectively (see $[\mathrm{V_I^{\cdot}}]$ profile in Figure S13d vs dark case in S13c). It follows that $|\bar{\phi}_B| > |\bar{\phi}_C|$ and, once again, the steady state electron and hole concentrations deviate from each other ($\bar{p} > \bar{n}$). This is in stark contrast with the case where no mobile ions are present, where $\bar{p} = \bar{n}$ is expected (see above).

As the concentration of electronic carriers becomes significant at large bias (particularly holes in this case), the electroneutrality condition in the bulk demands a slight change in mobile ionic concentration. Indeed, for the 0.1 sun case, $[\mathrm{V_I^{\cdot}}] \approx 9.88 \times 10^{18}\ cm^{-3}$ in the bulk of the active layer. Such value is lower than the set value of $[\mathrm{A'}] = 1 \times 10^{19}\ cm^{-3}$ by almost exactly the bulk value of $p$. The assumption that the charge stored in the two space charge regions at the interfaces on the active layer sides are the same gradually loses validity and Equation S42 can no longer be simplified.

Similar analysis performed on data obtained for input parameters where the contacts are not symmetric, in terms of doping density and/or band edge offset, is presented in Figure S14. Figure S14a combines the data from Figure S13c and S13d, used as reference. Figure S14b–d show the charge concentration profiles when such symmetry is broken by varying the properties of the HTM. Such changes lead to asymmetry in either the band edge offset between contact and active layer for the two interfaces and/or in the doping of the two contact ($N_A$ is the acceptor doping in the HTM and $N_D$ is the donor doing in the ETM). With respect to Figure S14a,

- in Figure S14b the HTM has a lower work function (leading to lower overall $\phi_{bi}$), but same doping (this means also $\Delta E_{V,1} > \Delta E_{C,2}$)
- In Figure S14c the HTM has a low work function as in S14b, but also lower doping than the symmetrical case, so that $\Delta E_{V,1} = \Delta E_{C,2}$ still holds.
- In Figure S14d, the same work function for the HTM as S14a is used (implying the same $\phi_{bi}$), but with lower doping ($\Delta E_{V,1} > \Delta E_{C,2}$)

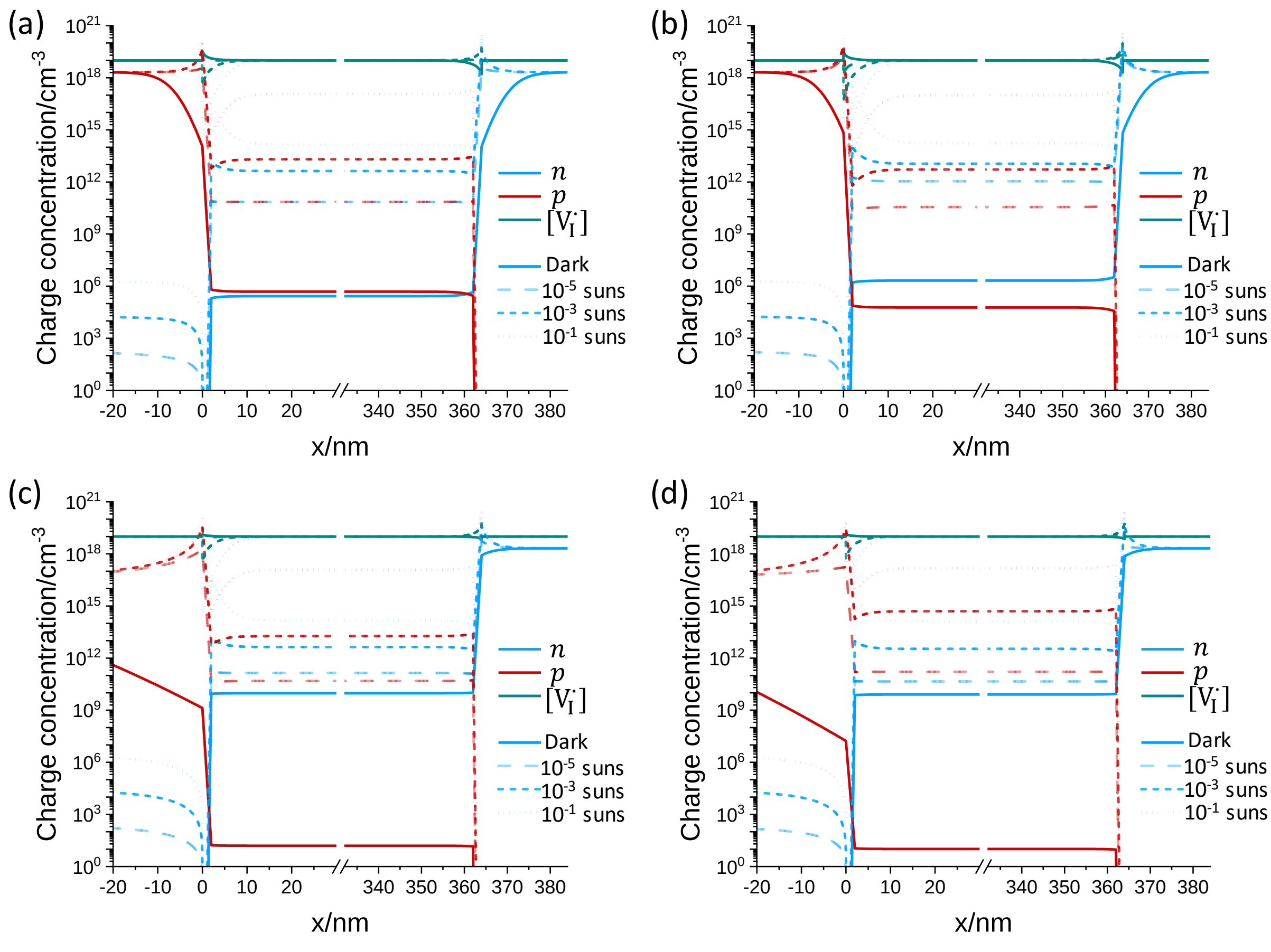

Figure S14. Steady-state charge concentrations evaluated for a solar cell at open circuit for different light intensity conditions. The panels refer to different properties of the contact layers. (a) Symmetrical situation also shown in Figure S13 ($\Delta E_{V,1} = \Delta E_{C,2} = 0.4\ eV$, $N_A = N_D = 2.1 \times 10^{18} cm^{-3}$). (b) $\Delta E_{V,1} = 0.5\ eV, \Delta E_{C,2} = 0.4\ eV$, $N_A = N_D = 2.1 \times 10^{18} cm^{-3}$. (c) $\Delta E_{V,1} = 0.5\ eV, \Delta E_{C,2} = 0.4\ eV$, $N_A = 2.1 \times 10^{18} cm^{-3} = N_D e^{-\frac{0.1 eV}{k_B T}}$. (d) $\Delta E_{V,1} = \Delta E_{C,2} = 0.4\ eV$, $N_A = 2.1 \times 10^{18}\ cm^{-3} = N_D e^{-\frac{0.1 eV}{k_B T}}$.

The results highlight that the two cases where asymmetry in doping between the HTM and ETM is considered show a significant difference in the equilibrium electronic concentrations, almost regardless of whether $\Delta E_{V,1} = \Delta E_{C,2}$ (Figure S14c) or $\Delta E_{V,1} > \Delta E_{C,2}$ (Figure S14d). Such difference is more pronounced than for the case involving a mismatch between $\Delta E_{V,1}$ and $\Delta E_{C,2}$ and symmetrical doping (Figure S14b). Low doping in the hole (electron) transport layer decreases its equilibrium depletion space charge capacitance. This results in an increase in the value of $\bar{\phi}_A$ ($\bar{\phi}_D$) which implies a more n- (p-) type character of the active layer. Under large bias, the active layer presents a mismatch in the bulk concentrations of electrons and holes ($\bar{p} \gg \bar{n}$ at 0.1 suns illumination) and space charge situation that are slightly different in each case but follow similar trend.

## 9. Approximated chemical capacitance network for mixed conductors under bias

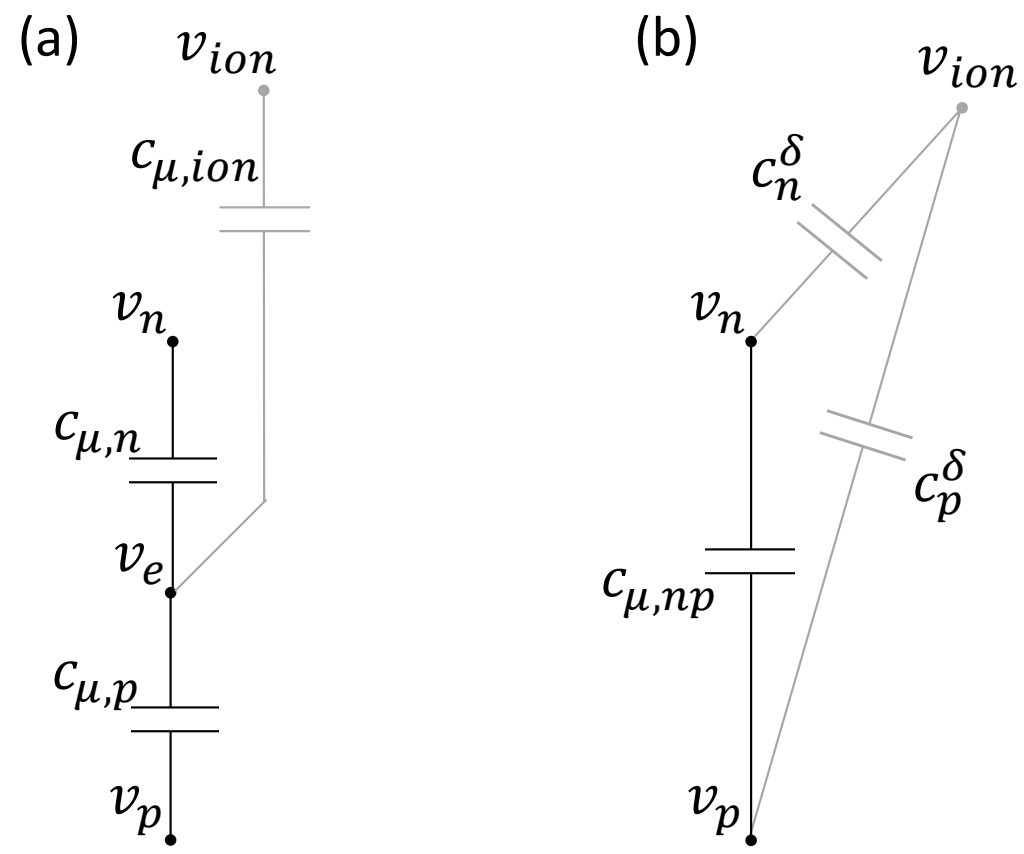

$$c_{\mu,np} = \frac{c_{\mu,n} c_{\mu,p}}{c_{\mu,n} + c_{\mu,ion} + c_{\mu,p}} \approx \frac{c_{\mu,n} c_{\mu,p}}{c_{\mu,ion}}$$

$$c_n^\delta = \frac{c_{\mu,n} c_{\mu,ion}}{c_{\mu,n} + c_{\mu,ion} + c_{\mu,p}} \approx c_{\mu,n}$$

$$c_p^\delta = \frac{c_{\mu,p} c_{\mu,ion}}{c_{\mu,n} + c_{\mu,ion} + c_{\mu,p}} \approx c_{\mu,p}$$

Figure S15. Y-Δ transformation for the chemical capacitance network in the circuit shown in Figure 2, neglecting the connection of electrostatic capacitors to the electrostatic rail. The chemical capacitors $c_n^\delta$ and $c_p^\delta$ are included in the network of Figure 5, while $c_{\mu,np}$ can be neglected, provided that $c_{\mu,ion} \gg c_{\mu,n}$ and $c_{\mu,ion} \gg c_{\mu,p}$. If that is the case, $v_{ion} \approx v_e$ is still valid in the bulk, and the $v_{ion}$ nodes can be connected to the electrostatic rail, as discussed in the main text and in Appendix C. The resulting network effectively accounts for the electronic chemical capacitance (storage of electrons and holes). For this reason, the $\hat{c}$ element used in Figure 4 is not required in Figure 5.

## 10. Impedance calculation using MATLAB codes

Supporting files with MATLAB codes in section 15 of this document include:

- Z_Voc_DD_ECM: this script can be run in combination with the Driftfusion software, repository: https://github.com/barnesgroupICL/Driftfusion/tree/2022-EA_SDP_EIS. [3] The code returns calculated impedance spectra evaluated at open circuit for solar cells using the drift-diffusion calculation, the complete transmission line (Figure S3) and the IC and MC-i approximated models. Which solution to calculate can be selected in the first section of the code, along with the details on frequency range, light intensities etc. The calculation time becomes long for input files containing a large number of mesh points. In addition, discontinuity in mesh spacing should be minimized at the interfaces between layers and junctions to improve correspondence between the drift-diffusion and the circuit model solution. Spectra in the main text and in this document that are obtained using this code refer to input files with the following thicknesses and number of points:
    - HTM: 200 nm, 107 points
    - Junction interface 1: 2nm, 50 points
    - Active layer: 360 nm, 186 nm
    - Junction interface 2: 2 nm, 50 points
    - ETH: 200 nm, 107 points

  Such mesh leads to small discontinuity in spacing at the interfaces and reasonable calculation time. In the input file, the values of the electrodes' work function Phi_left and Phi_right should be set to match the work function (E0) of the HTM and of the ETM respectively.

- Approx_IC_MCi: this script is a toy model for the IC and MC-i approximations introduced in this study and it can be run independently. The user can input values for the steady-state parameters describing the solar cell (bulk charge concentrations and mobilities, space charge potentials, electronic lifetimes and details on traps in the bulk and at interfaces). The interfacial capacitors are calculated based on the input space charge potentials. The frequency dependent small signal electrostatic potential and the total impedance are calculated based on the analysis presented in Appendix C. This is plotted and saved, along with the calculated impedance.

Figure S16–20 show spectra obtained with the 'Approx_IC_MCi' code. All these examples consider the same asymmetry in the contacts' dielectric properties and a fixed set of interfacial space charge potentials. By varying the bulk concentration of electrons and holes and the capture time constants in the bulk, at interface 1 and at interface 2, different low frequency impedance features are obtained.

Figure S16 shows Nyquist plots of impedance data where the IC approximation reproduces well the low frequency impedance. Here, the electronic concentration is small enough so that $c_n^\delta$ and $c_p^\delta$ are small compared with the interfacial capacitors, but they are large enough to induce a difference between the results obtained with the IC or the MC-i approximations at low frequencies.

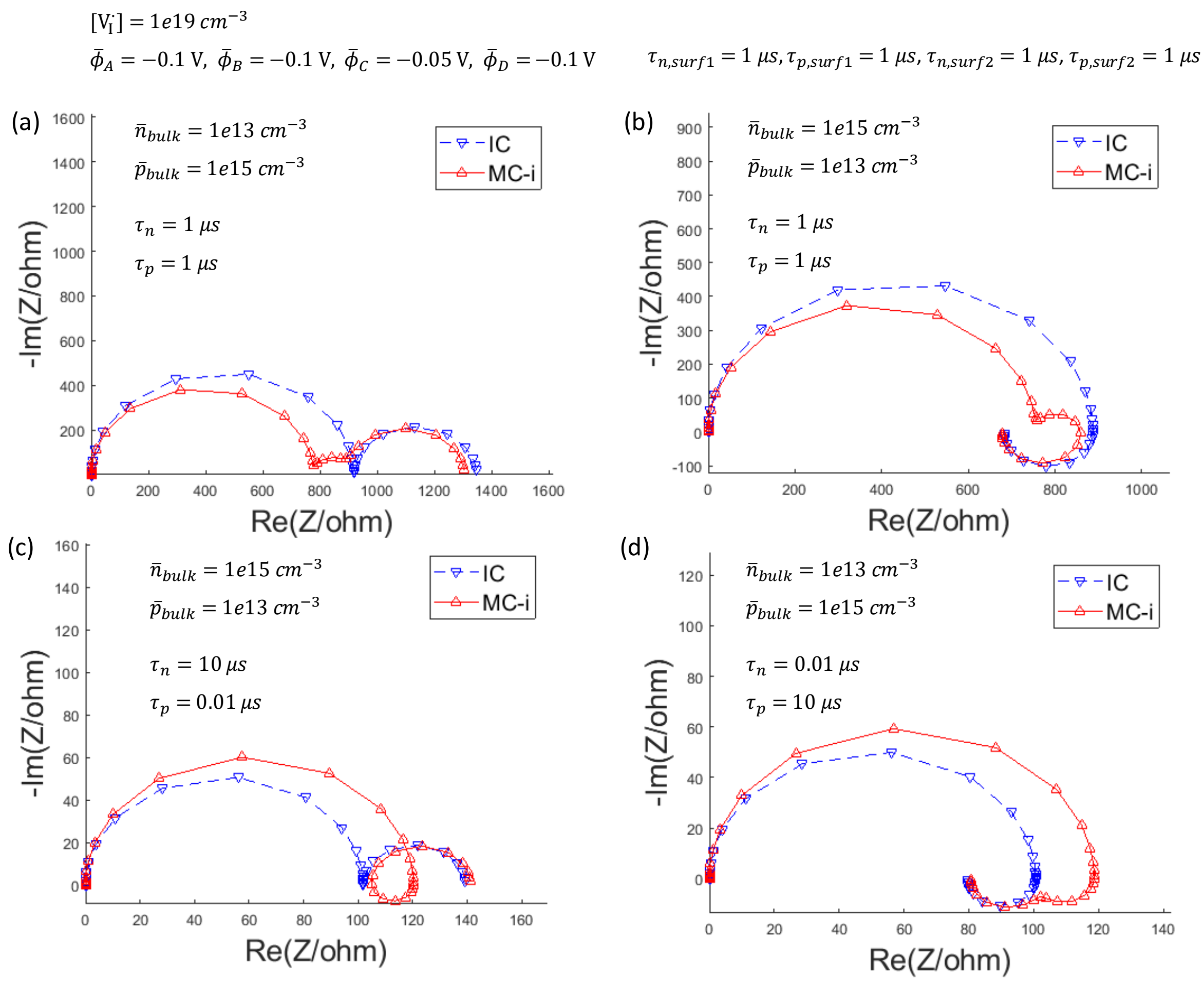

Figure S16. Calculated impedance (frequency range $1\ \text{MHz} - 10\ \text{mHz}$) for different recombination situations. In all cases, bulk recombination dominates, giving rise to two evident low frequency features. Within each panel, the electron and hole bulk concentrations and capture lifetimes are specified. The remaining parameters used for the calculations are shown on top of the figure and have been used for all panels. For the contacts, $N_A = N_D = 1e18\ cm^{-3}$, $\epsilon_{HTM} = 10$, $\epsilon_{ETM} = 20$ and for the active layer $\epsilon = 32$. All remaining parameters are the same as in Table S3.

Figure S17–20 show examples where $c_n^\delta$ or $c_p^\delta$ are comparable or even larger than the interfacial capacitors. In these cases, the IC approximation fails to reproduce the low frequency impedance (see also Appendix C). As shown in these examples, the code can be used to explore the sensitivity of the impedance calculated with IC and MC-i approximation to each input parameter, or as a base for impedance fitting procedures.

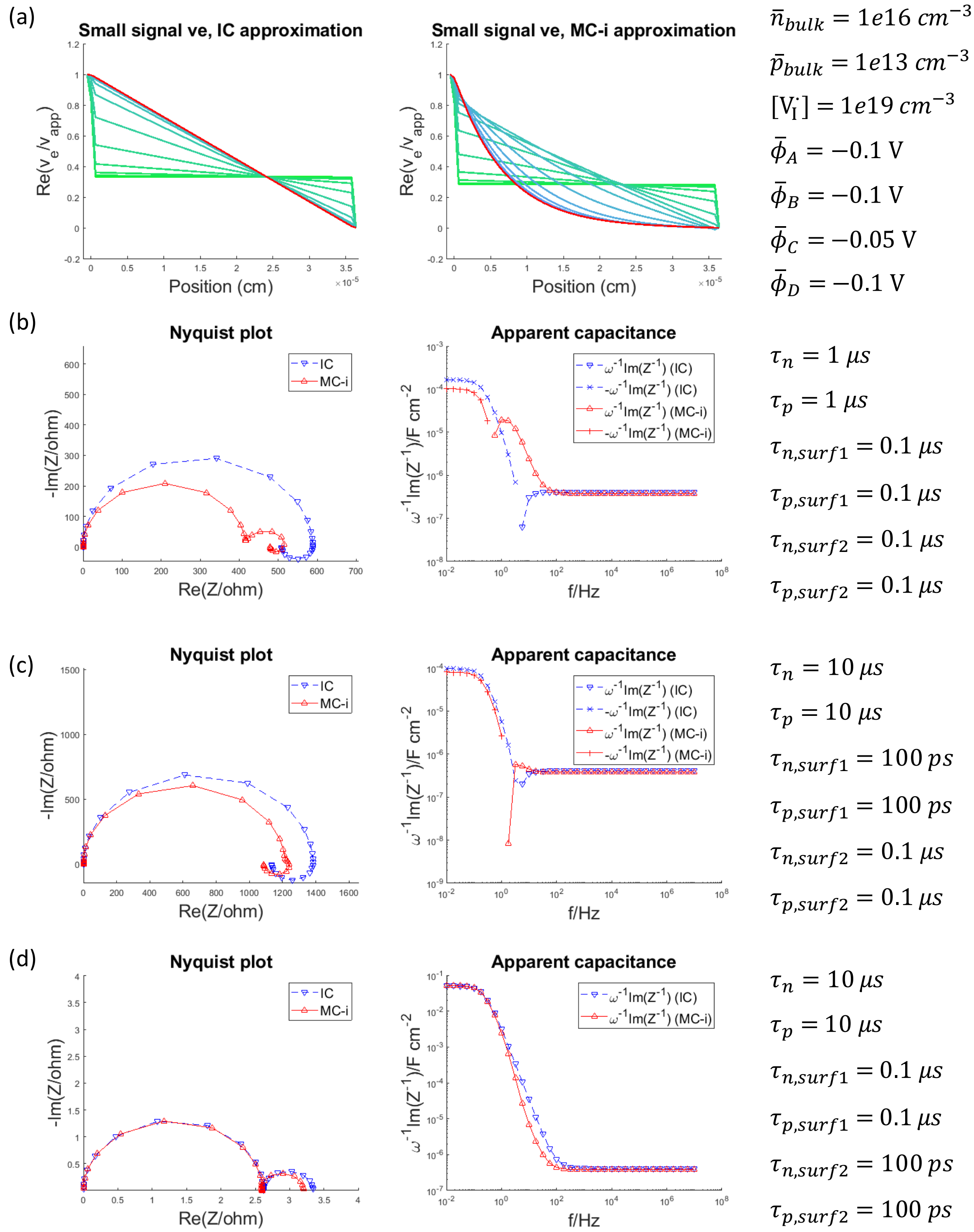

Figure S17. (a) Calculated small signal electrostatic potential for the case $\bar{n}_{bulk} \gg \bar{p}_{bulk}$. Calculated impedance (frequency range $1\ \mathrm{MHz} - 10\ \mathrm{mHz}$) for different recombination situations: (b) bulk recombination dominates; (c) surface recombination at interface 1 dominates; (d) surface recombination at interface 2 dominates. For the contacts, $N_A = N_D = 1e18\ cm^{-3}$, $\epsilon_{HTM} = 10$, $\epsilon_{ETM} = 20$ and for the active layer $\epsilon = 32$. All remaining parameters are the same as in Table S3.

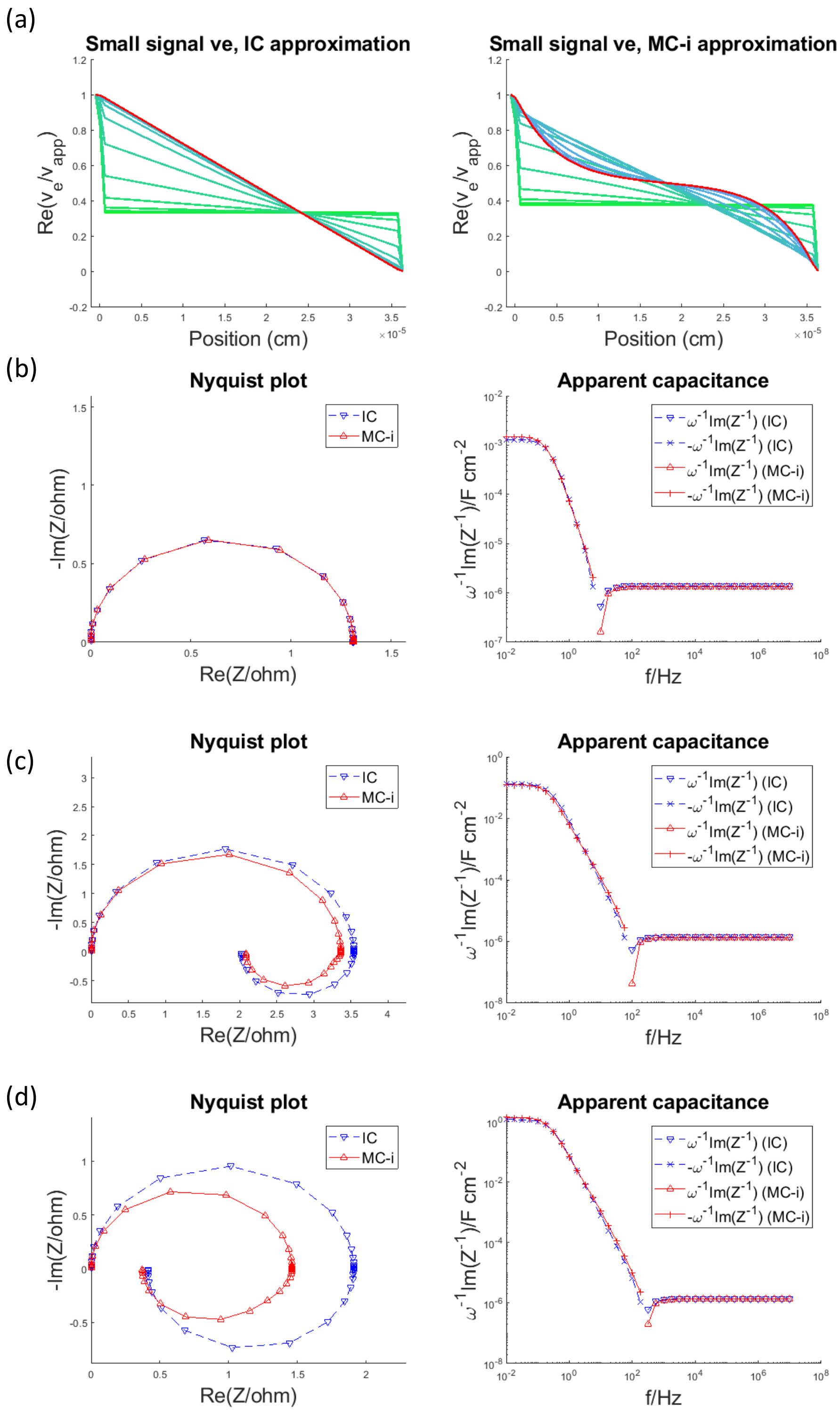

(a) $\bar{n}_{bulk} = 1e16\ cm^{-3}$, $\bar{p}_{bulk} = 1e16\ cm^{-3}$, $[\mathrm{V_I^{\cdot}}] = 1e19\ cm^{-3}$, $\bar{\phi}_A = -0.1$ V, $\bar{\phi}_B = -0.1$ V, $\bar{\phi}_C = -0.05$ V, $\bar{\phi}_D = -0.1$ V

(b) $\tau_n = 1\ \mu s$, $\tau_p = 1\ \mu s$, $\tau_{n,surf1} = 0.1\ \mu s$, $\tau_{p,surf1} = 0.1\ \mu s$, $\tau_{n,surf2} = 0.1\ \mu s$, $\tau_{p,surf2} = 0.1\ \mu s$

(c) $\tau_n = 10\ \mu s$, $\tau_p = 10\ \mu s$, $\tau_{n,surf1} = 100\ ps$, $\tau_{p,surf1} = 100\ ps$, $\tau_{n,surf2} = 0.1\ \mu s$, $\tau_{p,surf2} = 0.1\ \mu s$

(d) $\tau_n = 10\ \mu s$, $\tau_p = 10\ \mu s$, $\tau_{n,surf1} = 0.1\ \mu s$, $\tau_{p,surf1} = 0.1\ \mu s$, $\tau_{n,surf2} = 100\ ps$, $\tau_{p,surf2} = 100\ ps$

Figure S18. (a) Calculated small signal electrostatic potential for the case $\bar{n}_{bulk} = \bar{p}_{bulk}$. Calculated impedance (frequency range $1\ \mathrm{MHz} - 10\ \mathrm{mHz}$) for different recombination situations: (b) bulk recombination dominates; (c) surface recombination at interface 1 dominates; (d) surface recombination at interface 2 dominates. For the contacts, $N_A = N_D = 1e18\ cm^{-3}$, $\epsilon_{HTM} = 10$, $\epsilon_{ETM} = 20$ and for the active layer $\epsilon = 32$. All remaining parameters are the same as in Table S3.

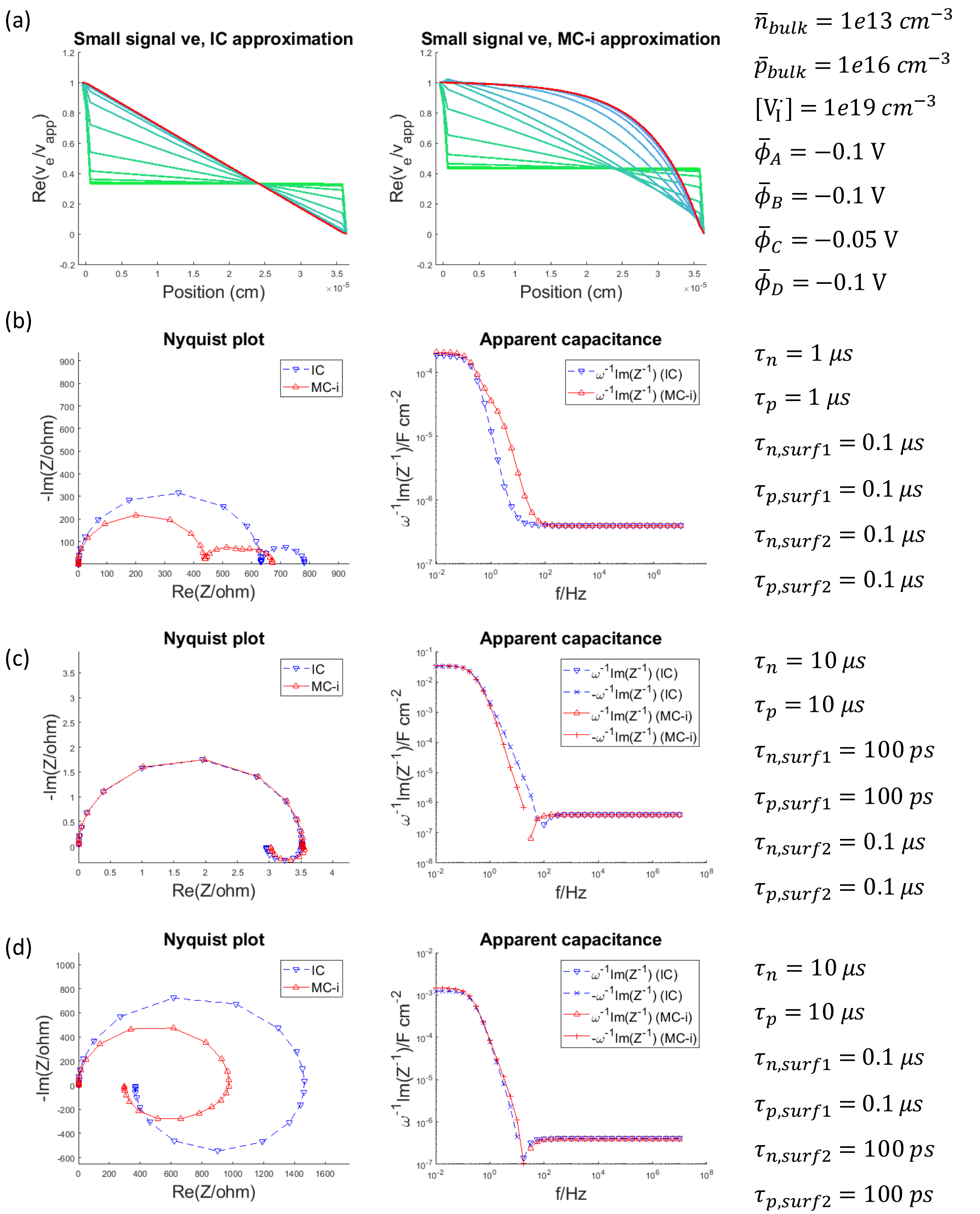

Figure S19. (a) Calculated small signal electrostatic potential for the case $\bar{n}_{bulk} \ll \bar{p}_{bulk}$. Calculated impedance (frequency range $1\ \mathrm{MHz} - 10\ \mathrm{mHz}$) for different recombination situations: (b) bulk recombination dominates; (c) surface recombination at interface 1 dominates; (d) surface recombination at interface 2 dominates. For the contacts, $N_A = N_D = 1e18\ cm^{-3}$, $\epsilon_{HTM} = 10$, $\epsilon_{ETM} = 20$ and for the active layer $\epsilon = 32$. All remaining parameters are the same as in Table S3.

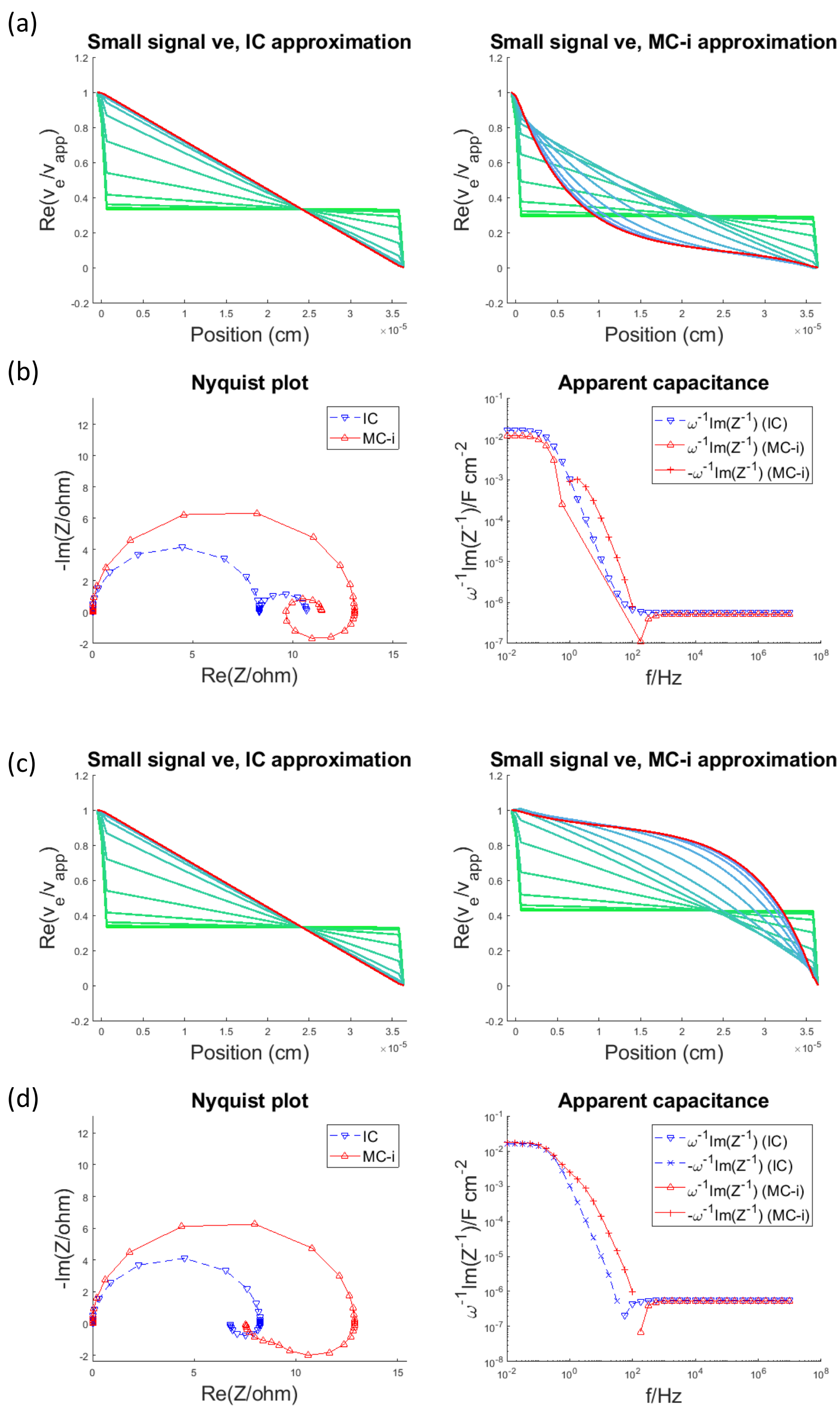

$\bar{n}_{bulk} = 1e16\ cm^{-3}$

$\bar{p}_{bulk} = 1e15\ cm^{-3}$

$[V_I^{\cdot}] = 1e19\ cm^{-3}$

$\bar{\phi}_A = -0.1$ V

$\bar{\phi}_B = -0.1$ V

$\bar{\phi}_C = -0.05$ V

$\bar{\phi}_D = -0.1$ V

$\tau_n = 10\ \mu s$

$\tau_p = 0.1\ \mu s$

$\tau_{n,surf1} = 1\ \mu s$

$\tau_{p,surf1} = 1\ \mu s$

$\tau_{n,surf2} = 1\ \mu s$

$\tau_{p,surf2} = 1\ \mu s$

$\bar{n}_{bulk} = 1e15\ cm^{-3}$

$\bar{p}_{bulk} = 1e16\ cm^{-3}$

$[V_I^{\cdot}] = 1e19\ cm^{-3}$

$\bar{\phi}_A = -0.1$ V

$\bar{\phi}_B = -0.1$ V

$\bar{\phi}_C = -0.05$ V

$\bar{\phi}_D = -0.1$ V

$\tau_n = 0.1\ \mu s$

$\tau_p = 10\ \mu s$

$\tau_{n,surf1} = 1\ \mu s$

$\tau_{p,surf1} = 1\ \mu s$

$\tau_{n,surf2} = 1\ \mu s$

$\tau_{p,surf2} = 1\ \mu s$

Figure S20. Calculated small signal electrostatic potential for the case (a) $\bar{n}_{bulk} \ll \bar{p}_{bulk}$ and (c) $\bar{n}_{bulk} \gg \bar{p}_{bulk}$. (b, d) Calculated impedance (frequency range $1\ \mathrm{MHz} - 10\ \mathrm{mHz}$) for situations where the majority carriers in the bulk presents significantly longer lifetime than the minority carriers, limiting the overall recombination (in all cases the bulk dominates recombination): (b) $\tau_n \gg \tau_p$; (d) $\tau_n \ll \tau_p$. As a result, the impedance feature associated with the bulk polarization process presents a negative apparent capacitance. For the contacts, $N_A = N_D = 1e18\ cm^{-3}$, $\epsilon_{HTM} = 10$, $\epsilon_{ETM} = 20$ and for the active layer $\epsilon = 32$. All remaining parameters are the same as in Table S3.

## 11. Input parameters for impedance calculations

The following input parameters are used to obtain the data displayed in the main text and in this document using the Driftfusion software:

- The drift-diffusion steady-state solution used to calculate the impedance data in Figure 6 and Figure S21 is obtained considering the input parameters listed in Table S3. Details on the mesh are discussed in section 10 of this document.
- The drift-diffusion steady-state solution used to calculate the impedance spectra in Figure 7 is obtained with the same parameters as Table S3 with the following modifications:
    - the IP, $E_t$ and $\phi_W$ of the HTM are all shifted to more negative values by 0.1 eV
    - the EAff, $E_t$ and $\phi_W$ of the ETM are all shifted to less negative values by 0.1 eV
    - For the simulation where either $u_{surf1}$ or $u_{surf2}$ dominates, the capture time constants are set to: $\tau_n = \tau_p = 10\ \mu s$ in the bulk, $\tau_n = \tau_p = 100\ ps$ in the junction layer at the recombination dominant interface and $\tau_n = \tau_p = 10\ ns$ in the junction layer at the other interface.
    - For simulations where $u_{bulk}$ dominates, the value of $\tau_n = \tau_p$ in Table S3 are used.
- The drift-diffusion steady-state solution displayed in Figure S13 refers to the same dataset as the ones in Table S3 with the following changes:
    - the IP, $E_t$ and $\phi_W$ of the HTM and of the ETM are all shifted to more negative values by 0.1 eV.
    - In addition, the EAff of the HTM and the IP of the ETM are also changed so that both contact layers have a bandgap of 3 eV.
    - Finally, both transport layers have a dielectric constant of 10, to obtain a symmetric device.
- Data in Figure S14a are obtained starting from the input parameters used for Figure S13, with the changes described in the discussion of the Figure.
- The data in Figure S24–28 are obtained using the same input parameters used for the results displayed in Figure 7 (see above), and by varying the electron and hole charge carrier mobilities.

In all cases, the work function of each electrode is set to the same value as the work function of the charge transport layer (ETM or ETM) which it is in contact with.

Table S3. Drift-diffusion simulation parameters used in the calculations shown in Figure 6 (main text) and in Figure S21.

| Parameter name | Symbol | HTM | Active layer | ETM | Unit |
|---|---|---|---|---|---|
| Layer thickness | $d$ | 200 | 360 | 200 | nm |
| Band gap | $E_g$ | 2.7 | 1.6 | 3.2 | eV |
| Relative dielectric constant | $\varepsilon_s$ | 2 | 32 | 10 | |
| Mobile ionic defect density | $N_{ion}$ | 0 | $10^{19}$ | 0 | $cm^{-3}$ |
| Ion mobility | $u_c$ | - | $10^{-10}$ | - | $cm^2 V^{-1} s^{-1}$ |
| Electron mobility | $u_e$ | 0.1 | 20 | 0.1 | $cm^2 V^{-1} s^{-1}$ |
| Hole mobility | $u_h$ | 0.1 | 20 | 0.1 | $cm^2 V^{-1} s^{-1}$ |
| Acceptor doping density | $N_A$ | $2.1 \times 10^{18}$ | - | - | $cm^{-3}$ |
| Donor doping density | $N_D$ | - | - | $2.1 \times 10^{18}$ | $cm^{-3}$ |
| Work function | $\phi_W$ | -4.8 | -4.6 | -4.2 | eV |
| Ionization potential | $IP$ | -4.9 | -5.4 | -7.3 | eV |
| Electron affinity | $EAff$ | -2.2 | -3.8 | -4.1 | eV |
| Effective density of states | $N_c$, $N_V$ | $10^{20}$ | $10^{19}$ | $10^{20}$ | $cm^{-3}$ |
| Radiative rate constant | $k_{rad}$ | $10^{-11}$ | $3.6\ 10^{-12}$ | $10^{-11}$ | $cm^{-3} s^{-1}$ |
| SRH trap energy | $E_t$ | $E_{VB}$+0.15 | $E_{CB}$-0.8 | $E_{CB}$-0.15 | eV |
| SRH time constants | $\tau_n$, $\tau_p$ | $10^{-9}$ | $10^{-6}$ | $10^{-9}$ | s |

## 12. Discussion and interpretation of bias dependent impedance

### *12.1 Impedance calculation with efficient electronic transport*

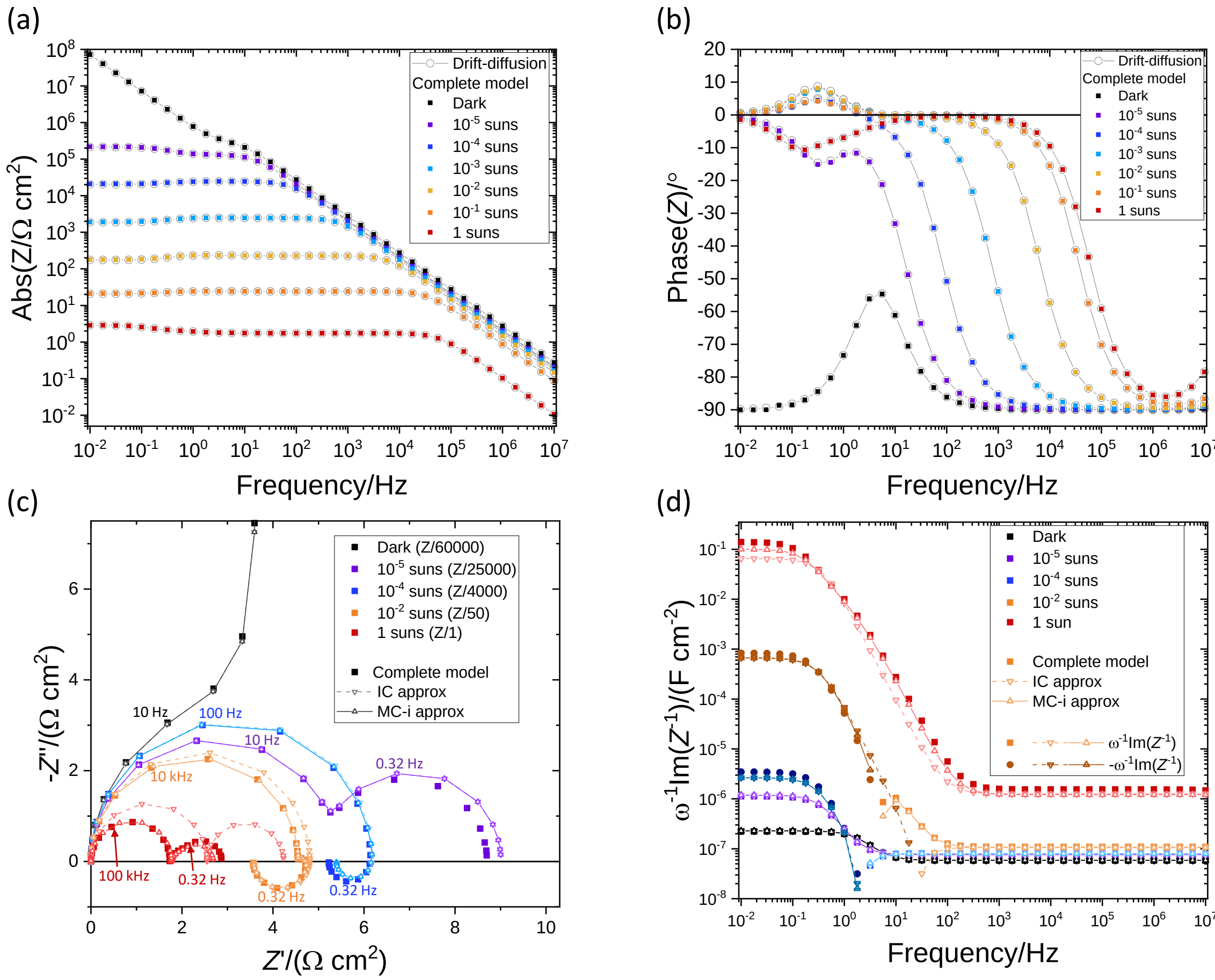

Figure S21. Impedance of a mixed conducting semiconductor based solar cell calculated at open circuit and for different bias light intensities using either drift-diffusion or the transmission line transistor based equivalent circuit model presented in this study (same data shown in Figure 6 in the main text). Bode plots of the (a) magnitude and (b) phase of the impedance (same as in Figure 6). (c) Nyquist representation and (d) apparent capacitance. The data obtained using the equivalent circuit model in Figure S3 (complete model) are compared with drift-diffusion simulation results in (a, b), while in (c, d) they are plotted along with the impedance evaluated using the IC and the MC-i approximated models described in Figure 4a, b and Figure 5.

Figure S21 displays the same data shown in Figure 6. Guided by the circuit model representations, the data in Figure S21, taken here as example, are discussed further to exemplify the relevance of electron-hole recombination processes on the observed impedance.

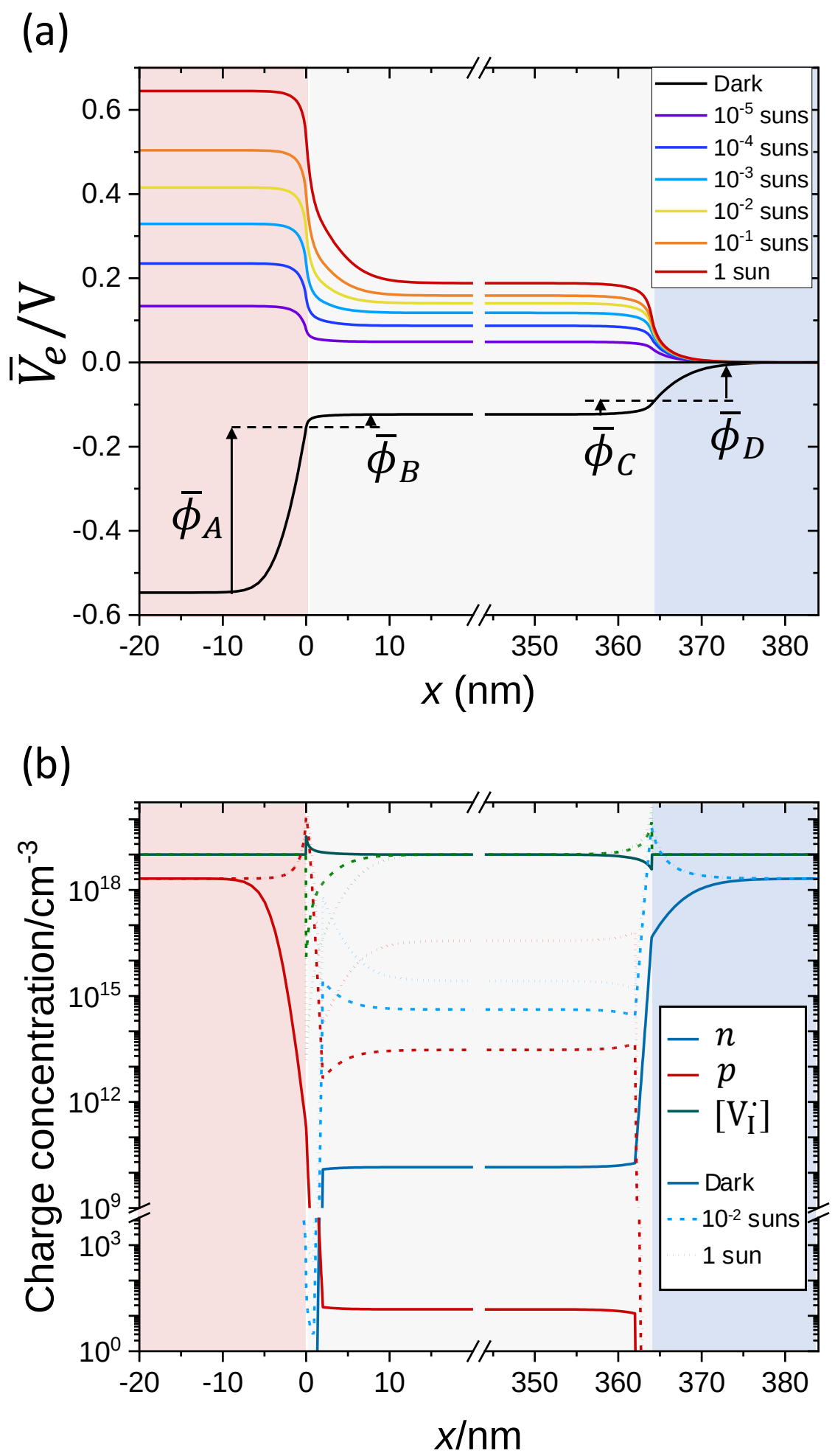

Figure S22. Steady-state properties of a mixed conducting (e.g. halide perovskite) solar cell at open circuit in the dark and under different light intensities obtained from drift-diffusion simulations. (a) Electrostatic potential and (b) charge concentrations as function of position in the device for selected conditions. $[V_I^\cdot]$ refers to the concentration of iodide vacancies, the mobile ionic species assumed in this example.

First, it is important to clarify the steady-state properties of the device under the different conditions, as they determine the recombination regime that is relevant at each position in the active layer (see FigureS2 and Table S1). Figure S22a illustrates the steady-state electrostatic potential $\bar{V}_e$ in the device for the different light intensities used to obtain the data in Figure S21. In the dark, the equilibrium space charge potentials add up to the built-in potential (a small deviation from the input value of $\phi_{bi} = 0.6$ V is present for the drift-diffusion solution in the dark). When varying the light intensity, and therefore the value of steady-state applied potential ($\bar{V}_{app} = V_{OC}$, $V_{OC}$ is the open circuit potential), two regimes are identified. For small values of the applied bias, the steady-state interfacial space charge potentials decrease in magnitude compared with the dark situation, while at large bias their values become negative and increase in magnitude. It follows that the electronic carriers that are injected at each contact (holes at interface 1 and electrons at interface 2) are accumulated at the interface while the other carrier is depleted, only in the dark and at low enough bias. At large enough bias (specifically when $\bar{V}_{app} > \phi_{bi}$), the opposite trend is found, as shown in Figure 22b. For the solar cell parameters considered here, already at the lowest light intensity $V_{OC} > \phi_{bi}$ and all interfacial space charge potentials are negative.

In addition, Figure S22b highlights a change from n-type to p-type of the bulk for increasing light intensity. This is due to a gradual decrease in the mobile ion bulk concentration, resulting from the ion redistribution at the interfaces. This is compensated by an increase in hole concentration in the bulk (see section 8 of this document for details). The steady-state charge concentration profiles have consequences on the dominant recombination transconductance in the device, as discussed below.

Before commenting on the behavior under large bias, it is important to discuss the contribution of ionic conduction to the impedance at low bias. This is significant in the dark, where a high frequency feature is essentially related with the ionic resistance $r_{ion,bulk}$ and the geometric capacitance of the device. [1] The contribution of $r_{ion,bulk}$ remains significant also at low electronic charge concentrations: at $10^{-5}$ suns, a low frequency feature with positive apparent capacitance is obtained, although the frequency dependent recombination current contributes towards a negative apparent capacitance behavior (see discussion below). The latter starts to dominate only at higher light intensities ($10^{-4}$ suns).

Figure S23 exemplifies the interpretation of the impedance features observed in Figure 6 and S21 further, by illustrating the real part of $v_e/v_{app}$, $Re(v_e/v_{app})$ (top row), the relevant recombination (trans)conductance profiles (middle row), and the resulting recombination current per unit volume based on Equations 16 and 17 in the main text (bottom row), for three light intensities. The inset of the bottom row displays the corresponding impedance spectra which, at low frequencies, are largely determined by the recombination response under large enough bias ($Z(\omega) \approx Z_{eon}(\omega) \approx \frac{v_{app}}{j_{rec,tot}(\omega)}$).

Similarly to the discussion in the main text, the $Re(v_e/v_{app})$ profiles highlight the changes in recombination driving force. The frequency dependent $Re\left(v_e/v_{app}\right)$ data for the $10^{-4}$ suns light intensity case displays the space charge polarization occurring at low frequencies. The (trans)conductance $g_{rec,p}$ is largest in this case, as holes are minority carriers across the active layer (see Table 1 and 2 in the main text). The resulting small signal net recombination current contributions per unit volume associated with the *npn* and *pnp* transistors are shown for two different frequencies, one lower and one higher than the frequency range over which the polarization occurs (see low frequency impedance feature in the inset). As the recombination associated with the *pnp* transistors integrated across the device is dominant, the frequency dependent change in $v_{rec,p}$ determines the shape of $Z(\omega)$ at low frequencies. In this specific case, a negative apparent capacitance (or inductive behavior) is observed, as also evident from Figure S21d.

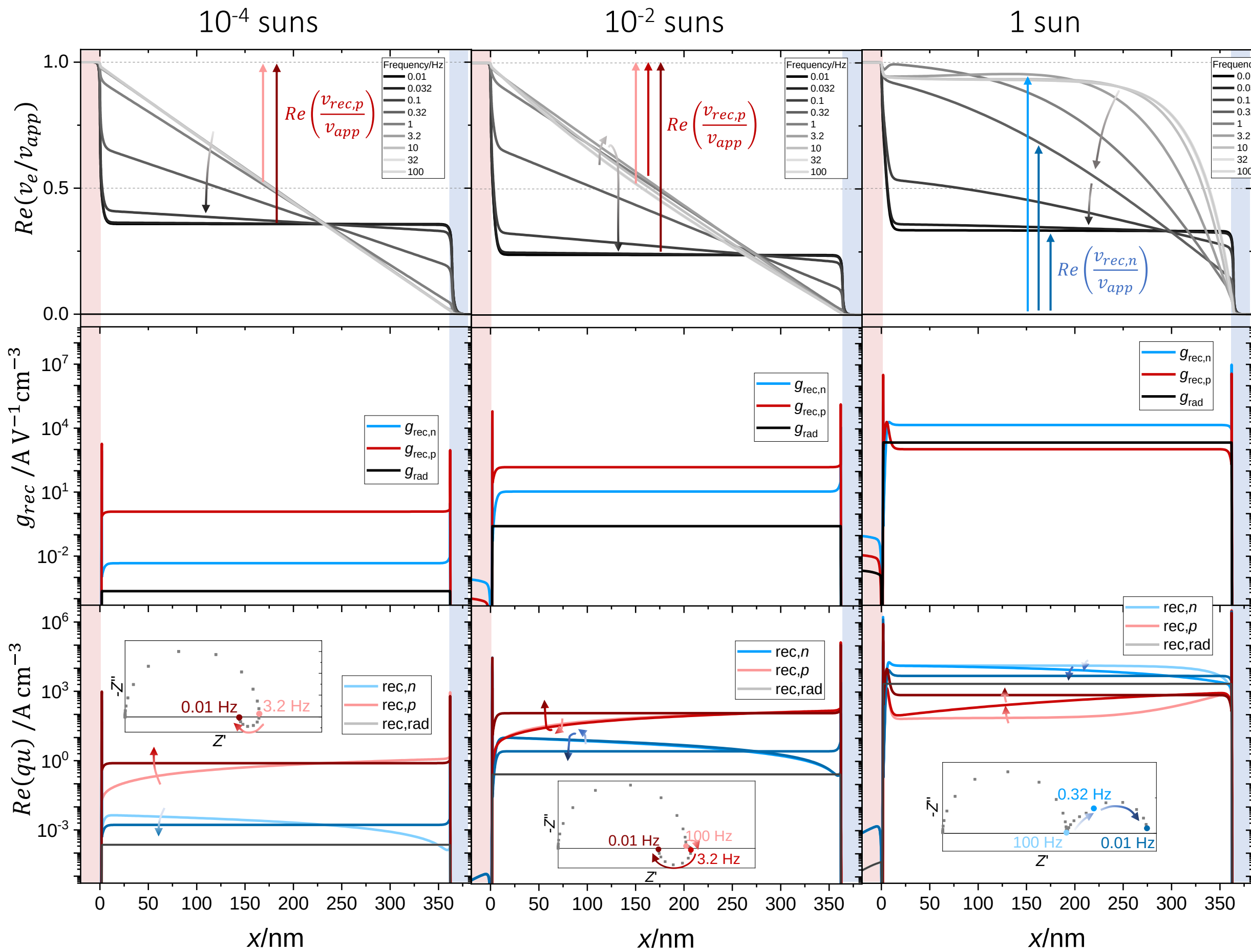

Figure S23. (top row) Real part of the normalized small signal electrostatic potential as a function of position in the device plotted for different frequencies and for the case of (left to right) $10^{-4}$, $10^{-2}$ and 1 sun illumination. The recombination driving force of the dominant mechanism is highlighted. Below each panel, (middle row) the corresponding recombination (trans)conductance profiles are shown as well as (bottom row) the resulting net recombination current components per unit volume. In the inset of the bottom graphs, the Nyquist plots of the corresponding impedance are shown, highlighting the low frequency feature(s) for which the net recombination terms are displayed.

At higher light intensities ($10^{-2}$ suns), the $Re(v_e/v_{app})$ profile in the active layer at high frequencies is influenced by the electronic charging of the chemical capacitors. This occurs already at very fast time scales, leading to a non-linear profile of the small signal electrostatic potential, which dictates the recombination voltages and current in the bulk and at interfaces at high frequencies. A linear profile of $v_e$ is restored at lower frequencies, at which ambipolar diffusion of electronic and ionic charges “can follow” the applied voltage perturbation. Such polarization occurs over time scales determined by the chemical diffusion of the neutral component (ionic + electronic) in the bulk of the active layer. This process gives rise to an additional impedance feature, which can be assigned to the change in recombination voltage due to the transition from a ‘curved’ to linear profile of $v_e$. In this example, the space charge polarization described above occurs at even lower frequencies.

Finally, the data referring to the high light intensity (1 sun) in Figure S23 (right column) show a qualitatively similar evolution of $v_e$ to the $10^{-2}$ suns data, with two clear differences. First, the curvature of $v_e$ at very high frequencies is much more pronounced, due to the even larger electronic charge concentration at such bias condition. A small 'overshoot' in $v_e$ is also observed during the polarization process. Secondly, such curvature has opposite sign for the calculated data at 1 sun compared with the $10^{-2}$ suns case. This is because these two scenarios involve electrons and holes, respectively, as electronic majority carriers in the bulk (see Figure S22b). This change in carrier type has also consequences on which recombination transconductance and recombination current is dominant. In this case, it results in a change of the sign from negative to positive for the apparent capacitance associated with the low frequency feature when going from the $10^{-2}$ suns to the 1 sun situation, respectively (see also Figure S21d).

From the discussion of the example in Figures S21, it becomes clear that several device properties (e.g., electronic charge carrier lifetimes, contact doping, energy level alignment) contribute to the 'multi-feature' impedance of a mixed conducting device, as also described in Ref. [5]. The discussion in this study allows the generalization of such picture also in terms of equivalent circuit models. Specifically, the analysis presented here emphasizes how the frequency dependence of the recombination voltage (driving force) associated with the dominant recombination process determines the overall low frequency impedance spectrum.

### *12.2 Impedance calculation with electronic transport limitation*

Figures S24–27 illustrate the effect of electronic transport limitations on the small signal response of the mixed conducting solar cell model system considered in this study. Based on the same input parameters used in the calculations displayed in Figure 7c, e and f, the effect of reducing the mobility of electrons and holes from a value of 20 $cm^2\ V^{-1}\ s^{-1}$ (left column) to values of 0.02 (middle column) and 0.002 $cm^2\ V^{-1}\ s^{-1}$ (right column) is shown. Situations where the dominant recombination contribution is localized at the interface between the active layer with contact 1 (Figure S24 and S25) or with contact 2 (Figure S26 and S27) are considered. For each input parameter set, results are calculated for open circuit and two light intensity conditions ($10^{-4}$ and $10^{-2}$ suns)

The data highlight the following:

- *Steady-state charge concentrations* (panels a, b and c). Reducing the electronic mobilities leads to increasingly larger variation in the concentration of the electronic minority carrier across the active layer thickness. This observation stresses that the approximation $v_p - v_n = v_{app}$ is no longer valid when the rate of transport is not sufficiently fast compared with the rate of recombination.
- *Frequency dependent small signal electrochemical and electrostatic potential profiles* (panels d, e and f). In situations where electronic charge transport is sluggish, frequency dependent changes in the local charge carrier concentrations are not only due to the frequency dependence of the $v_e$ profile, but also of the electrochemical potentials of electrons and holes. This is especially the case for the minority carriers, in this case holes (see different $v_p$ profiles depending on frequency). For very small values of mobilities, also the potential associated with the electronic majority carriers (electrons) presents a significant frequency dependence.
- *Impedance spectra* (panels g, h and i). As a result of the above, the IC and the MC-i approximations, as defined in the main text, cannot reproduce the impedance calculated with the complete model for the case of slow electronic transport. If the rate of transport becomes rate limiting to $j_{tot}$, its frequency dependence is reflected in the impedance features of the complete model.
Specifically, the impedance spectra in Figures S24–27 show that:

- The spectra obtained with the complete model highlight that the overall magnitude of the impedance increases for transport limited situations.
- The frequency dependent impedance cannot be explained by only analyzing the change in $v_p, v_n$ and $v_e$ at the position associated with the dominant recombination contribution. In these figures, recombination occurs predominantly at either interface 1 or interface 2. In these two regions, large variations of $v_e$ are observed during the space charge polarization process at low frequencies, while very limited changes occur during the ambipolar diffusion of ions and electronic charges occurring at intermediate frequencies. According to the discussion in the main text and in the previous section, a single low frequency impedance feature with characteristic time $\tau^{\perp}$ would be expected. Nevertheless, the data show instances of impedance spectra with two low frequency features (Figure S25i and S27i). This is explained based on the frequency dependent changes in charge carrier concentration in the bulk during the two polarization processes, which affect transport and therefore the overall device impedance (see also next point).
- The low frequency feature(s) in the impedance spectra of a mixed conducting solar cell can provide direct information on the nature and location of the dominant recombination mechanism only when assuming fast transport. For example, the "complete model" spectrum in Figure S25f would suggest *dominant bulk recombination* and a significantly longer capture time constant of the majority carrier compared to the one of minority carrier (in this case $\tau_n \gg \tau_p$, see also discussion of Figure S16d). This is in contrast with the actual input parameters used here ($\tau_n = \tau_p$, and *dominant recombination at interface 1*). The presence of the low frequency features with negative capacitance is due to the frequency dependent change in the local chemical potential of the minority carriers ($v_{\mu,p} = v_p - v_e$) and consequent change in charge carrier concentration in the bulk of the device when decreasing the frequency of the perturbation. In the complete equivalent circuit model, such changes in transport are due to the $j_\mu$ terms, which are gated by the local $v_{\mu,p}$. As a result, the small signal interfacial recombination current is modulated at time scale associated with the dynamics of the small signal potentials not only at the interface but also in the bulk of the device.
- An additional feature is observed at high frequencies and at large bias for situations where low electronic mobility is used in the calculation. This is evident in the Nyquist representation in the case of Figure S27i. This feature is associated with the characteristic time of electronic charging of the device. To analyze this point in more detail the apparent capacitance spectra are displayed in Figure S28. The high frequency capacitance matches the geometric capacitance at low enough bias. This is the case in Figure S28a and b. At large bias, the high frequency capacitance is expected to exceed the geometric capacitance due to non-negligible contribution from the effective electronic capacitance, $\hat{c}$ (see Appendix B and section 6.1 of this document). Figure S28c–f highlights that this electronic contribution becomes relevant only at frequencies that are slower than the relevant rate of transport in the active layer. Its effect is also evident in the change in $v_p$ and $v_n$ at high frequencies, which indicate the electronic charging of the active layer (100 kHz vs 100 Hz profiles in Figure S27f). Importantly, at such large frequencies, no significant ionic redistribution occurs. The value of $\hat{c}$ can be dominated by the bulk electronic chemical capacitance, $c_{\mu,eon,tot}$, or by an interfacial space charge capacitance (see Figure S6). The former case is relevant to situations where both $\bar{n}$ and $\bar{p}$ are large enough so that $L_{D,p} \ll L$ and $L_{D,n} \ll L$, implying $c_{\mu,eon,tot} > c_{g,tot}$ and $\hat{c} \approx c_{\mu,eon,tot}$. The latter case concerns scenarios where $\bar{n}$ and $\bar{p}$ are very different, *e.g.* $L_{D,p} > L$ and $L_{D,n} < L$ and the electronic contribution to the capacitance is associated with the electronic capacitance of the interface where the majority carrier is blocked (in this case, the series of $c_A$ and $c_B$). Note that the definition of single values for $L_{D,p}$ and $L_{D,n}$ loses its meaning in situations where significant changes in electronic charge concentration occur across the active layer, while the comparison of the capacitance contributions remains meaningful.

- Depending on whether $c_{\mu,eon,tot}$ or an interfacial capacitance dominates the value of $\hat{c}$, ambipolar transport of both electrons and holes, or transport of the majority carrier, respectively, determines the time scale of the additional high frequency feature. In both Figure S25 and S27, electrons are the electronic majority carriers and $L_{D,p} > L$ and $L_{D,n} < L$. Therefore, the electronic contribution to the capacitance displayed in Figure S28c and d can be attributed to the space charge capacitance due to electron accumulation at interface 1 on the active layer side. Indeed, the high frequency capacitance increases for decreasing frequencies from a value of $\omega^{-1}Im(Z^{-1}) \approx c_{g,tot} \approx 70\ nF\ cm^{-2}$ at very high frequency to $\omega^{-1}Im(Z^{-1}) \approx c_B \approx 150 - 300\ nF\ cm^{-2}$ ($c_A \gg c_B$ in this situation, so $c_1 \approx c_B$). The characteristic frequency describing this transition is in the order of $10\ kHz - 1\ MHz$ depending on the input electronic mobility. In Figure S28e and f the apparent capacitance of the device calculated for the 10 suns illumination case is displayed. At such large bias and for the high electronic mobility case, $\bar{n}$ and $\bar{p}$ are large enough so that $L_{D,p} \ll L$ and $L_{D,n} \ll L$. It follows that $\omega^{-1}Im(Z^{-1}) \approx c_{\mu,eon,tot} \approx 1 - 10\ \mu F\ cm^{-2}$.
- For situations where low electronic mobility is used as input, the electronic charge concentrations can vary significantly within the active leading to a drop in the value of $c_{\mu,eon,tot}$ and to a more complex capacitance spectrum.

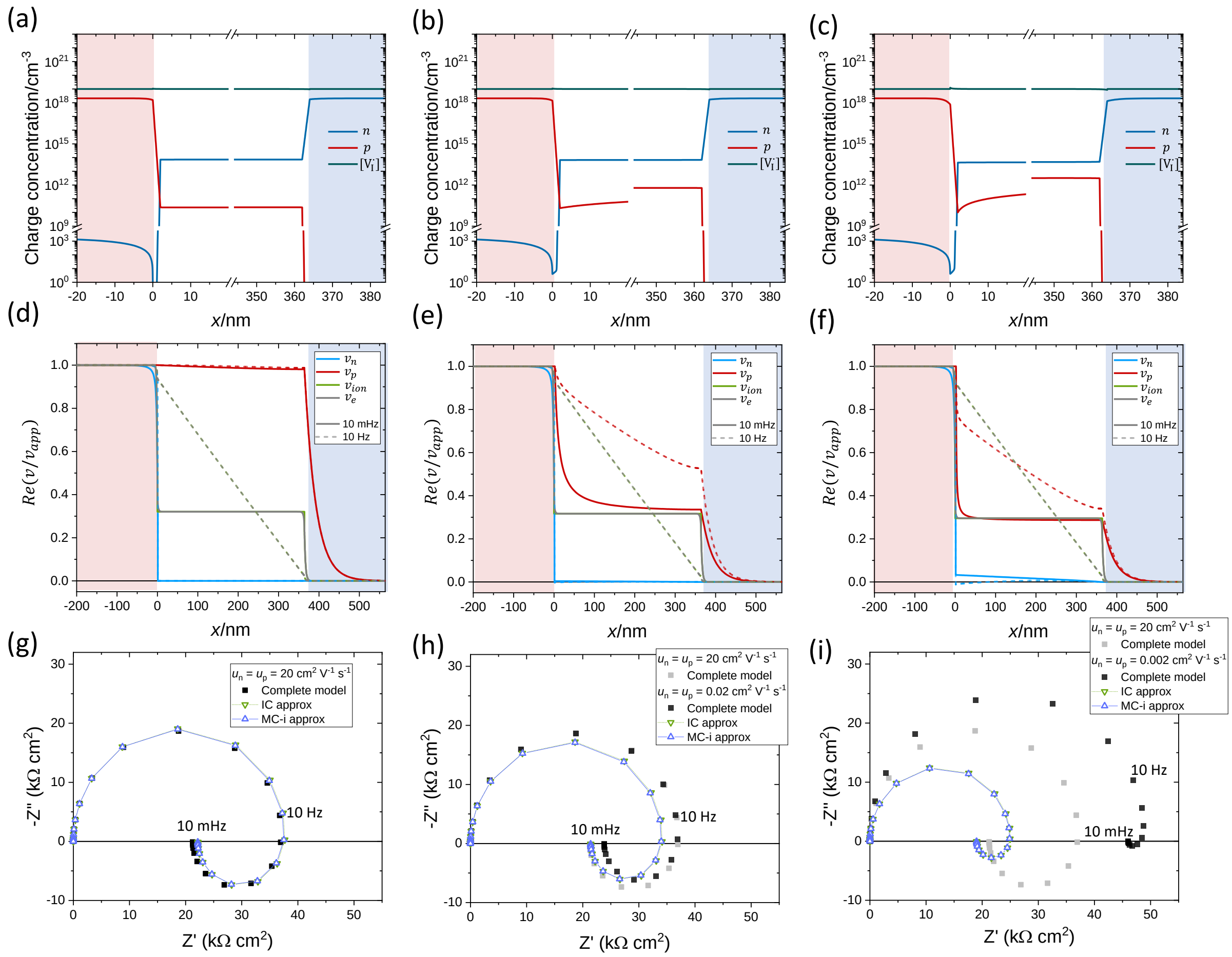

Figure S24. Analysis of the electrical response for a mixed conducting solar cell at open circuit considering different values of the electron and hole mobilities (left to right: 20, 0.02 and 0.002 $cm^2 V^{-1} s^{-1}$) under $10^{-4}$ suns illumination. Recombination at interface 1 is dominant. Top row: steady-state charge concentrations. Middle row: Real part of the normalized small signal electrostatic potential as a function of position in the device plotted for different frequencies. Bottom row: impedance spectra obtained with the complete model and with the IC and MC-i approximated models.

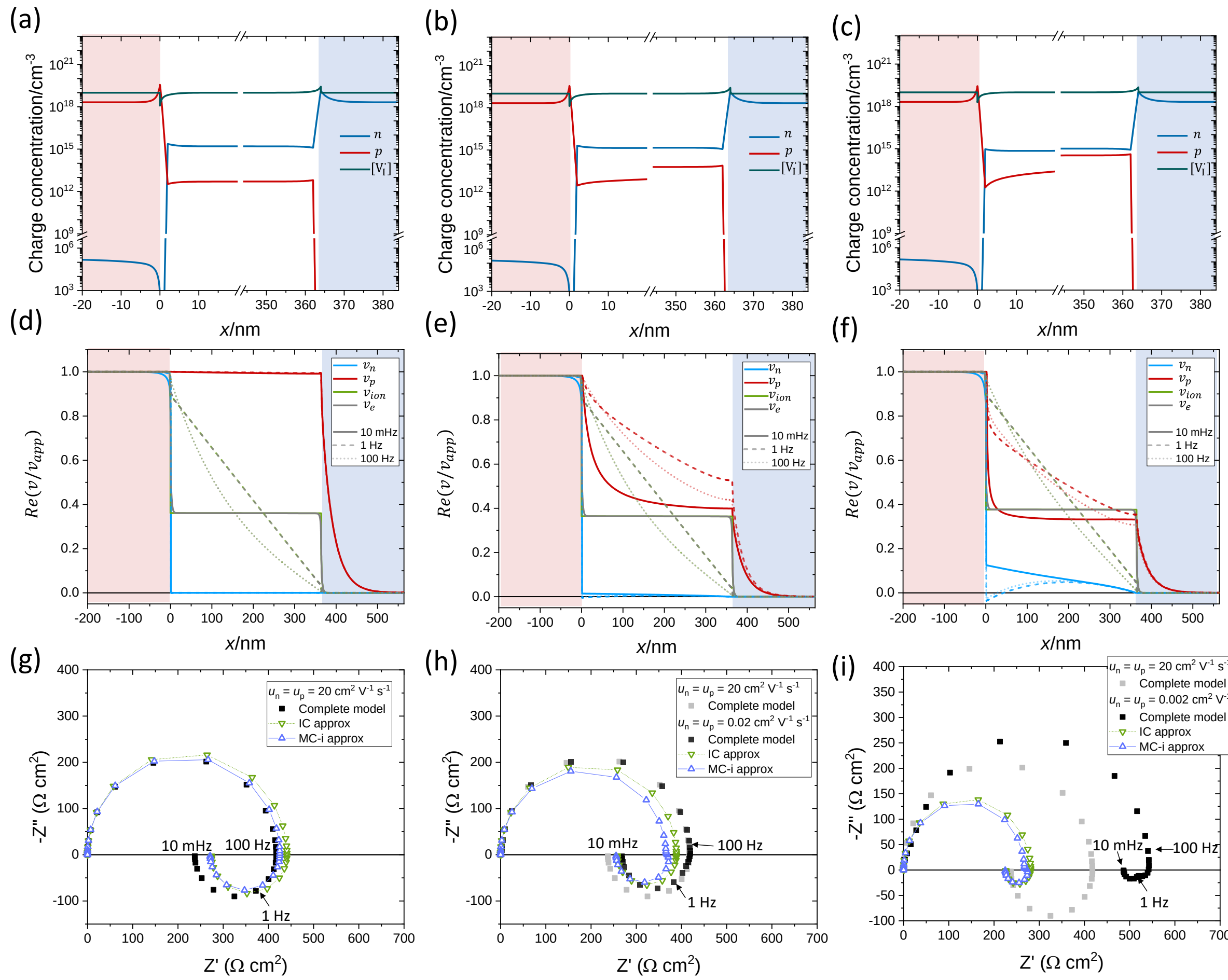

Figure S25. Analysis of the electrical response for a mixed conducting solar cell at open circuit considering different values of the electron and hole mobilities (left to right: 20, 0.02 and 0.002 $cm^2 V^{-1} s^{-1}$) under $10^{-2}$ suns illumination. Recombination at interface 1 is dominant. Top row: steady-state charge concentrations. Middle row: Real part of the normalized small signal electrostatic potential as a function of position in the device plotted for different frequencies. Bottom row: impedance spectra obtained with the complete model and with the IC and MC-i approximated models.

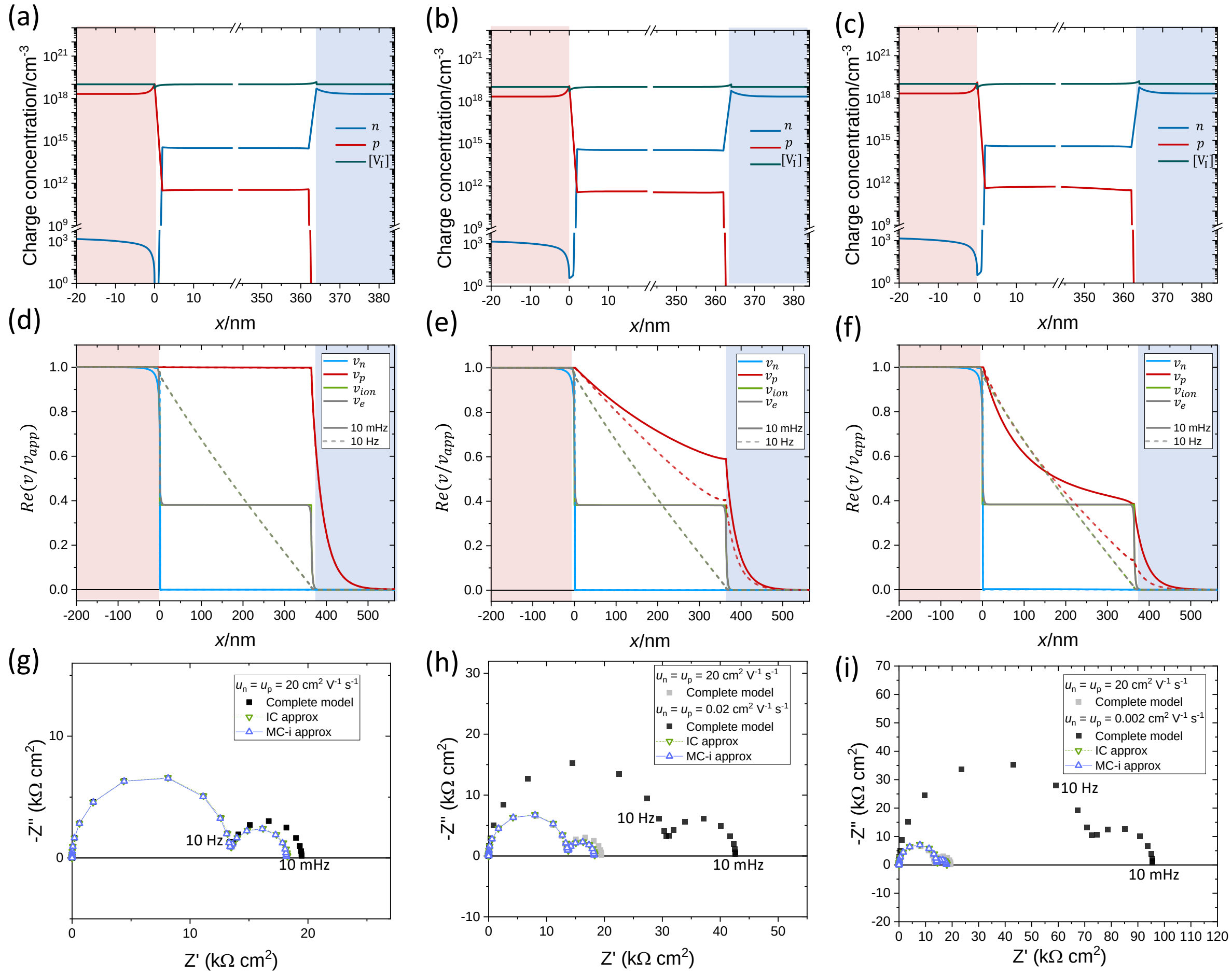

Figure S26. Analysis of the electrical response for a mixed conducting solar cell at open circuit considering different values of the electron and hole mobilities (left to right: 20, 0.02 and 0.002 $cm^2 V^{-1} s^{-1}$) under $10^{-4}$ suns illumination. Recombination at interface 2 is dominant. Top row: steady-state charge concentrations. Middle row: Real part of the normalized small signal electrostatic potential as a function of position in the device plotted for different frequencies. Bottom row: impedance spectra obtained with the complete model and with the IC and MC-i approximated models.

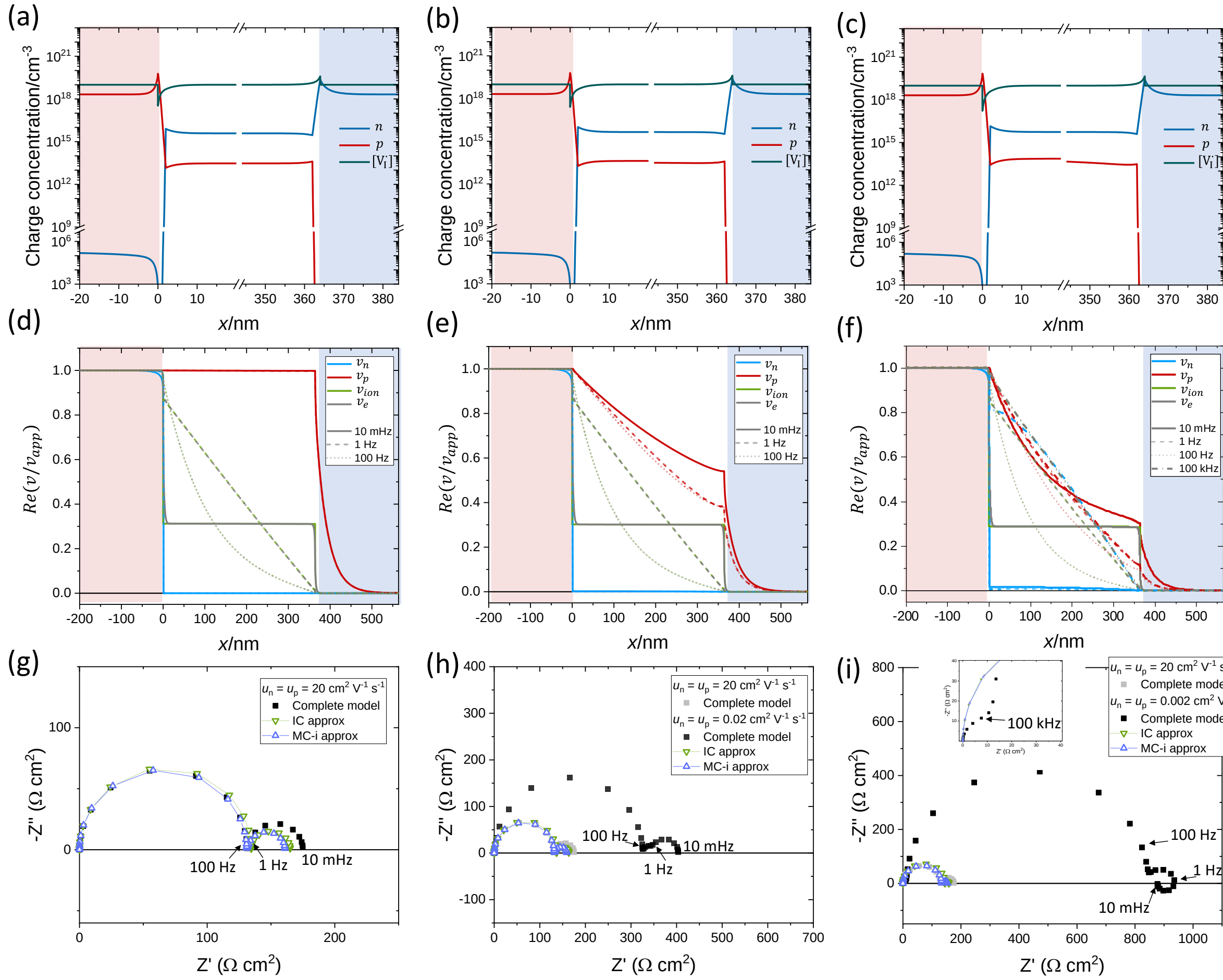

Figure S27. Analysis of the electrical response for a mixed conducting solar cell at open circuit considering different values of the electron and hole mobilities (left to right: 20, 0.02 and 0.002 $cm^2 V^{-1} s^{-1}$) under $10^{-2}$ suns illumination. Recombination at interface 2 is dominant. Top row: steady-state charge concentrations. Middle row: Real part of the normalized small signal electrostatic potential as a function of position in the device plotted for different frequencies. Bottom row: impedance spectra obtained with the complete model and with the IC and MC-i approximated models.

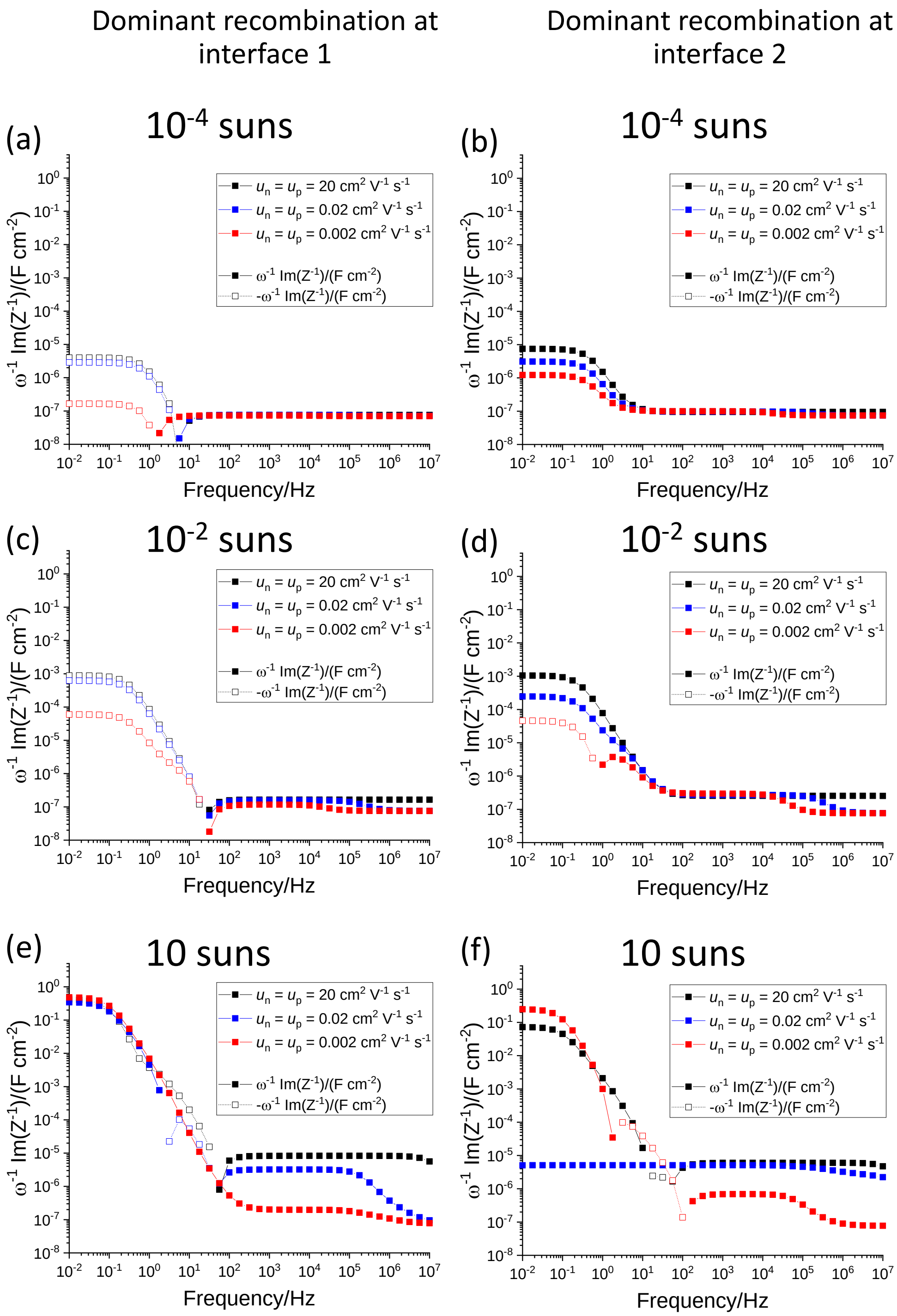

Figure S28. Apparent capacitance associated with the impedance data illustrated in (a) Figure S24g, h and i, (c) Figure S25g, h and i, (b) Figure S26g, h and i, and (d) Figure S27g, h and i. Left and right columns refer to the devices where recombination is dominant at interface 1 and interface 2, respectively. In (e) and (f), the apparent capacitance calculated for 10 suns illumination is illustrated.

## 13. Construction of the discretized equivalent circuit model

Two meshes are defined, $x_i$ ($i = 1,2, \dots\ N+1$) and $x_{i+1/2}$, where $x_{i+1/2} = (x_i + x_{i+1})/2$ ($i = 1, 2, \dots\ N$). In the construction of the transmission line model, the steady-state solution that is returned by the Driftfusion solver is defined on $x_{i+1/2}$ (in terms of electrostatic potential, $\bar{V}_{e,i+1/2}$, and charge concentrations, $\bar{n}_{i+1/2}$).

- Chemical capacitors elements of species $j$ are defined at each mesh point $x_{i+1/2}$ as $c_{\mu,j,i+1/2} = q\frac{n_{i+1/2,j}\Delta x_i}{V_{th}}$ consistent with their connection between electrochemical potential and electrostatic potential nodes in the same position. Here, $\Delta x_i = x_{i+1} - x_i$.
- Electrochemical transport resistors of species $j$ and electrostatic capacitors are defined on mesh $x_i$ as $r_{i,j} = \frac{\Delta x_{i+1/2}}{\sigma_{i,j}}$ and $c_{g,i} = \frac{\epsilon_{i+1/2}}{\Delta x_{i+1/2}}$, respectively, ($\Delta x_{i+1/2} = x_{i+3/2} - x_{i+1/2}$. Here, $\sigma_{i,j} = qn_{i,j}u_{i,j}$ is the partial conductivity of species $j$ ($u$ is the mobility) and $\epsilon_i$ is the dielectric constant of the material, both defined at position $x_i$.
- Voltage controlled current sources for species $j$ are defined on mesh $x_i$ as $j_{i,\mu,j} = g_{i,\mu,j}v_{i,\mu,j}$. Here, the transconductance $g_{i,\mu,j} = \frac{\bar{J}_{i,j}}{V_{th}}$ is defined based on the steady-state current associated with the *j*-th charge carrier. The change in chemical potential at position $x_i$ ($v_{i,\mu,j}$) is defined as the average between its values at the neighboring points of the voltage node mesh $v_{i,\mu,j} = \frac{v_{i+1/2,\mu,j}+v_{i-1/2,\mu,j}}{2}$.
- Recombination conductance and transconductance terms are defined at $x_{i+1/2}$ points, based on the expressions in the main text, multiplied by $\Delta x_i$ to obtain terms with A V$^{-1}$ cm$^{-2}$ units.

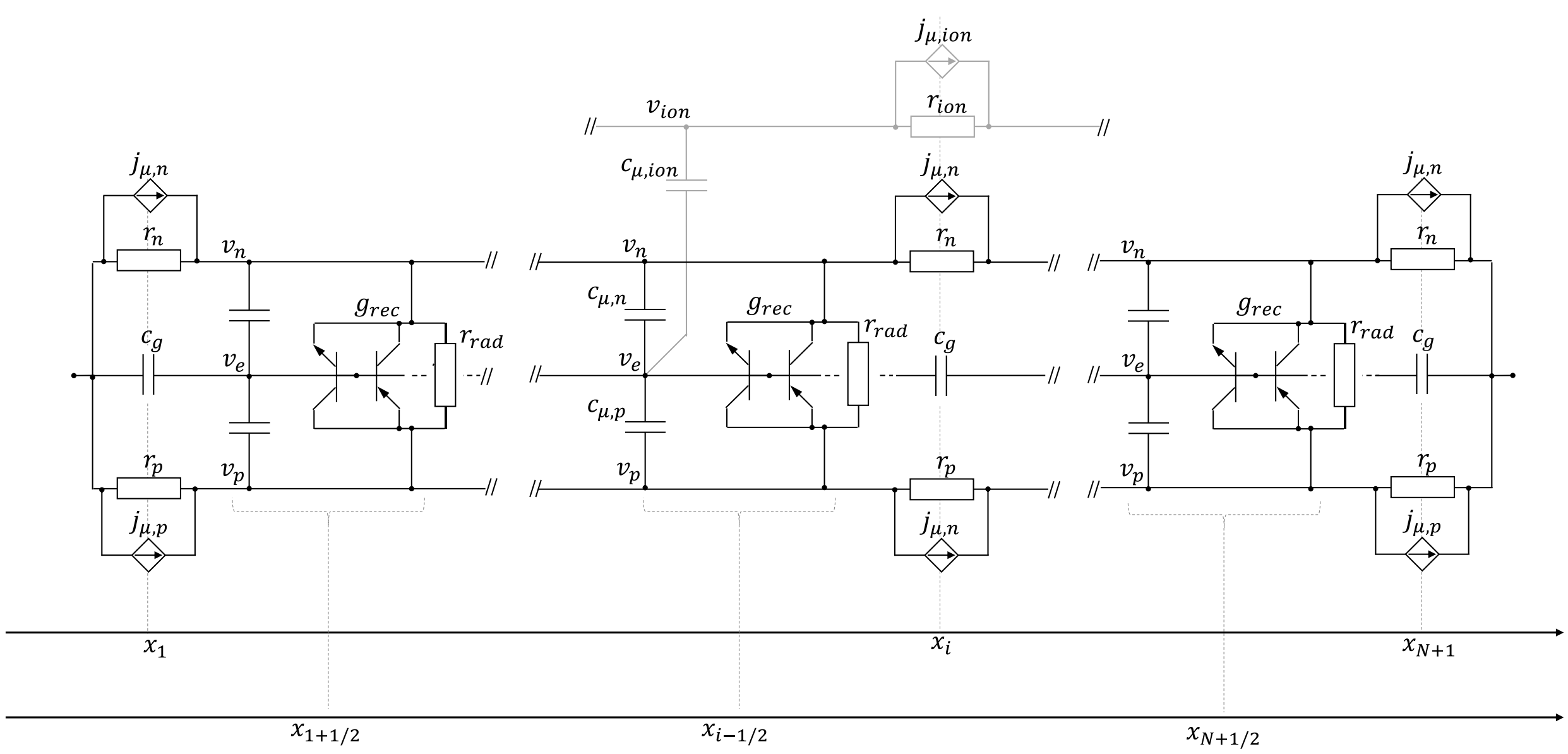

Figure S29. Discretized elements in the complete equivalent circuit model, displayed with reference to the two meshes used.

## 14. Supporting References

15. MATLAB codes

## MATLAB code 1: approximated impedance calculation Approx_IC_MCi

```
%This code is a toy model that computes the approximated impedance of a
%mixed conducting solar cell (IC and MC-i approximations) with ideally
%selective and ion blocking contacts: HTM/MIEC/ETM.
%It is based on input parameters describing properties related with charge
%carriers, space charge potentials, recombination. The calculation is
%performed in a selected frequency range and the results are plotted and
%saved in the selected directory.

%After setting the folder name and input parameters (lines 22-89), press Run.

%---------------------------------------------------------------------------
% LICENSE
% Copyright (C) 2024  Davide Moia
% Max Planck Institute for Solid State Research
% This program is free software: you can redistribute it and/or modify
% it under the terms of the GNU Affero General Public License as published
% by the Free Software Foundation, either version 3 of the License, or
% (at your option) any later version.
%---------------------------------------------------------------------------

%Select directory where to save data
Directory = cd;
newFolder = '\Approx_IC_MCi\';
if ~exist([Directory,newFolder], 'dir')
[void] = mkdir([Directory,newFolder]);
end
NewDirectory = [Directory newFolder];
%Select Filename for this calculation
Filename = 'Test';

%Frequencies used for the impedance calculations (log scaling by default)
N_freq = 37;
f_max = 1e7;
f_min = 1e-2;
f_all = logspace(log10(f_min),log10(f_max),N_freq)';
omega_all = 2*pi*f_all;

%Elementary charge and thermal voltage at 300 K
q = 1.6e-19;
Vth = 0.026;

%Parameters needed for the evaluation of the impedance using the
%approximated equivalent circuit models. Parameters followed by '_' are
%differential parameters, e.g. cmup_ = dcmup/dx

%Dielectric constant of the three layers
eps0 = 8.85e-14;
eps = 32*eps0;
eps_HTM = 20*eps0;
eps_ETM = 10*eps0;

%Electronic and ionic parameters
Eg = 1.6;                                %Bandgap (eV)
NcNv = 1e38;                             %Product of the effective DOS for VB and CB
of the active layer(cm^-6)
ni = sqrt(NcNv)*exp(-Eg/2/Vth);          %Intrinsic concentration (cm^-3)
NA_HTM = 1e18;                           %Acceptor doping HTM (left contact) (cm^-3)
ND_ETM = 1e18;                           %Donor doping ETM (right contact) (cm^-3)
N_VI_bulk = 1e19;                        %Mobile ion concentration in the active
layer (e.g. iodide vacancies in MAPI) (cm^-3)
mu_VI = 1e-10;                           %Mobile ion mobility (cm^2 V^-1 s^-1)
rion_ = 1/(q*N_VI_bulk*mu_VI);           %Differential ionic resitance rion_ =
drion/dx (ohms cm^-1)
```

```
cg_ = eps;                              %Differential geometric capacitance cg_ =
(dcg/d(x)^-1) (F cm^-1)
n_bulk = 1e13;                          %Electron bulk concentration (cm^-3)
p_bulk = 1e15;                          %Hole bulk concentration (cm^-3)
cmun_ = q*n_bulk/Vth;                   %Differential electron chemical capacitance
cmun_ = dcmun/dx (F cm^-3)
cdelta_n_ = cmun_;                      %Differential component chemical
capacitance associated with electrons (assuming ions are majority carriers, it is
equal to cmun) (F cm^-3)
cmup_ = q*p_bulk/Vth;                   %Differential hole chemical capacitance
cmup_ = dcmup/dx (F cm^-3)
cdelta_p_ = cmup_;                      %Differential component chemical
capacitance associated with holes (assuming ions are majority carriers, it is equal
to cmup) (F cm^-3)
L = 360e-7;                             %Thickness of the active layer (cm)
taun = 1e-6;                            %Recombination time constant electrons in
the bulk (s)
taup = 1e-6;                            %Recombination time constant holes in the
bulk (s)
n1 = ni;                                %SRH trap parameter in the bulk (cm^-3)
p1 = ni;                                %SRH trap parameter in the bulk (cm^-3)
taun_surf1 = 1e-6;                      %Surface recombination time constant
electrons at interface 1 (s)
taup_surf1 = 1e-6;                      %Surface recombination time constant holes
at interface 1 (s)
n1_surf1 = ni;                          %SRH trap parameter at interface 1 (cm^-3)
p1_surf1 = ni;                          %SRH trap parameter at interface 1 (cm^-3)
taun_surf2 = 1e-6;                      %Surface recombination time constant
electrons at interface 2 (s)
taup_surf2 = 1e-6;                      %Surface recombination time constant holes
at interface 2 (s)
n1_surf2 = ni;                          %SRH trap parameter at interface 2 (cm^-3)
p1_surf2 = ni;                          %SRH trap parameter at interface 2 (cm^-3)
dsurf = 1e-07;                          %Distance from interface 1 and 2 where
surface recombination is active (cm). dsurf/tau_surf is the surface recombination
velocity
krad = 1e-11;                           %Radiative constant

%The interfacial capacitors are defined based on the space charge potential
%at each interface (all phi are defined positive for a solar cell in the dark at
equilibrium, their sum is equal to the buil-it potential)
phi_A = -0.1;
phi_B = -0.1;
phi_C = -0.05;
phi_D = -0.1;

%Surface concetrations calculated based on bulk concentrations and space
%charge at the interfaces
n_surf1 = n_bulk*exp(-phi_B/Vth);
p_surf1 = p_bulk*exp(phi_B/Vth);
n_surf2 = n_bulk*exp(phi_C/Vth);
p_surf2 = p_bulk*exp(-phi_C/Vth);

%Debye lengths for each layer
L_Debye = sqrt(eps*Vth/q/(N_VI_bulk+n_bulk+p_bulk));
L_Debye_n = sqrt(eps*Vth/q/n_bulk);
L_Debye_p = sqrt(eps*Vth/q/p_bulk);
L_Debye_HTM = sqrt(eps_HTM*Vth/q/NA_HTM);
L_Debye_ETM = sqrt(eps_ETM*Vth/q/ND_ETM);

%Evaluating the space charge widths based on whether an accumulation or
%depletion situation is present (making sure that Lsc is not shorter than L_Debye)
if phi_B > 0
    Lsc_B = L_Debye;
    Lsc_C = max(L_Debye*sqrt(2*phi_C/Vth),L_Debye);
    Lsc_A = max(L_Debye_HTM*sqrt(2*phi_A/Vth),L_Debye_HTM);
    Lsc_D = max(L_Debye_ETM*sqrt(2*phi_B/Vth),L_Debye_ETM);
else
    Lsc_B = max(L_Debye*sqrt(2*(-phi_B)/Vth),L_Debye);
```

```
    Lsc_C = L_Debye;
    Lsc_A = L_Debye_HTM;
    Lsc_D = L_Debye_ETM;
end

%Calculation of interfacial capacitance. If a 0 space charge potential is
%input the capacitance is set to the Debye capacitance value of the
%relevant layer. Note that for c_B, the formula uses the opposite sign for
%the space charge potential phi_B, based on the its definition here.
if phi_A == 0
    c_A = sqrt(eps_HTM*q*NA_HTM/Vth);
else
    c_A = sign(phi_A)*sqrt(q*eps_HTM*NA_HTM/2/Vth)*(1-exp(-phi_A/Vth))/sqrt(exp(-
phi_A/Vth)+phi_A/Vth-1);
end
if phi_B == 0
    c_B = sqrt(eps*q*N_VI_bulk/Vth);
else
    c_B = sign(-phi_B)*sqrt(q*eps*N_VI_bulk/2/Vth)*(1-exp(-(-
phi_B)/Vth))/sqrt(exp(-(-phi_B)/Vth)+(-phi_B)/Vth-1);
end
if phi_C == 0
    c_C = sqrt(eps*q*N_VI_bulk/Vth);
else
    c_C = sign(phi_C)*sqrt(q*eps*N_VI_bulk/2/Vth)*(1-exp(-phi_C/Vth))/sqrt(exp(-
phi_C/Vth)+phi_C/Vth-1);
end
if phi_D == 0
    c_D = sqrt(eps_ETM*q*ND_ETM/Vth);
else
    c_D = sign(phi_D)*sqrt(q*eps_ETM*ND_ETM/2/Vth)*(1-exp(-phi_D/Vth))/sqrt(exp(-
phi_D/Vth)+phi_D/Vth-1);
end

%Defining the bulk in the active layer and its properties
Lbulk = L - Lsc_B -Lsc_C;
rion_bulk = rion_*Lbulk;
cg_bulk = cg_/Lbulk;
cg_tot = ((cg_/L)^-1+c_A^-1+c_D^-1)^-1;

%Interfacial capacitance at each side of the device
c1 = c_A*c_B/(c_A+c_B);
c2 = c_C*c_D/(c_C+c_D);

%Recombination transconductance due to bulk and interfacial traps
grec_n_bulk =
q/Vth*(n_bulk*p_bulk*(taun*(p_bulk+p1)+taup*n1))/(taun*(p_bulk+p1)+taup*(n_bulk+n1)
)^2;
grec_p_bulk =
q/Vth*(n_bulk*p_bulk*(taup*(n_bulk+n1)+taun*p1))/(taun*(p_bulk+p1)+taup*(n_bulk+n1)
)^2;
grad_bulk = (Vth/(q*krad*n_bulk*p_bulk))^-1;

grec_n_surf1 =
dsurf*q/Vth*(n_surf1*p_surf1*(taun_surf1*(p_surf1+p1_surf1)+taup_surf1*n1_surf1))/(
taun_surf1*(p_surf1+p1_surf1)+taup_surf1*(n_surf1+n1_surf1))^2;
grec_p_surf1 =
dsurf*q/Vth*(n_surf1*p_surf1*(taup_surf1*(n_surf1+n1_surf1)+taun_surf1*p1_surf1))/(
taun_surf1*(p_surf1+p1_surf1)+taup_surf1*(n_surf1+n1_surf1))^2;
grec_n_surf2 =
dsurf*q/Vth*(n_surf2*p_surf2*(taun_surf2*(p_surf2+p1_surf2)+taup_surf2*n1_surf2))/(
taun_surf2*(p_surf2+p1_surf2)+taup_surf2*(n_surf2+n1_surf2))^2;
grec_p_surf2 =
dsurf*q/Vth*(n_surf2*p_surf2*(taup_surf2*(n_surf2+n1_surf2)+taun_surf2*p1_surf2))/(
taun_surf2*(p_surf2+p1_surf2)+taup_surf2*(n_surf2+n1_surf2))^2;

vapp = 1;

%Definition of the mesh (linear mesh), constraining at least 20 mesh
```

```
%points within the shortest space charge region in the active layer
Nint = 20;
if Lsc_B<Lsc_C
    Lscmin = Lsc_B;
    Nint1 = Nint;
    Nint2 = floor(Nint1*Lsc_C/Lsc_B);
else
    Lscmin = Lsc_C;
    Nint2 = Nint;
    Nint1 = floor(Nint2*Lsc_B/Lsc_C);
end

%Recalculate L, to make it a multiple of xstep
L = Lscmin/(Nint-1)*floor(L/Lscmin*(Nint-1));
N = L/Lscmin*(Nint-1)+1;
xstep = L/(N-1);
x = 0:xstep:L;

%z is the position axis that spans the bulk only (at x = Lsc1, z = 0)
zstep = xstep;
z = x(Nint1:N-Nint2+1)-x(Nint1);

%xplot is defined to save and plot the data
x_plot = [-Lsc_A, 0, z + x(Nint1), L, L + Lsc_D];

%Matrices that will contain the impedance calculated with the
%approximated ECMs
Ztot_IC = zeros(N_freq,1);
Ztot_MCi = zeros(N_freq,1);

%-------------------------------------------------------------------------
%Approximated ECM with transmission line assuming low cmun and low cmup
%(neglecting their effect) ->IC approximation

%Eigenvalue for the calculation of the equivalent capacitance below
kappa_eone = sqrt((cmun_+cmup_)/cg_);
%Equivalent capacitance for the capacitive network associated with the
%electronic (eon) and electrostatic (e) contributions.
ceq =
(2*c_A*c_D*cmun_*cmup_+(c_A*c_D*(cmun_^2+cmup_^2)+(c_A+c_D)*cmun_*cmup_*(cmun_+cmup
_)*L)*cosh(kappa_eone*L)+...

kappa_eone*(c_A*cmun_*(cg_*cmun_+c_D*cmup_*L)+cg_*cmup_*(c_D*cmup_+cmun_*(cmun_+cmu
p_)*L))*sinh(kappa_eone*L))/...

((c_A+c_D)*(cmun_+cmup_)^2*cosh(kappa_eone*L)+cg_*kappa_eone^3*(c_A*c_D+cg_*(cmun_+
cmup_))*sinh(kappa_eone*L));

%Effective electronic capacitance. It is used in the IC approximation to
%improve the descripion of the high frequency capacitance.
ceon_eff = (ceq-cg_tot);

%--------------------------------------------------------------------------
%IC approximation
Zione_IC = zeros(length(omega_all),1);

figure(2)
close(2)

%Definition of matrix where the ve data are stored
ve_all_IC = zeros(length(x_plot), N_freq+1);
ve_all_IC(:,1) = x_plot;

for jj = 1:length(omega_all)

    omega = omega_all(jj);

    %Calculation of the small signal electrostatic potential
```

```
    ve_IC = vapp*c1/(c1+c2)*(1+1i*omega*rion_bulk*(cg_bulk+c2*(Lbulk-
z)/Lbulk))/(1+1i*omega*rion_bulk*(c1*c2/(c1+c2)+cg_bulk));

    ve1_IC = ve_IC(1)+(vapp-ve_IC(1))*c_A/(c_A+c_B);
    ve2_IC = ve_IC(end)*c_C/(c_C+c_D);

    %Used for comparing the data with other approximations
    ve_IC_plot = [vapp, ve1_IC, ve_IC, ve2_IC, 0];

    %Impedance of the ionic-electrostatic circuit in the IC model.
    Zione_IC(jj) =
rion_bulk/(1+1i*omega*cg_bulk*rion_bulk)+1/(1i*omega*c1*c2/(c1+c2));

    %Total impedance
    Ztot_IC(jj) = ((vapp/(vapp*1i*omega*ceon_eff + vapp*grad_bulk*L +...
        grec_n_bulk*sum(ve_IC)*zstep + grec_p_bulk*sum(vapp - ve_IC)*zstep +...
        (grec_n_surf1)*ve1_IC + (grec_p_surf1)*(vapp - ve1_IC)+...
        (grec_n_surf2)*ve2_IC + (grec_p_surf2)*(vapp - ve2_IC)))^-1+...
        Zione_IC(jj)^-1)^-1;

    %Plotting the real part of ve (assuming vapp = 1 here)
    figure(2)
    hold on
    plot(x_plot,real(ve_IC_plot),'Linewidth',2,'color',[jj/(length(omega_all)) 1-
jj/(length(omega_all)) 4*jj/(length(omega_all))*(1-jj/(length(omega_all)))])

    figure(4)
    subplot(2,2,1)
    hold on
    plot(x_plot,real(ve_IC_plot),'Linewidth',2,'color',[jj/(length(omega_all)) 1-
jj/(length(omega_all)) 4*jj/(length(omega_all))*(1-jj/(length(omega_all)))])

    ve_all_IC(:,jj+1) = real(ve_IC_plot');

end

C = 1/2/pi./f_all.*imag(Ztot_IC.^-1);
Z_r = real(Ztot_IC);
Z_i = imag(Ztot_IC);
Z_abs = sqrt(Z_r.^2 + Z_i.^2);
Z_phase = 180/pi*phase(Ztot_IC);

figure(1)
subplot(2,2,1)
hold on
plot(Z_r,-Z_i,'b--v')

figure(4)
subplot(2,2,3)
hold on
plot(Z_r,-Z_i,'b--v')

figure(1)
subplot(2,2,2)
hold on
loglog(f_all(C>0),C(C>0),'b--v',f_all(C<0),-C(C<0),'b--x')

figure(4)
subplot(2,2,4)
hold on
loglog(f_all(C>0),C(C>0),'b--v',f_all(C<0),-C(C<0),'b--x')

figure(1)
subplot(2,2,3)
hold on
loglog(f_all,Z_abs,'b--v')

figure(1)
subplot(2,2,4)
```

```
hold on
semilogx(f_all,Z_phase,'b--v')

filename = [NewDirectory, Filename,'_IC_Z.txt'];
fid = fopen(filename, 'w');
fprintf(fid,
'%s,%s,%s,%s,%s,%s,%s,%s,%s,%s\n','Freq(Hz)','Z_IC_r(ohm*cm^2)','Z_IC_i(ohm*cm^2)',
'Z_IC_abs(ohm*cm^2)','Z_IC_phase(ohm*cm^2)','Capacitance(F/cm^2)','Z_IC_r(ohm*cm^2)
','-Z_IC_i(ohm*cm^2)','Znorm_IC_r(ohm*cm^2)','-Znorm_IC_i(ohm*cm^2)');
fclose(fid);
MaxZ_r_IC = max(Z_r);
MaxZ_r_plot = MaxZ_r_IC;
MaxZ_i_plot = max(Z_i);

dlmwrite(filename,[f_all Z_r Z_i Z_abs Z_phase C Z_r -Z_i Z_r/MaxZ_r_IC -
Z_i/MaxZ_r_IC],'-append');

%-------------------------------------------------------------------------
%MC-i approximation

j_ionemu_MCi = zeros(N_freq,1);
j_rec_MCi = zeros(N_freq,1);

figure(3)
close(3)

%Definition of matrix where the ve data are stored
ve_all_MCi = zeros(length(x_plot), N_freq+1);
ve_all_MCi(:,1) = x_plot;

for jj = 1:length(omega_all)

    omega = omega_all(jj);

    %Calculation of the small signal electrostatic potential in the MC-i
    %approximation
    kappa_ione_np = sqrt(1i*omega*(cdelta_n_ + cdelta_p_)*rion_/(1 +
1i*omega*cg_*rion_));
    ve_MCi = vapp*(cdelta_p_/(cdelta_p_ + cdelta_n_) + ((1 -
c2/c1*cdelta_p_/cdelta_n_)*cosh(kappa_ione_np*z)*(1 +
1i*omega*rion_*cg_)*kappa_ione_np^2+1i*omega*rion_*((cdelta_n_+cdelta_p_)*(cosh(kap
pa_ione_np*(Lbulk-z))-
cosh(kappa_ione_np*z))+kappa_ione_np*c2*(sinh(kappa_ione_np*(Lbulk-z))-
cdelta_p_/cdelta_n_*sinh(kappa_ione_np*z)))))/...

((cdelta_p_/cdelta_n_+1)*(1i*omega*rion_*((cdelta_n_+cdelta_p_)*(cosh(kappa_ione_np
*Lbulk)-
1)+c2*kappa_ione_np*sinh(kappa_ione_np*Lbulk))+(1+1i*omega*cg_*rion_)*kappa_ione_np
*(kappa_ione_np+c2/c1*kappa_ione_np*cosh(Lbulk*kappa_ione_np)+(cdelta_p_+cdelta_n_)
/c1*sinh(kappa_ione_np*Lbulk)))));

    %Interfacial electrostatic potentials
    ve1_MCi = c_B/(c_A+c_B)*ve_MCi(1)+vapp*c_A/(c_A+c_B);
    ve2_MCi = c_C/(c_C+c_D)*ve_MCi(end);

    %Current associated with ionic-electrostatic and electronic chemical
    %contributions
    j_ionemu_MCi(jj) = (vapp-ve_MCi(1))*1i*omega*c1 + sum(vapp-
ve_MCi)*1i*omega*cdelta_p_*xstep;
    %Current associated with recombination in the bulk and at the
    %interfaces
    j_rec_MCi(jj) = grec_n_bulk*sum(ve_MCi)*zstep + grec_p_bulk*zstep*sum(vapp-
ve_MCi)+(ve1_MCi)*grec_n_surf1 +(ve2_MCi)*grec_n_surf2+(vapp-ve1_MCi)*grec_p_surf1
+(vapp-ve2_MCi)*grec_p_surf2 + vapp*grad_bulk*L;

    %Total impedance
    Ztot_MCi(jj) = vapp/(j_ionemu_MCi(jj) + j_rec_MCi(jj));

    ve = [1, real(ve1_MCi), real(ve_MCi), real(ve2_MCi), 0];
```

```
    %Plotting the real part of ve (assuming vapp = 1 here)
    figure(3)
    hold on
    plot(x_plot,ve,'Linewidth',2,'color',[jj/(length(omega_all)) 1-
jj/(length(omega_all)) 4*jj/(length(omega_all))*(1-jj/(length(omega_all)))])

    figure(4)
    subplot(2,2,2)
    hold on
    plot(x_plot,ve,'Linewidth',2,'color',[jj/(length(omega_all)) 1-
jj/(length(omega_all)) 4*jj/(length(omega_all))*(1-jj/(length(omega_all)))])

    ve_all_MCi(:,jj+1) = real(ve');

end

%Plotting impedance data
C =1/2/pi./f_all.*imag(Ztot_MCi.^-1);
Z_r = real(Ztot_MCi);
Z_i = imag(Ztot_MCi);
Z_abs = sqrt(Z_r.^2 + Z_i.^2);
Z_phase = 180/pi*phase(Ztot_MCi);

figure(1)
subplot(2,2,1)
hold on
plot(Z_r,-Z_i,'r-^')
subplot(2,2,2)
hold on
semilogx(f_all(C>0),C(C>0),'r-^',f_all(C<0),-C(C<0),'r-+')
subplot(2,2,3)
hold on
loglog(f_all,Z_abs,'r-^')
subplot(2,2,4)
hold on
loglog(f_all,Z_phase,'r-^')

figure(4)
subplot(2,2,3)
hold on
plot(Z_r,-Z_i,'r-^')
subplot(2,2,4)
hold on
semilogx(f_all(C>0),C(C>0),'r-^',f_all(C<0),-C(C<0),'r-+')

%Scaling of axes and axes' labels
MaxZ_r_MCi = max(Z_r);
dlmwrite(filename,[f_all Z_r Z_i Z_abs Z_phase C Z_r -Z_i Z_r/MaxZ_r_MCi -
Z_i/MaxZ_r_MCi],'-append');
MaxZ_r_plot = max(MaxZ_r_plot, MaxZ_r_MCi);
MaxZ_i_plot = max(MaxZ_i_plot, max(Z_i));

figure(1)
subplot(2,2,1)
set(gca, 'XScale', 'lin', 'YScale', 'lin');
title('Nyquist plot')
xlabel('Re(Z/ohm)')
ylabel('-Im(Z/ohm)')
xlim([0 1.2*MaxZ_r_plot])
ylim([-1.2*MaxZ_i_plot 1.2*(MaxZ_r_plot-MaxZ_i_plot)])
legend('IC','MC-i')
subplot(2,2,2)
set(gca, 'XScale', 'log', 'YScale', 'log');
title('Apparent capacitance')
xlabel('f/Hz')
ylabel('\omega^-^1Im(Z^-^1)/F cm^-^2')
subplot(2,2,3)
set(gca, 'XScale', 'log', 'YScale', 'log');
```

```
title('Bode - magnitude')
xlabel('f/Hz')
ylabel('Abs(Z/ohm cm^2)')
subplot(2,2,4)
set(gca, 'XScale', 'log', 'YScale', 'lin');
title('Bode - phase')
xlabel('f/Hz')
ylabel('Phase(Z)/deg')

figure(2)
title('Small signal ve, IC approximation','FontSize', 20)
xlabel('Position (cm)','FontSize', 20)
ylabel('Re(v_e/v_a_p_p)','FontSize', 20)
xlim([-2*Lsc_A L + 2*Lsc_D])
ylim([-0.2 1.2])

figure(3)
title('Small signal ve, MC-i approximation','FontSize', 20)
xlabel('Position (cm)','FontSize', 20)
ylabel('Re(v_e/v_a_p_p)','FontSize', 20)
xlim([-2*Lsc_A L + 2*Lsc_D])
ylim([-0.2 1.2])

figure(4)
subplot(2,2,3)
set(gca, 'XScale', 'lin', 'YScale', 'lin');
title('Nyquist plot','FontSize', 20)
xlabel('Re(Z/ohm)','FontSize', 20)
ylabel('-Im(Z/ohm)','FontSize', 20)
xlim([0 1.2*MaxZ_r_plot])
ylim([-1.2*MaxZ_i_plot 1.2*(MaxZ_r_plot-MaxZ_i_plot)])
legend('IC','MC-i','FontSize', 15)
subplot(2,2,4)
set(gca, 'XScale', 'log', 'YScale', 'log');
title('Apparent capacitance','FontSize', 20)
xlabel('f/Hz','FontSize', 20)
ylabel('\omega^-^1Im(Z^-^1)/F cm^-^2','FontSize', 20)
% legend('\omega^-^1Im(Z^-^1) (IC)','-\omega^-^1Im(Z^-^1) (IC)','\omega^-^1Im(Z^-
^1) (MC-i)','-\omega^-^1Im(Z^-^1) (MC-i)','FontSize', 15)
subplot(2,2,2)
title('Small signal ve, MC-i approximation','FontSize', 20)
xlabel('Position (cm)','FontSize', 20)
ylabel('Re(v_e/v_a_p_p)','FontSize', 20)
subplot(2,2,1)
title('Small signal ve, IC approximation','FontSize', 20)
xlabel('Position (cm)','FontSize', 20)
ylabel('Re(v_e/v_a_p_p)','FontSize', 20)
subplot(2,2,1)
xlim([-2*Lsc_A L + 2*Lsc_D])
ylim([-0.2 1.2])
subplot(2,2,2)
xlim([-2*Lsc_A L + 2*Lsc_D])
ylim([-0.2 1.2])
set(gcf, 'Units', 'Normalized', 'OuterPosition', [0, 0.04, 0.65, 0.96]);

%Saving the impedance data
filename = [NewDirectory, Filename,'_MCi_Z.txt'];
fid = fopen(filename, 'w');
fprintf(fid,
'%s,%s,%s,%s,%s,%s,%s,%s,%s,%s\n','Freq(Hz)','Z_MCi_r(ohm*cm^2)','Z_MCi_i(ohm*cm^2)
','Z_MCi_abs(ohm*cm^2)','Z_MCi_phase(ohm*cm^2)','Capacitance(F/cm^2)','Z_MCi_r(ohm*
cm^2)','-Z_MCi_i(ohm*cm^2)','Znorm_MCi_r(ohm*cm^2)','-Znorm_MCi_i(ohm*cm^2)');
fclose(fid);

%Saving real part of ve for both approximations
filename = [NewDirectory, Filename,'_real_ve_IC.txt'];
fid = fopen(filename, 'w');
fclose(fid);
dlmwrite(filename,ve_all_IC,'-append');
```

```
filename = [NewDirectory, Filename,'_real_ve_MCi.txt'];
fid = fopen(filename, 'w');
fclose(fid);
dlmwrite(filename,ve_all_MCi,'-append');

% Saving the figures
cd(NewDirectory)
figure(1)
print(gcf, '-dtiff', 'Approx_Impedance.tiff');
figure(2)
print(gcf, '-dtiff', 'Real_ve_IC.tiff');
figure(3)
print(gcf, '-dtiff', 'Real_ve_MC-i.tiff');
figure(4)
print(gcf, '-dtiff', 've_Z_IC_MC-i.tiff');
cd ..
```

**MATLAB code 2: calculated impedance from drift-diffusion, complete model and approximated models Z_Voc_DD_ECM**

```
%------------------------------------------------------------------------
%What does the code do?

%     1) The code evaluates the steady-state solution of a semiconductor based
%     device with or without mobile ions (1 mobile ionic species considered
%     here) defined as an input file of Driftfusion at open circuit under dark
%     and under defined light conditions using Driftfusion
%     (Calado et al. J Comput Electron 21, 960–991 (2022).
https://doi.org/10.1007/s10825-021-01827-z).
%
%     2) It then proceeds to evaluate the impedance of the device under such
%     steady-state conditions, using different methods:
%     -> Drift-diffusion simulation of the impedance experiment (Driftfusion
package used)
%     -> Complete transmission line model (analytically equivalent to the drift-
diffusion model)
%     -> Approximated transmission line models for low and for large bias
%
%     3) Finally, it saves txt files with the solutions of the calculations
%     as well as other information for further analysis (small signal potentials vs
x,
%     jrec vs x, transconductances, steady-state solution from Driftfusion,
%     and the solution to the impedance for the drift diffusion (DD), for
%     the complete transmission line model and the approximated models (IC and MC-
i)

%How to set up the simulation?

%     To run this code you need this version of the Driftfusion software
%     https://github.com/barnesgroupICL/Driftfusion/tree/2022-EA_SDP_EIS
%     Simply save this script in the same working folder containing the Driftfusion
scripts

%How to run the code?

%     In the first section of the code (line 64-90) you can enter these details:
%     folder containing the input files to process, which simulations you want to
run,
%     light intensities and frequencies for the impedance calculations.
%     -> Set the folder name containing the input files that need to be processed
%     (the code runs the simulations for all of such input files).
%     This folder should be in the 'Input_files' folder
%     -> Set the name of the folder where the results will be saved
%     (additional information on the input files are then added to such name
%     depending on the input file)
%     The array WhichZ allows one to indicate which impedance to run
%     Enter the light intensities used to determine the steady-state solution
%     and the frequency range for the impedance calculations
%     -> Run

%------------------------------------------------------------------------
% LICENSE
% Copyright (C) 2024  Davide Moia
% Max Planck Institute for Solid State Research
% This program is free software: you can redistribute it and/or modify
% it under the terms of the GNU Affero General Public License as published
% by the Free Software Foundation, either version 3 of the License, or
% (at your option) any later version.
% For details on the license of the software Driftfusion, see
% Driftdiffusion code
%------------------------------------------------------------------------

%Initialise Driftfusion
initialise_df
```

```
%Select folder with input files to use for calculations
%All input files contained in this folders will be processed one at a time.
%In this case, this folder is inside the 'Input_files' folder of Driftfusion
input_files_folder = 'InputFolderName';

D = dir(['Input_files/',input_files_folder]);

%Results for each input file will be saved in a separate folder in the
%current directory. All folders start with the same name (here defined as
%Sim_name), followed by the input_file name. Change this name before running
%the code to avoid overwriting data
Sim_name = '\DD_ECM_IC_MCi_';

%Here one can tell the code which impedance to calculate
%(1: do calculate; 0: do not calculate). In all cases the steady-state
%solution from DriftFusion will bbe used.
%The first element in the array refers to the drift-diffusion calculation.
%The second to the complete transmission line equivalent circuit model.
%The third to the IC and MC-i approximated transmission line equivalent circuit
models.
WhichZ = [1,1,1];

%Range of light intensities to consider (units are suns). An array of
%intensity with log scaling is created below. Dark is also calculated
%automatically
Intensity_low = 1e-5;
Intensity_high = 1e-1;
Nintensities = 3;

%Frequencies used for the impedance calculations (log scaling by default)
N_freq = 37;
f_max = 1e7;
f_min = 1e-2;

%--------------------------------------------------------------------------
%Start of the code

%Go through all input files in the selected directory
for ll = 3:length(D)
    %Try-catch is used here so that if one of the calculations gives an
    %error the code goes through all input files in the folder anyway
    try
        %Current Input file name
        input_file_name = D(ll).name;

        %Directory and folder where to save the solutions
        Directory = cd;
        newFolder = [Sim_name,input_file_name,'\'];
        if ~exist([Directory,newFolder], 'dir')
        [void] = mkdir([Directory,newFolder]);
        end
        NewDirectory = [Directory newFolder];

        %Import csv file, make a copy to place in the folder of the calculation
        input_csv = ['Input_files/',input_files_folder,'/',input_file_name];
        source = fullfile('Input_files/',input_files_folder,'/',input_file_name);
        destination = fullfile(NewDirectory,input_file_name);
        copyfile(source,destination)
        par = pc(input_csv);
        soleq = equilibrate(par);
        soleqi = stabilize(soleq.ion);

        %Find the steady-state solution of the device starting from the equilibrium
        %solution soleqi (the one with mobile ions) and applying a range of light
        %intensities. Specifically, a number (4th parameter) of light intensities
        %ranging between the 2nd input to the 3rd input parameter in suns is
        %considered. The last input specifies whether a dark solution should be
        %included (true) or not (false)
```

```
        [structs_OC, Vocs] =
genIntStructsVoc(soleqi,Intensity_low,Intensity_high,Nintensities,true);

        %Extract the steady-state solution in the dark
        Sol_dark = structs_OC{1,1};

        %Define the array of light intensities
        Intensities =
logspace(log10(Intensity_low),log10(Intensity_high),Nintensities);

        %Save the open circuit potentials
        filename = [NewDirectory, 'Vocs.txt'];
        fid = fopen(filename, 'w');
        fprintf(fid, '%s,%s\n','Intensity(suns)','Voc(V)');
        fclose(fid);
        dlmwrite(filename,[Intensities', Vocs(2:end)'],'-append');

        %---------------------------------------------------------------------------
        %Evaluate the impedance using Driftfusion if WhichZ(1) is 1

        if WhichZ(1) == 1
            %Voltage amplitude, chosen small for linearity, but also
            %computation time
            Vac = 2e-3;

            %Drift-diffusion impedance calculation for all conditions.
            %IS_oc contains all solutions
            IS_oc = IS_script(structs_OC, f_max, f_min, N_freq, Vac, false, true,
false);

            %Saves all the Driftfusion impedance solutions and plots the
            %results
            filenamedark = [NewDirectory, 'DD_Z_dark.txt'];
            fid = fopen(filenamedark, 'w');
            fprintf(fid,
'%s,%s,%s,%s,%s,%s,%s,%s,%s,%s\n','Freq(Hz)','ZDD_r(ohm*cm^2)','ZDD_i(ohm*cm^2)','Z
DD_abs(ohm*cm^2)','ZDD_phase(ohm*cm^2)','Capacitance(F/cm^2)','ZDD_r(ohm*cm^2)','-
ZDD_i(ohm*cm^2)','ZDD_r_norm(ohm*cm^2)','-ZDD_i_norm(ohm*cm^2)');
            fclose(fid);

            for jj = 1:Nintensities+1
                freq = IS_oc.Freq(jj,:)';
                ZDD_r = IS_oc.impedance_re(jj,:)';
                ZDD_i = IS_oc.impedance_im(jj,:)';
                Capacitance = 1./(2*pi*freq).*imag((ZDD_r+1i*ZDD_i).^-1);
                ZDD = ZDD_r+1i*ZDD_i;
                ZDD_abs = abs(ZDD);
                MaxZ = max(ZDD_r);
                ZDD_phase = 180/pi*phase(ZDD);

                figure(1)
                subplot(2,2,1)
                hold on
                plot(ZDD_r,-ZDD_i,'g*')

                subplot(2,2,2)
                hold on

loglog(freq(Capacitance>0),Capacitance(Capacitance>0),'g*',freq(Capacitance<0),-
Capacitance(Capacitance<0),'g*')

                subplot(2,2,3)
                hold on
                loglog(freq,ZDD_abs,'g*')

                subplot(2,2,4)
                hold on
                semilogx(freq,-ZDD_phase,'g*')
```

```
                if jj == 1

dlmwrite(filenamedark,[freq,ZDD_r,ZDD_i,ZDD_abs,ZDD_phase,Capacitance,ZDD_r,-
ZDD_i,ZDD_r/MaxZ,-ZDD_i/MaxZ],'-append');
                else
                    filename = [NewDirectory, 'DD_Z_', num2str(Intensities(jj-
1)),'suns.txt'];
                    fid = fopen(filename, 'w');
                    fprintf(fid,
'%s,%s,%s,%s,%s,%s,%s,%s\n','Freq(Hz)','ZDD_r(ohm*cm^2)','ZDD_i(ohm*cm^2)','ZDD_abs
(ohm*cm^2)','ZDD_phase(ohm*cm^2)','Capacitance(F/cm^2)','ZDD_r(ohm*cm^2)','-
ZDD_i(ohm*cm^2)');
                    fclose(fid);

dlmwrite(filename,[freq,ZDD_r,ZDD_i,ZDD_abs,ZDD_phase,Capacitance,ZDD_r,-ZDD_i],'-
append');
                end
            end
        end

        %-----------------------------------------------------------------------------
        %Calculation of the impedance using the equivalent circuit model by
        %calling the function MixedConductorTransmissionLine_DDinput.

        Zdark = MixedConductorTransmissionLine_DDinput(f_max, f_min, N_freq,
Sol_dark, NewDirectory, [input_file_name(1:end-4),'_Dark'], WhichZ);

        Zbias = zeros(N_freq, Nintensities);
        for jj = 1:Nintensities
            Zbias(:,jj) = MixedConductorTransmissionLine_DDinput(f_max, f_min,
N_freq, structs_OC{1,jj+1}, NewDirectory, [input_file_name(1:end-
4),'_',num2str(Intensities(jj)),'sun'], WhichZ);
        end

    catch
    end
end
% Set the axes and labels of Figure 1
figure(1)
subplot(2,2,1)
set(gca, 'XScale', 'lin', 'YScale', 'lin');
title('Nyquist plot')
xlabel('Re(Z/ohm)')
ylabel('-Im(Z/ohm)')

subplot(2,2,2)
set(gca, 'XScale', 'log', 'YScale', 'log');
title('Apparent capacitance')
xlabel('f/Hz')
ylabel('Apparent capacitance/F cm^2')

subplot(2,2,3)
set(gca, 'XScale', 'log', 'YScale', 'log');
title('Bode - magnitude')
xlabel('f/Hz')
ylabel('Abs(Z/ohm cm^2)')

subplot(2,2,4)
set(gca, 'XScale', 'log', 'YScale', 'lin');
title('Bode - phase')
xlabel('f/Hz')
ylabel('-Phase(Z)/deg')

function Z0 = MixedConductorTransmissionLine_DDinput(f_max, f_min, N_freq, sol,
NewDirectory, Filename, WhichZ)

%The input sol is the steady-state solution and is used to extract input
%parameters and the value of equivalent circuit model elements.
```

```
%Import parameters from drift-diffusion solution
par = sol.par;

%Constants
q = par.e;
kB = par.kB*q;
eps0 = 8.85e-14;

%Temperature
T = par.T;

%Thermal voltage at room temperature
Vth = kB*T/q;

%N is the number of ihalf mesh points in the drift diffusion solution and the same
number that is used in the impedance calculation
N = par.pcum(5);

%The mesh for the complete transmission line model calculation corresponds to the
%mesh in the whole device, including junctions, ETM and HTM
x = sol.x;                     %Position array
dx = x(2:end)-x(1:end-1);      %Intervals array
x_ihalf = par.x_ihalf;         %Position array, mid-points in the x array (this is
used in the calculation below as the nodes in the equivalent circuit)
dx_ihalf = [x_ihalf(1), x_ihalf(2:end)-x_ihalf(1:end-1), x(end)-x_ihalf(end)];
%Interval array for the x_ihalf array

%Index referring to the middle of the ETM and HTM (used below to reference the
electrostatic potential and evaluating space charge potentials)
i_midETM = floor((par.pcum(5)+par.pcum(4))/2);
i_midHTM = floor(par.pcum(2)/2);

%Indices for interfacial positions and mid-active layer. These will be used
%below in the approximated ECM calculations
i_Lhalf = floor((par.pcum(4)+par.pcum(1))/2);   %Middle of active layer
i_1_A = par.pcum(1)+1;                          %Interface 1 on HTM side
i_1_B = par.pcum(2)+1;                          %Interface 1 on active layer side
i_2_C = par.pcum(3)+1;                          %Interface 2 on active layer side
i_2_D = par.pcum(4)+1;                          %Interface 2 on ETM side

%Thickness of the active layer (including junctions)
L = par.dcum0(5)-par.dcum0(2);

%Thickness of the HTM (used below to reference the position x, when saving the
solution)
x0 = par.dcum0(2);

%Saving steady-state solution: Ve(referenced to Ve in the middle of the ETM), n, p
and cion
filename = [NewDirectory, Filename,'_SteadyState.txt'];
fid = fopen(filename, 'w');
fprintf(fid, '%s,%s,%s,%s,%s\n','x(nm)','Ve(V)','n(cm^-3)','p(cm^-3)','[VI](cm^-
3)');
fclose(fid);
dlmwrite(filename,[x'-x0, sol.u(end,:,1)'-sol.u(end,i_midETM,1), sol.u(end,:,2)',
sol.u(end,:,3)', sol.u(end,:,4)'],'-append');

n = sol.u(end,:,2);            %Electron concentration
p = sol.u(end,:,3);            %Hole concentration
N_VI = sol.u(end,:,4);         %Ion concentration

n_ihalf = (n(2:end)+n(1:end-1))/2;          %Electron concentration at x_ihalf
p_ihalf = (p(2:end)+p(1:end-1))/2;          %Hole concentration at x_ihalf
N_VI_ihalf = (N_VI(2:end)+N_VI(1:end-1))/2; %Ion concentration at x_ihalf

%Material parameters
%x_ihalf are the voltage nodes, x are the nodes where the transport
%resistors and geometric capacitors are defined.
```

```
%See Calado et al. J Comput Electron 21, 960-991 (2022).
https://doi.org/10.1007/s10825-021-01827-z
ni_ihalf = par.dev_ihalf.ni;      %Intrinsic concentration
mu_p = par.dev.muh;               %Hole mobility
mu_n = par.dev.mue;               %Electron mobility
mu_VI = par.dev.mucat;            %Ion mobility
eps = par.dev.epp;                %Reltive permittivity
krad_ihalf = par.dev_ihalf.B;     %Radiative recombination constant
taun_ihalf = par.dev_ihalf.taun;  %Recombination time constant electrons
taup_ihalf = par.dev_ihalf.taup;  %Recombination time constant holes
n1_ihalf = par.dev_ihalf.nt;      %SRH trap parameter
p1_ihalf = par.dev_ihalf.pt;      %SRH trap parameter

[J, j, xmesh] = dfana.calcJ(sol);   %Current components frm solution
Jnx = [J.n(end,1), (J.n(end,2:end)+J.n(end,1:end-1))/2, J.n(end)];
Jpx = [J.p(end,1), (J.p(end,2:end)+J.p(end,1:end-1))/2, J.p(end)];
Jcx = [J.c(end,1), (J.c(end,2:end)+J.c(end,1:end-1))/2, J.c(end)];

rn = dx_ihalf./(q*n.*mu_n);       %Electron transport resistors
rp = dx_ihalf./(q*p.*mu_p);       %Hole transport resistors
rion = dx_ihalf./(q*N_VI.*mu_VI);%Ion transport resistors
cmun = q*dx.*n_ihalf/Vth;         %Electron chemical capacitors
cmup = q*dx.*p_ihalf/Vth;         %Hole chemical capacitors
cmuion = q*dx.*N_VI_ihalf/Vth;    %Ion chemical capacitors
cg = eps0*eps./dx_ihalf;          %Electrostatic capacitors

%Elements used at the boundaries of the complete transmission line model
rn_L = rn(1);                      %Electron resistor at the left boundary
rn_R = rn(end);                    %Electron resistor at the right boundary
rp_L = rp(1);                      %Hole resistor at the left boundary
rp_R = rp(end);                    %Hole resistor at the right boundary
rion_L = rion(1);                  %Ion resistor at the left boundary
rion_R = rion(end);                %Ion resistor at the right boundary
cg_L = cg(1);                      %Electrostatic capacitor at the left boundary
cg_R = cg(end);                    %Electrostatic capacitor at the right boundary

%Removing the first and last transport resistors and geometric capacitor
%from the array used in the bulk transmission line calculation. These are
%used for the boundary conditions, see above.
rn = rn(2:end-1);
rp = rp(2:end-1);
rion = rion(2:end-1);
cg = cg(2:end-1);

%Volumetric recombination and generation transconductance (A V^-1 cm^-3)
grec_n_vol =
q/Vth*(n_ihalf.*p_ihalf.*(taun_ihalf.*(p_ihalf+p1_ihalf)+taup_ihalf.*n1_ihalf))./(t
aun_ihalf.*(p_ihalf+p1_ihalf)+taup_ihalf.*(n_ihalf+n1_ihalf)).^2;
ggen_n_vol =
q/Vth*taun_ihalf.*ni_ihalf.^2.*p_ihalf./(taun_ihalf.*(p_ihalf+p1_ihalf)+taup_ihalf.
*(n_ihalf+n1_ihalf)).^2;
grec_p_vol =
q/Vth*(n_ihalf.*p_ihalf.*(taup_ihalf.*(n_ihalf+n1_ihalf)+taun_ihalf.*p1_ihalf))./(t
aun_ihalf.*(p_ihalf+p1_ihalf)+taup_ihalf.*(n_ihalf+n1_ihalf)).^2;
ggen_p_vol =
q/Vth*taup_ihalf.*ni_ihalf.^2.*n_ihalf./(taun_ihalf.*(p_ihalf+p1_ihalf)+taup_ihalf.
*(n_ihalf+n1_ihalf)).^2;

%Discretized recombination and generation transconductance elements
%(A V^-1 cm^-2)
grec_n = grec_n_vol.*dx;
ggen_n = ggen_n_vol.*dx;
grec_p = grec_p_vol.*dx;
ggen_p = ggen_p_vol.*dx;

%Transconductance terms for transport
g_n = Jnx/Vth;
g_p = Jpx/Vth;
g_ion = Jcx/Vth;
```

```
g_n_L = g_n(1);
g_n_R = g_n(end);
g_p_L = g_p(1);
g_p_R = g_p(end);
g_ion_L = g_ion(1);
g_ion_R = g_ion(end);

g_n = g_n(2:end-1);
g_p = g_p(2:end-1);
g_ion = g_ion(2:end-1);

%Interfacial recombination at left boundary (set to 0 when using input from
%drift-diffusion) These surfaces in the equations below are in fact the
%boundaries between ETM, HTM and the electrodes
grec_surf1_n = 0;
ggen_surf1_n = 0;
grec_surf1_p = 0;
ggen_surf1_p = 0;

%Interfacial recombination at right boundary (set to 0 when using input from drift-
diffusion)
grec_surf2_n = 0;
ggen_surf2_n = 0;
grec_surf2_p = 0;
ggen_surf2_p = 0;

%Discretised radiative recombination resistors and conductors
rrad = Vth./(q*krad_ihalf.*n_ihalf.*p_ihalf)./dx;

%Volumetric radiative conductance (A V^-1 cm^-3)
grad_vol = (rrad.^-1)./dx;

%Save files with steady-state recombination (trans)conductance per unit
%volute
filename = [NewDirectory, Filename,'_Trans-conductance.txt'];
fid = fopen(filename, 'w');
fprintf(fid, '%s,%s,%s,%s\n','x(nm)','grec_n(S cm^-3)','grec_p(S cm^-3)','grad(S
cm^-3)');
fclose(fid);
dlmwrite(filename,[x_ihalf'-x0, grec_n_vol', grec_p_vol', grad_vol'],'-append');

%Frequency and angular frequency arrays
f_all = logspace(log10(f_min),log10(f_max),N_freq)';
omega_all = 2*pi*f_all;

if WhichZ(2) == 1

    %Applied potential (can be any value ~=0 as the linearized equations are
solved)
    vapp = 1;

    %Potential of all nodes in the circuit (electronic, ionic and electrostatic
rails)
    vn_x = zeros(N,1);
    vp_x = zeros(N,1);
    vion_x = zeros(N,1);
    ve_x = zeros(N,1);

    %Initial guess is an array with all elements = 0
    vi_trial = [vn_x; vp_x; vion_x; ve_x];

    %Impedance array
    Z0 = zeros(length(omega_all),1);

    %Collecting the small signal ve for some frequencies (nf_pd is the number of
frequencies per decade)
    nf_pd = 2;
    nf = (log10(f_max)-log10(f_min))*nf_pd + 1;
```

```
    ve_some_r = zeros(length(x_ihalf), nf+1);
    ve_some_r(:,1) = x_ihalf'-x0;
    ve_some_i = zeros(length(x_ihalf), nf+1);
    ve_some_i(:,1) = x_ihalf'-x0;
    f_some = zeros(nf,1);
    jj = 1; %Index used for these matrices

    %Collecting the small signal recombination driving forces (real and imag) as
function
    %of frequencies for 5 positions (x = 0, L/4, L/2, 3*L/4, L)
    i_0p25 = find(x_ihalf'-x0>L/4);
    i_0p25 = i_0p25(1);
    i_0p75 = find(x_ihalf'-x0>L*3/4);
    i_0p75 = i_0p75(1);
    vrecn_f = zeros(N_freq, 11);
    vrecp_f = zeros(N_freq, 11);
    vrecrad_f = zeros(N_freq, 11);

    %Evaluation of impedance at each frequency for each frequency value
    %using the complete model
    for h =1:length(omega_all)
        omega = omega_all(h);
        options = optimoptions('fsolve','Display','off');
        vsol = fsolve(@(vi) Vfunction(vi), vi_trial, options);
        Z0(h) = vapp/((vapp-vsol(1))/rn_L + (vapp-vsol(N+1))/rp_L + (vapp-
vsol(2*N+1))/rion_L + (vapp-vsol(3*N+1))*1i*omega*cg_L);

        %Evaluating the recombination driving forces as function of
        %position
        vrec_n = vsol(3*N+1:4*N)-vsol(1:N);
        vrec_p = vsol(N+1:2*N)-vsol(3*N+1:4*N);
        vrec_rad = vsol(N+1:2*N)-vsol(1:N);

        vrecn_f(h,:) = [f_all(h), real(vrec_n(i_1_B)), -imag(vrec_n(i_1_B)),
real(vrec_n(i_0p25)), -imag(vrec_n(i_0p25)), real(vrec_n(i_Lhalf)), -
imag(vrec_n(i_Lhalf)), real(vrec_n(i_0p75)), -imag(vrec_n(i_0p75)),
real(vrec_n(i_2_C)), -imag(vrec_n(i_2_C))];
        vrecp_f(h,:) = [f_all(h), real(vrec_p(i_1_B)), -imag(vrec_p(i_1_B)),
real(vrec_p(i_0p25)), -imag(vrec_p(i_0p25)), real(vrec_p(i_Lhalf)), -
imag(vrec_p(i_Lhalf)), real(vrec_p(i_0p75)), -imag(vrec_p(i_0p75)),
real(vrec_p(i_2_C)), -imag(vrec_p(i_2_C))];
        vrecrad_f(h,:) = [f_all(h), real(vrec_rad(i_1_B)), -imag(vrec_rad(i_1_B)),
real(vrec_rad(i_0p25)), -imag(vrec_rad(i_0p25)), real(vrec_rad(i_Lhalf)), -
imag(vrec_rad(i_Lhalf)), real(vrec_rad(i_0p75)), -imag(vrec_rad(i_0p75)),
real(vrec_rad(i_2_C)), -imag(vrec_rad(i_2_C))];

        %For some frequencies, the small signal potentials, the recombination
driving forces and
        %recombination currents are saved in separate text files
        if mod(log10((f_all(h))^nf_pd),1)<0.01

            %Saving the small signal potentials at the selected frequency
            filename = [NewDirectory,
Filename,'_v_real',num2str(f_all(h)),'Hz.txt'];
            fid = fopen(filename, 'w');
            fprintf(fid,
'%s,%s,%s,%s,%s,%s\n','Index','x(cm)','Re(vn)(V)','Re(vp)(V)','Re(vion)(V)','Re(ve)
(V)');
            fclose(fid);
            dlmwrite(filename,[(1:N)', x_ihalf'-x0, real(vsol(1:N)),
real(vsol(N+1:2*N)), real(vsol(2*N+1:3*N)),real(vsol(3*N+1:4*N))],'-append');
            filename = [NewDirectory,
Filename,'_v_imag',num2str(f_all(h)),'Hz.txt'];
            fid = fopen(filename, 'w');
            fprintf(fid,
'%s,%s,%s,%s,%s,%s\n','Index','x(cm)','Im(vn)(V)','Im(vp)(V)','Im(vion)(V)','Im(ve)
(V)');
            fclose(fid);
```

```
            dlmwrite(filename,[(1:N)', x_ihalf'-x0, imag(vsol(1:N)),
imag(vsol(N+1:2*N)), imag(vsol(2*N+1:3*N)),imag(vsol(3*N+1:4*N))],'-append');

            %Saving the recombination potential and current (real part)
            filename = [NewDirectory,
Filename,'_vrec_jrec_',num2str(f_all(h)),'Hz.txt'];
            fid = fopen(filename, 'w');
            fprintf(fid,
'%s,%s,%s,%s,%s,%s,%s\n','x(cm)','Re(vrec_n)(V)','Re(vrec_p)(V)','Re(vrec_rad)(V)',
'Re(jrec_n)(A/cm^3)','Re(jrec_p)(A/cm^3)','Re(jrec_rad)(A/cm^3)');
            fclose(fid);
            dlmwrite(filename,[x_ihalf'-x0, real(vrec_n), real(vrec_p),
real(vrec_rad), real(vrec_n).*grec_n_vol', real(vrec_p).*grec_p_vol',
real(vrec_rad).*grad_vol'],'-append');
            ve_some_r(:,jj+1) = real(vsol(3*N+1:4*N));
            ve_some_i(:,jj+1) = imag(vsol(3*N+1:4*N));
            f_some(jj) = f_all(h);
            jj = jj+1;
        end
    end

    %Saving all small signal ve for all selected frequencies in one file
    %The list of selected frequencies is stored in a 'SelectedFreq' txt file
    dlmwrite([NewDirectory, Filename,'_ve_selectedFreq_real.txt'],ve_some_r);
    dlmwrite([NewDirectory, Filename,'_ve_selectedFreq_imag.txt'],ve_some_i);
    dlmwrite([NewDirectory, Filename,'_selectedFreq.txt'],f_some);
    dlmwrite([NewDirectory, Filename,'_vrecn_vs_f.txt'],vrecn_f);
    dlmwrite([NewDirectory, Filename,'_vrecp_vs_f.txt'],vrecp_f);

    %Calculation of the apparent capacitance
    Capparent = real(-1i./(omega_all.*Z0));
    %The calculated impedance is saved in a text file named
    %(Filename)_Z.txt
    Z0_r = real(Z0);
    Z0_i = imag(Z0);
    Z0_abs = sqrt(Z0_r.^2 + Z0_i.^2);
    Z0_phase = 180/pi*phase(Z0);
    filename = [NewDirectory, Filename,'_CompleteModel_Z.txt'];
    fid = fopen(filename, 'w');
    fprintf(fid,
'%s,%s,%s,%s,%s,%s,%s,%s,%s,%s\n','Freq(Hz)','ZECM_r(ohm*cm^2)','ZECM_i(ohm*cm^2)',
'ZECM_abs(ohm*cm^2)','ZECM_phase(ohm*cm^2)','Capacitance(F/cm^2)','ZECM_r(ohm*cm^2)
','-ZECM_i(ohm*cm^2)','ZECMnorm_r(ohm*cm^2)','-ZECMnorm_i(ohm*cm^2)');
    fclose(fid);
    MaxZ = max(Z0_r);
    dlmwrite(filename,[omega_all/2/pi Z0_r Z0_i Z0_abs Z0_phase Capparent Z0_r -
Z0_i Z0_r/MaxZ -Z0_i/MaxZ],'-append');

    %Figure 1 has four panels plotting impedance in a Nyquist plot,the apparent
    %capacitance, the magnitude and the phase of the impedance as a function of
    %frequency
    figure(1)
    subplot(2,2,1)
    hold on
    plot(real(Z0),-imag(Z0),'ks')
    title('Nyquist plot')
    xlabel('Re(Z)')
    ylabel('-Im(Z)')

    subplot(2,2,2)
    hold on

loglog(omega_all(Capparent>0)/2/pi,abs(Capparent(Capparent>0)),'ks',omega_all(Cappa
rent<0)/2/pi,-Capparent(Capparent<0),'k+')
    title('Effective capacitance')
    xlabel('f (Hz)')
    ylabel('C*')

    subplot(2,2,3)
```

```
    hold on
    loglog(omega_all/2/pi,abs(Z0),'ks')
    title('Abs(Z)')
    xlabel('f (Hz)')
    ylabel('Abs(Z)')

    subplot(2,2,4)
    hold on
    semilogx(omega_all/2/pi,-angle(Z0)*180/pi,'ks')
    title('Phase(Z)')
    xlabel('f (Hz)')
    ylabel('-Phase(Z)')
else
    Z0 = 0;
end

%Calculate impedance with approximated models if WhichZ(3) is 1
if WhichZ(3) == 1

    %------------------------------------------------------------------------
    %Parameters needed for the evaluation of the impedance using the
    %approximated equivalent circuit models. Parameters followed by '_' are
    %differential parameters, e.g. cmup_ = dcmup/dx
    N_VI_ave = N_VI(i_Lhalf);
    mu_VI_ave = mu_VI(i_Lhalf);
    rion_ = 1/(q*N_VI_ave*mu_VI_ave);
    cg_ = eps(i_Lhalf)*eps0;
    n_ave = n(i_Lhalf);
    p_ave = p(i_Lhalf);
    %Differential capacitors
    cmun_ = q*n_ave/Vth;
    cdelta_n_ = cmun_;
    cmup_ = q*p_ave/Vth;
    cdelta_p_ = cmup_;
    %Open circuit and built in potential (valid only if Vapp = Voc)
    Voc = sol.par.Vapp;
    Vbi = sol.par.Vbi;

    %Because of the interfacial layers between the contact and the active
    %layer there are different ways to define the space charge potentials
    %phi_A,B,C,D. Here, the various definitions are listed and only one is
    %assigned to the phi_A,B,C,D used in the calculations of space charge
    %widths and capacitance. Note that in the nomenclature Phi_x_y, x
    %indicates which space charge being considered, and y which side of
    %the junction at the interface is taken for the evaluation of the space
    %charge potential e.g. phi_A_B evaluates the phi_A space charge
    %potential based on the value of the electrostatic potential at the
    %active layer/junction interface. In the case of phi_A_AB, an average
    %of the potential across the junction is taken

    %Space charge potential evaluated as the average within the interfacial
    %region defined between the active and the contact layers. Reasonable
    %if the recombination current is roughly homogeneous in this layer
%     phi_A_AB = mean([sol.u(end,i_1_A,1),sol.u(end,i_1_B,1)]) -
sol.u(end,i_midHTM,1);
%     phi_B_AB = sol.u(end,i_Lhalf,1) -
mean([sol.u(end,i_1_A,1),sol.u(end,i_1_B,1)]);
%     phi_C_CD = mean([sol.u(end,i_2_C,1),sol.u(end,i_2_D,1)]) -
sol.u(end,i_Lhalf,1);
%     phi_D_CD = sol.u(end,i_midETM,1) -
mean([sol.u(end,i_2_C,1),sol.u(end,i_2_D,1)]);

    %Space charge potential evaluated at the beginning of the interfacial
    %layer coming from the active layer side.
%     phi_A_B = sol.u(end,i_1_B,1) - sol.u(end,i_midHTM,1);
%     phi_B_B = sol.u(end,i_Lhalf,1) - sol.u(end,i_1_B,1);
%     phi_C_C = sol.u(end,i_2_C,1) - sol.u(end,i_Lhalf,1);
%     phi_D_C = sol.u(end,i_midETM,1) - sol.u(end,i_2_C,1);
```

```
    %Space charge potential evaluated at the beginning of the junction
    %layer coming from the contact layer side. This is more reasonable in
    %situations where mobile ions are allowed in the junction

    phi_A_A = sol.u(end,i_1_A,1) - sol.u(end,i_midHTM,1);
    phi_B_A = sol.u(end,i_Lhalf,1) - sol.u(end,i_1_A,1);
    phi_C_D = sol.u(end,i_2_D,1) - sol.u(end,i_Lhalf,1);
    phi_D_D = sol.u(end,i_midETM,1) - sol.u(end,i_2_D,1);

    %Here one can decide how to define the space charge potentials that
    %will be used for the calculation of the interfacial capacitance
    phi_A = phi_A_A;
    phi_B = phi_B_A;
    phi_C = phi_C_D;
    phi_D = phi_D_D;

    %Debye length assuming cations are the only mobile species and they are
    %the majority carriers
    L_Debye = sqrt(eps0*eps(i_Lhalf)*Vth/q/par.Ncat(3));

    %Space charge widths at interface 1 and 2
    if Voc <= Vbi
        %Making sure that Lsc is not shorter than L_Debye
        Lsc_B = L_Debye;
        Lsc_C = max(L_Debye*sqrt(2*phi_C/Vth),L_Debye);
    else
        %Making sure that Lsc is not shorter than L_Debye
        Lsc_B = max(L_Debye*sqrt(2*(-phi_B)/Vth),L_Debye);
        Lsc_C = L_Debye;
    end

    %Small signal space charge capacitance for each space charge region
    if phi_A == 0
        c_A = sqrt(eps0*eps(i_midHTM)*q*par.NA(1)/Vth);
    else
        c_A = sign(phi_A)*sqrt(q*eps0*eps(i_midHTM)*par.NA(1)/2/Vth)*(1-exp(-
phi_A/Vth))/sqrt(exp(-phi_A/Vth)+phi_A/Vth-1);
    end
    if phi_B == 0
        c_B = sqrt(eps0*eps(i_Lhalf)*q*par.Ncat(3)/Vth);
    else
        c_B = sign(-phi_B)*sqrt(q*eps0*eps(i_Lhalf)*par.Ncat(3)/2/Vth)*(1-exp(-(-
phi_B)/Vth))/sqrt(exp(-(-phi_B)/Vth)+(-phi_B)/Vth-1);
    end
    if phi_C == 0
        c_C = sqrt(eps0*eps(i_Lhalf)*q*par.Ncat(3)/Vth);
    else
        c_C = sign(phi_C)*sqrt(q*eps0*eps(i_Lhalf)*par.Ncat(3)/2/Vth)*(1-exp(-
phi_C/Vth))/sqrt(exp(-phi_C/Vth)+phi_C/Vth-1);
    end
    if phi_D == 0
        c_D = sqrt(eps0*eps(i_midETM)*q*par.ND(5)/Vth);
    else
        c_D = sign(phi_D)*sqrt(q*eps0*eps(i_midETM)*par.ND(5)/2/Vth)*(1-exp(-
phi_D/Vth))/sqrt(exp(-phi_D/Vth)+phi_D/Vth-1);
    end

    %Define bulk thickenss and capacitance based on the space charge widths and
total thickness of the
    %active layer
    Lbulk = L - Lsc_B -Lsc_C;
    rion_bulk = rion_*Lbulk;
    cg_bulk = cg_/Lbulk;
    cg_tot = ((cg_/L)^-1+c_A^-1+c_D^-1)^-1;

    %Total interfacial small signal capacitors at interface 1 and 2
    c1 = c_A*c_B/(c_A+c_B);
    c2 = c_C*c_D/(c_C+c_D);
```

```
    %For the bulk, the transconductance per unit volume in the middle of
    %the active layer is taken and is integrated below in the impedance
calculation.
    %This is an acceptable approximation if the concentration of electrons and
holes is almost
    %constant across the bulk
    grec_n_bulk = grec_n_vol(i_Lhalf);
    grec_p_bulk = grec_p_vol(i_Lhalf);
    grad_bulk = grad_vol(i_Lhalf);
    %For the surface transconductances (in the juctions), the integration happens
here, as
    %only one value of ve is considered for each interfaces in the
    %simplified IC and MC-i models
    grec_n_surf1 = sum(grec_n(i_1_A:i_1_B));
    grec_p_surf1 = sum(grec_p(i_1_A:i_1_B));
    grec_n_surf2 = sum(grec_n(i_2_C:i_2_D));
    grec_p_surf2 = sum(grec_p(i_2_C:i_2_D));

    %This value of vapp is set to 1, it is irrelevant in the calculation of
    %the impedance below. It used only for readability of the math expression
    vapp = 1;

    %Definition of the mesh (linear mesh), constraining at least 20 mesh
    %points within the shortest space charge width in the active layer
    Nint = 20;
    if Lsc_B<Lsc_C
        Lscmin = Lsc_B;
        Nint1 = Nint;
        Nint2 = floor(Nint1*Lsc_C/Lsc_B);
    else
        Lscmin = Lsc_C;
        Nint2 = Nint;
        Nint1 = floor(Nint2*Lsc_B/Lsc_C);
    end

    %Recalculate L, to make it a multiple of xstep
    L = Lscmin/(Nint-1)*floor(L/Lscmin*(Nint-1));
    N = L/Lscmin*(Nint-1)+1;
    xstep = L/(N-1);
    x = 0:xstep:L;

    %z is the position axis that spans the bulk only (at x = Lsc1, z = 0)
    zstep = xstep;
    z = x(Nint1:N-Nint2+1)-x(Nint1);

    %xplot is defined to save and plot the data
    x_plot = [0, z + x(Nint1), L];

    %Matrices that will contain the impedance calculated with the
    %approximated ECMs
    Z_IC = zeros(N_freq,1);
    Z_MCi = zeros(N_freq,1);

    %-------------------------------------------------------------------------
    %IC approximation

    %Eigenvalue for the calculation of the equivalent capacitance below
    kappa_eone = sqrt((cmun_+cmup_)/cg_);
    %Equivalent capacitance for the capacitive network associated with the
    %electronic (eon) and electrostatic (e) contributions.
    ceq =
(2*c_A*c_D*cmun_*cmup_+(c_A*c_D*(cmun_^2+cmup_^2)+(c_A+c_D)*cmun_*cmup_*(cmun_+cmup
_)*L)*cosh(kappa_eone*L)+kappa_eone*(c_A*cmun_*(cg_*cmun_+c_D*cmup_*L)+cg_*cmup_*(c
_D*cmup_+cmun_*(cmun_+cmup_)*L))*sinh(kappa_eone*L))/...

((c_A+c_D)*(cmun_+cmup_)^2*cosh(kappa_eone*L)+cg_*kappa_eone^3*(c_A*c_D+cg_*(cmun_+
cmup_))*sinh(kappa_eone*L));

    %Effective electronic capacitance. It is used in the IC approximation to
```

```
    %improve the descripion of the high frequency capacitance.
    ceon_eff = (ceq-cg_tot);

    %Impedance of the ionic-electrostatic circuit branch
    Zione_IC = zeros(length(omega_all),1);

    %Refresh figure 2
    figure(2)
    close(2)

    for jj = 1:length(omega_all)

        omega = omega_all(jj);

        %IC Approximation
        ve_IC = vapp*c1/(c1+c2)*(1+1i*omega*rion_bulk*(cg_bulk+c2*(Lbulk-
z)/Lbulk))/(1+1i*omega*rion_bulk*(c1*c2/(c1+c2)+cg_bulk));

        %Calculation of the small signal electrostatic potential at
        %interface 1 and 2
        ve1_IC = ve_IC(1)+(vapp-ve_IC(1))*c_A/(c_A+c_B);
        ve2_IC = ve_IC(end)*c_C/(c_C+c_D);

        ve_IC_plot = [real(ve1_IC), real(ve_IC), real(ve2_IC)];

        %Impedance of the IC model. Note that the
        %solution without distributed elements can also be evaluated in this case
        Zione_IC(jj) =
rion_bulk/(1+1i*omega*cg_bulk*rion_bulk)+1/(1i*omega*c1*c2/(c1+c2));

        %Total impedance calculated with the IC model
        Z_IC(jj) = ((vapp/(vapp*1i*omega*ceon_eff + vapp*grad_bulk*L +...
            grec_n_bulk*sum(ve_IC)*zstep + grec_p_bulk*sum(vapp - ve_IC)*zstep +...
            (grec_n_surf1)*ve1_IC + (grec_p_surf1)*(vapp - ve1_IC)+...
            (grec_n_surf2)*ve2_IC + (grec_p_surf2)*(vapp - ve2_IC)))^-1+...
            Zione_IC(jj)^-1)^-1;

        %Plot the real part of the small signal (normalized) electrostatic
        %potential as function of frequency
        figure(2)
        hold on
        plot(x_plot,real(ve_IC_plot),'.','color',[jj/(length(omega_all)) 1-
jj/(length(omega_all)) 4*jj/(length(omega_all))*(1-jj/(length(omega_all)))])
        title('Small signal ve/vapp, low electronic chemical capacitance')
        xlabel('Position (cm)')
        ylabel('Re(ve/vapp)')

    end

    figure(1)
    subplot(2,2,1)
    hold on
    plot(real(Z_IC),-imag(Z_IC),'b--v')

    C = 1/2/pi./f_all.*imag(Z_IC.^-1);
    Z_r = real(Z_IC);
    Z_i = imag(Z_IC);
    Z_abs = sqrt(Z_r.^2 + Z_i.^2);
    Z_phase = 180/pi*phase(Z_IC);

    figure(1)
    subplot(2,2,2)
    hold on
    loglog(f_all(C>0),C(C>0),'b--v',f_all(C<0),-C(C<0),'b--x')

    figure(1)
    subplot(2,2,3)
    hold on
    loglog(f_all,Z_abs,'b--v')
```

```
    figure(1)
    subplot(2,2,4)
    hold on
    semilogx(f_all,-Z_phase,'b--v')

    filename = [NewDirectory, Filename,'_IC_Z.txt'];
    fid = fopen(filename, 'w');
    fprintf(fid,
'%s,%s,%s,%s,%s,%s,%s,%s,%s,%s\n','Freq(Hz)','Z_IC_r(ohm*cm^2)','Z_IC_i(ohm*cm^2)',
'Z_IC_abs(ohm*cm^2)','Z_IC_phase(ohm*cm^2)','Capacitance(F/cm^2)','Z_IC_r(ohm*cm^2)
','-Z_IC_i(ohm*cm^2)','Znorm_IC_r(ohm*cm^2)','-Znorm_IC_i(ohm*cm^2)');
    fclose(fid);
    MaxZ = max(Z_r);
    dlmwrite(filename,[f_all Z_r Z_i Z_abs Z_phase C Z_r -Z_i Z_r/MaxZ -
Z_i/MaxZ],'-append');

    %-----------------------------------------------------------------------
    %MC-i approximation

    %Array for the ionic-electrostatic-electronicchemical contribution
    j_ionemu_MCi = zeros(N_freq,1);
    %Array for the recombination current
    j_rec_MCi = zeros(N_freq,1);

    %Refresh figure 3
    figure(3)
    close(3)

    for jj = 1:length(omega_all)

        omega = omega_all(jj);

        %Calculation of the small signal electrostatic potential in the MC-i
        %approximation
        kappa_ione_np = sqrt(1i*omega*(cdelta_n_ + cdelta_p_)*rion_/(1 +
1i*omega*cg_*rion_));
        ve_MCi = vapp*(cdelta_p_/(cdelta_p_ + cdelta_n_) + ((1 -
c2/c1*cdelta_p_/cdelta_n_)*cosh(kappa_ione_np*z)*(1 +
1i*omega*rion_*cg_)*kappa_ione_np^2+1i*omega*rion_*((cdelta_n_+cdelta_p_)*(cosh(kap
pa_ione_np*(Lbulk-z))-
cosh(kappa_ione_np*z))+kappa_ione_np*c2*(sinh(kappa_ione_np*(Lbulk-z))-
cdelta_p_/cdelta_n_*sinh(kappa_ione_np*z))))/...

((cdelta_p_/cdelta_n_+1)*(1i*omega*rion_*((cdelta_n_+cdelta_p_)*(cosh(kappa_ione_np
*Lbulk)-
1)+c2*kappa_ione_np*sinh(kappa_ione_np*Lbulk))+(1+1i*omega*cg_*rion_)*kappa_ione_np
*(kappa_ione_np+c2/c1*kappa_ione_np*cosh(Lbulk*kappa_ione_np)+(cdelta_p_+cdelta_n_)
/c1*sinh(kappa_ione_np*Lbulk)))));

        %Calculation of the small signal electrostatic potential at
        %interface 1 and 2
        ve1_MCi = c_B/(c_A+c_B)*ve_MCi(1)+vapp*c_A/(c_A+c_B);
        ve2_MCi = c_C/(c_C+c_D)*ve_MCi(end);

        ve_MCi_plot = [real(ve1_MCi), real(ve_MCi), real(ve2_MCi)];

        %Current associated with ionic-electrostatic and electronic chemical
        %contributions
        j_ionemu_MCi(jj) = (vapp-ve_MCi(1))*1i*omega*c1 + sum(vapp-
ve_MCi)*1i*omega*cdelta_p_*xstep;
        %Current associated with recombination in the bulk and at the
        %interfaces
        j_rec_MCi(jj) = grec_n_bulk*sum(ve_MCi)*zstep + grec_p_bulk*xstep*sum(vapp-
ve_MCi)+(ve1_MCi)*grec_n_surf1 +(ve2_MCi)*grec_n_surf2+(vapp-ve1_MCi)*grec_p_surf1
+(vapp-ve2_MCi)*grec_p_surf2 + vapp*grad_bulk*L;

        %Total impedance
        Z_MCi(jj) = vapp/(j_ionemu_MCi(jj) + j_rec_MCi(jj));
```

```
        %Plotting the real part of ve (assuming vapp = 1 here)
        figure(3)
        hold on
        plot(x_plot,ve_MCi_plot,'.','color',[jj/(length(omega_all)) 1-
jj/(length(omega_all)) 4*jj/(length(omega_all))*(1-jj/(length(omega_all)))])
        title('Small signal ve, MC-i approximation')
        xlabel('Position (cm)')
        ylabel('Re(ve/vapp)')
    end

    figure(1)
    subplot(2,2,1)
    hold on
    plot(real(Z_MCi),-imag(Z_MCi),'r-^')

    C = 1/2/pi./f_all.*imag(Z_MCi.^-1);
    Z_r = real(Z_MCi);
    Z_i = imag(Z_MCi);
    Z_abs = sqrt(Z_r.^2 + Z_i.^2);
    Z_phase = 180/pi*phase(Z_MCi);

    figure(1)
    subplot(2,2,2)
    hold on
    loglog(f_all(C>0),C(C>0),'r-^',f_all(C<0),-C(C<0),'r-+')

    figure(1)
    subplot(2,2,3)
    hold on
    loglog(f_all,Z_abs,'r-^')

    figure(1)
    subplot(2,2,4)
    hold on
    semilogx(f_all,-Z_phase,'r-^')

    filename = [NewDirectory, Filename,'_MCi_Z.txt'];
    fid = fopen(filename, 'w');
    fprintf(fid,
'%s,%s,%s,%s,%s,%s,%s,%s,%s,%s\n','Freq(Hz)','Z_MCi_r(ohm*cm^2)','Z_MCi_i(ohm*cm^2)
','Z_MCi_abs(ohm*cm^2)','Z_MCi_phase(ohm*cm^2)','Capacitance(F/cm^2)','Z_MCi_r(ohm*
cm^2)','-Z_MCi_i(ohm*cm^2)','Znorm_MCi_r(ohm*cm^2)','-Znorm_MCi_i(ohm*cm^2)');
    fclose(fid);
    MaxZ = max(Z_r);
    dlmwrite(filename,[f_all Z_r Z_i Z_abs Z_phase C Z_r -Z_i Z_r/MaxZ -
Z_i/MaxZ],'-append');
end

%----------------------------------------------------------------------
function Residual = Vfunction(vx)
%Function to find the values of vn, vp, vion, ve for which Jtrial = 0
%used in the complete model

Jtrial = zeros(4*N,1);

Jtrial(1) = 1e10*((vapp-vx(1))/rn_L - (vx(1)-vx(2))/rn(1) + g_n_L*(vx(3*N+1)-
vx(1))/2 - g_n(1)*(vx(3*N+1)-vx(1)+vx(3*N+2)-vx(2))/2 - (vx(1)-
vx(3*N+1))*1i*omega*cmun(1) - (vx(3*N+1)-vx(N+1))*(ggen_surf1_n + ggen_n(1)) +
(vx(3*N+1)-vx(1))*(grec_surf1_n + grec_n(1))  - (vx(1)-vx(3*N+1))*(ggen_surf1_p +
ggen_p(1)) + (vx(N+1)-vx(3*N+1))*(grec_surf1_p + grec_p(1)) + (vx(N+1)-
vx(1))/rrad(1));
Jtrial(N) = 1e10*(-vx(N)/rn_R + (vx(N-1)-vx(N))/rn(N-1) + g_n(N-1)*(vx(4*N-1)-vx(N-
1)+vx(4*N)-vx(N))/2 - g_n_R*(vx(4*N)-vx(N))/2 - (vx(N)-vx(4*N))*1i*omega*cmun(N) -
(vx(4*N)-vx(2*N))*(ggen_surf2_n + ggen_n(N)) + (vx(4*N)-vx(N))*(grec_surf2_n +
grec_n(N))  - (vx(N)-vx(4*N))*(ggen_surf2_p + ggen_p(N)) + (vx(2*N)-
vx(4*N))*(grec_surf2_p + grec_p(N)) + (vx(2*N)-vx(N))/rrad(N));

for k=2:N-1
```

```
    Jtrial(k) = 1e10*((vx(k-1)-vx(k))/rn(k-1) - (vx(k)-vx(k+1))/rn(k) + g_n(k-
1)*(vx(3*N+k-1)-vx(k-1)+vx(3*N+k)-vx(k))/2 - g_n(k)*(vx(3*N+k)-vx(k)+vx(3*N+k+1)-
vx(k+1))/2 - (vx(k)-vx(k+3*N))*1i*omega*cmun(k) - (vx(3*N+k)-vx(N+k))*ggen_n(k) +
(vx(3*N+k)-vx(k))*grec_n(k) - (vx(k)-vx(3*N+k))*ggen_p(k) + (vx(N+k)-
vx(3*N+k))*grec_p(k) + (vx(N+k)-vx(k))/rrad(k));
end

Jtrial(N+1) = 1e10*((vapp-vx(N+1))/rp_L - (vx(N+1)-vx(N+2))/rp(1) + g_p_L*(vx(N+1)-
vx(3*N+1))/2 - g_p(1)*(vx(N+1)-vx(3*N+1)+vx(N+2)-vx(3*N+2))/2 - (vx(N+1)-
vx(3*N+1))*1i*omega*cmup(1) + (vx(3*N+1)-vx(N+1))*(ggen_surf1_n + ggen_n(1)) -
(vx(3*N+1)-vx(1))*(grec_surf1_n + grec_n(1))  + (vx(1)-vx(3*N+1))*(ggen_surf1_p +
ggen_p(1)) - (vx(N+1)-vx(3*N+1))*(grec_surf1_p + grec_p(1)) - (vx(N+1)-
vx(1))/rrad(1));
Jtrial(2*N) = 1e10*(-vx(2*N)/rp_R + (vx(2*N-1)-vx(2*N))/rp(N-1) + g_p(N-1)*(vx(2*N-
1)-vx(4*N-1)+vx(2*N)-vx(4*N))/2 - g_p_R*(vx(2*N)-vx(4*N))/2 - (vx(2*N)-
vx(4*N))*1i*omega*cmup(N) + (vx(4*N)-vx(2*N))*(ggen_surf2_n + ggen_n(N)) -
(vx(4*N)-vx(N))*(grec_surf2_n + grec_n(N))  + (vx(N)-vx(4*N))*(ggen_surf2_p +
ggen_p(N)) - (vx(2*N)-vx(4*N))*(grec_surf2_p + grec_p(N)) - (vx(2*N)-
vx(N))/rrad(N));

for k=N+2:2*N-1
    Jtrial(k) = 1e10*((vx(k-1)-vx(k))/rp(k-N-1) - (vx(k)-vx(k+1))/rp(k-N) + g_p(k-
N-1)*(vx(k-1)-vx(2*N+k-1)+vx(k)-vx(2*N+k))/2 - g_p(k-N)*(vx(k)-vx(2*N+k)+vx(k+1)-
vx(2*N+k+1))/2 - (vx(k)-vx(k+2*N))*1i*omega*cmup(k-N) + (vx(2*N+k)-vx(k))*ggen_n(k-
N) - (vx(2*N+k)-vx(k-N))*grec_n(k-N) + (vx(k-N)-vx(2*N+k))*ggen_p(k-N) - (vx(k)-
vx(2*N+k))*grec_p(k-N) - (vx(k)-vx(k-N))/rrad(k-N));
end

Jtrial(2*N+1) = 1e10*((vapp-vx(2*N+1))/rion_L - (vx(2*N+1)-vx(2*N+2))/rion(1) +
g_ion_L*(vx(2*N+1)-vx(3*N+1))/2 - g_ion(1)*(vx(2*N+1)-vx(3*N+1)+vx(2*N+2)-
vx(3*N+2))/2 - (vx(2*N+1)-vx(3*N+1))*1i*omega*cmuion(1));
Jtrial(3*N) = 1e10*(-vx(3*N)/rion_R + (vx(3*N-1)-vx(3*N))/rion(N-1) + g_ion(N-
1)*(vx(3*N-1)-vx(4*N-1)+vx(3*N)-vx(4*N))/2 - g_ion_R*(vx(3*N)-vx(4*N))/2 -
(vx(3*N)-vx(4*N))*1i*omega*cmuion(N));

for k=2*N+2:3*N-1
    Jtrial(k) = 1e10*((vx(k-1)-vx(k))/rion(k-2*N-1) - (vx(k)-vx(k+1))/rion(k-2*N) +
g_ion(k-2*N-1)*(vx(k-1)-vx(N+k-1)+vx(k)-vx(N+k))/2 - g_ion(k-2*N)*(vx(k)-
vx(N+k)+vx(k+1)-vx(N+k+1))/2 - (vx(k)-vx(k+N))*1i*omega*cmuion(k-2*N));
end

Jtrial(3*N+1) = 1e10*((vapp-vx(3*N+1))*1i*omega*cg_L - (vx(3*N+1)-
vx(3*N+2))*1i*omega*cg(1) + (vx(1)-vx(3*N+1))*1i*omega*cmun(1) + (vx(N+1)-
vx(3*N+1))*1i*omega*cmup(1) + (vx(2*N+1)-vx(3*N+1))*1i*omega*cmuion(1));
Jtrial(4*N) = 1e10*(-vx(4*N)*1i*omega*cg_R + (vx(4*N-1)-vx(4*N))*1i*omega*cg(N-1) +
(vx(N)-vx(4*N))*1i*omega*cmun(N) + (vx(2*N)-vx(4*N))*1i*omega*cmup(N) + (vx(3*N)-
vx(4*N))*1i*omega*cmuion(N));

for k=3*N+2:4*N-1
    Jtrial(k) = 1e10*((vx(k-1)-vx(k))*1i*omega*cg(k-3*N-1) - (vx(k)-
vx(k+1))*1i*omega*cg(k-3*N) + (vx(k-3*N)-vx(k))*1i*omega*cmun(k-3*N) + (vx(k-2*N)-
vx(k))*1i*omega*cmup(k-3*N) + (vx(k-N)-vx(k))*1i*omega*cmuion(k-3*N));
end

Residual = Jtrial;

end
end
```